\documentclass[11pt,a4paper]{book}
\usepackage{epsfig}
\usepackage{graphicx}
\usepackage[figuresright]{rotating} 
\usepackage{multirow}
\usepackage[tight,hang]{subfigure}
\usepackage{amsmath}
\usepackage{amssymb}
\usepackage{amsfonts}
\usepackage{amsxtra}
\usepackage{amstext}
\usepackage{amsbsy}
\usepackage[latin1]{inputenc} 
\usepackage{wasysym} 
\usepackage{array} 
\usepackage{tabularx} 
\usepackage{eurosym} 
\usepackage{subfigure}
\usepackage{wrapfig}
\usepackage{caption}
\usepackage{setspace}
\usepackage{longtable}
\usepackage{pifont} 
\usepackage{fix-cm} 
\usepackage{upgreek} 
\usepackage[all]{xy} 
\usepackage{dsfont}
\usepackage{bbold} 

\usepackage[NoDate]{currvita} 
\let\oldsqrt\sqrt
\def\sqrt{\mathpalette\DHLhksqrt}
\def\DHLhksqrt#1#2{\setbox0=\hbox{$#1\oldsqrt{#2\,}$}\dimen0=\ht0
\advance\dimen0-0.2\ht0
\setbox2=\hbox{\vrule height\ht0 depth -\dimen0}%
{\box0\lower0.4pt\box2}}
\newcommand{\bs}[1]{\boldsymbol{#1}}
\newcommand{\SSS}{\bs{\mathcal{S}}}

\usepackage{fancyhdr}
\usepackage{makeidx}
\usepackage{graphics}

\usepackage{natbib,natbibspacing}

\usepackage[breaklinks,pagebackref=true]{hyperref}	
\hypersetup{
    unicode=false,          
    pdftoolbar=true,        
    pdfmenubar=true,        
    pdffitwindow=false,     
    pdfstartview={FitH},    
    pdftitle={Spectropolarimetry of Sunspot Penumbrae},    
    pdfauthor={Morten Franz},     
    pdfsubject={PhD Thesis},   
    pdfcreator={Morten Franz},   
    pdfproducer={Morten Franz}, 
    pdfkeywords={Sunspots} {Penumbra} {Spectropolarimetry} {HINODE}, 
    pdfnewwindow=true,      
    colorlinks=true,       
    linkcolor=red,          
    citecolor=blue,        
    filecolor=green,      
    urlcolor=cyan,           
    citebordercolor={0 1 0}, 
    linktocpage=false,
    raiselinks=false
}

\newcommand{\captionfonts}{\small}

\makeatletter  
\long\def\@makecaption#1#2{%
  \vskip\abovecaptionskip
  \sbox\@tempboxa{{\captionfonts #1: #2}}%
  \ifdim \wd\@tempboxa >\hsize
    {\captionfonts #1: #2\par}
  \else
    \hbox to\hsize{\hfil\box\@tempboxa\hfil}%
  \fi
  \vskip\belowcaptionskip}
\makeatother   

\def\arcsec{$^{\prime\prime}$}
\makeatletter
\def\@makechapterhead#1{%
  {\parindent \z@\raggedleft
    \ifnum \c@secnumdepth >\m@ne
    {\scalebox{2}{\Huge\bfseries\sffamily{\thechapter}}}
    \fi
    \rule{\columnwidth}{0.9pt} \par
    {\Huge \bfseries\sffamily #1\par}
    \nobreak
    \vskip 70\p@
    }}

\def\@makeschapterhead#1{%
  {\parindent \z@ \raggedleft
    {\Huge \bf #1\par}
    \nobreak
    \vskip 70\p@
    }}
\makeatother
\newcommand{\clearemptydoublepage}
   {\newpage\thispagestyle{empty}\cleardoublepage}
\makeindex 

\begin{document}

\frontmatter
\pagenumbering{roman}
\singlespacing

\thispagestyle{empty}
\begin{center}

\vspace{-5cm}

\begin{Huge}
\textbf{
Morten Franz
}
\end{Huge}

\vspace{0.8cm}

\begin{Huge}
\textbf{\fontsize{35}{40}\selectfont
Spectropolarimetry\\
of Sunspot Penumbrae\\}
\end{Huge}

\vspace{0.6cm}

\begin{Large}
\textbf{
A Comprehensive Study of the Evershed Effect\\
Using High Resolution Data from the\\
Space-Borne Solar Observatory HINODE\\
}
\end{Large}

\vspace{1.5cm}

\includegraphics[width=\textwidth]{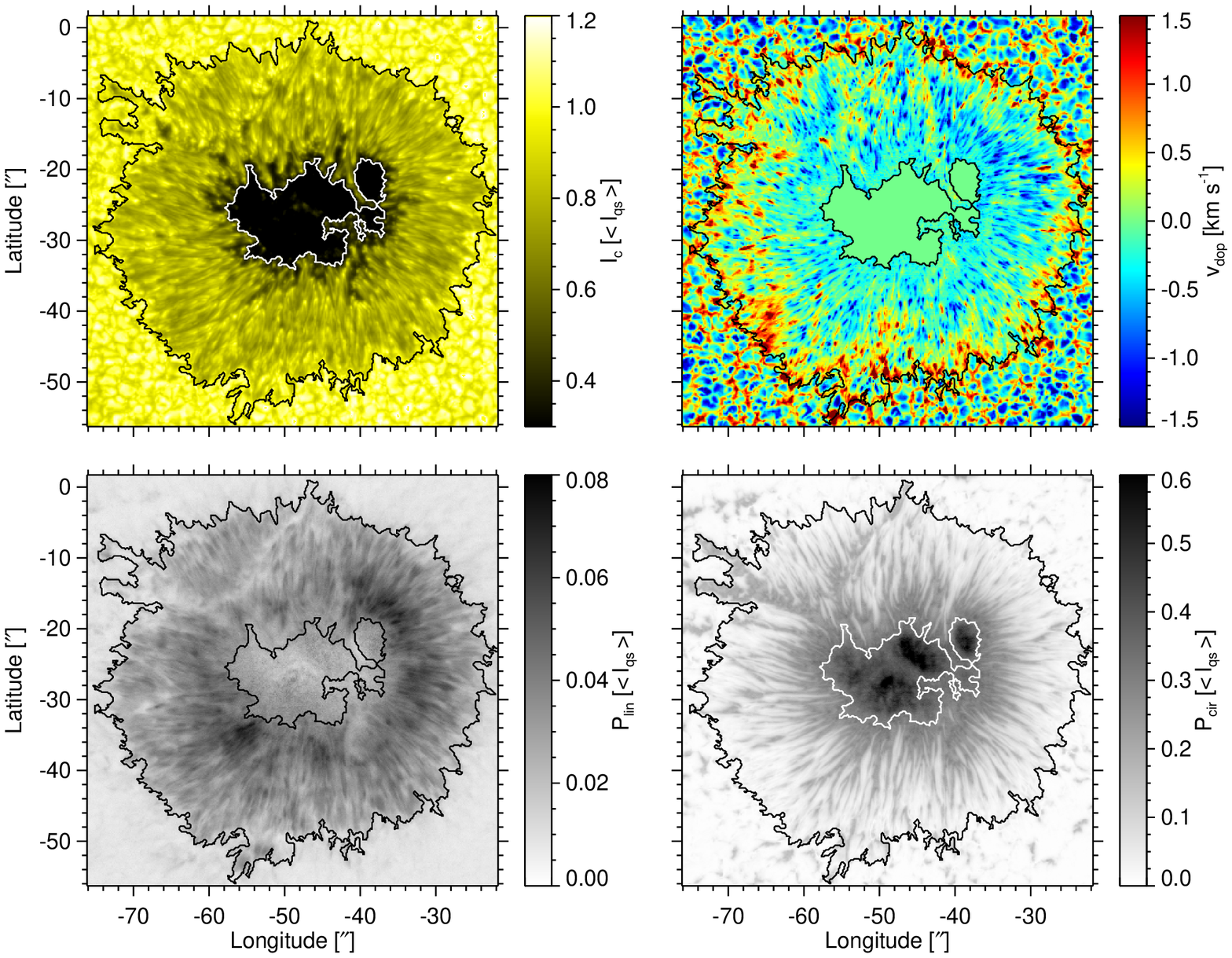}

\end{center}

\newpage
\clearpage
\thispagestyle{empty}

\vspace*{16.5cm}
\begin{flushleft}
\hrule
\vspace*{0.05cm}
\hrule
\vspace*{0.2cm}
	{\bf{Cover Image:}}\,\,\,\, A sunspot at the center of the solar disk observed by HINODE on January 5$^{\rm{th}}$ 2007. The panels show clockwise: Continuum Intensity, Doppler Velocity, and the inverse of Total Circular as well as Total Linear Polarization.
\vspace*{0.2cm}
\hrule
\vspace*{0.05cm}
\hrule
\end{flushleft}


\newpage
\thispagestyle{empty}
\begin{center}

\vspace{0.3cm}

\begin{Huge}
\textbf{\fontsize{35}{40}\selectfont
Spectropolarimetry\\
of Sunspot Penumbrae\\}
\end{Huge}

\vspace{0.7cm}

\begin{Large}
\textbf{
A Comprehensive Study of the Evershed Effect\\
Using High Resolution Data from the\\
Space-Borne Solar Observatory HINODE\\
}
\end{Large}
\vspace{0.7cm}

\includegraphics[height=8cm]{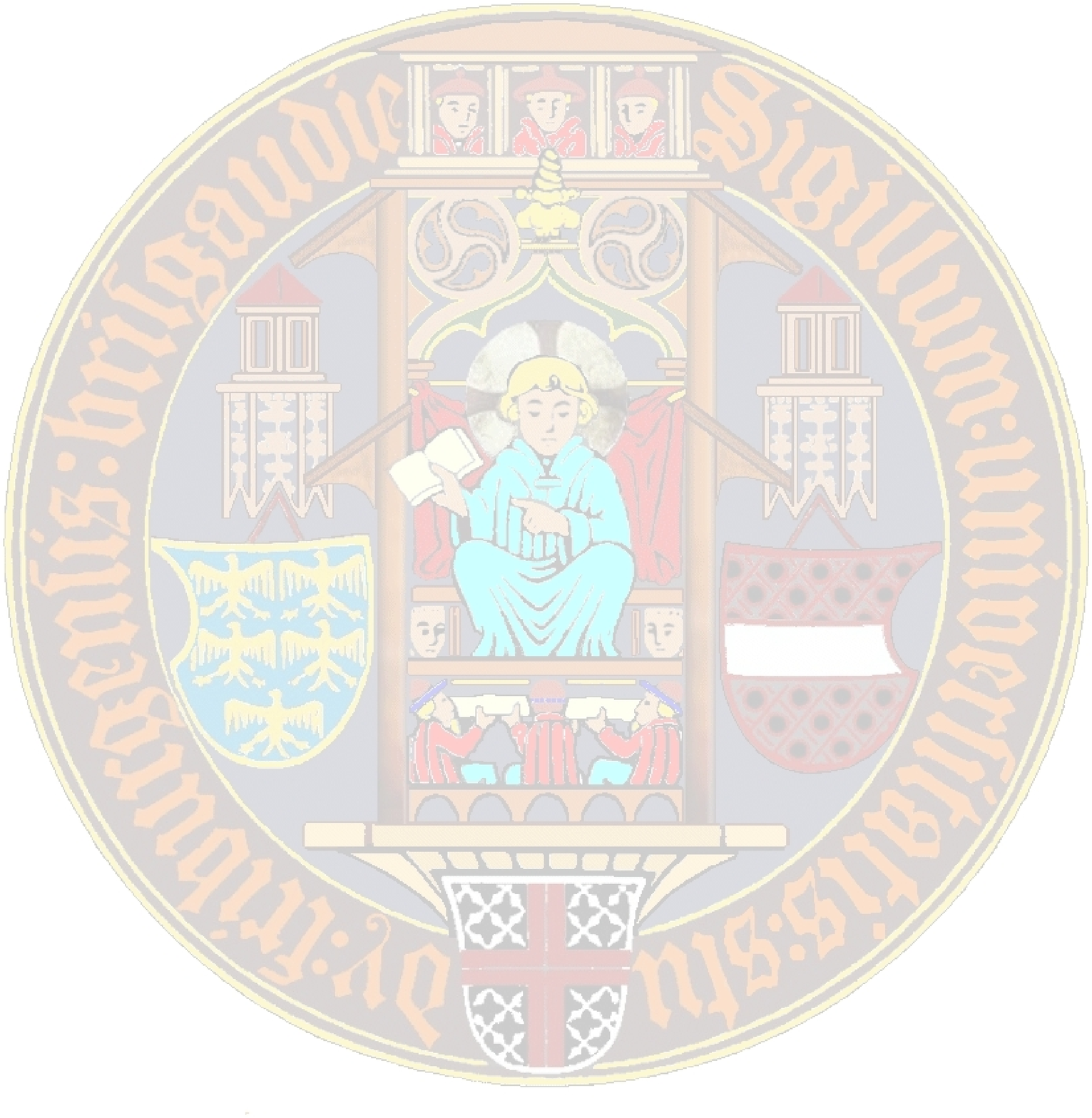}

\vspace{-5.1cm}
\begin{Large}
Inaugural-Dissertation zur Erlangung des Doktorgrades\\
der Fakult\"at f\"ur Mathermatik und Physik\\
der Albert-Ludwigs-Universit\"at Freiburg im Breisgau\\
\end{Large}

\vspace{3.5cm}

\begin{Huge}
\textbf{
Morten Franz\\
}
\end{Huge}

\vspace{0.8cm}

\includegraphics[height=4cm]{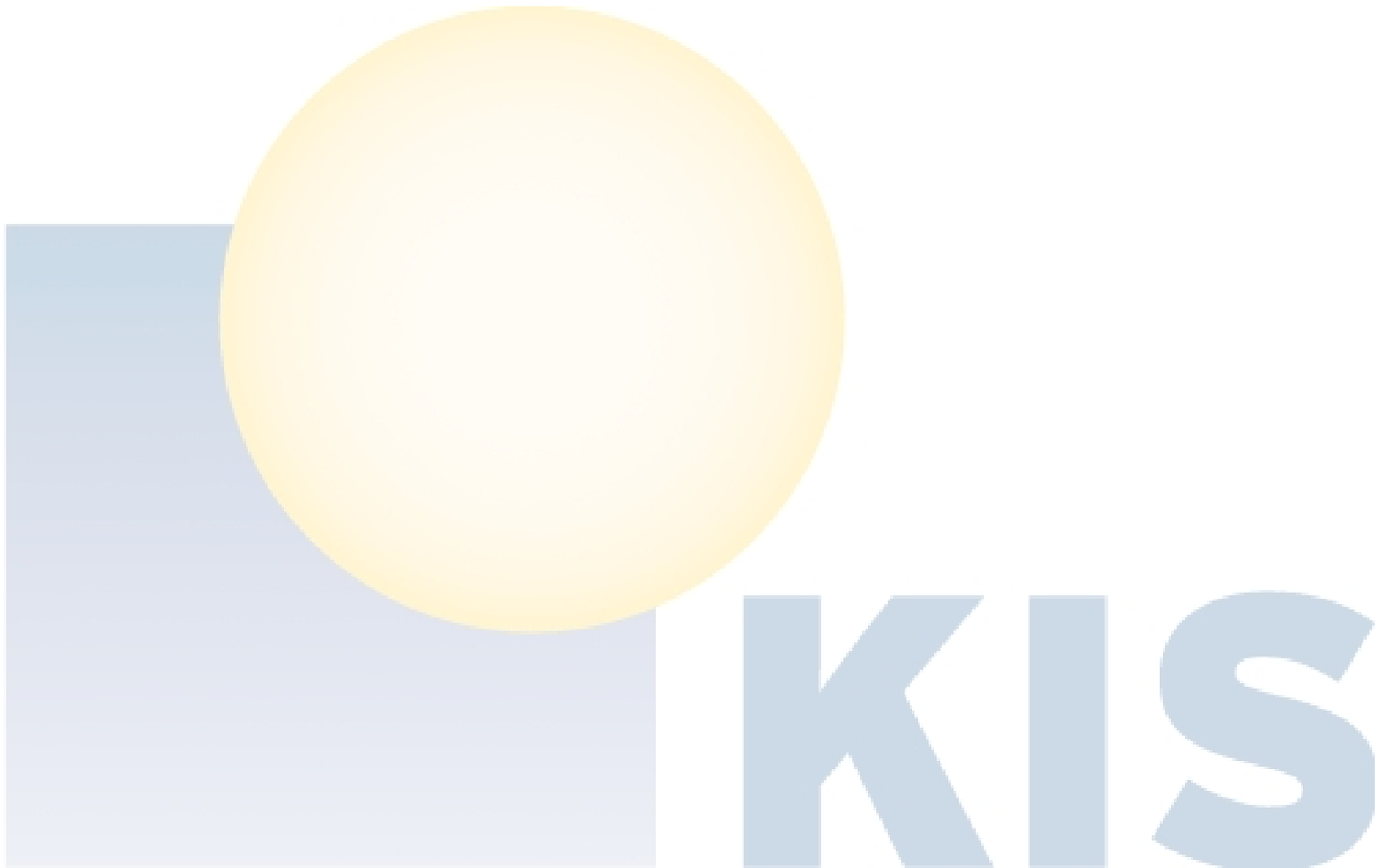}

\vspace{-2.6cm}

\begin{LARGE}
\textbf{Kiepenheuer Institut\\
f\"ur Sonnenphysik\\}
\end{LARGE}
\vspace{1.2cm}
\begin{LARGE}
\textbf{May 2011\\}
\end{LARGE}

\end{center}

\newpage
\clearpage
\thispagestyle{empty}

\newpage
\clearpage
\thispagestyle{empty}

\vspace*{17.5cm}

\begin{table}[h!]
\begin{center}
	\begin{tabular}{ll}
		\hline
		\hline	
		{Dekan:} & {Prof. Dr. Kay K\"onigsmann}\\
		{Referent:} & {Prof. Dr. Wolfgang Schmidt}\\
		{Korreferent:} & {Prof. Dr. Svetlana Berdyugina}\\
		{Disputation:\,\,\,\,\,\,\,\,\,\,\,\,\,\,\,\,\,\,\,\,\,\,\,\,\,\,\,\,\,\,\,\,\,\,\,\,\,\,\,\,\,\,\,\,\,\,\,\,\,\,\,\,\,\,\,\,\,\,\,\,\,\,\,\,\,\,\,\,\,\,\,\,\,\,\,\,\,\,\,\,\,\,\,\,\,} & {21.06.2011}\\
		\hline
		\hline
	\end{tabular}
\end{center}
\end{table}

\chapter*{Publications and Conference Contributions
\label{ch:publication}
\markboth{Publications and Conference Contributions}{}}
\addcontentsline{toc}{chapter}{{\indent{}}Publications}

\section*{Publication\footnote{Contributions marked with \ding{253} have been or will be published in the context of this thesis.} in Peer Reviewed Journals}

\begin{small}

\begin{itemize}\itemsep0.1pt

\item[\ding{253}] {{\bf M. Franz} J. Borrero, and R. Schlichenmaier, {\sf "Reversal of NCP in penumbra at large heliocentric angles"}, (2011), {\it{in preparation}}}

\item[\ding{253}] {{\bf M. Franz} and R. Schlichenmaier, {\sf "Opposite Polarities in sunspot penumbrae"}, (2011), {\it{in preparation}}}

\item {A. Prokhorov, J. Bruls, {\bf M. Franz} and S. Berdyugina, {\sf "Comparison of simulation of solar grannulation with IMaX observation"}, (2011), {\it{in preparation}}}

\item {{\bf M. Franz}, B. Fischer and M. Walther, {\sf "Probing structure and phase-transitions in molecular crystals by terahertz time-domain spectroscopy"}, Journal of Molecular Structure, (2011), {\it{submitted}}}

\item {O. Steiner, {\bf M. Franz}, N. Bello Gonz{\'a}lez, C. Nutto, R. Rezaei, V. Mart{\'{\i}}nez Pillet, J. A. Bonet, J. C. del Toro Iniesta, V. Domingo, S. K. Solanki, M. Kn{\"o}lker, W. Schmidt, P. Barthol and A. Gandorfer, {\sf "Detection of Vortex Tubes in Solar Granulation from Observations with SUNRISE"}, The Astrophysical Journal Letters Vol. 723, pp. L180, (2010) \href{http://adsabs.harvard.edu/abs/2010ApJ...723L.180S}{URL}}

\item {M. Roth, {\bf M. Franz}, N. Bello Gonz{\'a}lez, V. Mart{\'{\i}}nez Pillet, J. A. Bonet, A. Gandorfer, P. Barthol, S. K. Solanki, T. Berkefeld, W. Schmidt, J. C. del Toro Iniesta, V. Domingo and M. Kn{\"o}lker, {\sf "Surface Waves in Solar Granulation Observed with SUNRISE"}, The Astrophysical Journal Letters Vol. 723, pp. L175, (2010) \href{http://adsabs.harvard.edu/abs/2010ApJ...723L.175R}{URL}}

\item {T. Riethm{\"u}ller, S. K. Solanki, V. Mart{\'{\i}}nez Pillet, J. Hirzberger, A. Feller, J. Antonio Bonet, N. Bello Gonz{\'a}lez, {\bf M. Franz}, M. Sch{\"u}ssler, P. Barthol, T. Berkefeld, J. C. del Toro Iniesta, V. Domingo, A. Gandorfer, M. Kn{\"o}lker and W. Schmidt, {\sf "Bright Points in the Quiet Sun as Observed in the Visible and Near-UV by the Balloon-borne Observatory SUNRISE"}, The Astrophysical Journal Letters Vol. 723, pp. L169, (2010) \href{http://adsabs.harvard.edu/abs/2010ApJ...723L.169R}{URL}}

\item {N. Bello Gonz{\'a}lez, {\bf M. Franz}, V. Mart{\'{\i}}nez Pillet, J. A. Bonet, S. K. Solanki, J. C. del Toro Iniesta, W. Schmidt, A. Gandorfer, V. Domingo, P. Barthol, T. Berkefeld, and M. Kn{\"o}lker, {\sf "Detection of Large Acoustic Energy Flux in the Solar Atmosphere"}, The Astrophysical Journal Letters Vol. 723, pp. L134, (2010) \href{http://adsabs.harvard.edu/abs/2010ApJ...723L.134B}{URL}}

\item {S. K. Solanki, P. Barthol, S. Danilovic, A. Feller, A. Gandorfer, J. Hirzberger, T. Riethm{\"u}ller, M. Sch{\"u}ssler, J. A. Bonet, V. Mart{\'{\i}}nez Pillet, J. C. del Toro Iniesta, V. Domingo, J. Palacios, M. Kn{\"o}lker, N. Bello Gonz{\'a}lez, T. Berkefeld, {\bf M. Franz}, W. Schmidt and A. M. Title, {\sf "SUNRISE: Instrument, Mission, Data, and First Results"}, The Astrophysical Journal Letters Vol. 723, pp. L127, (2010) \href{http://adsabs.harvard.edu/abs/2010ApJ...723L.127S}{URL}}

\item[\ding{253}] {{\bf M. Franz} and R. Schlichenmaier, {\sf "Center to limb variation of penumbral Stokes V profiles"}, Astronomische Nachrichten Vol. 331, pp. 570, (2010) \href{http://adsabs.harvard.edu/abs/2010AN....331..570F}{URL}}

\item[\ding{253}] {{\bf M. Franz} and R. Schlichenmaier, {\sf "The velocity field of sunspot penumbrae. I. A global view"}, Astronomy \& Astrophysics Vol. 508, pp. 1453, (2009) \href{http://adsabs.harvard.edu/abs/2009A&A...508.1453F}{URL}}

\item {{\bf M. Franz}, B. Fischer and M. Walther, {\sf "The Christiansen effect in terahertz time-domain spectra of coarse-grained powders"}, Applied Physics Letters Vol. 92, 021107, (2008) \href{http://apl.aip.org/resource/1/applab/v92/i2/p021107\_s1}{URL}}
\end{itemize}

\section*{Proceedings$^1$}

\begin{itemize}\itemsep0.1pt

\item {O. Steiner, {\bf M. Franz}, N. Bello Gonz{\'a}lez, C. Nutto, R. Rezaei, V. Mart{\'{\i}}nez Pillet, J. A. Bonet Navarro, J. C. del Toro Iniesta, V. Domingo, S. K. Solanki, M. Kn{\"o}lker, W. Schmidt, P. Barthol, and A. Gandorfer, {\sf "Detection of Vortex Tubes in Solar Granulation from Observations with Sunrise"}, Astronomical Society of the Pacific Conference Series -- HINODE 4 -- Unsolved problems and recent insights, (2011), {\it{in press}}}

\item {S. K. Solanki, P. Barthol, S. Danilovic, A. Feller, A. Gandorfer, J. Hirzberger, S. Jafarzadeh, A. Lagg,  T. Riethm{\"u}ller, M. Sch{\"u}ssler, T. Wiegelmann, J. A. Bonet, M. Mart{\'{\i}}nez Gonz{\'{a}}lez, V. Mart{\'{\i}}nez Pillet, E. Khomenko, L. Yelles Chaouche, J. C. del Toro Iniesta, V. Domingo, J. Palacios, M. Kn{\"o}lker, N. Bello Gonz{\'a}lez, J.M. Borrero, T. Berkefeld, {\bf M. Franz}, M. Roth, W. Schmidt, O. Steiner, and A. M. Title, {\sf "First results from the Sunrise mission"}, Astronomical Society of the Pacific Conference Series -- HINODE 4 -- Unsolved problems and recent insights,  (2011), {\it{in press}}}

\item {S. K. Solanki, P. Barthol, S. Danilovic, A. Feller, A. Gandorfer, J. Hirzberger, A. Lagg, T. Riethm{\"u}ller, M. Sch{\"u}ssler, T. Wiegelmann, J. A. Bonet, V. Mart{\'{\i}}nez Pillet, E. Khomenko, J. C. del Toro Iniesta, V. Domingo, J. Palacios, M. Kn{\"o}lker, N. Bello Gonz{\'a}lez, J. M. Borrero, T. Berkefeld, {\bf M. Franz}, M. Roth, W. Schmidt, O. Steiner and A. M. Title, {\sf "The Sun at high resolution: first results from the Sunrise mission"}, Proceedings IAU Symposium No. 273 -- Physics of Sun and Star Spots,  (2011), {\it{in press}}}

\item[\ding{253}] {{\bf M. Franz}, R. Schlichenmaier and W. Schmidt, {\sf "Small-Scale Velocities in Sunspot Penumbrae
"}, Astrophysics and Space Science Proceedings Vol. 510, pp. 510, (2010) \href{http://adsabs.harvard.edu/abs/2010mcia.conf..510F}{URL}}

\item[\ding{253}] {{\bf M. Franz} and R. Schlichenmaier, {\sf "Spectral Analysis of Sunspot Penumbrae Observed with Hinode
"}, Astronomical Society of the Pacific Conference Series Vol. 415, pp. 369, (2009) \href{http://adsabs.harvard.edu/abs/2009ASPC..415..369F}{URL}}

\item[\ding{253}] {R. Schlichenmaier and {\bf M. Franz}, {\sf "The Small-Scale Flow Field of a Sunspot Penumbra"}, 12th European Solar Physics Meeting, Freiburg, Germany, (2008) \href{http://adsabs.harvard.edu/abs/2008ESPM...12.2.28S}{URL}}

\item {B. M. Fischer, {\bf M. Franz}, and D. Abbott, {\sf "T-ray biosensing: a versatile tool for studying low-frequency intermolecular vibrations"}, Proceedings of SPIE Vol. 6416, 64160U, (2006) \href{http://spiedigitallibrary.org/proceedings/resource/2/psisdg/6416/1/64160U\_1?isAuthorized=no}{URL}}

\item {{\bf M. Franz}, B. Fischer, D. Abbott and Hanspeter Helm, {\sf "Terahertz Study of Chiral and Racemic Crystals"}, Joint 31st International Conference on IRMMW-THz, Shanghai, China, (2006) \href{http://ieeexplore.ieee.org/xpl/freeabs\_all.jsp?arnumber=4222172}{URL}}

\item {B. M. Fischer, {\bf M. Franz} and D. Abbott, {\sf "THz spectroscopy as a versatile tool for investigating crystalline structures"}, Joint 31st International Conference on IRMMW-THz, Shanghai, China, (2006) \href{http://ieeexplore.ieee.org/xpls/abs\_all.jsp?arnumber=4222304}{URL}}

\end{itemize}

\section*{Talks and Other Contributions$^1$}
\begin{itemize}\itemsep0.1pt

\item[\ding{253}] {{\bf M. Franz}, J. Borrero and R. Schlichenmaier, {\sf "Crossover Profiles tracing Opposite Polarities in Sunspot Penumbrae"}
HINODE 4 -- Unsolved problems and recent insights, Palermo, Italy, (2010) \href{http://www.astropa.unipa.it/hinode4/Hinode4.html}{URL}}

\item[\ding{253}] {{\bf M. Franz}, R. Schlichenmaier and W.Schmidt, {\sf "Spectral Analysis of Sunspot  Penumbrae"}, USO Summer School "Solar Magnetism", Dwingeloo, Netherlands, (2009) \href{http://www.astro.uu.nl/~rutten/uso-dwingeloo-2009/Home.html}{URL}}

\item[\ding{253}] {{\bf M. Franz}, R. Schlichenmaier and W.Schmidt, {\sf "Observation of Sunspot Penumbrae with Hinode"}, DPG Fr\"uhjahrstagung, Freiburg, Germany, (2008) \href{http://www.dpg-verhandlungen.de/2008/freiburg/ep.pdf}{URL}}

\item {{\bf M. Franz}, B. Fischer and M. Walther, {\sf "THz Spectroscopy on Crystalline Biomolecules"}, 1st THz Frischlinge Meeting, Freiburg, Germany, (2007)}

\end{itemize}
\end{small}

\newpage
\clearpage
\thispagestyle{empty}

\vspace*{15cm}

\begin{flushright}
{\it{When I was a little kid my mother told me not to stare into the Sun.\\
So once when I was six I did.\\
The doctors didn't know if my eyes would ever heal.\\
I was terrified, alone in that darkness.\\
Slowly, daylight crept in through the bandages, and I could see.\\
But something else had changed inside of me.}}\\

\bigskip

Maximilian Cohen, ${\bf{\pi}}$ $-$ faith in chaos
\end{flushright}

\begingroup
\hypersetup{linkcolor=black}
\makeatletter \renewcommand{\@dotsep}{10000} \makeatother
\tableofcontents
\clearemptydoublepage
\endgroup

%


\chapter*{Abstract
\markboth{Abstract}{}}
\label{ch:abstract}
\addcontentsline{toc}{chapter}{{\indent{}}Abstract}

This research project focuses on the structure of sunspot penumbrae and on the Evershed Effect (EE). Even though the EE has been known for more than one hundred years, its driving mechanism remains an issue of debate until the present day. High resolution spectropolarimetric data obtained by the space-borne observatory HINODE is used to characterize the small-scale ($\approx 240$~km) penumbral magnetic field as well as the vertical and horizontal component of the EE.

After an introduction to the Sun and its magnetic activity, sunspots are characterized phenomenologically, the state of the art of penumbral modeling is summarized and the theoretical background to spectropolarimetry is provided. The HINODE observatory is sketched and a range of techniques that allow for an absolute wavelength calibration as well as a measure of solar plasma velocities in the deep photosphere are compared.

The results demonstrate that the penumbral velocity field differs significantly from that of the quiet Sun. Morphological studies yield elongated upflow channels in the inner penumbra and round downflows in the outer penumbra. These flows are identified as the sources and the sinks of the EE. What is remarkable is the high plasma velocity in these sinks, which is much larger when compared to the quiet Sun. Furthermore, an extraordinary high zenith angle was found for the penumbral downflows. The small-scale velocity field within penumbral filaments is investigated both by statistics and case studies. The outcome of these surveys confirms the predictions of penumbral flux-tube models. The high quality of HINODE data allows the investigation of bright penumbral downflows as well as two families of penumbral filaments. Observations of sunspots close to the solar limb are used to review the horizontal component of the EE.

The study of the asymmetries of Stokes profiles shows that the penumbral plasma flow is concentrated in the deep photosphere and that its amplitude di\-minishes much faster with height when compared to the quiet Sun. Another important result is the unequivocal confirmation of the magnetized character of the horizontal and the vertical EE. In contrast to previous studies, this analysis proves that the sinks of the EE are filled with a magnetic field of opposite polarity. Additionally, the influence of atmospheric parameters on the asymmetries of Stokes profiles is explored within the framework of a two-layer model atmosphere and by means of spectral inversion. Interestingly, it is only the polarity of the gradients with height of the magnetic field strength that causes the sign of the total net circular polarization in the center side penumbra.

\makeatletter
\renewcommand\subsubsection{\newpage\par%
  \thispagestyle{plain}%
  \global\@topnum\z@
  \@afterindentfalse
  \secdef\@chapter\@schapter
}
\makeatother

\subsubsection*{Zusammenfassung
\markboth{Zusammenfassung}{}}
\label{ch:zusammenfassung}


Die vorliegende Arbeit befasst sich mit der Struktur der Penumbra von Sonnenflecken und dem Evershed Effekt (EE). Obwohl der EE seit \"uber einhundert Jahren bekannt ist, sind seine urs\"achlichen Mechanismen bisher nicht vollst\"andig gekl\"art. Mittels hochaufge\-l\"oster spektropolarimetrischer Daten des Satellitenobser\-vatoriums HINODE werden sowohl das kleinskalige ($\approx 240$~km) penumbrale Magnetfeld, als auch die Vertikal- und Horizontalkomponenten des EE untersucht.

Am Anfang wird der Aufbau und die magnetische Aktivit\"at der Sonne beschrie\-ben. Eine Reihe von theoretischen Modellen zur Erkl\"arung des EE werden diskutiert und die theoretischen Grundlagen der Spektropolarimetrie werden zusammengefasst. 
Das Beobachtungsinstrument wird skizziert und eine Reihe von Methoden zur absoluten Wellenl\"angenkalibrierung sowie zur Untersuchung von solaren Materiestr\"o\-mung\-en in der tiefen Photosph\"are werden miteinander verglichen.

Die Forschungsresultate zeigen, dass sich das penumbrale Geschwindigkeitsfeld signifikant von dem der ruhigen Sonne unterscheidet. Morphologische Studien ergeben elongierte Aufstr\"omungen in der inneren und runde Abstr\"omungen in der \"au{\ss}eren Penumbra, welche als Quellen und Senken des EE interpretiert werden. F\"ur die Senken des EE konnte ein au{\ss}erordentlich gro{\ss}er Zenitwinkel nachgewiesen werden. Weiterhin ist anzumerken, dass die Plasmageschwindigkeit in den penumbralen Abstr\"omungen weit gr\"o{\ss}er ist als in der ruhigen Sonne. Das kleinskalige Geschwindigkeitsfeld innerhalb penumbraler Filamente wird statistisch und anhand von Fallstudien untersucht, wobei die Vorhersagen von Flu{\ss}r\"ohrenmodellen best\"atigt werden. Des Weiteren werden helle penumbrale Abstr\"omungen beschrie\-ben und zwei Klassen von penumbralen Filamenten 
identifiziert. Beobachtungen von Flecken am Rand der Sonnenscheibe werden benutzt, um die Horizontalkomponente des EE zu untersuchen.

Anhand von asymmetrischen Stokes Profilen wird gezeigt, dass die penumbralen Plasmastr\"omungen vornehmlich in der unteren Photosph\"are vorliegen und dass deren Amplitude, im Gegensatz zur ruhigen Sonne, schnell mit der H\"ohe abf\"allt. Es wird nachgewiesen, dass sowohl die Horizontal- als auch die Vertikalkomponente des EE magnetisiert sind. Hierbei wird im Gegensatz zu vorherigen Studien demonstriert, dass die Senken des EE ein Magnetfeld gegens\"atzlicher Polarit\"at aufweisen. Der Einfluss atmosph\"arischer Parameter auf Asymmetrien wird im Rahmen eines Zwei-Schichten-Modells und mittels Spektralinversionen untersucht. Die Ergebnisse zeigen dass es nur die Polarit\"at des H\"ohengradienten der Magnetfeldst\"arke ist, welche das Vorzeichen der totalen Netto-Zirkolarpolarisation auf der zentrumsseitigen Penumbra bestimmt.

\mainmatter

\chapter{Introduction}
\label{ch:Intro}

The Sun plays a vital role 
for virtually all lifeforms on earth, and its observation by humans is probably as old as mankind. Prehistoric cave drawings in Lascaux (France) indicate that people put the Sun and other celestial bodies into context with their daily life as early as 15.000 BC \citep{Rappen:2004a,Rappen:2004b}. The development of agriculture made it important to forecast the annual seed and harvest times, increasing the need for solar observation. Constructions like Stonehenge, the Pyramids of Gizeh and artifacts like the Nebra sky disk show that our ancestors had a precise knowledge of the yearly occurrence of the summer and winter solstice as well as the equinoxes in-between. Astronomy and mathematics were used by early civilizations to define a detailed calendar predicting phenomena such as the variation in the length of daylight over a solar year, the first and last visible risings of different planets over a period of decades and the prediction of solar and lunar eclipses. One of the oldest historical evidences of an astronomical incident is the solar eclipse recorded 2137 BC in China \citep{1980SoPh...66..187W}, and Aristarchus of Samos (310 BC - 230 BC), a supporter of the heliocentric system, estimated the distance between the Earth, the Moon and the Sun more than two thousand years ago \citep{Heath:1913}. 

In the following millennia, numerous scientists, e.g.\,Nicholas Copernicus, Galileo Galilei, Isaac Newton, William Herschel, Joseph von Fraunhofer, Lord Kelvin, Hermann von Helmholtz, Albert Einstein, Sir Arthur Eddington, Subrahmanyan Chandrasekhar and Hans Bethe, contributed to our modern picture of the Sun. Their work was aided by a steady improvement of scientific instruments and a fruitful exchange between the different branches of science. 
Famous examples of this interdisciplinary work include: a) Spectroscopic observation of sunlight, which led to the discovery of the hitherto unknown element helium. 
b) The long lasting controversy between geological and astrophysical measurements about the age of the Sun, which was not settled until the discovery of nuclear fusion \citep{vonWeizsacker1938,1939PhRv...55..434B}. c) The discrepancy between the flux of solar neutrinos measured at earth and theoretical predictions, which finally led to the postulation of neutrino oscillations \citep{1985YaFiz..42.1441M}. Besides evident facts such as its impact on the terrestrial weather and climate \citep{2010CliPa...6..723J}, the Sun still influences our daily life in many ways. Modern communication via satellites, for example, is very prone to solar activity, and solar storms can cause power failures.

The Sun's stellar classification is G2V, accounting for its surface tem\-pe\-rature of 5780 K and its location within the main sequence of the Hertzsprung-Russel-Diagram \citep{1909AN....179..373H,1914PA.....22..275R}. At an age of 4.86$\cdot10^9$ years, it has completed about half of the main-sequence evolution. The Sun has a diameter of about 1.4$\cdot10^{9}$ m, a mass of 2$\cdot10^{30}$ kg and a luminosity of 3.8$\cdot10^{26}$~W \citep{Unsoeld:2002}. On Earth, it appears as the brightest object in the sky ($\rm{mag}= -$26.74), even though its absolute magnitude is only $+$4.83. About 75\% of its mass consists of hydrogen H, while the rest is mostly He. Less than 2\% consist of heavier elements, including carbon (C), nitrogen (N), oxygen (O), neon (Ne),  calcium (Ca), iron (Fe) and others. This composition is typical for a Population I star of the galactic disk; Population II stars from the halo of the Milky Way contain much less heavy elements. Out of the $10^{23}$ estimated stars in the universe \citep{2010Natur.468..940V}, the Sun is exceptional only in its proximity to Earth. This vicinity makes it a unique laboratory to test computer models and theoretical predictions,  
since it is the only star for which surface phenomena can be resolved and studied in detail. 

\section{Scope and Organization of this Work}

Sunspots are one of the most prominent solar features. Under certain conditions, these dark dots on the solar disk can be seen with the naked eye. Even though records of sunspots have existed since two millennia \citep{1987A&AS...70...83W}, it was not until the 16$^{\rm{th}}$ century that they were investigated systematically. 
While some astronomers considered them as solar features, others believed that they are caused by exosolar phenomena. 
An important observation was made by \citet{1908ApJ....28..315H}, who discovered strong magnetic fields within sunspots and proposed that these fields are caused by strong currents due to a cyclonic motion of plasma around the spot. In contrast to this idea, \citet{Evershed:1909p167} found spectroscopic evidence for a radial outflow of plasma in the penumbrae of sunspots. 

The aim of this work is to study the small-scale structure of this so-called Evershed effect and to compare the results with predictions of penumbral models by using data of high spectral and spatial resolution obtained by HINODE. 

Chapter~\ref{ch:sun} provides a brief introduction to the Sun, while the current knowledge of sunspots is summarized in Chapter~\ref{ch:sunspots}. Chapters~\ref{ch:specpol} and \ref{ch:data} elaborate on the method used in this study from a theoretical and experimental perspective. Chapters~\ref{ch:center} and \ref{ch:limb} describe the investigation of the horizontal and the vertical components of the Evershed effect, while Chapters \ref{ch:asym} and \ref{ch:ncp} study the gradients with height of the penumbral magnetic and velocity field. In conclusion, Chapter~\ref{ch:outlook} evaluates the quality of penumbral models in the light of HINODE observations.

A short summary of the contents is provided in the beginning of each Chapter. 
For convenience, hyperlinks are used throughout the electronic version of this text, e.g. for sections, figures, tables and references. Additionally, references may be accessed directly using the corresponding URL in the bibliography.

\chapter{The Structure of the Sun}
\label{ch:sun}
The structure of the Sun is the result of a balance of forces (mainly between gravity and pressure), the energy balance between its generation in the core and its loss at the solar surface as well as the stationary energy transport in-between. Section \ref{sec:intern} reviews the inner body of the Sun, while the atmospheric layers, including a range of observable features, are summarized in Section \ref{sec:atmos}.

\begin{figure}[h!]
	\centering
		\includegraphics[width={0.9\textwidth}]{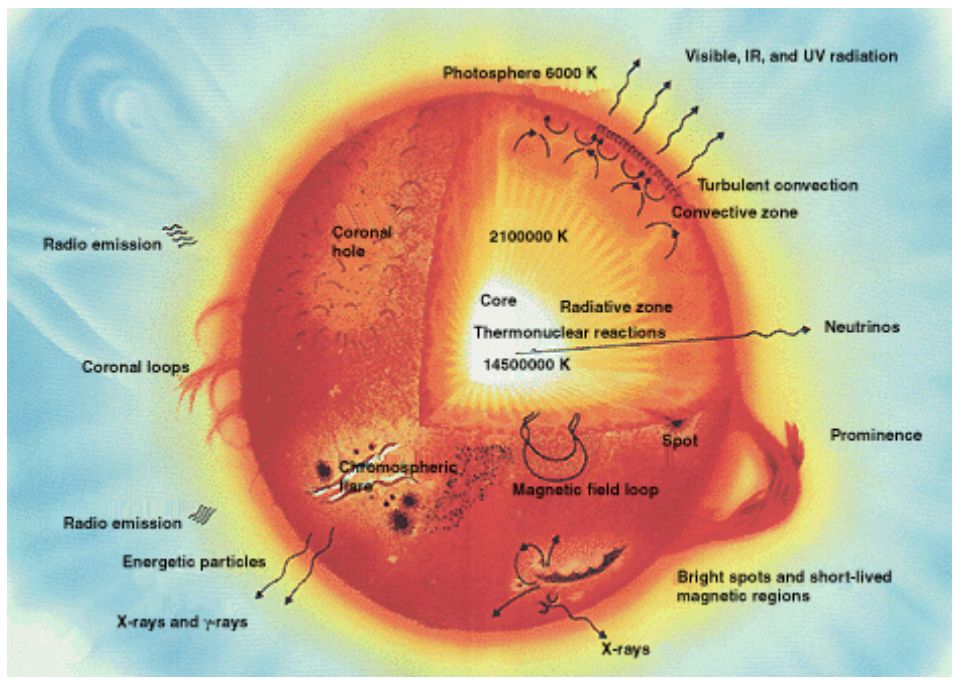}
		\caption{Sketch of the structure of the Sun. Adopted from \citet{wiki:sun}.}
		\label{fig:sun_cut}
\end{figure}

\section{The Solar Interior}
\label{sec:intern}

The solar model governing its internal structure has to obey physical laws such as the conservation of mass and momentum, and it has to describe the balance and transport of energy. Theoretical considerations yield four first order differential and four constitutive equations. To solve the differential equations, certain boundary conditions are assumed and then modified in an iterative process until they provide reasonable observable values \citep{2004suin.book.....S}. The solution to these equations yields, for example, the distribution of temperature (T), pressure and density ($\rho$) with solar radius \citep{1996Sci...272.1286C}. 

\paragraph{The Core:} In the solar core H is converted into He by nuclear fusion. This process is described by the CNO-tricycle \citep{vonWeizsacker1938,1939PhRv...55..434B} or by different proton-proton chains \citep{1938PhRv...54..248B}. Measurements of the neutrino spectrum show that the latter process, which converts 7$\permil$ of the mass of the participating particles into energy, is dominant in the Sun. The fusion rate is in an equilibrium: On the one hand, a higher rate would increase T and cause an expansion of the core, which would decrease T again. On the other hand, a lower fusion rate would cause the core to cool and shrink, increasing the fusion rate and reverting it to its current level.

\paragraph{Source of Energy:} The source of the solar energy supply has long been the issue of debate. One theory proposed gravitational contraction: The Sun is collapsing slowly, thereby converting gravitational potential energy into heat and light. Taking the current solar luminosity into account, the so-called Kelvin-Helmholtz (KH) timescale would be around 3$\cdot10^7$ years, which was considered a good estimate of the age of the Sun in the 19$^{\rm{th}}$ century. In the 20$^{\rm{th}}$ century, however, geological evidence yielded an age of the Earth of the order of 4$\cdot10^9$ years. The resulting conflict between the geological timescale and the KH timescale was not settled until it was accepted that nuclear fusion provides another source of energy within stars. If it is assumed that 1$\permil$ of the mass of the Sun is converted into radiant energy, the nuclear time scale is of the order of 10$^{10}$ years, i.e. the present age of the universe.

\paragraph{Radiation Zone:} H and He are fully ionized within the radiation zone. Free-free absorption, bremsstrahlung and electron scattering cause the mean free path of photons of a few millimeters. With increasing radius, the gradient in T and $\rho$ yields a cooler and thinner, hence less opaque, solar plasma. On average, the photons thus  
diffuse towards the outer layers, transporting the energy by thermal radiation. Estimates of the photon travel time range between 1.7$\cdot10^5$ years, if a random walk is assumed \citep{1992ApJ...401..759M}, to 1.7$\cdot10^7$ years, if thermal adjustment of the Sun is taken as a measure \citep{2003SoPh..212....3S}.

\paragraph{Convection Zone:} Within the convection zone, T drops rapidly and allows the formation of neutral H an He. The reconnection consumes a large fraction of the free electrons, increasing the opacity until the gradient of T becomes larger than the adiabatic gradient. This triggers convective instabilities, and energy is transported by moving plasma that reaches the surface within months \citep{2004suin.book.....S}.

\section{The Atmospheric Layers}
\label{sec:atmos}

The surface of the Sun is commonly defined as the atmospheric layer at which the solar plasma changes from opaque to transparent, or more precise, where the opacity of the solar plasma at 500~nm ($\tau_{500}$) corresponds to unity. This transition depends on the wavelength (solar absorption lines) and the position on the solar disk (limb darkening effect). The fact that $\tau=1$ occurs at different geometrical height within an absorption line provides an important tool to study the thermodynamic state\footnote{In the local thermodynamic equilibrium, collisions within the plasma distribute the energy equally among the degrees of freedom of the constituent particles. Thus temperature may be defined as a thermodynamic quantity. If the density decreases, collisions occur less frequently and the temperature becomes a kinetic quantity that may assume separate values for different directions or different particle species, e.g. ions and electrons. It is necessary to keep this in mind when comparing photospheric with chromospheric or coronal temperatures.} 
of different atmospheric layers. Theoretical models, which have to obey observational constrains, yield a distribution of T within the atmosphere \citep{1981ApJS...45..635V}. 
It is remarkable that T decreases to 4100 K at a height of 500~km above $\tau_{500}=1$, but then increases again, reaching several 10$^6$ K in the corona. 

\paragraph{Photosphere:} Convective plumes transport the hot plasma from below into the photosphere, where it cools radiatively. The top of these plumes are seen as bright elements, so-called granules. Like B\'enard cells, the granules resemble a honeycomb structure of hexagonal prisms. They have a diameter of around 1000~km and a lifetime of approximately 10~minutes. However, in observations with a high spatial resolution, the granules show a substructure \citep{2010ApJ...723L.180S}. Between the granules, where the cooler plasma sinks back below the surface, multiply connected intergranular lanes appear. This granular pattern is called the quiet Sun (QS). 

Supergranules are the giant version of granules with diameters of 2$\cdot10^4$~km across and are best seen in synoptic Doppler maps of the solar disk. Individual supergranules last for one or two days and have flow speeds of about 0.5 km s$^{-1}$.

\begin{figure}[h!]
	\centering
		\includegraphics[width={0.69\textwidth}]{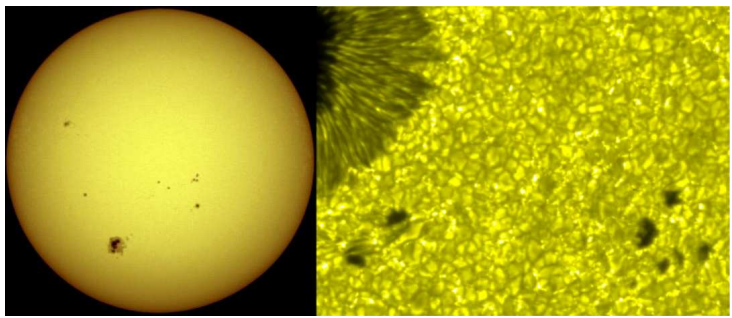}
		\caption{Left: White light picture of the solar disk. Adopted from \citet{nasa:photos1}. Right: HINODE high resolution image at 555.1 nm. Adopted from \citet{nasa:photos2}}
		\label{fig:photos}
\end{figure}

The QS appears darker at the solar limb when compared to the center of the disk. This is because the line of sight (LOS) penetrates the photosphere under an angle which is largest at the solar limb. Thus, the path through the atmosphere is increased both in length and opacity, causing the limb to appear darker since the photons stem from higher and cooler photospheric regions.

Magnetic flux concentrations alter the granular pattern, and form intergranular bright points, faculae or pores. If the pore is at least partially surrounded by a filamentary and less dark ring, it is called a sunspot\footnote{This definition has been criticized by \citet{1981phss.conf....7M}, but will be used throughout this work.} \citep{Loughhead:1964}.

\paragraph{Chromosphere:} During the totality of a solar eclipse, the chromosphere appears as a deep red ring (emission of H$_{\alpha}$ at 656.2~nm) around the lunar disk. The chromosphere is the coolest layer of the solar atmosphere. Its magnetic field forms hammock-like structures, suspending plasma above the surface. Depending on whether these thread-like strands are seen on the disk or at the limb, they are called filaments or prominences respectively. Other features, i.e. plage, are often seen around sunspots. The web-like pattern at the edges of supergranular cells is called the chromospheric network and the highly dynamic magnetic fields filled with luminous gas moving up and down within 10~minutes are referred to as Spicules.

\paragraph{Corona:} The outmost layer of the atmosphere, i.e. the corona, is so hot that not only H and He, but also C, Ni and O are completely ionized. Heavier trace elements, i.e. Fe and Ca, are highly ionized and cause the emission line corona. Prominent explanations for the coronal T, which seem at odds with the second law of thermodynamics, involve reconnection of magnetic field lines as well as magneto-acoustic and Alfv\'en waves. Coronal feature, mainly caused by the magnetic field, are: Coronal loops, solar flares, helmet streamers, polar plumes or coronal holes.

\begin{figure}[h!]
	\centering
		\includegraphics[width={0.8\textwidth}]{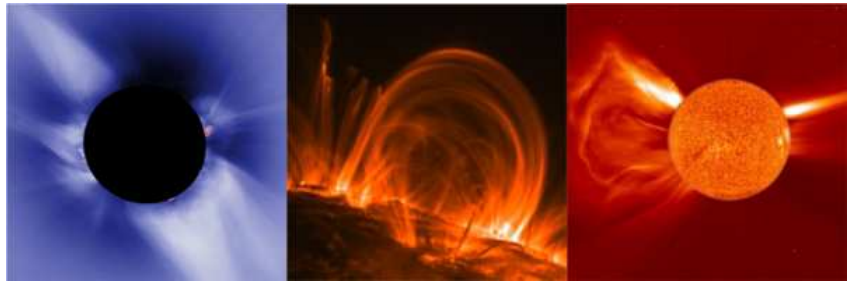}
		\caption{Left: White light corona during the totality of a solar eclipse. Adopted from \citet{nasa:corona1}. Center: Image taken in Fe IX/X at 17.1~nm which corresponds to a temperature of 1$\cdot10^6$~K. Adopted from \citet{nasa:corona2}. Right: White light image of the SOHO coronograph. The occulted disk is filled with an image taken by SOHO in He II at 30.4~nm, which corresponds to a temperature\protect\footnotemark[3]\, of 7$\cdot10^4$~K. Adopted from \citet{nasa:corona3}.}
		\label{fig:corona}
\end{figure}

\addtocounter{footnote}{+1}\footnotetext{Usually it is the collision with the electrons from the local environment that excite or ionize atoms and cause solar spectral lines. In the case of He II, non-local electrons from the hotter coronal regions cause this ionization. The temperature is inferred from model calculations similar to the VAL-C model and does not refer to the kinetic energy of the electrons, e.g. in Fe XII.}

\chapter{Sunspots}
\label{ch:sunspots}

Section~\ref{sec:cycle} provides a summary of the properties of sunspots during the solar cycle as well as wide-spread ideas for the generation of magnetic fields. Part of it is based on the work of \citet{uso:summerschool2009} and \citet{2009SSRv..144..351V}. The following two Sections (\ref{sec:theo_umbra}~and~\ref{sec:theo_penum}) draw a more detailed picture on sunspots, following the extensive review by \citet{Solanki:2003p2072}. The focus lies on small-scale features and their dynamical behavior in as well as around the umbra and the penumbra. Section~\ref{sec:theo_life} summarizes the current knowledge about the formation and decay of sunspots. This chapter is concluded in Section~\ref{sec:evershed} with a discussion of different physical mechanisms explaining the Evershed flow as well as a review of the state of the art of penumbral models, including their limitations and shortcomings.

\begin{figure}[h!]
	\centering
		\includegraphics[width={0.90\textwidth}]{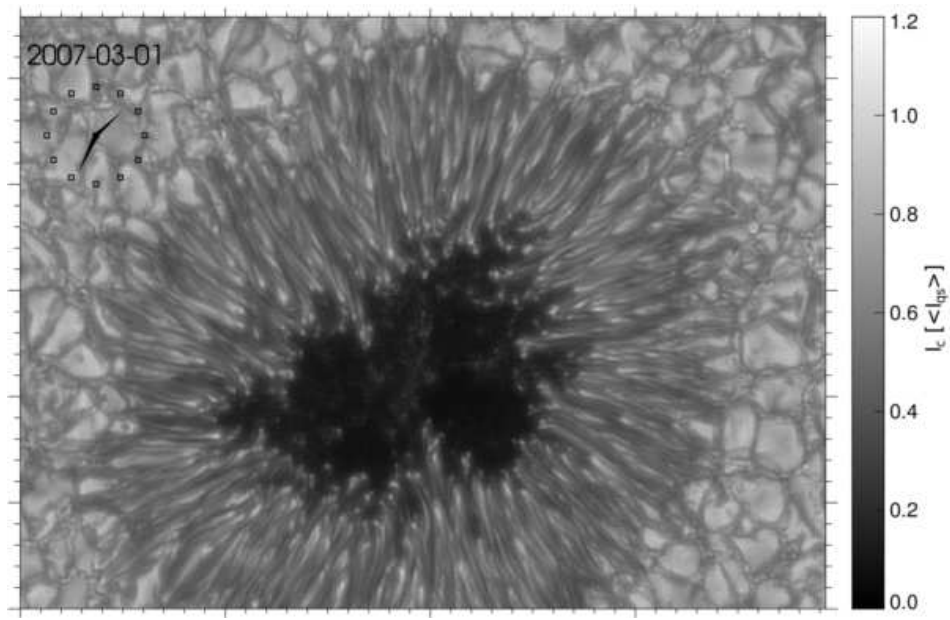}
		\caption{HINODE BFI image of a sunspot taken in the blue continuum at 450.5 nm with a spatial resolution of 0\arcsec.2. The image has been post-processed to increase the contrast and make the small-scale features more apparent. Tick-marks are in seconds of arc.}
		\label{fig:spotcont}
\end{figure}

\section{Global Properties and Periodicity}
\label{sec:cycle}

The invention of the telescope in the early 17$^{\rm{th}}$ century led to a systematic investigation of the Sun. Daily observation of sunspots and other solar features showed that the equatorial plane of the Sun rotates roughly 20\% faster than the polar regions  \citep{Scheiner:1630,1861AN.....55..289S}. 
A closed expression for this differential rotation was first given by \citet{Carrington:1863}. Modern observation of magnetic features yield 
$\omega(\rm{l})=14.38-1.95\,\rm{sin}^2\,(\rm{l})-2.17\,\rm{sin}^4\,(\rm{l})$ for the rotational speed of the solar latitudes (l).

\begin{figure}[h!]
	\centering
		\includegraphics[width={0.75\textwidth}]{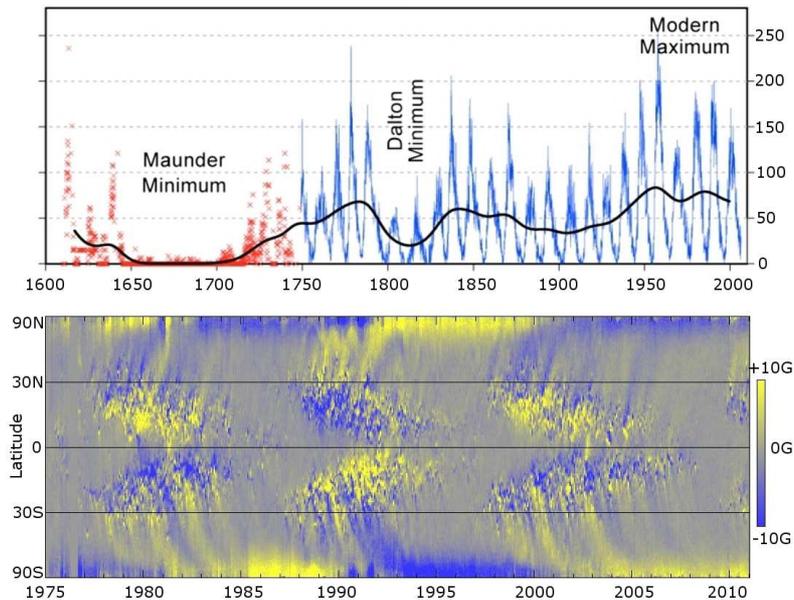}
		\caption{Top: Number of apparent sunspots on the Sun (blue) and sporadic observations (red). Adopted from \citet{wiki:cycle}. Bottom: Modern butterfly diagram of the last three solar cycles. Adopted from \citet{solardyn}.}
		\label{fig:spotnumblat}
\end{figure}

\paragraph{Sunspot cycle:} 
\noindent\citet{1844AN.....21..233S} 
was among the first to report on a 11 year cycle of the apparent number of sunspots. This cycle was confirmed by \citet{1850MiZur...1....3W}, 
who counted sunspots together with active regions. The waxing and waning of this so-called Wolf number is visible in the top panel of Fig.~\ref{fig:spotnumblat}. Cycles with a large number of spots (during the last 50 years) and almost no spots at all (during the Maunder minimum) have been measured. To what extent this variation of solar activity might influence the terrestrial climate, i.e. result in ice ages or cause global warming, is still under debate \citep{2010CliPa...6..723J}.

\citet{1858MNRAS..19....1C} 
noted that sunspots appear at progressively lower latitudes as the solar cycle evolves. This was confirmed by \citet{1883AN....105..169S}
, who visualized this effect by calculating the solar area occupied by spots in a certain time interval and plotted this versus the solar latitude. The result can be seen in the lower panel of Fig.~\ref{fig:spotnumblat}. The spots appear around 30$^{\circ}$ north and south of the equator in the beginning and emerge close to the equator at the end of the sunspot cycle. The region between $\pm$30$^{\circ}$ latitude is called the activity belt because sunspots usually do not appear at larger latitudes. Since the distinct pattern in the activity belt resembles the shape of the wings of a butterfly, Sp\"orer's plot is often referred to as the butterfly diagram\footnote{In addition to the original butterfly diagram, the lower panel of Fig.~\ref{fig:spotnumblat} shows the polarity of the magnetic fields. Sp\"orer was not able to measure solar magnetic fields, but only the sunspot area as a function of latitude.}. The Wolf number reaches its maximum in the middle of the cycle when the majority of sunspots appear at $\pm$15$^{\circ}$ latitude.

After the discovery of solar magnetic fields by \citet{1908ApJ....28..315H}, 
it was recognized that sunspots are only the most prominent manifestations of solar magnetism and can be used as a proxy of the latter. Since areas of increased magnetic activity -- i.e. active regions (ARs) -- typically have a bipolar structure and sunspots are always located within such regions, they often appear in binary groups. The western or preceding (p) spot of such a group is usually larger and the first to be formed, while the eastern or following (f) spot appears later, frequently splits into several components and disappears sooner. 

\begin{figure}[h!]
	\centering
		\includegraphics[width={0.775\textwidth}]{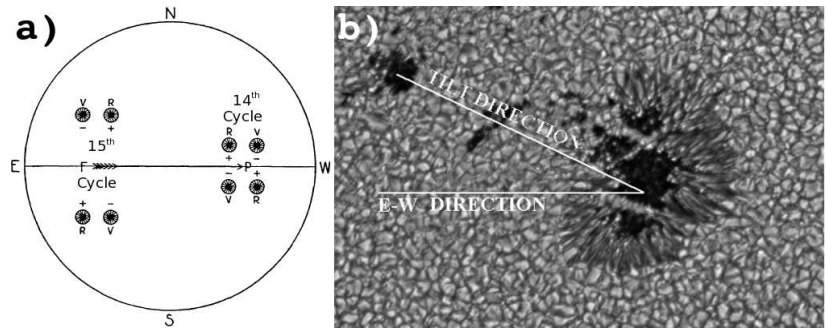}
		\caption{Left: Hales law states that the polarity of the p- and f-spots is opposite in different hemispheres and opposite within one hemisphere and different cycles. Adopted from \citet{1919ApJ....49..153H}. Right: The tilt angle between the axis of a bipolar AR and the equator is described by Joy's law. Adopted from \citet{solardyn}.}
		\label{fig:joyhale}
\end{figure}

\paragraph{Hale's law:}
\noindent\citet{1925ApJ....62..270H} 
not only reported that the polarity of the magnetic field is opposite in p- and f-spots, 
but also found that the magnetic field in binary ARs is of opposite polarity in both hemispheres as well as in subsequent sunspot cycles. This behavior, which is called Hale's law, is shown in the left panel of Fig.~\ref{fig:joyhale}%
. During the 14$^{\rm{th}}$ cycle, the polarity of p-spots was negative in the northern hemisphere, while the respective spot in the southern hemisphere showed positive polarity. In the 15$^{\rm{th}}$ cycle, this pattern was just the opposite.

Hale's law does not only apply to sunspots and AR, but also to the polarity of the average magnetic field in the respective hemispheres. This phenomenon is depicted in the lower panel of Fig.~\ref{fig:spotnumblat}%
. During the maximum of the 22$^{\rm{nd}}$ cycle, ($\approx$1980), the northern hemisphere showed more positive polarity, while the southern hemisphere was dominated by negative polarity. During the maximum of the 23$^{\rm{rd}}$ cycle ($\approx$1991), this configuration was reversed, while in the 24$^{\rm{th}}$ cycle ($\approx$2002), it was the same as in the 22$^{\rm{nd}}$. In conclusion, the period of a solar cycle -- i.e. the time until the magnetic field in one hemisphere shows the same polarity again -- actually amounts to 22 years\footnote{The polarity of magnetic features at the poles does not change simultaneously with the polarity of sunspots in the activity belts, but approximately at the maximum of the sunspot cycle.} \citep{1938QB525.H13......}. 

Bipolar regions that obey Hale's law are referred to as Hale oriented. Anti-Hale orientated AR occur preferentially in the end of a sunspot cycle, when magnetic flux with the configuration from the previous cycle emerges close to the equator, while flux emerging in higher latitudes already belongs to the present cycle.

\paragraph{Joy's law:} Careful analysis of a large number of bipolar sunspots led to the conclusion that, throughout the cycle, f-spots appears at higher latitudes when compared to the position of the p-spots -- cf. right panel in Fig \ref{fig:joyhale} and \citet{1919ApJ....49..153H}. 
This behavior, as well as the fact that the tilt angle between the axis of the bipole and the equator becomes larger for increasing latitude, is called Joy's law. More recent studies indicate that it is rather the distance between the spots within the bipolar ARs which is correlated with the tilt angle \citep{1995ApJ...438..463F}.

\paragraph{The Babcock Model:} A conceptual model explaining the evolution of the magnetic field during the 22-year solar cycle was put forward by \citet{1961ApJ...133..572B}. 

\begin{figure}[h!]
	\centering
		\includegraphics[width={\textwidth}]{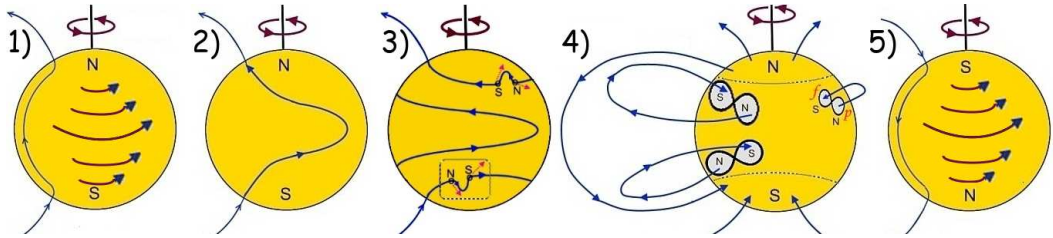}
		\caption{Magnetic field configuration during the solar cycle according to \citet{1961ApJ...133..572B}.}
		\label{fig:babcock}
\end{figure}

1) The magnetic field is an axisymmetric dipole, in which the lines of force lie in meridional planes and loop out from the north pole (positive polarity). They cross the equatorial plane at some distance and re-enter the Sun at the south pole (negative polarity). Only latitudes larger than $\pm55^{\circ}$ show magnetic activity. 

2) As the magnetic field lines are frozen\footnote{A combination of the laws by Ohm, Amp\`ere and Gau{\ss} yields the induction equation, containing a conductive and a diffusive term. Since the conductivity of the solar plasma is orders of magnitude higher than its diffusivity, the latter can be neglected. In other words, the time scale for the magnetic field to diffuse through the solar plasma is so large that the field lines appear to be attached to (or frozen into) the plasma itself.} into the solar plasma, the differential rotation of the Sun shears the poloidal field into a toroidal configuration \citep{Bullard:1954}, thereby amplifying its initial strength thousandfold.

3) This amplification is maximal around $\rm{l}=\pm30^{\circ}$. If flux tubes with Kilogau{\ss} field strength are obtained, they become buoyant and start to rise in the form of an $\Omega$-loop \citep{1955ApJ...121..491P}. 
When they erupt through the surface, they form a bipolar AR with opposite magnetic polarities, reversing their orientation across the equator. The drift of AR towards the equator during the sunspot cycle is a consequence of the differential rotation of the Sun. 

4) The reversal of the poloidal field is due to the systematic inclination of ARs. The p-polarity moves towards the equator, where it neutralizes with the opposite polarity from the other hemisphere, while the f-polarity drifts towards the nearest pole, where it eventually reverses the polarity of the polar field\footnote{\citet{1969ApJ...156....1L} 
interpreted the mean flux transport as the combined effect of the dispersal of magnetic elements by a random walk process and the asymmetry in the flux emergence as stated by Joy's law. He included the flux transport in a quantitative, closed kinematic model for the solar cycle called the Babcock-Leighton model.}.

5) A poloidal field configuration of reversed polarity is obtained after 11 years. Analogues to steps 2), 3) and 4) complete the whole 22-year magnetic cycle.

\begin{wrapfigure}{r}{0.5\textwidth}
	\vspace{-23pt}
  	\begin{center}
  		\includegraphics[width={0.5\textwidth}]{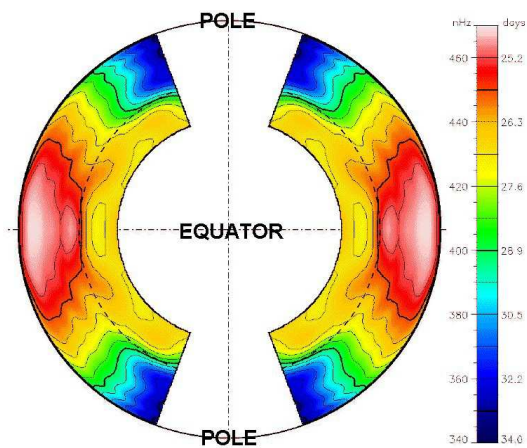}
  	\end{center}
	 \vspace{-15pt}
  		\caption{Differential rotation in the solar convection zone. The equator rotates faster than the poles. Adopted from \citep{diffrot}.}
	\label{fig:diffrot}
  	\vspace{-5pt}
\end{wrapfigure}

\paragraph{Generation of Magnetic Fields:} It is assumed that solar magnetic fields are generated by a dynamo process ope\-rating in the tachocline\footnote{The tachocline is a thin shell at the base of the solar convection zone, where the latitudinal differential rotation interferes with the solid rotation of the solar radiative core \citep{2005AN....326..208G}.}. 
Elaborated dynamo models try to combine the induction equation with the coupled mass, momentum and energy relations for the plasma, to obtain a dynamo equation. However, since the tachocline cannot be measured directly by existing helioseismologic techniques \citep{2009SSRv..144..351V}, 
all mo\-dels must rely both on theoretical considerations and on boundary conditions inferred from observations.

Common to all dynamo models are the so-called $\alpha$- and $\omega$-effects. The $\omega$-effect describes the sheer of an initial poloidal field into a toroidal configuration, as well as its resulting amplification, by the differential rotation of the Sun. The reversed transformation, from the toroidal back into a poloidal configuration, is more difficult to describe. \citet{1955ApJ...122..293P} showed that the plasma within the convection zone is subject to Coriolis forces which induce helicity in such a way that the zonal magnetic field gains a meridional component. 
As a result of this $\alpha$-effect, rising magnetic elements carry a poloidal field component opposite to the present cycle \citep{1970ApJ...162..665P}. 

Even though the details of the dynamo process are still under debate, substantial progress has been made by modifying the dynamo equations to account for the meridional circulation. This flow influences the configuration of the global magnetic fields during a solar cycle \citep{1995A&A...303L..29C,1999ApJ...518..508D}, and calculations with an advective dynamo model have shown that it aids the transformation of the toroidal field back into a poloidal configuration with opposite polarity at the end of the solar cycle \citep{2001ApJ...551..536D,2001ApJ...559..428D}

\section{The Umbra}
\label{sec:theo_umbra}

\begin{wrapfigure}{r}{0.54\textwidth}
	\vspace{-32pt}
  	\begin{center}
  		\includegraphics[width={0.54\textwidth}]{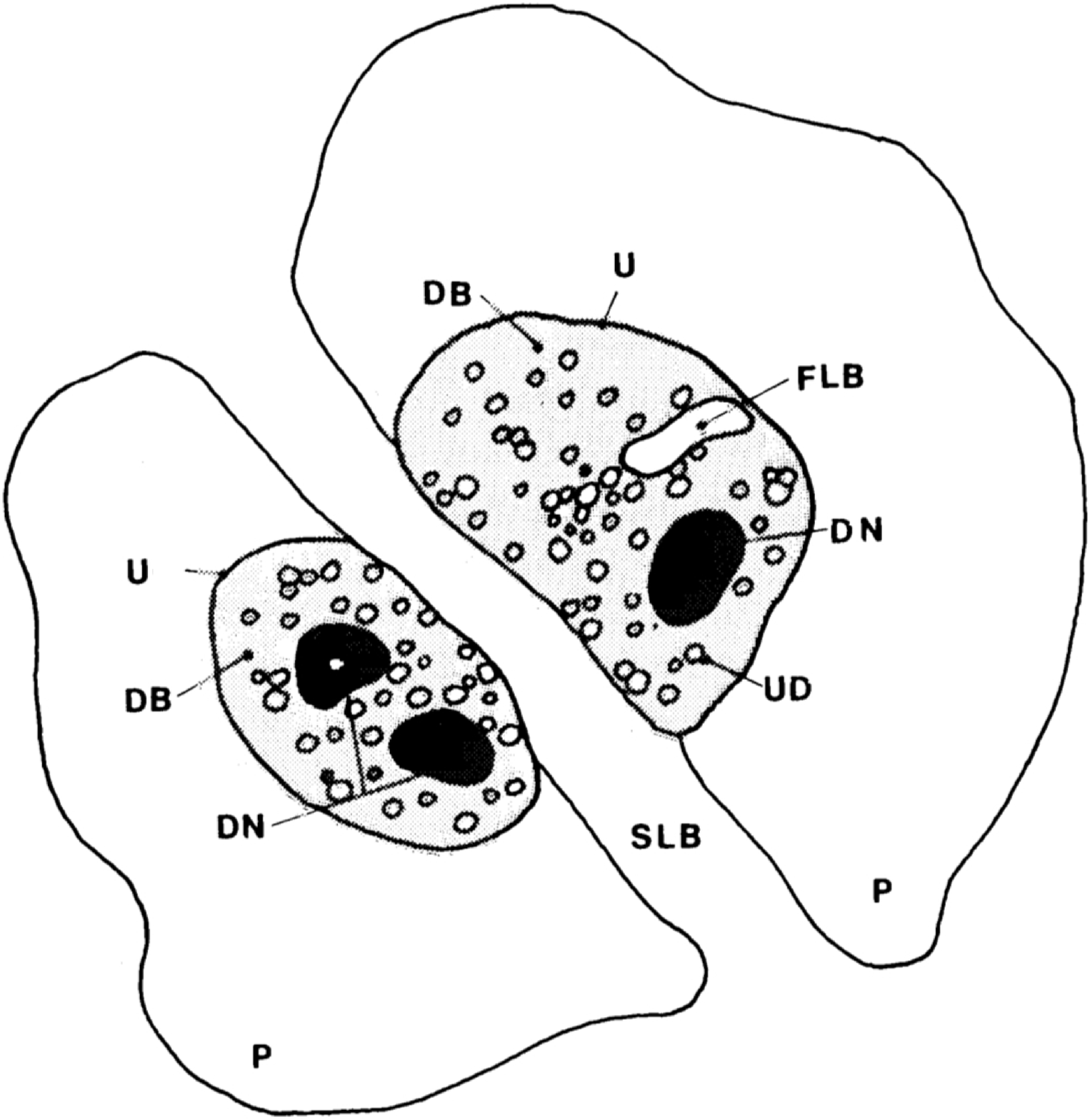}
  	\end{center}
	 \vspace{-15pt}
  		\caption{Schematic drawing of umbral features. P: penumbra, U: umbra, SLB: strong light bridge, FLB: faint light bridge, UD: umbral dot, DB: diffuse background, DN: dark nucleus. Adopted from \citet{1993ApJ...415..832S}.}
	\label{fig:umbra}
  	\vspace{-10pt}
\end{wrapfigure}

In the umbra, convective motions are suppresses by the strong magnetic field, and radiative losses cannot be balanced by energy form the solar interior.  Thus, the umbra is cooler than the surrounding QS and appears dark in continuum observations. 
The umbral brightness is not uniform, but exhibits cellular variations of intensity. The dark nuclei (DN) cover about 10\% to 20\% of the umbral area and have a size and temperature of approximately 1.\arcsec5 and 3500 K. 
They are the darkest part of the umbra and show, depending on wavelength, only 5\% to 30\% of the continuum intensity of the QS \citep{1991BAICz..42..250S}. In contrast to the diffuse background, 
DN show almost no variation in brightness \citep{1991Natur.350...45L}.

\paragraph{Umbral Dots:} Small and bright intrusions, called umbral dots (UDs), cover 3\% to 10\% of the umbral area, but contribute 10\% to 20\% to its brightness \citep{1993ApJ...415..832S}. Usually, a distinction is made between peripheral umbral dots (PUDs), which move from the outer umbra to its center, and central umbral dots (CUDs), which remain fixed and are slightly darker than PUD \citep{1986A&A...156..347G}. There is evidence that UDs are elevated with respect to the umbral background and that they are about 700~K to 1000~K hotter than the DN \citep{Sutterlin:1998p4334,Tritschler:2002p4342}. Reports of the lifetime and size of UDs range from 3 to 80 minutes \citep{1986A&A...160...51K,1992SoPh..137..215E,2008A&A...492..233R} and from 0.\arcsec1 to 0.\arcsec8 respectively \citep{1981A&A....99..111K,1997ApJ...490..458R,Sobotka:2009p4021}.

It is assumed that umbral dots are due to convection in field free gaps below the surface \citep{Parker:1979p4022}. This idea is supported by the simulation of \citet{Schussler:2006p189} in which UDs are caused by an altered mode of magnetoconvection in regions of weak magnetic field below the umbra. The hot plasma in these regions provides an energy reservoir for convective motions, which not only cause the bright UDs at the surface, but also decrease the field strength within them. The predicted Doppler velocity pattern in and around UDs was confirmed in observation \citep{Ortiz:2010p4018}. Additionally, the upflow within UDs shifts the $\tau=1$ level into cooler atmospheric regions. This explains the dark lane across the UD, which is visible in observation with a resolution better than 0.{\arcsec}2 \citep{Bharti:2007p4019,Sobotka:2009p4021}.

\paragraph{Light Bridges:} Long bright structures crossing the dark umbrae of sunspots are called light bridges (LBs). They are classified according to their fine structure and brightness. \citet{1997ASPC..118..155S} distinguishes between: Granular LBs, which harbor cells that are similar but smaller than granulation cells in the QS, and filamentary LBs, which look like intrusions of penumbral filaments. Both types of LBs may appear either as a strong or a faint feature, separating the umbra or being part of it. Strong granular LBs usually evolve into regions of regular granulation splitting the spot, while faint filamentary LBs seem to be associated with PUDs. It is an established fact that LBs harbor convective flows and contain a weaker (reduction of 1.5 kG) and more inclined (zenith angle of 5$^{\circ}$ to 30$^{\circ}$) magnetic field 
\citep{1997ApJ...490..458R,2008ApJ...672..684R,2003ApJ...589L.117B,2006A&A...453.1079J}. With sufficient spatial resolution, a dark lane is visible along the main axis of the LB. It is approximately 0.\arcsec5 wide, produces small barb-like extensions to the sides and is elevated above the umbral background \citep{2003ApJ...589L.117B,2004SoPh..221...65L}.

\paragraph{Wilson Depression:}%
\noindent \citet{1774RSPT...64....1W} first noted that the limb-ward penumbra appears broader when compared to its center-side part, if a sunspot is observed at the edge of the solar disk. This phenomenon is interpreted as a geometrical effect. Due to the depression of the umbra, the sunspot forms a dip resembling the shape of a funnel in the solar surface. 

Today it is accepted that the magnetic field causes this depression, because lateral pressure balance requires that the gas pressure, hence density and opacity, in the umbra is lower when compared to the QS. Furthermore, the cool umbral atmosphere is per se more transparent, since the H$^-$ bound-free opacity -- the major contribution to photospheric opacity -- is very sensitive to temperature. Thus, depending on the size of the spot, the umbral surface ($\tau_{500} = 1$) is located 500 to 800 km below\footnote{In the penumbra, the complex filamentary structure makes it difficult to convert a $\tau$ scale, e.g. from inversion results, into a geometrical height scale. This is because strong jumps of the $\tau_{500} = 1$ surface occur within distances of less than 1\arcsec\,\,\citep{Puschmann:2010p171}.} that of the QS \citep{Loughhead:1964,2004suin.book.....S}. 

\section{The Penumbra}
\label{sec:theo_penum}

The penumbra is a semi-dark structure that surrounds the umbra at least partially. In observations with a resolution better than 1\arcsec, bright and dark fibrils, which are elongated and radially aligned, become visible \citep{Schwarzschild:1959p2298,Loughhead:1964}. Depending on their location within the penumbra, these penumbral filaments (PFs) have a width ranging from 0.\arcsec2 to 0.\arcsec8 and a length between 0.\arcsec5 and 2\arcsec\,\,or longer, e.g.  \citep{Danielson:1961p2347,Denker:1998p4363}. There is evidence that some of them show dark, thread like features, i.e. dark-cores, similar to the features found in certain LBs \citep{Scharmer:2002p3495,BellotRubio:2007p4028}. PFs have a lifetime from 10 minutes to 4 hours \citep{Danielson:1961p2347,Loughhead:1964}, and different values have been reported for their brightness, ranging from 30\% to 70\% of the intensity of the average QS for the dark structures and from 70\% to 100\% for the bright ones \citep{Muller:1973p4408,Denker:1998p4363,Tritschler:2002p4342}. Note, however, that the terms bright and dark have only a local significance, since bright structures in one part of the penumbra may be darker than dark structures in another region.

\paragraph{Penumbral Grains:} Bright features often appear at the head, i.e. the umbral side, of the bright PFs. The bright heads are referred to as penumbral grains (PG), and it has been argued that the bright PFs are actually PGs, each with a less bright comet-shaped tail \citep{Muller:1973p3980,Muller:1973p4408,Denker:2008p4315}. It is useful to distinguish between PGs located in the inner penumbra and PGs located in the outer penumbra, since different lifetimes (3 hours vs. 40 minutes) have been reported. Furthermore, PGs move radially at different speeds and directions (inwards at 0.3~km~s$^{-1}$ to 1~km~s$^{-1}$ vs. outwards at 0.5~km~s$^{-1}$ to 0.7~km~s$^{-1}$) \citep{Muller:1973p3980,Shine:1987p4495,Sobotka:1995p4510,Sobotka:1999p4524,Sobotka:2001p4536}.

\paragraph{The Evershed Effect:} In spectroscopic observations of sunspots, \citet{Evershed:1909p167} found that photospheric lines are blueshifted in the center side and redshifted in the limb side penumbra. He interpreted this shift as a radial and outward directed flow of plasma, the Evershed flow (EF).

However, the Evershed effect is not only a simple displacement of the spectral line, but includes a broadening, an asymmetry, and in extreme cases a doubling of the line \citep{Bumba:1960p212,Holmes:1961p2361}. The interpretation of these asymmetries is not distinct: On the one hand, they are seen as evidence for two lateral displaced velocity fields in the dark and bright PFs \citet{Schroter:1965p2368}, while on the other hand, there is evidence that they are due to changing Doppler velocities with height \citep{Maltby:1964p233}\footnote{\citet{StJohn:1913p179} was the first to report on a variation of v$_{\rm{dop}}$ with $\tau$, including a reversal of the flow direction in the chromosphere.}. The first concept is in accordance with the idea of plasma motion occurring along individual PFs \citep{Loughhead:1964}, while the second scenario is able to explain the center-to-limb variation of the maximum amplitude of the EF as reported by \citet{Michard:1951p2367}.

The EF seems to stop abruptly at the white light boundary of the spot \citep{Wiehr:1992p3002,Title:1993p216,1999A&A...349L..37S}. Contradicting observations, e.g. \citep{Rimmele:1995p2411}, could be explained by a partial continuation of that flow along the canopy into the chromosphere \citep{Solanki:1994p4039,Rezaei:2006p157}. The EF is not steady, but velocity packages, called Evershed clouds (EC), propagate radially outward with a repetitive but irregular behavior on a timescale of 10 to 15 minutes \citep{1994ApJ...430..413S,1994A&A...290..972R}. ECs evolve and remain coherent until they go through the outer penumbral border, where they seem to vanish (type I) or continue into the moat region (type II) \citep{2007A&A...475.1067C}. It has been suggested that EC are a precursor of moving magnetic features (MMFs) \citep{2006ApJ...649L..41C}.

The EF is structured on small scales, and reports point to a anticorrelation between Doppler velocity and continuum intensity \citep{Stellmacher:1971p240,Title:1993p216,1994ApJ...430..413S,Hofmann:1994p2568,WestendorpPlaza:2001p1784,Langhans:2005p9}. This correlation, however, is difficult to interpret 
as only observations with high spatial resolution \citep{Wiehr:1992p3002}, or those of lines forming at a comparable height can be used for such a study \citep{Wiehr:1994p412,Rimmele:1995p2411}. Newer observations reveal a more complicated picture, in which the correlation reverses its polarity in the penumbra \citep{Ichimoto:2007p178} or exists only on a local scale \citep{Schlichenmaier:2005p219}.

\paragraph{The Moat Flow:} Using feature tracking techniques, a radial outward directed flow can be measured in the periphery of sunspots. This so-called moat flow (MF) develops after the formation of the spot, and its velocity ranges from 0.5~km~s$^{-1}$ to 1~km~s$^{-1}$. The MF extends $1\cdot10^7$ m to $2\cdot10^7$ m into the QS for small spots and roughly twice the spot radius for large spots \citep{1973SoPh...28...61H,1988SoPh..115...43B,1997ApJ...490..458R,2008ApJ...679..900V}. Contrary to the EF, which is a surface phenomenon, helioseismic techniques provide evidence that the MF continues with speeds of 1 km s$^{-1}$ for $3\cdot10^7$ m and seems to be present in depths of 2000 km \citep{2000JApA...21..339G}.

\paragraph{Magnetic Field Configuration:}

Magnetic fields in sunspots were first measured by \citet{1908ApJ....28..315H}. On scales larger than 2\arcsec, their distribution is relatively smooth and can be approximated by a flux tube. 
However, the strong intensity jump 
between umbra and penumbra is not evident in the magnetic field strength. 
Depending on the size of the spot, the field is strongest (2~kG to 3.7 kG) in the center of the umbra, drops monotonously to values from 1.4~kG to 2.2~kG at the umbral-penumbral border, reaches 0.7 to 1 kG at the penumbra-QS boundary and drops rapidly in strength beyond that line
, e.g. \citet{1953ZA.....31..273M,Beckers:1969p188,1982SoPh...80..251B,Adam:1990p3428,BellotRubio:2003p206}.

\begin{figure}[h!]
	\centering
		\includegraphics[width={\textwidth}]{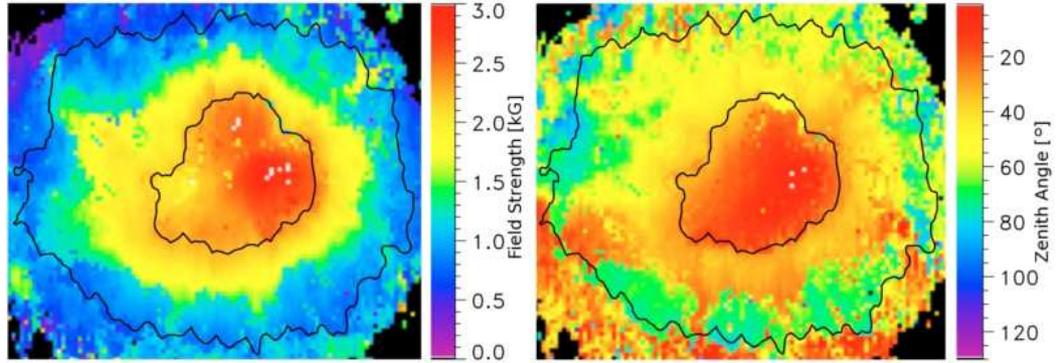}
		\caption{Average magnetic field strength (left) and inclination (right) within a sunspot. Adopted from \citet{BellotRubio:2004p4357}.}
		\label{fig:spotmagconf}
\end{figure}

The strongest field is usually associated with the DN in the umbra, where it is almost vertical. The azimuthally averaged inclination increases with radial distance, reaching 70$^{\circ}$ to 80$^{\circ}$ at the outer penumbra \citep{Kawakami:1983p3487,Adam:1990p3428,Lites:1990p2995,Keppens:1996p308,WestendorpPlaza:2001p1784}. These findings are at odds with the results from all the Doppler measurements that show that the EF is parallel to the solar surface and imply a larger inclination of the magnetic field.

In their pioneering work, \citet{Beckers:1969p188} found a larger zenith angle of the magnetic field in dark PFs than in the bright ones. But it was accepted only later that the penumbral magnetic field shows large azimuthal variations of the inclination on scales of less than 2\arcsec. This scenario explains the contradictory results, 
if it is assumed that the EF occurs along the more horizontal field lines.

High resolution observations \citep{Degenhardt:1991p2736,Schmidt:1992p2732,Lites:1993p2386,WestendorpPlaza:1997p238,Wiehr:2000p2574} and more recent theoretical studies; e.g. \citet{MartinezPillet:2000p223}, confirm the existence of this so-called "spine-intra-spine" structure \citep{BellotRubio:2003p177} of the magnetic field, which has been incorporated into more advanced penumbral models, e.g the "fluted" penumbra \citep{Title:1993p216} or the "uncombed" penumbra \citep{Solanki:1993p210}.

\paragraph{Subsurface Structure:} Helioseismic techniques may be used to infer the subsurface structure of sunspots, since their thermal and magnetic inhomogeneities change the phase and amplitude of solar oscillations \citep{1999ApJ...510..494L}. Some results indicate the presence of converging collar flows at 4000 km below the spot, which turns into a downflow and then, at greater depths, into an outflow \citep{2002AN....323..186K}. However, results are ambiguous and do not yet allow to draw definite conclusions on the subsurface structure of the magnetic field in sunspots, e.g. differentiate between the spaghetti or cluster model, \citep{Moradi:2010p3329}.

\paragraph{Canopy:} Observation of different absorption lines can be used to infer the decrease of magnetic field strength with height. Within the visible part of the spot, average rates of 0.3~G~km$^{-1}$ to 0.6~G~km$^{-1}$ have been found \citep{Loughhead:1964,2001ApJ...547.1130W}. As a result of the expansion with height, the field continues beyond the white light boundary in the form of an almost horizontal canopy in the middle and upper photospheres \citep{1980SoPh...68...49G} 
and in a more vertical shape in the chromosphere, where it forms the so-called superpenumbra \citep{1968SoPh....5..489L}.

\paragraph{Moving Magnetic Features:} 
MMFs may be distinguished according to their magnetic and velocity properties: Some MMFs tend to appear in pairs of opposite polarity (type I). Other MMFs are unipolar, have the same (type II) or opposite (type III) polarity compared to the polarity of the spot and move with speeds similar to the MF, or faster with up to 2~km~s$^{-1}$ \citep{1973SoPh...28...61H,1998ApJ...492..402R,2007ApJ...659..812K}. Some MMFs are related to the canopy and follow the orientation of the superpenumbral fibrils \citep{2003A&A...399..755Z}, while others are related to the decay of sunspots. 

\section{Formation and Decay of Sunspots}
\label{sec:theo_life}

\begin{wrapfigure}{r}{0.32\textwidth}
  	\vspace{-23pt}
  	\begin{center}
  		\includegraphics[width={0.32\textwidth}]{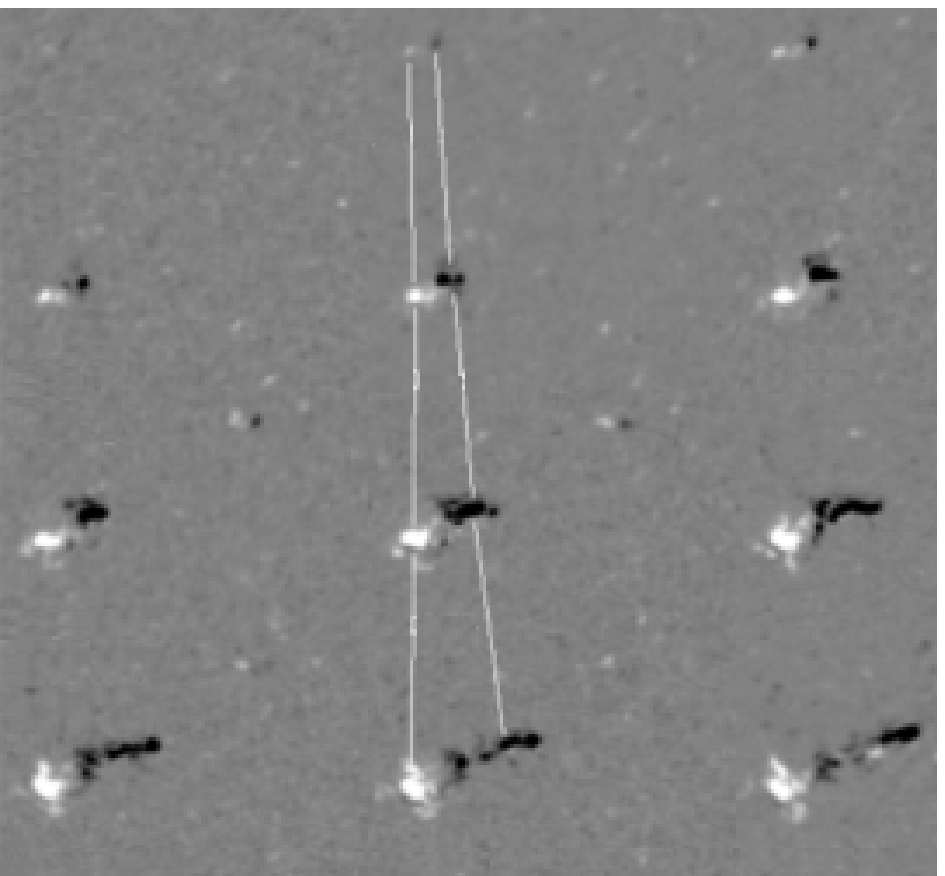}
  	\end{center}
		\vspace{-15pt}
  		\caption{Asymmetric evolution of an AR. Adopted from \citet{uso:summerschool2009}.}
	\label{fig:fluxemergence2}
	\vspace{-20pt}
\end{wrapfigure}

Today, it is accepted that sunspots and other photopheric magnetic features are caused by magnetic flux tubes. There is evidence that toroidal flux tubes, produced by the $\omega$-effect, reside at the bottom of the tachocline, where the degree of subadiabaticity is just right to store them for a substantial fraction of a solar cycle \citep{1993GApFD..72..209F}. 
\citet{1955ApJ...121..491P} showed that they become buoyant and rise in form of an $\Omega$-loop, if they contain magnetic fields in the Kilogau{\ss} regime.

\paragraph{Observation:} %
\noindent \citet{1987ARA&A..25...83Z} distinguishes between: a) Large ARs, which appear within the activity belts, live for several weeks and contain sunspots, pores, plague and faculae. b) Small ARs, which can be observed for a couple of days, do not contain spots, but pores and smaller magnetic features. c) Ephemeral ARs, that do exist only for hours and may emerge at high latitudes. Furthermore, ARs show asymmetries in size and lifetime -- i.e. within ARs, p-spots are larger and live longer when compared to f-spots --  as well as in the divergent motions during their emergence -- i.e. p-spots moves faster westward than f-spots move eastward (cf. Fig.~\ref{fig:fluxemergence2}).

Observations indicate that the flux in ARs builds up as the result of many small magnetic elements of opposite polarity appearing in the photosphere. 
They move apart with velocities of up to 2 km s$^{-1}$, while new flux continues to emerge near the polarity inversion line. The orientation of the emerging field is aligned along the axis connecting the two polarities. The accumulation of flux in both polarities leads to the appearance of pores and to the formation of sunspots if pores merge \citep{1985SoPh..100..397Z,1987ARA&A..25...83Z}. Pores are associated with redshifts, and it is not clear whether this is due to material draining from the emerging loop or due to convective collapse \citep{Parker:1978p3679,Spruit:1979p3687}.

\begin{wrapfigure}{r}{0.40\textwidth}
  	\vspace{-18pt}
  	\begin{center}
  		\includegraphics[width={0.40\textwidth}]{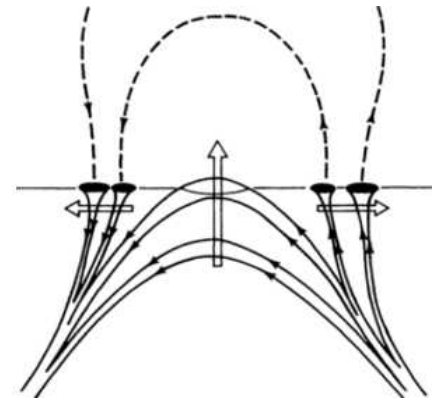}
  	\end{center}
	  	\vspace{-15pt}
  		\caption{Cartoon of flux emergence. Adopted from \citet{1985SoPh..100..397Z}.}
	\label{fig:fluxemergence}
		  	\vspace{-5pt}
\end{wrapfigure}

\paragraph{Models and Simulations of Flux Emergence:} In the heuristic model of \citet{1978SoPh...60..213Z,1985SoPh..100..397Z}, flux rises through the convection zone, but fragments below the surface. A collection of $\Omega$-loops, which are connected to the same roots, pe\-ne\-trate the surface (cf. Fig~\ref{fig:fluxemergence}). After the topmost loops have emerged, their photospheric footpoints separate, causing an increasingly vertical field. The coalescence of the vertical flux in each polarity leads to the formation of pores and sunspots. In other models \citep{1978ApJ...222..357P}, it is not the coalescence of the vertical flux, but the hydrodynamic attraction of the individual and rising fragments of the flux tube that lead to the formation of pores and sunspots \citep{1978ApJ...222..357P}.

The exact process of flux emergence is still not well understood, and results from simulations are contradictory.
\citet{1995ApJ...441..886C} 
showed that the conservation of angular momentum during the ascent of the flux tube results in a larger inclination of the magnetic field in the p- than in the f-footpoint. The emergence of such a deformed $\Omega$-loop yields divergent motions of p- and f-spots. The conservation of angular momentum induces a retrograde (eastward) plasma flow in the flux tube, increasing the magnetic pressure and concomitantly the lifetime of the p-spot \citep{1993ApJ...405..390F}.

By contrast, the simulation of \citet{2008ApJ...676..680F} shows that Coriolis forces cause asymmetric stretching, which in turn yields higher field strengths in the p-leg of the rising flux tube. This results in a more buoyant p-leg with less inclined fields as well as in larger and more stable p-spots. The asymmetry in sunspot proper motions is explained by the faster ascent of the p-leg \citep{2009SSRv..144..351V}.

Note that, according to model calculations, only a considerable twist of the flux tube conserves its integrity while it rises through the convection zone \citep{1996ApJ...464..999L,1998ApJ...492..804E}. Observational evidence of twisted flux tubes has been found by, e.g. \citet{1996ApJ...462..547L,2003ApJ...594.1033N}. A highly twisted flux tube could be the reason for "knotted" $\delta$-sunspots \citep{1991SoPh..136..133T} or ARs alternating between Hale and non-Hale orientation \citep{2000ApJ...544..540L,2009SSRv..144..351V}.

\paragraph{Formation of Sunspots:} 
Above a critical size, pores start to develop penumbral structures. In proto-spots, partial penumbrae do not completely surround the umbra, but appear within hours \citep{Loughhead:1964}. The penumbra grows sector after sector, starting at the side that points away from the opposite polarity of the AR \citep{2010A&A...512L...1S}. It is interesting that a newly developed penumbral sector already harbors the EF and is indistinguishable from a more mature filament \citep{1998ApJ...507..454L}. Sunspot sizes range from 5\arcsec\,\,to 80\arcsec\,\,or even more \citep{Loughhead:1964}, and their lifetime is linearly related to their maximum size \citep{1997SoPh..176..249P}.

\begin{figure}[h!]
	\centering
		\includegraphics[width={0.98\textwidth}]{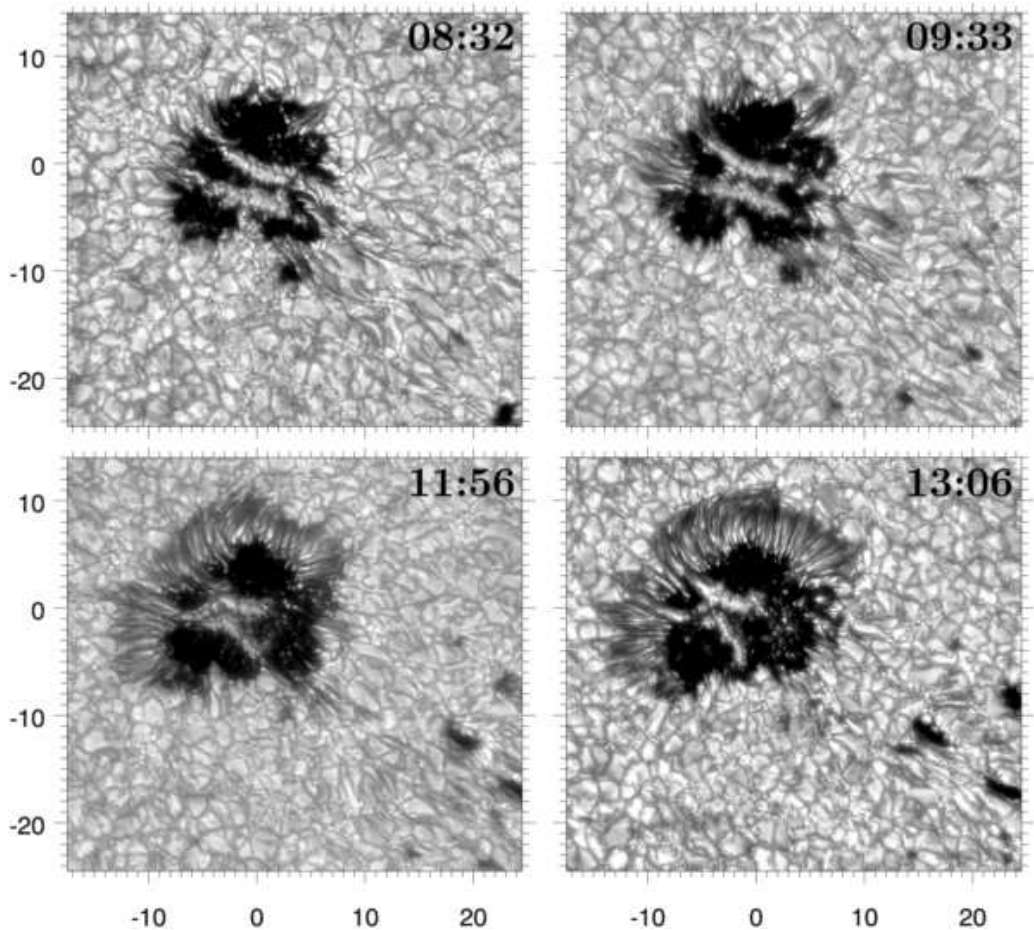}
		\caption{Formation of a proto-sunspot including a penumbra from pores observed with the VTT. Adopted from \citet{2010A&A...512L...1S}.}
		\label{fig:spotform}
\end{figure}

\paragraph{Spread of Flux:} Magnetic fields of sufficiently low strength are moved around by the turbulent motion of granulation. The process of magnetic flux expulsion, for example, concentrates them in the intergranular lanes \citep{1963ApJ...138..552P,1964RSPTA.256...99W} as well as on the borders of supergranular cells \citep{1964ApJ...140.1120S}. The meridional flow sweeps small-scale fields of predominately opposite polarity to the nearest pole, where they cancel with the existing polarity and finally create a poloidal field of opposite polarity during the next cycle (cf. Section \ref{sec:cycle}).

\paragraph{Flux Cancellation:} The most common way of magnetic flux removal, both in the QS and in ARs, is flux cancellation, e.g. \citet{1985AuJPh..38..929M,1993SSRv...63....1S}. During this process, magnetic features of opposite polarity approach each other, merge and disappear. Different interpretations of such events involve the submergence of magnetic loops \citep{1984ApJ...287..404R} and reconnection above or below the photosphere, e.g. \citet{1987ARA&A..25...83Z}. Other processes leading to flux removal involve the fragmentation of the flux tube, e.g. by Rayleigh-Taylor \citep{1979A&A....71...79S} or by fluting \citep{1975SoPh...40..291P} instabilities and subsequent diffusion.

\paragraph{Sunspot Decay:}  It is highly probable that the processes leading to the decay of sunspots also operate during the formation phase, but they become apparent only afterwards \citep{1981phss.conf....7M}. Since it is intrinsically easier to observe a sunspot during its decay phase, there are many more reports  on this process, and extensive studies have been performed especially on the decay of sunspot area. It has been reported that sunspots with an irregular shape \citep{1982SoPh...81...25R}, extensive bright umbral structures \citep{1968ARA&A...6..135Z}, large proper motion \citep{1992SoPh..137...51H} as well as sunspots occurring in higher latitudes \citep{1995SoPh..157..389L} suffer from a higher decay rate. For the time dependency of the decay of sunspot area, linear \citep{1963BAICz..14...91B}, exponential \citep{1997SoPh..176..249P} or lognormal \citep{1993A&A...274..521M} relationships have been proposed, but a definite conclusion has not been reached \citep{1993A&A...274..521M, 2008SoPh..250..269H}. The form of the decay curves allows to differentiate between different theoretical models. A linear decay law, for example, can be explained by Ohmic diffusion across the current sheet between the sunspot and the QS \citep{1972SoPh...26...52G}. Other models assume a turbulent diffusion front that erodes the flux tube forming the sunspot and favor a parabolic or quadratic decay law \citep{1997ApJ...485..398P}. This idea is supported by recent observations of the MF and (especially unipolar) MMFs outside the white light boundary of the spot, which are believed to remove flux from the sunspot \citep{1973SoPh...28...61H,2007ApJ...659..812K,2008ApJ...681.1677K,2008ApJ...686.1447K}. 

\section{Penumbral Models}
\label{sec:evershed}

Penumbral models have to explain a range of observational features such as the EF, PFs and the small-scale configuration of the magnetic field. The concept of \citet{Danielson:1961p2347,Danielson:1961p2350} was an early attempt. He argued in favor of horizontal magnetic field lines in the penumbra, which allows the EF to occur parallel\footnote{A flow that occurs perpendicular to the magnetic field lines will be suppressed by Lorentz forces acting on the highly conductive solar plasma.} to the lines of force. In his model, the presence of the magnetic field alters the convective motions, resulting in elongated cells which form the PFs. Bright PFs were identified as hot and rising tubes of force, while the dark PFs correspond to sinking and cold tubes of force. This scenario explained the observation of  \citet{Beckers:1969p188} but did not give a reason for the EF. \citet{Galloway:1975p2886} refined this model and assumed that the convective roll motion leads to an increase of magnetic field strength in the dark PFs. Together with the overall pressure balance, the excess of magnetic pressure drives an outward flow within the dark PFs and an inward flow within the bright PFs. This concept was already used by \citet{Schroter:1965p168}, however with reversely directed flows, to explain penumbral line asymmetries.

Even though 
the quality of penumbral models has increased tremendously in recent years, a definite conclusion has neither been reached on the question of the underlying structure of the penumbra nor on the driving mechanism of the EF. In the following, the state of the art of penumbral models is discussed including their limitations and shortcomings.

\subsection{Siphon Flows and Turbulent Pumping}
\label{sec:siphon}

\begin{figure}[h!]
	\centering
		\includegraphics[width={\textwidth}]{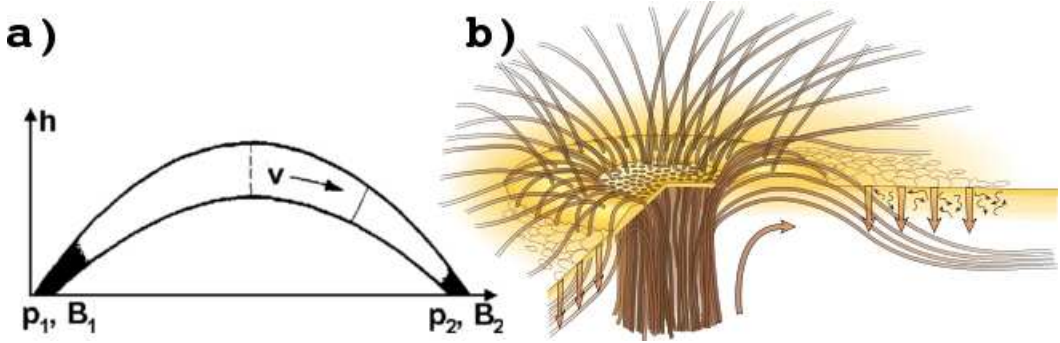}
		\caption{Left: Plasma flow along a flux tube driven by the pressure imbalance at the footpoints. Adopted from \citet{Meyer:1968p2870}. Right: Sketch of the penumbral magnetic field that is kept down by turbulent pumping outside the spot. Adopted from \citet{Thomas:2002p3290}.}
		\label{fig:siphonflow}
\end{figure}

The first proposal for a driving mechanism of the EF was the siphon flow introduced by \citet{Meyer:1968p2879,Meyer:1968p2870} and then updated by \citet{1981phss.conf..359S}. These authors conducted magnetohydrodynamical studies on a flux tube 
that forms an $\Omega$-loop 
(cf. left side of Figure~\ref{fig:siphonflow}). If B$_1$($\rm{h}=0$)~$<$~B$_2$($\rm{h}=0$), that is if the magnetic field strength B$_2$ 
of the second footpoint at an arbitrary geometrical height h exceeds the strength of the first footpoint, then the magnetic pressure $\rm{p(z)_{\rm{mag}}}=\frac{\rm{B}^2}{2\mu_0}$ at the second footpoint is also higher when compared to the first one. As a result of the lateral pressure balance, i.e. $\rm{p(z)_{\rm{tot}}}=\rm{p(z)_{\rm{gas}}}+\rm{p(z)_{\rm{mag}}}=\rm{const}$, in the solar atmosphere, $\rm{p(z)_{\rm{gas}}}$ at the first footpoint will be higher when compared to $\rm{p(z)_{\rm{gas}}}$ at the second one. Thus, a flow along the tube will be maintained by this imparity.

In a next step, one footpoint was positioned inside the penumbra, while the other was located in another spot or a field concentration outside the penumbra. This construction was justified by theoretical studies that showed that the field strength in small-scale magnetic features may surpass penumbral values under certain physical conditions -- i.e. convective collapse \citep{Parker:1978p3679,Spruit:1979p3687}. Furthermore, \citet{Meyer:1968p2870} proposed that the inverse EF is a consequence of a second family of flux tubes. They contain reverse flows, because they have one footpoint in the inner penumbra resulting in B$_1 < \rm{B}_2$. Since these flux tubes reach higher atmospheric layers, the observed velocities are much lower.

Siphon flows along isolated flux tubes have been studied at different levels of complexity to include realistic values of plasma-$\beta$ \citep{Thomas:1988p3726,Thomas:1990p3729} -- which is around unity in the penumbra -- as well as radiative losses of the moving plasma inside the tube \citep{Degenhardt:1991p2736,Montesinos:1993p3728}.

\paragraph{Drawbacks of Siphon Flows:} An important issue in these models is the length of the loop, which is shorter than the width of the penumbra \citep{Thomas:1990p3729}, especially if the apex of the former is located in the photosphere. Even though the length of the loop can be increased if the field strength inside the flux tube is lowered \citep{Degenhardt:1991p2736} or if the loop is embedded in an ambient field \citep{Thomas:1993p3733}, the length of the loop remains smaller than the width of a typical penumbra. To overcome this problem, it has been suggested \citep{delToroIniesta:2001p3738} that the EF takes place in small loops that exist at different radii.

Another challenge of siphon flow models arises from observations of downflows within or at the outer edge of the penumbra, where the plasma stream dips back into the solar surface \citep{WestendorpPlaza:1997p238,Schlichenmaier:2000p225,BellotRubio:2003p206,Franz:2009p515}. This implies that the outer footpoint has a lower field strength than the inner one. Thus, contrary to observation, the plasma flow would be directed towards the umbra. 

These difficulties were tackled by \citet{Montesinos:1997p3731} using observations of return flow in isolated magnetic elements just outside the spot boundary \citep{Boerner:1992p3740} as an argument for a mechanism that would enhance the magnetic field strength inside the penumbra. Other proposals \citep{2002AN....323..303S,2003A&A...399..755Z,SainzDalda:2008p1689} invoke a sea serpent like structure of the flux tube that returns and reappears within the penumbra, but ends outside the spot in an intense magnetic element. As a result, the siphon flow would be driven by the pressure difference between the end points in the inner penumbra and the magnetic element outside the spot.

\paragraph{Flux Pumping:} The siphon flow models mentioned so far postulate an uncombed penumbral magnetic field and assume flux tubes returning to the solar surface without giving any physical reason for that. \citet{Thomas:2002p3290,Thomas:2002p3903} proposed a scenario -- see also \citet{Weiss:2004p3904} -- in which the accumulation process of flux leads to an increase of inclination of the magnetic field lines along the outer boundary of a protospot. Above a critical angle, convectively driven fluting instability\footnote{\citet{Thomas:2002p3903} argue that this instability is different from the fluting instability of a flux tube in an adiabatic environment as described by \citet{1977MNRAS.179..741M}.} sets in, which bends some of the outermost field lines down to the solar surface, resulting in the interlocked geometry of the penumbral magnetic field (cf. right panel in Fig.~\ref{fig:siphonflow}). These horizontal fields cannot only be advected, but also be kept below the photosphere by turbulent pumping, which overcompensates buoyancy and magnetic curvature forces of the flux tubes.

This scenario is attractive in so far as it explains the hysteresis\footnote{Some sunspots contain less magnetic flux than large pores \citep{2010mcia.conf..229T}.} of sunspots since the penumbra, once developed, will be maintained even if the sunspot decays. However, turbulent pumping has been investigated only in idealized three-dimensional numerical simulation of granulation \citep{Brummell:2008p3943} until today, and has not yet been confirmed by observations of penumbral formation. Furthermore, it is not clear which type of solution is realized in the penumbra. Siphon models are steady state solutions, while the penumbral structure is highly dynamic. It needs to be investigated how these models evolve over time and whether their solutions remain stable.

\subsection{Buoyant Flux Tubes}
\label{sec:buoyant}

\begin{figure}[h!]
  \begin{center}
\includegraphics[width=0.95\textwidth]{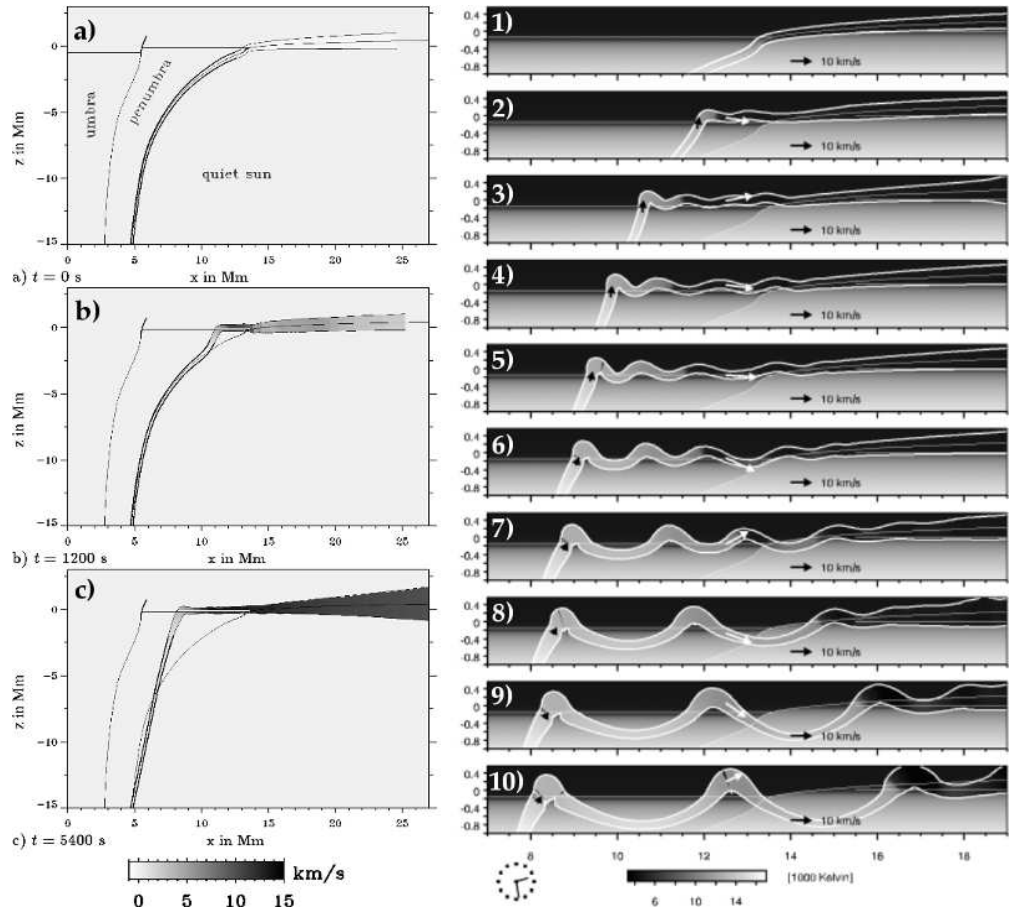}
  \end{center}
  \caption{Left Column: Rise of magnetic flux bundle through the photosphere. Adopted from \citet{Schlichenmaier:1998p2295}. Right Column: Development of flux tube with sea serpent structure through "convective overshoot". Adopted from \citet{2002AN....323..303S}.}
  \label{fig:rolftubes}
\end{figure}

An alternative idea to the siphon flow mechanism 
considers the penumbra, including the Evershed effect, as a result of the convective interchange of flux tubes \citep{1981phss.conf..359S,Schmidt:1991p3974,1992sto..work..139J,Jahn:1994p3975}\footnote{\citet{1981phss.conf..359S} assumed the penumbra to be shallow, but it was recognized by \citet{1987rfsm.conf..219S} and approved by \citet{Solanki:1993p3978} that the penumbra is not a surface phenomenon.}. 
This scenario has been investigated by means of numerical simulations \citep{Schlichenmaier:1998p2295,Schlichenmaier:1998p2296} and is illustrated in the left column of Fig.~\ref{fig:rolftubes}.

\paragraph{Simulation:} The initial condition of the simulation is depicted in panel a) of Fig.~\ref{fig:rolftubes}. A magnetic flux bundle resides at the magnetopause\footnote{The current sheet between the QS and the penumbra.} and is in contact with the field free plasma below. The cooler plasma inside the flux tube is heated, expands, gains buoyancy and, due to the superadiabatically stratification of the penumbral plasma, rises towards the surface. Once it reaches the photosphere, the convectively stable layers of the penumbral atmosphere reduce the buoyancy and the tube comes to rest in a horizontal position -- cf. panel b) in Fig.~\ref{fig:rolftubes}. Since the background magnetic field is more vertical, an uncombed geometry is obtained. 

The outermost part of the flux tube reaches the surface first, but with time, parts of the tube closer to the umbra penetrate the surface and allow hot plasma to rise. Due to the stratification of the plasma below the penumbra, the tube has to expand while it rises. In combination with the total pressure balance between tube and environment, the magnetic pressure inside the tube is lowered, while the gas pressure has to increase. As the outer end of the tube does not rise, the respective gas pressure is always lower when compared to the pressure at the inner point. This imbalance results in an outward directed plasma flow inside the tube. If the tube is optically thick, this flow can be observed as the EF -- cf. panel c) in Fig.~\ref{fig:rolftubes}. 

Contrary to the ideas of \citet{Jahn:1994p3975}, these simulations show that the tube does not submerge anymore, but stays in the photosphere as long as the outflow continues. This is because an equilibrium is reached between two reversely directed forces operating at the inner end of the tube: The magnetic forces pulls the tube into deeper layers, while the centrifugal force, caused by the momentum of the plasma, acts in the opposite direction.

During the ascent of the tube, the magnetic pressure inside changes, which also alters the Alfv\'en speed. In other words: The closer it is to the umbra, the deeper below the penumbra the plasma rises from, and thus, the lower the Alfv\'en speed inside the tube at that position is. 
If the flow speed of the plasma surpasses the Alfv\'en speed, the magnetic tension is no longer sufficient to bend the flow into the horizontal, and the plasma convectively overshoots into the atmosphere. The plasma is decelerated by buoyancy in this convectively stable region, returns and causes the tube to form a standing wave -- cf. panels 2) $-$ 5) in the right column of Fig.~\ref{fig:rolftubes}. During the evolution of the tube, the Alfv\'en speed inside drops further, thereby increasing the amplitude of the wave. Eventually, the minimum of the first wave will enter the superadiabatically stratification below the penumbra and is dragged further down. While the trough of the wave sinks, the inner apex migrates towards the umbra and the outer apex migrates towards the QS -- cf. panels 6) $-$ 10) in the right column of Fig.~\ref{fig:rolftubes}.

\paragraph{Observational Evidences:} Within the context of this model, PGs are interpreted as hot plasma rising inside the flux tube. With time, parts of the tube closer to the umbra reach the surface, and cause the PGs to move radially inwards. 
The outward migration of PGs in the outer penumbra is due to the sea serpent structure of the flux tube \citep{2002AN....323..303S,2003ASPC..286..211S}. In the inner penumbra, the hot plasma rising from below bends towards the horizontal, is accelerated radially outwards and cools on its way towards the outer penumbra, causing the EF. 

Radiative transfer calculations, which treat the radiative losses of the hot plasma in an isothermal atmosphere, are able to model the intensity pattern of bright PFs \citep{Schlichenmaier:1999p3981}. \citet{RuizCobo:2008p3982} explain the dark-cores of bright PFs as an opacity effect of the magnetic flux tube embedded in a stronger ambient field together with a stratified atmosphere. 

If the bright PF is identified with the hot part of the horizontal flux tube, it is not expected to show strong Doppler shifts because the temperature dependency of the H$^-$ opacity moves the $\tau=1$ level to higher and cooler atmospheric regions outside the flow channel. However, if the gas cools on its way to the outer penumbra, the $\tau=1$ level eventually drops below the top of the flux tube, and the outflowing gas causes a Doppler shift of an absorption line. This could explain the anticorrelation between Doppler velocity and continuum intensity. Furthermore, the flow channel is elevated with respect to the penumbral background, which resembles the geometry of bright PFs inferred from observation \citep{Schmidt:2004p3724}. 

It has be shown qualitatively that the flux tube model reproduces the Stokes V asymmetries of Fe I 630.15 nm and Fe I 1564.8 nm \citep{Schlichenmaier:2002p220}. Finally, the azimuthal variation of total net circular polarization can be reproduced if the effects of anomalous dispersion are taken into account \citep{Schlichenmaier:2002p722,Muller:2002p3983}.

\paragraph{Shortcomings:} Despite the ability of the simulation to explain this broad range of observation, the existence of a thin flux tube is an ad hoc assumption. Only a single flux tube is simulated, while the influence of neighboring flux tubes is not accounted for. Furthermore, the model is 1-dimensional and the properties of the background atmosphere remain unaffected by the presence of the flux tube. In a more realistic scenario the curvature forces of the background field, which wraps around the flux tube, would have to be considered.

\subsection{Convection in Penumbral Gaps}
\label{sec:gaps}

The gappy penumbral model \citep{Spruit:2006p2} completely avoids the concept of flux tubes and postulates that certain regions of the penumbra, i.e. the gaps, are dominated by the kinetic energy of the plasma \citep{Spruit:2010p239}. In these regions, which are identified with the bright PFs, convection does not only transfer energy from below to the surface, but also pushes aside the penumbral magnetic field. Similar to the scenario of UDs (cf. Section~\ref{sec:theo_umbra}), the upflows inside the gap will move the $\tau=1$ surface into cooler atmospheric regions, causing the dark lane in the center of the bright PF. Support for this model comes from: a) Observations \citep{Muller:1973p3980,2007ASPC..369...71S} which show a transition between UDs or LBs on the one side and PFs on the other side and b) simulations of LBs \citep{2006ASPC..354..353N} which show that the dark lanes running along the axis of LBs are caused by the same opacity effect as in UDs.

\begin{figure}[h!]
  \begin{center}
\includegraphics[width=\textwidth]{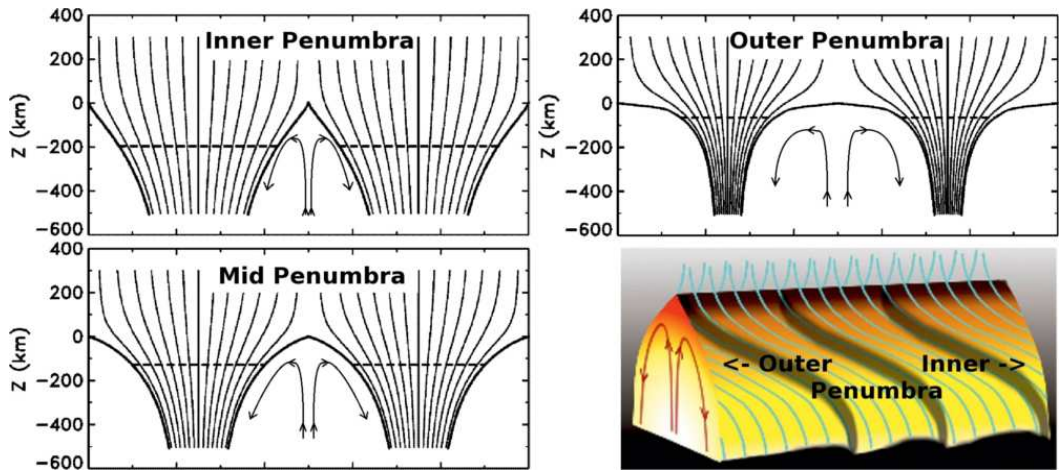}
  \end{center}
  \caption{Left and top right: Penumbral magnetic field configuration in the azimuth vs. height plane for the gappy model in the inner, mid and outer penumbra. Adopted from \citet{Spruit:2006p2}. Bottom right: Perspective view of a penumbral gap with fluted $\tau=1$ surface. Adopted from \citet{Spruit:2010p239}}
  \label{fig:gaps}
\end{figure}

Fig.~\ref{fig:gaps} shows the configuration of the magnetic field in the z $-$ $\phi$ (geometrical height $-$ azimuth) plane at different penumbral radii. The magnetic field wraps around the gap, causing large gradients with height. 
In the inner penumbra where the gaps are close to each other, they show a cusp like structure at the top, following the z $-$ $\phi$ configuration of the magnetic field. This is because the gas pressure inside the gap is balanced by the magnetic pressure of the field around. 

Towards the outer penumbra, the field free regions become larger, and the cusp vanishes. This is a result of the geometry of the magnetic field in radial direction \citep{Scharmer:2006p25}, which is not shown in the sketch -- i.e. B$_{\rm{r}}$ is the magnetic field component projected on the axis perpendicular to the plane of Fig.~\ref{fig:gaps}. B$_{\rm{r}}$ varies in azimuthal direction because the inclination of the magnetic field fluctuates with respect to the local vertical, i.e. it is larger above than between the gaps. The amplitude of this fluctuation depends on the radial position and is largest in the outer penumbra. In other words: Since B$_{\rm{r}}$ increases with radius, especially above the gap, it will eventually surpass the average magnetic field strength and balance the gas pressure inside the gap, causing the cusp to disappear. Due to this configuration, the convection inside the gap can be altered, and the plasma is forced radially outwards resulting in the EF. 

The larger inclination of the magnetic field above the gap decreases the vertical field strength at the umbral side of the gap. Therefore, it is easier for the hot plasma to ascend, and the gap is pushed open opposite to the direction of the outflow along the horizontal field lines. The migration will eventually come to an end, if the gap penetrates the more vertical umbral field. This could be an explanation for the inward migration of PGs and the appearance of PUDs \citep{Scharmer:2008p11}.

\paragraph{Advantages:} The idea of convection throughout the penumbra is appealing as it easily explains the penumbral brightness. Furthermore, it provides an elegant explanation for the apparent twist of PFs \citep{Ichimoto:2007p4024}. The magnetic field causes fluting instabilities of the $\tau=1$ surface of the gap, which yields corrugations appearing with the same inclination as the field. These corrugations in turn increase the surface area and allow the plasma to cool more effectively causing PFs with inclined striation. Their movement is explained as a so-called barber-pole effect caused by convective downflows in the gap \citep{Spruit:2010p239}.

\paragraph{Problems:} The gappy penumbral model has been criticized for its underlying assumption of field free regions in the penumbra, \citep{Borrero:2008p143,Puschmann:2010p171}. Furthermore, the proposed correlation of continuum intensity and Doppler velocity \citep{Jurcak:2008p1548} and the expected morphology of the vertical flow field in PFs \citep{Franz:2009p515,Puschmann:2010p171} contradict present observations. 

On the basis of numerical simulations of PFs, \citet{Scharmer:2008p11} and \citet{Scharmer:2009p3010} argue that the EF is caused by overturning convection inside the gap. Even though the plasma at the top of the gap is deflected radially outwards by the magnetic field above, it is not aligned with the magnetic field -- cf. \citet{Heinemann:2007p184}. Furthermore, an EF present in the gap is unmagnetized, which is at odds with the results of e.g. \citet{Rezaei:2006p157} and Section~\ref{sec:implications}.

Other issues involve the submergence of the EF in the outer penumbra, which requires the magnetic field lines to dip back below the solar surface (cf. Section~\ref{sec:hiddenopp}) as well as the distribution of the total net circular polarization, which arises from the gappy model and has yet to be compared to observation.

\subsection{Alternative Proposals}
\label{sec:alternativ}

Siphon flows, buoyant flux tubes and penumbral gaps are not the only proposals to explain the penumbra and the EF.
Alternative scenarios are not exclusive of each other, and they sometimes differ only slightly from the concepts introduced above. For the sake of completeness, these ideas shall be briefly mentioned: 

\paragraph{Falling Flux Tubes:} This scenario was introduced by \citet{Wentzel:1992p4160} and assumes an impulsive, but temporal upflow along slightly inclined magnetic field lines, causing an inversion of density with height. This configuration, describing the magnetic field in the inner penumbra, creates a Rayleigh-Taylor instability which causes the flux tube to fall over. The surplus material inside the flux tube is drained along the horizontal part of the tube in the penumbra, thereby causing an episodic EF on a concave path. This model 
gives an explanation for the ragged border between the umbra and penumbra \citep{Solanki:1994p4039}, but the concave path of the EF contradicts observation \citep{Rimmele:1995p4029,Schlichenmaier:2000p225,Schmidt:2000p226}.

\paragraph{Micro Structured Magnetic Atmospheres (MISMAS):} The MISMA hypothesis assumes that the entire surface of the Sun, including the penumbra, is structured on scales far below the resolution limit of any telescope available today. This hypothesis was introduced by \citet{2010mcia.conf..210S} to explain the observed asymmetries in Stokes V profiles. While a range of observational features can be explained by this model \citep{SanchezAlmeida:2009p170,SanchezAlmeida:2010p426}, it presumes a large number of free parameters, which makes it difficult to obtain reliable experimental evidence and almost impossible to draw definite conclusions.

\paragraph{Collective Phenomenon:} The scenario developed by \citet{Ryutova:2008p4027}, in which the entire sunspot is modeled as an ensemble of interlaced flux tubes, is a rather theoretical approach. The varying inclinations of the dense conglomerate of twisted, interlaced flux tubes is a result of ongoing reconnection processes causing peripheral filaments to branch out at different heights, arching downward to the photosphere and eventually becoming horizontal.

\paragraph{Simulation of Radiative and Turbulent Magnetoconvection:} Computers may be used to solve the magnetohydrodynamical equations and model convection under the influence of largely inclined penumbral magnetic fields 
\citep{Heinemann:2007p184,2009ASPC..416..461R,Kitiashvili:2009p174}. Due to the restricted computational capacities, however, the domains are not large enough to host entire sunspots. Therefore, periodic boundary conditions, which cause artifacts, are used at the edge of the computational box, making it difficult to obtain structures similar to PFs. \citet{2009ASPC..415..351R} was the first to successfully model a domain which hosted one ore even two sunspots, showing a developed penumbra. These simulations are currently under investigation \citep{2008ESPM...12.2.28S,2009Sci...325..171R,Borrero:2010p135,2010arXiv1011.0981R,2011ApJ...729....5R} and will provide a new perspective on the underlying structure of the penumbra and the nature of the EF.

\chapter{Spectropolarimetry}
\label{ch:specpol}

This Chapter provides the theoretical background necessary to understand the technique of spectropolarimetry. To this end, the Stokes formalism is introduced in Section~\ref{sec:pollight}, which also provides information on how the polarization state of light can be measured. Section~\ref{sec:zeeman} explains the splitting and polarization of spectral lines due to the Zeeman effect. Section~\ref{sec:radtrans} shows how radiative transfer in a magnetized atmosphere can be formally described and reviews physical processes leading to the shape of solar absorption lines. Finally, Section~\ref{sec:inversion} explains how spectropolarimetric measurements can be used to derive the physical parameters that define the characteristics of the solar atmosphere.

\section{Description and Measurement of Polarized Light} 
\label{sec:pollight}
The observation and interpretation of polarization signals in suitable Frauenhofer 
lines is of uttermost importance for the investigation of solar and stellar magnetic fields \citep{1992soti.book.....S}. With the publication of \citet{1947ApJ...105..424C}, the Stokes formalism \citep{stokes1852} has become the standard representation of polarized radiation in astronomy \citep{solanki1987PhDT}. Its advantage lies in its 
description of arbitrarily polarized light via observable quantities, i.e. intensity.

\paragraph{Stokes Formalism:} If a quasi-monochromatic\footnote{This is the superposition of monochromatic waves with various amplitudes distributed over a frequency range $\scriptstyle\Delta\textstyle\nu$ centered at $\nu_0$ with $\nu_0 \gg \scriptstyle\Delta\textstyle\nu$.} beam of light is parametrized in Cartesian coordinates for which the z-axis is chosen in the direction of propagation, it is possible to write the vibrations of the electric field vector ($\xi_{\rm{x}}, \xi_{\rm{y}}$) as:

\begin{eqnarray}
\xi_{{\rm{x}}}(\rm{t}) &=& \hat{\xi}_{\rm{x}} \cos(\omega \rm{t} - \varphi_{\rm{x}}) \nonumber \\
\xi_{\rm{y}}(\rm{t}) &=& \hat{\xi}_{\rm{y}} \cos(\omega \rm{t} - \varphi_{\rm{y}}) 
\end{eqnarray}

\noindent where $\omega$ is the circular frequency of the vibration. $\hat{\xi}_{\rm{x}}$ and $\hat{\xi}_{\rm{y}}$ as well as $\varphi_{\rm{x}}$ and $\varphi_{\rm{y}}$ represent the amplitude and phase of the electric field vector in x and y directions.

\noindent With $\langle \,\, \rangle$ denoting the time average, 
the Stokes parameters are defined as:

\begin{eqnarray}
\rm{I}\,& \equiv & \langle \, \hat{\xi}^2_{\rm{x}} \, \rangle + \langle \, \hat{\xi}^2_{\rm{y}} \, \rangle \\
\rm{Q}& \equiv & \langle \, \hat{\xi}^2_{\rm{x}} \, \rangle - \langle \, \hat{\xi}^2_{\rm{y}} \, \rangle \\
\rm{U}& \equiv & 2 \langle \, \hat{\xi}_{\rm{x}} \, \hat{\xi}_{\rm{y}} \cos(\varphi_{\rm{x}} - \varphi_{\rm{y}}) \, \rangle \\
\rm{V}& \equiv & 2 \langle \, \hat{\xi}_{\rm{x}} \, \hat{\xi}_{\rm{y}} \sin(\varphi_{\rm{x}} - \varphi_{\rm{y}}) \, \rangle
\end{eqnarray}

\noindent The Stokes parameters are often written in vector form, i.e. {\bf{S}}=(I,Q,U,V)$^{\top}$ and completely characterize the polarization state of light:

\begin{eqnarray*}
 \bullet & \rm{I} & \textrm{Represents the total intensity.} \\
 \bullet & \rm{Q} & \textrm{Describes the amount of linearly polarized light.} \\
 \bullet & \rm{U} & \textrm{Is the linear polarization measured at an angle of 45$^{\circ}$ to that of Q.} \\
 \bullet & \rm{V} & \textrm{Accounts for the circularly polarized radiation}.
\end{eqnarray*}

\noindent The degree of polarization (P$_{\rm{tot}}$) can be written as:

\begin{equation}
 \rm{P_{\rm{tot}}}  \equiv \sqrt{\frac{\rm{Q}^2+\rm{U}^2+\rm{V}^2}{\rm{I}^2}}\,\,\,\,\,\text{with}\,\,\,\,\, 0 \le \rm{P}_{\rm{tot}} \le 1
\end{equation}

\noindent with its two extrema of total unpolarization, i.e. $\rm{Q}=\rm{U}=\rm{V}=0$, and complete polarization, i.e. $\rm{P_{\rm{tot}}}=1 \Leftrightarrow \rm{I}^2=\rm{Q}^2+\rm{U}^2+\rm{V}^2$.

\begin{figure}[h!]
\begin{center}
    \includegraphics[width=0.7\textwidth]{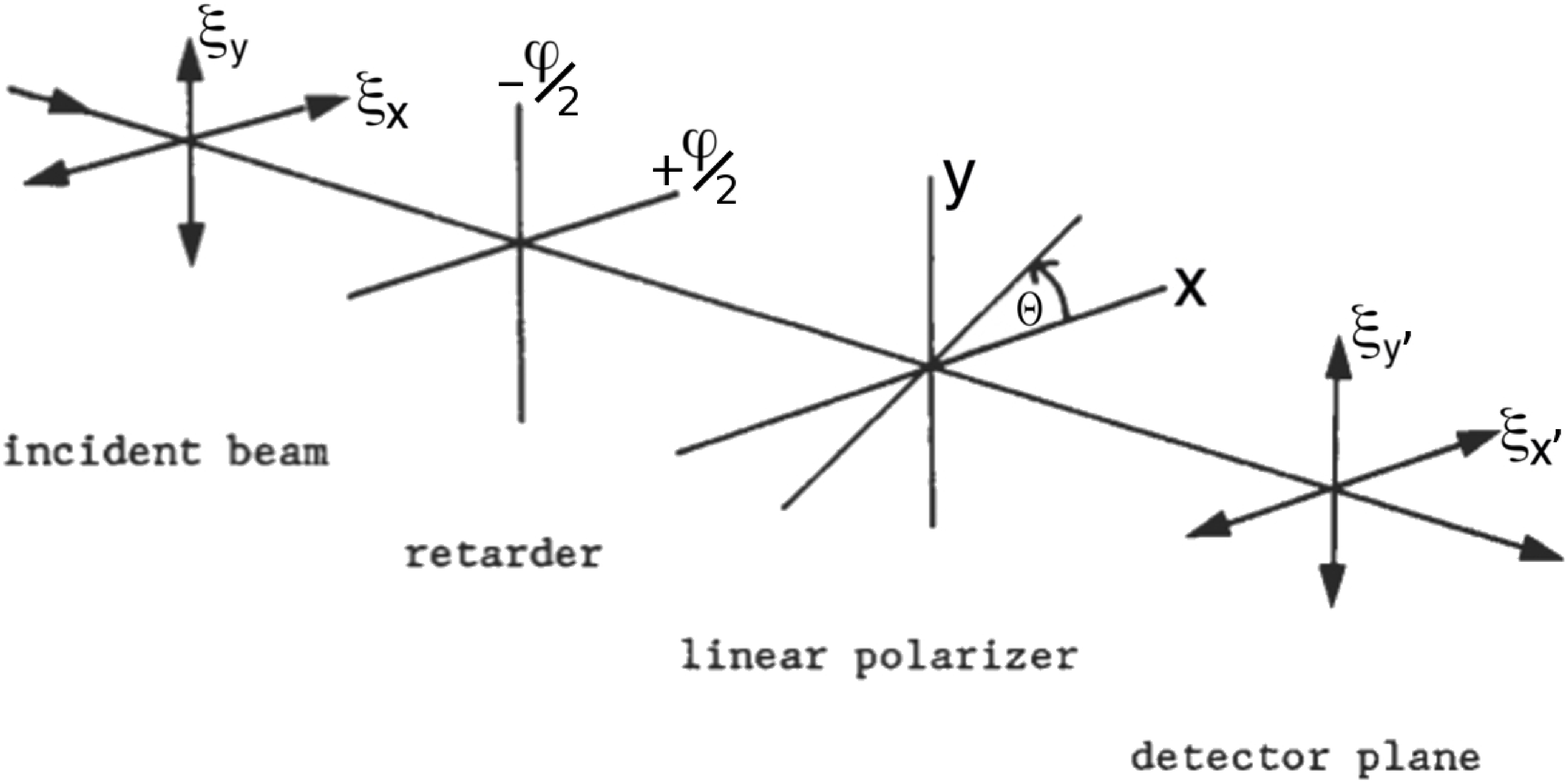}
\caption{Optical setup to measure the Stokes parameters. Adopted from \citet{collet}.}
\label{fig:polret}
\vspace{-20pt}
\end{center}
\end{figure}

\paragraph{Measurement of Stokes Parameters:} There is a range of techniques to determine the full set of Stokes parameters \citep{collet}. Nevertheless, they all rely on measurements of the intensity of light which has passed through adequate optical components. The basic principle shall be described by means of the classical method (cf. Fig.~\ref{fig:polret}), where the light passes through a quarter wave retarder followed by a polarizer.

\noindent In the framework of the M\"uller calculus, the incoming Stokes vector ({\bf{S}}) is transformed into the outgoing Stokes vector ({\bf{S'}}):

\begin{equation}
\mathbf{S'}=\mathbf{M}_{\rm{pol}}\mathbf{M}_{\rm{ret}}\mathbf{S}
\end{equation}

\noindent with {\bf{M}}$_{\rm{ret}}$ being the M\"uller matrix of the retarder with its fast axis along the x-plane:

\begin{equation}
\mathbf{M}_{\rm{ret}}= \left(\begin{array}{cccc}  
1 & 0 & 0 & 0 \\
0 &  1 & 0 & 0 \\
0 & 0 &  \cos \varphi & -\sin \varphi \\
0 & 0 & \sin \varphi &  \cos \varphi \\
\end{array}  \right)
\end{equation}

\noindent while {\bf{M}}$_{\rm{pol}}$ describes a polarizer with its transmission axis set at an angle $\Theta$:

\begin{equation}
\mathbf{M}_{\rm{pol}}= \left(\begin{array}{cccc}  
1 & \cos 2 \Theta & \sin 2 \Theta & 0 \\
\cos2 \Theta &  \cos^2 2 \Theta & \sin 2 \Theta \cos 2 \Theta & 0 \\
\sin 2 \Theta & \sin 2 \Theta \cos 2 \Theta &  \sin^2 2 \Theta & 0 \\
0 & 0 & 0 &  0 \\
\end{array}  \right)
\end{equation}

\noindent The intensity, in fact the only measurable quantity, is a linear combination of the four Stokes parameters \citep{collet}: 

\begin{equation}
\rm{I}({\Theta, \varphi}) = \frac{1}{2}(\rm{I}+\rm{Q} \cos 2 \Theta + \rm{U} \sin 2 \Theta \cos \varphi - \rm{V} \sin 2 \Theta \sin \varphi)
\label{eq:stokextract}
\end{equation}

\noindent Equation~\ref{eq:stokextract} allows to determine all Stokes parameters from a range of intensity measurements $\rm{I}({\Theta, \varphi})$:

\begin{eqnarray}
\rm{I} & = & \rm{I}({0, 0}) + \rm{I}({90^{\circ}, 0}) \\
\rm{Q} & = & \rm{I}({0, 0}) - \rm{I}({90^{\circ}, 0}) \\
\rm{U} & = & \rm{I}({45^{\circ}, 0}) - \rm{I}({135^{\circ}, 0}) \\
\rm{V} & = & \rm{I}({45^{\circ}, 90^{\circ}}) - \rm{I}({135^{\circ}, 90^{\circ}})  
\end{eqnarray}

For the spectropolarimeter onboard HINODE, a slightly different approach was chosen (cf. Section~\ref{sec:polarimeter}). Here, the light passes through a quarter-wave retarder, rotating at an angular frequency $\omega$, followed by a linear polarizer. Even though Equation~\ref{eq:stokextract} changes in this setup, the basic principle is the same.

\section{Zeeman Effect}
\label{sec:zeeman}

\citet{zeeman1897a, zeemannature} realized that the Na I D line broadens in the presence of a magnetic field and shows polarization peculiarities in the line-wings. 
Subsequent experiments \citep{zeeman1897b,zeeman1897c} confirmed that the magnetic field splits the line into differently polarized components. This so-called normal Zeeman effect was explained by Lorentz using the classical theory of electrodynamics. However, to understand the so-called anormalous Zeeman effect, i.e. the split into more than three components, quantum-mechanical concepts are required.

\begin{figure}[h!]
\begin{center}
    \includegraphics[width=0.6\textwidth]{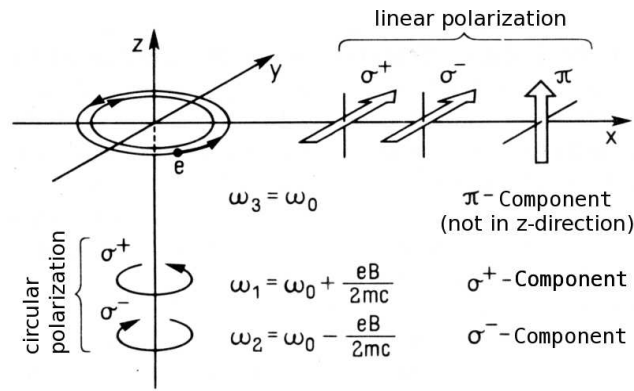}
\caption{Lorentz's explanation for the Zeeman-effect. A magnetic field in z-direction alters the rotational frequencies of e$^-$ in the x-y plane, while it has no impact on the oscillation in z-direction. Measurements parallel to the magnetic field yield two circularly polarized components, while perpendicular to it, the projection of the oscillation of the e$^-$ is measurable as three linearly polarized components. Adopted from \citet{MayerKuckuk}.}
\label{fig:zeeman1}
\end{center}
\end{figure}

\paragraph{Classical treatment:}
In the Lorentian theory, the electron (e$^-$) may oscillate in all three dimensions. In a coordinate system as in Fig.~\ref{fig:zeeman1}, the motion may be simplified to a linear oscillation in z-direction and a circular motion\footnote{A circular motion can be seen as the superposition of oscillations in the x- and in the y-direction} in the x-y plane. If an external magnetic field is applied in z-direction, a Lorentz force, in turn balanced by the centripetal force, will act on the e$^-$ $\propto$ {\bf{v}}$\times${\bf{B}}. Depending on their direction of motion, the rotational frequency for the e$^-$ is thus altered.

Observation in x-direction yields three linearly polarized lines: Two $\sigma$-components shifted by $\pm \scriptstyle \Delta \textstyle \omega$ and one unshifted $\pi$-component. Observation along the magnetic field lines results in two circularly polarized lines shifted by $\pm \scriptstyle \Delta \textstyle \omega$. The $\pi$-component is missing, since a dipole may not radiate along its axis of oscillation.

\paragraph{Quantum-Mechanical Interpretation:} To obtain the energy states of an multi-e$^-$ atom, it is helpful to consider the e$^-$ as subject to the kinetic forces and electrostatic interaction with an averaged electrostatic field (nucleus and other e$^-$) of spherical symmetry. Additionally, the spin-orbit coupling, which interlinks the spin (S) of the e$^-$ with the orbital angular momentum (L) into the total angular momentum (J), has to be considered.


If an external magnetic field is applied, the spherical symmetry is broken and the Hamiltonian has to be modified to account for the interaction of the magnetic moment of the e$^-$ with the external field. If a) Russell-Saunders coupling is valid, b) the coupling to the magnetic moment of the nucleus is neglected\footnote{This interaction is three orders of magnitude smaller than the spin-orbit interaction.} and c) the coupling of the e$^-$ to the external field is small\footnote{For the strength of solar magnetic fields, this is a reasonable approximation.} when compared to the spin-orbit interaction, the Hamiltionian ($\mathcal{H}$) can be written as \citep{solanki1987PhDT}:


\begin{eqnarray}
\mathcal{H} & = & \mathcal{H}_0 + \mathcal{H}_1 \nonumber \\
& = & \mathcal{H}_0 + \frac{\rm{e}}{2\rm{m}_{\rm{e}}\rm{c}}(\mathcal{L}+2\mathcal{S}){\mathbf{B}}
\label{eq:hamilton}
\end{eqnarray}

\noindent with charge (e) and mass of the e$^-$ (m$_{\rm{e}}$) as well as the speed of light (c) and the magnetic field vector ($\mathbf{B}$). $\mathcal{L}$ and $\mathcal{S}$ are the orbital and the spin angular momentum operators having eigenvalues of $\hbar\sqrt{\rm{L}(\rm{L}+1)}$ and $\hbar\sqrt{\rm{S}(\rm{S}+1)}$ with $\rm{L}\in\mathds{Z}$ and 2$\cdot$$\rm{S \in \mathds{Z}}$ as well as $\hbar=\rm{h}/2\pi$ with h representing Planck's constant .

Applying first-order time independent perturbation theory, it can be shown that the (2J+1)-fold degeneracy\footnote{The eigenvalues of the total angular momentum operator $\mathcal{J}=\mathcal{L}+\mathcal{S}$ are $\hbar\sqrt{\rm{J}(\rm{J}+1)}$ with 2$\cdot$$\rm{J}\in\mathds{Z}$} of each energy level disappears due to the splitting into magnetic sublevels with energy:

\begin{eqnarray}
\rm{E}_{\rm{J,M}} & = & \rm{E}_{\rm{J}}+\frac{\rm{e\hbar}}{2\rm{m}_{\rm{e}}\rm{c}}\rm{gMB} \nonumber \\
&=&\rm{E}_{\rm{J}}+\mu_0\rm{gMB}
\label{eq:energy}
\end{eqnarray}

\noindent with E$_{\rm{J}}$ being the energy of the atomic level in the absence of the magnetic field (E$_{\rm{J}}$ is an eigenvalue of H$_0$),  $\mu_0$ the Bohr magneton, B the strength of the magnetic field and $\hbar$M the eigenvalues of J$_{\rm{z}}$, i.e. the projection of the total angular momentum in the direction of the external magnetic field, with $\rm{J}\in\{-\rm{J}, -\rm{J}+1, ... , \rm{J}-1, \rm{J}\}$. Since LS-coupling is assumed to be valid, the Land\'e factor g may be written as:

\begin{equation}
\rm{g} \equiv 
 \begin{cases} 
 	1+\frac{\rm{J}(\rm{J}+1)+\rm{S}(\rm{S}+1)-\rm{L}(\rm{L}+1)}{2\rm{J}(\rm{J}+1)} & \text{if J} \not= 0 \\
     0 & \text{Otherwise}
\end{cases}
\end{equation}

\paragraph{Zeeman Pattern:} Spectral lines appear if transitions between two atomic states (l and u with E$_{\rm{l}}<\rm{E}_{\rm{u}}$) are triggered, whereby the selection rules for electric dipole\footnote{Higher multipole moments shall not be considered here, since they lead to weaker lines.} radiation have to be obeyed:

\begin{eqnarray}
\scriptstyle\Delta\displaystyle\rm{J} &=& \rm{J}_{\rm{u}}- \rm{J}_{\rm{l}} =0,\pm1\\
\scriptstyle\Delta\displaystyle\rm{M} &=& \rm{M}_{\rm{u}}- \rm{M}_{\rm{l}} =0,\pm1\\
\rm{M}=0 &\rightarrow& \rm{M}=0 \,\,\, \nexists \,\,\, \text{if} \,\,\, \scriptstyle\Delta\displaystyle\rm{J} = 0
\label{eq:energy2}
\end{eqnarray}

\noindent Transitions with $\scriptstyle\Delta\textstyle\rm{M}=0$ correspond to the $\pi$-component in the Zeeman pattern, while $\scriptstyle\Delta\textstyle\rm{M}=\pm1$ yields the $\sigma^{\pm}$-components. 
The shift of the components with respect to the unsplit line ($\lambda_0$) reads:

\begin{eqnarray}
\scriptstyle\Delta\displaystyle\lambda = \lambda-\lambda_0 & = &\frac{\rm{e} \lambda^2_0 \rm{B}}{4\pi\rm{m}_{\rm{e}}\rm{c}}(\rm{g}_{\rm{l}}\rm{M}_{\rm{l}}- \rm{g}_{\rm{u}}\rm{M}_{\rm{u}}) \nonumber \\
& = & \lambda_{\rm{B}} (\rm{g}_{\rm{l}}\rm{M}_{\rm{l}}- \rm{g}_{\rm{u}}\rm{M}_{\rm{u}})
\end{eqnarray}



\begin{figure}[h!]
\begin{center}
    \includegraphics[width=0.8\textwidth]{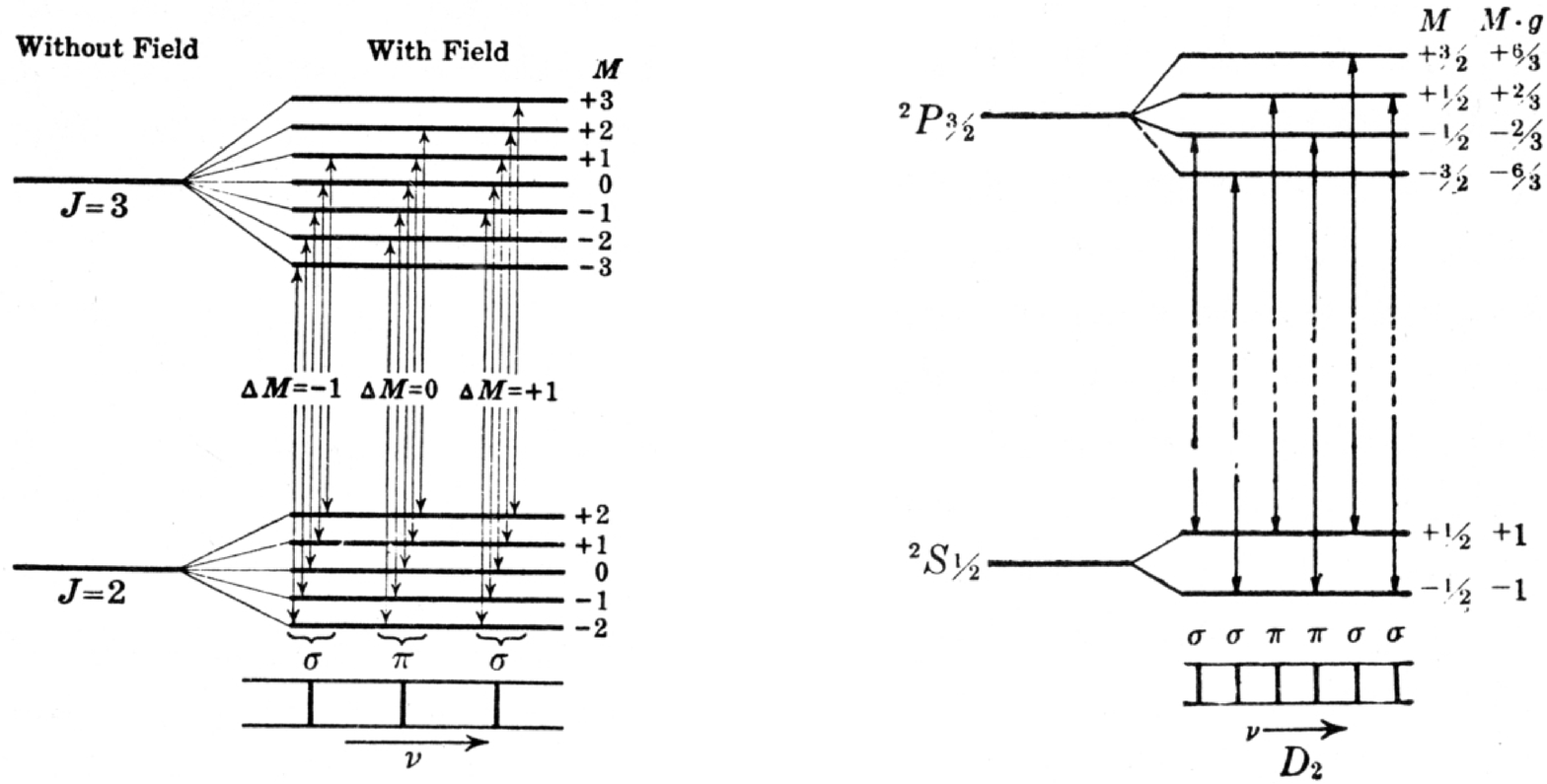}
\caption{Grotrian diagrams with transitions leading to the normal (left) and anormalous (righ) Zeeman effect. Adopted from \citet{herzberg}.} 
\label{fig:zeeman2}
\end{center}
\end{figure}

\noindent If $\rm{g}_{\rm{l}}=\rm{g}_{\rm{u}}$, i.e. the transition occurs between two levels with equal g-factors, a normal Zeeman pattern is obtained\footnote{If $\rm{J}=0 \rightarrow \rm{J}=1$, the multiplicity of $\rm{J}=1$ allows only for three transitions, too.}, since the separation between two consecutive sublevels M is the same for both levels J (cf. left panel of Fig.~\ref{fig:zeeman2}).

\noindent In general, $\rm{g}_{\rm{l}} \not= \rm{g}_{\rm{u}}$ and the spectral line is not a triplet anymore. Nevertheless, the denotation $\pi$- and $\sigma^{\pm}$-component is kept (cf. right panel of Fig.~\ref{fig:zeeman2}). In analogy to the previously introduced g-factor, an effective Land\'e factor\footnote{$\rm{g}_{\rm{eff}}$ is independent of the coupling scheme \citep{1982SoPh...77..285L}.} may be defined \citep{shenblair}:
\begin{eqnarray}
\rm{g}_{\rm{eff}} \equiv \frac{1}{2}(\rm{g}_{\rm{l}} + \rm{g}_{\rm{u}})+\frac{1}{4}(\rm{g}_{\rm{l}} - \rm{g}_{\rm{u}})[\rm{J}_{\rm{l}}(\rm{J}_{\rm{l}}+1)-\rm{J}_{\rm{u}}(\rm{J}_{\rm{u}}+1)]
\label{eq:efflande}
\end{eqnarray}

\noindent with $\scriptstyle\Delta\textstyle\lambda= \pm \lambda_{\rm{B}}  \rm{g}_{\rm{eff}}$ being the wavelength shift 
of the $\sigma^{\pm}$-components.

\begin{figure}[h!]
\begin{center}
    \includegraphics[width=\textwidth]{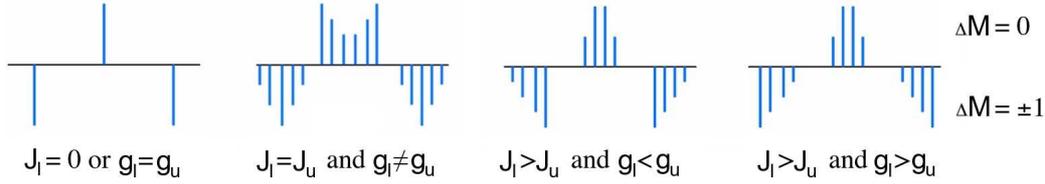}
\caption{Zeeman pattern for various relations between g and J between the lower (l) and upper (u) state of a transition. The leftmost pattern depicts the normal Zeeman effect, while the others are examples of the anomalous  effect. Adopted from \citet{svetalecture}.}
\label{fig:zeeman3}
\end{center}
\end{figure}

The relative intensities of the Zeeman components depend on the transition probability between the individual sublevels participating in the transition. These probabilities are given by 
the vector coupling coefficients\footnote{They are also called Clebsh-Gordan or Wigner coefficients and may be calculated using the expression for the 3-j symbols \citep{degl2004polarization}.}, 
reflecting the strength with which the electric dipole operator couples the sublevels. 
In contrast to the shifts of the Zeeman components, their relative intensities are independent of the coupling scheme. Fig.~\ref{fig:zeeman3} illustrates a range of Zeeman patterns for various relations between g and J. The graphical convention is such that the $\pi$- and $\sigma^{\pm}$- components point up- and downwards respectively, while the axis represents $\lambda$.

\section{Radiative Transfer}
\label{sec:radtrans}

To interprete solar and stellar spectra, it is necessary to understand how the interaction between light and matter influences the transport of energy through the atmosphere of stars. Because of the abundance of literature on this topic, e.g. \citet{1992soti.book.....S,ruttenlecture,delToro,2004suin.book.....S,degl2004polarization,svetalecture} and of course \citet{mihalas1978stellar} as well as \citet{gray2005observatio}, this Section reviews only the basic concepts and explains radiative transfer in a magnetized atmosphere using the Stokes Formalism.

\paragraph{Energy Transfer:} A basic quantity in the theory of radiative transfer is the intensity (I) of the radiation field. For a certain frequency ($\nu$), I$_{\nu}$ changes through the atmosphere:

\begin{eqnarray}
\rm{dI}_{\nu}& = & ( \epsilon_{\nu} - \kappa_{\nu} \rm{I}_{\nu} ) \rm{ds}
\label{eq:Ichange}
\end{eqnarray}

\noindent due to emission ($\epsilon$) and absorption ($\kappa$) processes 
occurring along the path (s).

If it is assumed that the light passes through an atmosphere of plane-parallel stratification, an optical depth scale can be defined:


\begin{eqnarray}
\rm{d} \tau_{\nu}= - \kappa_{\nu} \mu \rm{ds}
\label{eq:tau}
\end{eqnarray}

\noindent where 
$\mu = \cos \, \Theta$ measures the deviation of the LOS from the local vertical, i.e. the heliocentric angle. A combination of Equations~\ref{eq:Ichange} and \ref{eq:tau} yields:

\begin{eqnarray}
\mu \frac{\rm{dI}_{\nu}}{\rm{d}\tau_{\nu}} = \rm{I}_{\nu} - \frac{\epsilon_{\nu}}{\kappa_{\nu}} = \rm{I}_{\nu} - \mathcal{S}_{\nu}
\label{eq:radtrans}
\end{eqnarray}

\noindent with $\mathcal{S}$ being the so-called source function. %
%
%
If $\mathcal{S}$ is known, Equation~\ref{eq:radtrans} can be used to predict the emergent intensity. It is, however, more interesting to infer $\mathcal{S}$ from a measurement of I as a function of $\nu$ and $\mu$ (cf. Section~\ref{sec:inversion}).

\paragraph{Absorption and Emission:} Processes leading to $\epsilon_{\nu}$ and $\kappa_{\nu}$ are, for example, transitions between two energy levels of an atom in the stellar atmosphere\footnote{Other processes such as collisional excitation and de-excitation of atoms or molecular transitions are described in \citet{svetalecture} and shall not be discussed here.}. These transitions, which may be described by the Einstein coefficients, occur as a) spontaneous~$\equiv~\rm{A}_{\rm{ul}}$ and b) stimulated~$\equiv~\rm{B}_{\rm{ul}}$ emission as well as c) photo absorption~$\equiv~\rm{B}_{\rm{lu}}$. If $\rm{n}_{\rm{l}}$ and $\rm{n}_{\rm{u}}$ represent the number densities of atoms in the lower and upper state respectively, $\epsilon_{\nu}$ and $\kappa_{\nu}$ can be written as \citep{mihalas1978stellar}: 

\begin{eqnarray}
\epsilon_{\nu} & = & \frac{\rm{h}\nu}{4\pi}\rm{n}_{\rm{u}}\rm{A}_{\rm{ul}}
\label{eq:emeinst}
\\
\kappa_{\nu} & = & \frac{\rm{h}\nu}{4\pi} \left ( \rm{n}_{\rm{l}}\rm{B}_{\rm{lu}} -\rm{n}_{\rm{u}}\rm{B}_{\rm{ul}} \right)
\label{eq:abseinst}
\end{eqnarray}

\noindent Combining Equations~\ref{eq:emeinst} and \ref{eq:abseinst} with the relations for the Einstein coefficients:

\begin{equation}
\frac{\rm{B}_{\rm{ul}}}{\rm{A}_{\rm{ul}}} = \frac{\rm{c}^2}{2\rm{h}\nu^3} \,\,\,\,\, \text{and}  \,\,\,\,\, \frac{\rm{B}_{\rm{ul}}}{\rm{B}_{\rm{lu}}} = \frac{\rm{w}_{\rm{l}}}{\rm{w}_{\rm{u}}}  
\end{equation}

\noindent for which $\rm{w}_{\rm{l}}$ and $\rm{w}_{\rm{u}}$ represent the statistical weighting of the levels participation in the transition, Equation~\ref{eq:radtrans} can be rewritten as \citep{mihalas1978stellar}:

\begin{eqnarray}
\mu \frac{\rm{dI}_{\nu}}{\rm{d}\tau} & = & \rm{I}_{\nu} - \mathcal{S}_{\nu} \nonumber \\
& = & \rm{I}_{\nu} - \frac{\rm{w}_{\rm{u}}\rm{A}_{\rm{ul}}}{\rm{w}_{\rm{l}}\rm{B}_{\rm{lu}} - \rm{w}_{\rm{u}}\rm{B}_{\rm{ul}}} \nonumber \\
& = & \rm{I}_{\nu} - \frac{2\rm{h}\nu^3}{\rm{c}^2}\frac{1}{\rm{n}_{\rm{l}}\rm{w}_{\rm{u}} \, / \,  \rm{n}_{\rm{u}}\rm{w}_{\rm{l}}-1}
\label{eq:radtrans2}
\end{eqnarray}

\paragraph{Line Profiles:} In the solar spectrum, absorption lines appear if $\nu$ matches the resonance frequency ($\nu_0$) of an atomic transition. This results in a tremendous variation of $\kappa_{\nu}$ within a small frequency interval. For absorption lines, $\kappa_{\nu}$ may be written as a combination of absorption coefficients from the continuum\footnote{The continuum opacity of stars is caused by e.g. free-free transitions, scattering or the ionization of H$^-$ \citep{svetalecture}.} ($\kappa_{\nu}^{\rm{cont}}$) and from the line ($\kappa_{\nu}^{\rm{line}}$). Defining $\eta_0$ as the line to continuum absorption coefficient, $\kappa_{\nu}$ adopts the following form:


\begin{eqnarray}
\kappa_{\nu} & = & \left (1+ \frac{\kappa_{\nu}^{\rm{line}}}{\kappa_{\nu}^{\rm{cont}}} \right ) \kappa^{\rm{cont}}_{\nu} = \left (1+\eta_0 \right ) \kappa_{\nu}^{\rm{cont}}
\end{eqnarray}

\noindent To understand the behavior of $\kappa_{\nu}^{\rm{line}}$ in the vicinity of the resonance frequency, it is helpful to remember mechanisms leading to the spectral broadening of a line.

The lifetime of an atom in an excited state is limited by spontaneous emission $\scriptstyle\Delta\textstyle\rm{t} = \rm{A}^{-1}_{\rm{ul}}$. In accordance with the energy-time uncertainty principle, this leads to a certain spread in energy and hence frequency, i.e. $\scriptstyle\Delta\textstyle \rm{E} = \rm{h} \scriptstyle\Delta\textstyle \nu= \hbar / \scriptstyle\Delta\textstyle \rm{t} = \hbar \gamma^{\rm{rad}}$. This natural broadening always occurs and defines an emission\footnote{Additional assumptions have to be made, e.g. LTE, to explain absorption profiles.} probability distribution around $\nu_0$, resembling a Lorentz profile \citep{svetalecture}:

\begin{equation}
\Lambda(\scriptstyle\Delta\displaystyle\nu)=\frac{\gamma^{\rm{rad}}\, / \, 4\pi}{(\scriptstyle\Delta\displaystyle\nu)^2 + (\gamma^{\rm{rad}}\, / \, 4\pi)^2}
\end{equation}

In general, the natural line broadening is small when compared to collisional broadening mechanisms like Van-der-Waals broadening, i.e. interaction with H, or Stark broadening, i.e. interaction with free e$^-$. However, since a) the frequency dependency of these processes are given by Lorentz distributions too, and b) the latter is invariant to convolution, the resulting profile will be a Lorentz profile.

Due to the Doppler effect (cf.~Equation~\ref{eq:doppler}) random velocities of the atoms cause a shift of $\nu_0$. If their motion is purely thermal, i.e. if the approximation of LTE holds, the line profile is Doppler broadened:

\begin{equation}
\Gamma(\scriptstyle\Delta\displaystyle\nu) = \frac{\rm{1}}{\sqrt{\pi}\scriptstyle\Delta\textstyle\nu_{\scriptscriptstyle\rm{D}\textstyle}} \rm{e}^{-(\nu^2\,/\,\nu_{\rm{D}}^2)}
\label{eq:dopplerprof}
\end{equation}

\noindent where $\scriptstyle\Delta\displaystyle\nu_{\scriptscriptstyle\rm{D}\displaystyle} = \textstyle\nu_0 \rm{c}^{-1} \sqrt{2\rm{k}_0\rm{T}\, / \, \rm{m}_{\rm{A}} + \rm{v}_{\rm{mic}}^2 } $ is the Doppler width, k$_0$ Boltzmann's constant, m$_{\rm{A}}$ the mass of the atom and v$_{\rm{mic}}$ the microturbulence velocity accounting for random plasma motion on scales smaller than the mean free path of the photons.

In general, both Doppler and collisional broadening occurs, and $\Lambda(\scriptstyle\Delta\displaystyle\nu)$ as well as $\Gamma(\scriptstyle\Delta\displaystyle\nu)$ have to be folded, resulting in the so-called Voigt profile (cf. Equation~\ref{eq:voigt}) where $\Gamma(\scriptstyle\Delta\displaystyle\nu)$ dominates the core of the line and $\Lambda(\scriptstyle\Delta\displaystyle\nu)$ prevails in the wings.

\paragraph{Local Thermodynamic Equilibrium (LTE):} LTE implies that a single value of T is sufficient to describe the thermodynamic state of a volume with a radius smaller than the thermalization length \citep{2004suin.book.....S}. As a consequence, the velocity of the particles along the LOS follow a Maxwellian distribution leading to the Doppler broadened line profile of Equation~\ref{eq:dopplerprof}. The degree of ionization is given by the Saha equation, and the populations of the energy levels participating in the transition can be computed via the Boltzmann equation \citep{ruttenlecture}.





The most important simplification, however, is that $\mathcal{S}$ becomes an isotropic black-body radiation described by the Kirchhoff-Planck function like \citep{2004suin.book.....S}:

\begin{equation}
\mathcal{S}_{\nu} \equiv \rm{B}_{\nu}(\rm{T}) = \frac{2\rm{h}\nu^3}{\rm{c}^2} \frac{1}{{\rm{e}^{\rm{h}\nu\,/\,\rm{k}_0\rm{T}}-1}}
\end{equation}

\noindent An effective mechanism to impose LTE is the collision between the particles. Due to its relatively high density, the photosphere\footnote{LTE does neither exist in the upper parts of the chromosphere, nor in the corona.} closely satisfies this condition.

\paragraph{Radiative Transfer in a Magnetized Atmosphere:} %
\noindent \citet{1956PASJ....8..108U} was the first to derive Equation~\ref{eq:radtrans} in the presence of a magnetic field by means of classical electrodynamics. Equation~\ref{eq:radtrans} was then extended by \citet{Rachkovsky1962a,Rachkovsky1962b} to account for anomalous dispersion effects and by \citet{1969SoPh....9..372B,1969SoPh...10..262B} to treat arbitrary Zeeman patterns. Finally, \citet{1983SoPh...85....3L}
managed to deduce Equation~\ref{eq:radtrans} on the basis of more general quantum mechanical principles.

Using Stokes calculus, Equation~\ref{eq:radtrans} can be written as \citep{1974SoPh...35...11W}:

\begin{equation}
\mu \frac{\rm{d}\mathbf{I}_{\nu}}{\rm{d}\tau_{\nu}}  = \mathcal{K} (\mathbf{I}_{\nu} - \SSS_{\nu})
\label{eq:radtransstok}
\end{equation}

\noindent where the intensity of the radiation field is described by the Stokes vector, i.e. ${\mathbf{I}_{\nu}=\mathbf{S}=(\text{I,Q,U,V})^{\top}}$. Since LTE is assumed, $\SSS_{\nu}=(\text{B$_{\nu}$(T)},0,0,0)^{\top}$, i.e. the radiation field is unpolarized and can be represented by the Kirchhoff-Planck function via a specific T of the atmosphere at a certain $\tau$ level. 

The so-called propagation or absorption matrix ($\mathcal{K}= \kappa^{\rm{cont}}_{\nu} \mathbb{1} + \kappa_{\nu}^{\rm{line}} \tilde{\mathcal{K}}$) is defined in analogy to $\kappa_{\nu}$ as the sum of continuum ($\kappa^{\rm{cont}}_{\nu}$) and line ($\kappa^{\rm{line}}_{\nu}$) absorption, with $\mathbb{1}$ representing the identity matrix. $\tilde{\mathcal{K}}$ is the line absorption matrix which accounts for absorption, dichroism and dispersion phenomena in the following way:

\begin{equation}
\tilde{\mathcal{K}} = \underbrace {\left(\begin{array}{cccc}  
\eta_{\rm{I}} & 0 & 0 & 0 \\
0 &  \eta_{\rm{I}} & 0 & 0 \\
0 & 0 &  \eta_{\rm{I}} & 0 \\
0 & 0 & 0 &  \eta_{\rm{I}} \\
\end{array}  \right) }_{\displaystyle\rm{absorption}}
+ \underbrace {\left(\begin{array}{cccc}  
0 &  \eta_{\rm{Q}} &  \eta_{\rm{U}} &  \eta_{\rm{V}} \\
\eta_{\rm{Q}} & 0 & 0 & 0 \\
\eta_{\rm{U}} & 0 & 0 & 0 \\
\eta_{\rm{V}} & 0 & 0 & 0 \\
\end{array}  \right) }_{\displaystyle\rm{dichroism}}
+ \underbrace {\left(\begin{array}{cccc}  
0 & 0 & 0 & 0 \\
0 & 0 & \rho_{\rm{V}} & -\rho_{\rm{U}} \\
0 & -\rho_{\rm{V}} & 0 & \rho_{\rm{Q}} \\
0 & \rho_{\rm{U}} & -\rho_{\rm{Q}}  & 0 \\
\end{array}  \right) }_{\displaystyle \rm{dispersion}}
\nonumber
\end{equation}

\noindent In contrast to the absorption matrix, which attenuates the intensity of the beam of light without affecting its polarization, the dichroism matrix reflects absorption depending on the polarization state. While the intensity of the beam is altered by these two phenomena, the magneto-optical effect, expressed by the dispersion matrix, causes a redistribution of energy among the different states of polarization, i.e. Faraday rotation and Faraday pulsation \citep{delToro}.

\begin{figure}[h!]
\begin{center}
    \includegraphics[width=\textwidth]{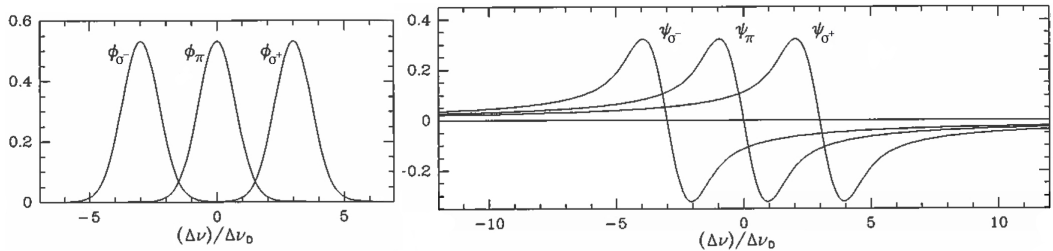}
\caption{Absorption (left) and dispersion (right) profiles for the $\pi$ and $\sigma^{\pm}$ components of the Zeeman triplet. Adopted from \citet{degl2004polarization}.}
\label{fig:absdisp}
\end{center}
\end{figure}

According to \citet{1992soti.book.....S}, the elements\footnote{See \citet{degl2004polarization} for a derivation of the elements of $\tilde{\mathcal{K}}$ in the framework of quantum mechanics.} of $\tilde{\mathcal{K}}$ are expressed as: 
\begin{eqnarray}
\eta_{\scriptscriptstyle\rm{I}\,} & = & \frac{\eta_0}{2} \left\{ \phi_{\pi} \, \sin^2 \theta + \frac{1}{2} \, [\phi_{\sigma^+} + \phi_{\sigma^-}] (1 +  \cos^2\theta) \right\} + 1 \\
\eta_{\scriptscriptstyle\rm{Q}} & = & \frac{\eta_0}{2} \left\{ \phi_{\pi} -  \frac{1}{2} \, [ \phi_{\sigma^+} + \phi_{\sigma^-}] \right\} \, \sin^2\theta \, \cos 2 \varphi  \\
\eta_{\scriptscriptstyle\rm{U}} & = & \frac{\eta_0}{2} \left\{ \phi_{\pi} -  \frac{1}{2} \,[\phi_{\sigma^+} + \phi_{\sigma^-}] \right \} \, \sin^2\theta \, \sin 2 \varphi  \\
\eta_{\scriptscriptstyle\rm{V}} & = & \frac{\eta_0}{2} \, [\phi_{\sigma^-} - \phi_{\sigma^+}] \,\cos \theta \\
\label{eq:rhoq}
\rho_{\scriptscriptstyle\rm{Q}} & = & \frac{\eta_0}{2} \left\{ \psi_{\pi} -  \frac{1}{2}\,[\psi_{\sigma^+} + \psi_{\sigma^-}] \right\} \, \sin^2\theta \, \cos 2 \varphi \\
\label{eq:rhou}
\rho_{\scriptscriptstyle\rm{U}} & = & \frac{\eta_0}{2} \left\{ \psi_{\pi} -  \frac{1}{2}\,[\psi_{\sigma^+} + \psi_{\sigma^-}] \right \} \, \sin^2\theta \, \sin 2 \varphi \\
\label{eq:rhov}
\rho_{\scriptscriptstyle\rm{V}} & = & \frac{\eta_0}{2} \, [\psi_{\sigma^-} - \psi_{\sigma^+}] \, \cos \theta
\label{eq:Kelements}
\end{eqnarray}

\noindent with $\phi_{\pi, \sigma^{\pm}}$ and $\psi_{\pi, \sigma^{\pm}}$ being the profile functions at the wavelength position of the $\pi$ and  $\sigma^{\pm}$ components of the Zeeman pattern\footnote{In the most simple case, these are the ${\pi}$ and $\sigma^{\pm}$ components of the Zeeman line triplet.} (cf.~Fig.~\ref{fig:absdisp}). If the Voigt ($\mathcal{V}$) and Faraday-Voigt ($\mathcal{F}$) function are used and reduced variables accounting for radiative damping (a) and frequency shifts due to the Doppler as well as Zeeman effect ($\tilde{\nu}$) are introduced, $\phi_{\pi, \sigma^{\pm}}$ and $\psi_{\pi, \sigma^{\pm}}$ read \citep{degl2004polarization}:

\begin{eqnarray}
\phi_{\pi, \sigma^{\pm}} (\rm{a}, \nu) \propto \mathcal{V} (\rm{a}, \tilde{\nu}) & \equiv &\frac{\rm{a}}{\pi}\int_{-\infty}^{+\infty} \rm{e}^{-\rm{y}^2}\frac{1}{(\tilde{\nu}-\rm{y})^2+ \rm{a}^2}{\rm dy}\\
\label{eq:voigt}
\psi_{\pi, \sigma^{\pm}} (\rm{a}, \nu)  \propto \mathcal{F} (\rm{a}, \tilde{\nu}) & \equiv & \frac{1}{\pi}\int_{-\infty}^{+\infty}\rm{e}^{-\rm{y}^2}\frac{\tilde{\nu}-\rm{y}}{(\tilde{\nu}-\rm{y})^2+\rm{a}^2}{\rm dy}
\label{eq:faradvoigt}
\end{eqnarray}

\section{Synthesis and Inversion of Stokes Profiles}
\label{sec:inversion}

According to \citet{1985SoPh...97..239L}, the formal solution of Equation~\ref{eq:radtransstok} can be written as: 

\begin{equation}
\mathbf{I}(\tau_1)  = \mathcal{O}(\tau_1,\tau_0)\mathbf{I}(\tau_0) - \int _{\tau_0}^{\tau_1} \mathcal{O} (\tau_1,\tau)  \mathcal{K} (\tau) \SSS({\tau})\rm{d}\tau
\label{eq:radtransstokeval}
\end{equation}

\noindent where $\mathcal{O}(\tau,\tau')$ is the so-called evolution operator that transforms the Stokes vector between two semi-infinite slabs 
of the atmosphere, i.e. $\mathbf{I} (\tau) = \mathcal{O}(\tau,\tau') \mathbf{I}(\tau')$.

If $\mathcal{K}$ is known, it is straightforward to evaluate Equation~\ref{eq:radtransstokeval}. This is the so-called synthesis of Stokes profiles by radiative transfer calculation in an atmosphere of known characteristics. 

The opposite and more interesting approach, i.e. inferring $\mathcal{K}$ from a measurement of $\mathbf{I}_{\nu} = \mathbf{S}_{\nu}$, is called inversion. Unfortunately, this approach is extremely difficult. To calculate $\mathcal{K}$, it is inter alia necessary to know the population densities of the atomic levels participating in the transition. In LTE, they are given by the Saha-Boltzmann equation, but generally depend on the radiation field itself. This self-similarity makes it necessary to solve Equation~\ref{eq:radtransstokeval} iteratively.

Even if LTE can be assumed, there is no analytical expression for $\mathcal{O}$ in general. Thus, Equation~\ref{eq:radtransstokeval} has to be solved numerically. This has been done using Feautrier's method \citep{1964CR....258.3189F}, Runge-Kutta algorithms \citep{1974SoPh...35...11W}, the Diagonal Element Lambda Operator approach \citep{1989ApJ...339.1093R}, Hermitian strategies \citep{1998ApJ...506..805B}, rotations in a Minkowski-like space \citep{1999A&A...350.1089L} or neural networks \citep{2001A&A...378..316C}.

\paragraph{Milne$-$Eddington Approximation:} The Unno-Rachkovsky solution is one of the rare exceptions where Equation~\ref{eq:radtransstokeval} can be evaluated analytically. It is obtained for an atmospheric model, which obeys the following assumptions \citep{degl2004polarization}:

\begin{itemize}
\item{The atmosphere is plane parallel, semi-infinite, and LTE applies.}
\item{All the APs affecting $\mathcal{K}$, i.e. $\eta_0$, a, v$_{\rm{dop}}$, B , $\gamma$ and $\varphi$, 
are constant with $\tau$.}
\item{The Planck function varies linearly with temperature, i.e. $\rm{B}_{\nu}=\rm{B}^0_{\nu}(1+\alpha\tau)$, resulting in linear variation of $\SSS_{\nu}$ with $\tau$.}
\end{itemize}

\begin{figure}[h!]
\begin{center}
    \includegraphics[width=\textwidth]{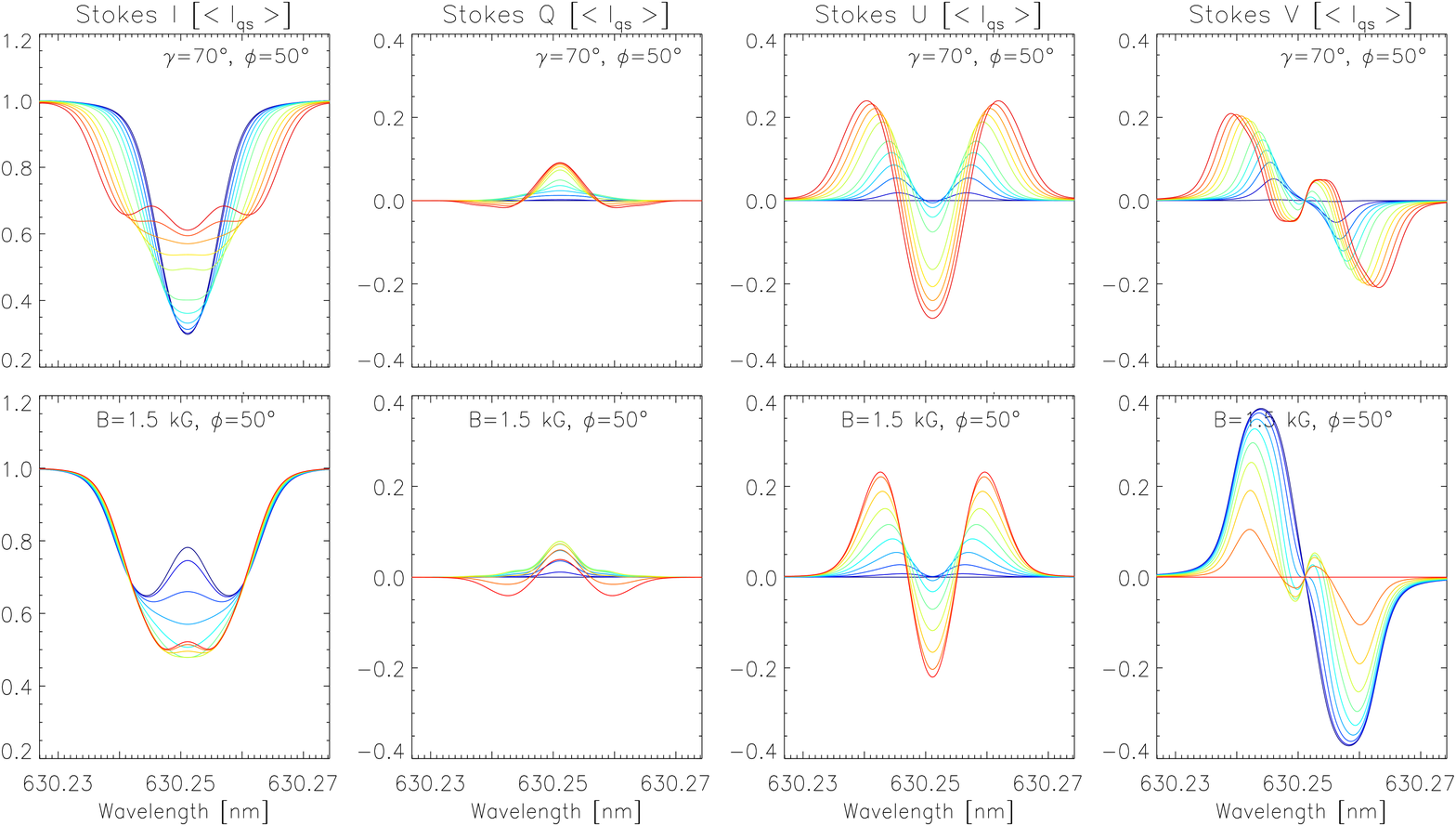}
\caption{Stokes profiles in a Milne-Eddington type atmosphere for a variation of atmospheric parameters. Top row: Magnetic field strength (B) increases from 0~G (blue) to 2.3~kG (red), while zenith angle ($\gamma$) and azimuth ($\varphi$) of the magnetic field  were set to 70$^{\circ}$ and 50$^{\circ}$ respectively. Bottom row: $\gamma$ increases from 0$^{\circ}$ (blue) to 90$^{\circ}$ (red), while B and $\varphi$ were set to 1.5~kG and 50$^{\circ}$ respectively. Adopted from \citet{nazaretstok}.}
\label{fig:varstok}
\end{center}
\end{figure}

\noindent A medium fulfilling these requirements is called a Milne-Eddington atmosphere. Fig.~\ref{fig:varstok} illustrates how the Stokes parameter change within such an atmospheric model if e.g. the strength or the inclination of the magnetic field varies. 

A drawback of the Unno-Rachkovsky solution is the fact that all the APs remain constant throughout the atmosphere by definition. Since the APs change drastically throughout the penumbral photosphere (cf.~Chapter~\ref{ch:asym} and \ref{ch:ncp}), the quality of the physical characteristics of the atmosphere obtained by the Milne-Eddington approximation is,  however, limited.

\paragraph{Stokes Inversion Based on Response Functions (SIR):} SIR \citep{1992ApJ...398..375R} is a package for the synthesis\footnote{In synthesis mode, SIR evaluates Equation~\ref{eq:radtransstok} for a given $\mathcal{K}$, i.e. a given set of APs.} and inversion of spectral lines in the presence of magnetic fields. In contrast to the Milne-Eddington approximation, SIR defines the APs on a grid in optical depth, i.e. $\tau \in \{\tau_0, \, ... \,,\tau_{\rm{i}} \}$,  and allows them to vary with $\tau$ if necessary.

In inversion mode, the algorithm modifies the APs of an initial guess model until the synthetic ($\rm{I}^{\rm{syn}}$) and observed ($\rm{I}^{\rm{obs}}$) profiles converge. Numerically, this is done by minimizing $\chi^2$, i.e. the merit function:

\begin{equation}
\chi^2 \equiv \sum_{\rm{j}=1}^4 \sum_{\rm{k}=1}^{\rm{N}} \left[ \rm{I}_{\rm{j}}^{\rm{obs}} (\lambda_{\rm{k}}) - \rm{I}_{\rm{j}}^{\rm{syn}}(\lambda_{\rm{k}}) \right] ^2 \frac{\rm{w}^2_{\rm{j}}}{\sigma^2_{\rm{}}}
\label{eq:merit}
\end{equation}

\noindent where $\rm{j} \in \{1,2,3,4 \}$ and represents the four components of the Stokes vector. $\rm{k} \in \{1,\,...\,,N \}$ and accounts for the wavelength at which the spectrum was measured. $\sigma$ represents the uncertainty in the observations, and w is a factor which allows for arbitrary weighting of each Stokes parameter \citep{sirmanual}. 

The minimization of Equation~\ref{eq:merit} is carried out iteratively using a Levenberg-Marquardt algorithm \citep{press2007numerical}, which makes use of the partial derivatives of the Stokes parameter with respect to the APs. For the success of the inversion, it is crucial that these derivatives can be expressed in terms of so-called response functions, which describe the resulting change of of the synthetic Stokes profiles due to a perturbation of any AP at a specific $\tau$-level \citep{delToro}. 

In order to reduce the number of free parameters, the perturbations of the depth-dependent APs are calculated only for a number of points (nodes) of the $\log (\tau)$ grid in which the atmosphere is discretized. Each AP is represented by a different set of nodes. The perturbations of the various APs in all the remaining grid points are approximated by a linear or cubic-spline interpolation at the nodes \citep{sirmanual}. Usually, SIR carries out a range of iteration-sets (cycles) in which the number of nodes (free parameters) increases.

\paragraph{Modification to the SIR Code:} In its original version, the SIR code allowed for unphysical values of the APs. 
These solutions were obtained if Stokes profiles with large asymmetries were inverted using a limited number of free parameters, i.e. a linear gradient of the APs with $\tau$. If for example $-50^{\circ} < \gamma  < 580 ^{\circ}$ for $-4 < \rm{log}(\tau) < 0$, the emergent profile shows large asymmetries, even though only a linear gradient with $\tau$ is assumed. However, such a rotational behavior of the magnetic field is a numerical artifact and very unlikely to occur on the Sun.

Therefore, the SIR code was modified, and the variation of the APs was restricted to the following intervals:

\begin{itemize}
\item{For the magnetic field strength : $\rm{B}  > 0 $}
\item{For the inclination of the magnetic field : $0^{\circ} < \gamma  < 180 ^{\circ}$}
\item{For the azimuth of the magnetic field : $0^{\circ} < \gamma  < 360 ^{\circ}$}
\end{itemize}

\noindent Afterwards, the code was recompiled under UNIX, Sun OS and Mac OS systems, using the Intel Fortran compiler. Extensive tests have shown a stable performance without any peculiarities.

\chapter{Data Acquisition and Calibration}
\label{ch:data}

In Section~\ref{sec:hinode_obs}, the space observatory HINODE is introduced, while Section~\ref{sec:polarimeter} gives a detailed explanation of the spectropolarimeter. Section~\ref{sec:reduc} provides a definition of maps of observable quantities as well as a discussion of data reduction issues. Various techniques to determine solar velocities from spectral line shifts are discussed in Section~\ref{sec:measuredoppler}. Two procedures for absolute wavelength calibration are compared in Section~\ref{sec:vcalib}, before the results of this Chapter are summarized in Section~\ref{sec:concl_data}.

\paragraph{Space-borne Observations:} The opacity of the terrestrial atmosphere already demands for space-borne observations in the extreme ultraviolet (EUV) and X-ray regimes to the electromagnetic spectrum (EMS). However, movements of air layers of Earth's atmosphere also cause intermittent distortions in easier accessible parts of the EMS, and lead to a deterioration of the imaging quality. This problem is known as seeing and has been tackled in various ways. Using Adaptive Optics, distortions of wavefronts are analyzed, and the surface of certain telescope mirrors is deformed in real-time to compensate for the seeing. Image quality may be further improved by post-processing of data, e.g. Multi-Object Multi-Frame Blind Deconvolution, Phase-Diversity Reconstruction, or Speckle Imaging. Using these techniques, it is nevertheless still difficult to perform e.g. spectropolarimetry with a signal to noise level suitable to study weak magnetic fields. In space-borne observations, the problem of seeing is avoided altogether, though at the cost of instrument lifetime and a lack of maintenance possibilities. Furthermore, space observatories are more expensive and extremely difficult to modify for additional needs once they have been deployed.

\section{The HINODE Observatory}
\label{sec:hinode_obs}

The space-borne observatory HINODE was launched at 21.32 UT on September 22$^{\rm{nd}}$ 2006. 
It is designed to address scientific questions concerning the heating mechanisms of the corona, the driving mechanism of solar flares as well as the creation of solar magnetic fields. For this purpose, HINODE is equipped with three main instruments: The solar optical telescope (SOT), the X-ray telescope (XRT) and the EUV imaging spectrometer (EIS). The spacecraft circles Earth each 98 minutes in a sun-synchronous orbit with an apogee of 694.23~km and a perigee of 672.24~km. 

Due to the inclination of 98.1$^{\circ}$ between the terrestrial plane of the equator and the orbit of the satellite, the latter precesses around Earth at a rate of 1$^{\circ}$ per day, which allows an uninterrupted observation of the Sun during nine months a year. Around the summer solstice on the northern hemisphere, HINODE experiences an eclipse season during which the Sun is occulted by Earth for a maximum of ten minutes during each orbit \citep{2007SoPh..243....3K}.

\paragraph{Extreme Ultraviolet Imaging Spectrometer:} The EIS is an off-axis parabolic telescope with a focal length of 1.9~m, a mirror diameter of 0.15 m and a toroidal diffraction grating in a normal incidence optical layout. The multi-layer mo\-lyb\-de\-num-silicon coatings have high reflectivity in two wavelength bands at 19$\pm$2 nm and 27$\pm$2~nm. These are observed simultaneously by two back-illuminated CCDs with a spectral resolution of 4000 and 4600, respectively. Many EUV emission lines from the transition region, the corona and flares are contained in these wavelength ranges, and up to 25 spectral windows can be selected. Two-dimensional maps are obtained in a raster observation with different slit widths, i.e.~1\arcsec\,\,and 2\arcsec\,\,in the narrow mode and 40\arcsec\,\,and 266\arcsec\,\,in the wide mode. The center of the field of view (FOV) can be changed by $\pm$890\arcsec\,\,in the east-west direction, which allows both to observe high-altitude regions of the corona at the limb and to see the region near the limb when the nominal observing region of HINODE is located near the center of the Sun \citep{2007SoPh..243...19C,2007SoPh..243....3K}.

\paragraph{X-Ray Telescope:} The XRT is a high resolution Wolter-I grazing incidence telescope with an aperture of 0.34~m and a focal length of 2.7~m. It consists of the X-ray optics, a focal plane mechanism (nine filters and a shutter) to choose a wavelength band between 0.2~nm and 20~nm as well as a back-illuminated CCD. Additionally, the XRT contains optics to focus visible light at 430.5~nm on the same detector to aid the co-alignment of XRT and SOT images. The CCD is mounted on an adjustable stage with a stroke of $\pm$1~mm along the optical axis to allow refocusing in orbit. The image plane of the grazing incidence optics is extremely curved, and the focus adjustment allows to choose either on-axis maximum resolution with rapid off-axis degradation (Gaussian focus) or a focus position that provides resolution as uniform as possible over a larger FOV \citep{2007SoPh..243....3K}.

\begin{figure}[h!]
	\centering
		\includegraphics[width={\textwidth}]{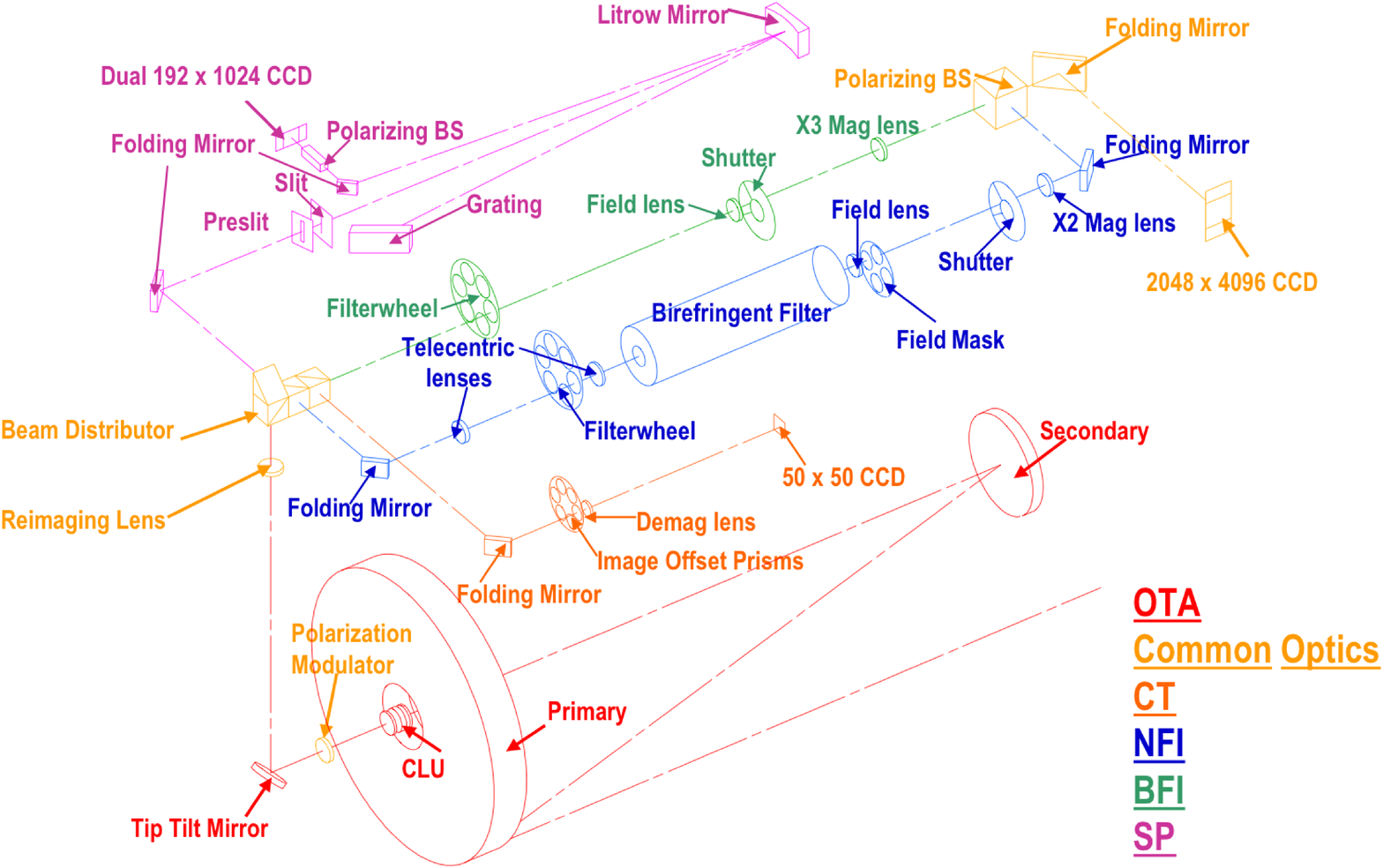}
		\caption{Optical layout of HINODE SOT. Adopted from \citet{2008SoPh..249..167T}}
		\label{fig:hinode}
\end{figure}


\paragraph{The Solar Optical Telescope Assembly:} The SOT consists of the optical telescope assembly (OTA) and the focal plane package (FPP). The OTA is an aplanatic Gregorian f/9 telescope with a clear aperture of 0.5~m that operates at the diffraction limit, achieving a spatial resolution from 0.\arcsec2 to 0.\arcsec3 between 388~nm and 668~nm. It was designed to minimize chromatic aberration and instrumental polarization. 
A heat dump mirror is installed in the primary focus to reflect unused sunlight 
back into space. At the secondary focus, a field stop limits the FOV to 361\arcsec\,\,by 197\arcsec. To ensure a compact design and to deliver parallel light to the FPP, a collimating lens unit with a focal length of 0.37~m is placed at the center of the primary mirror. Located behind the primary mirror, the polarizing modulator unit (PMU) is followed by the piezo-driven tip-tilt mirror. The latter ensures image stability as high as 0.\arcsec01 (RMS) in the horizontal and the vertical direction \citep{2008SoPh..249..197S,2008SoPh..249..221S,2008SoPh..249..167T}.

\paragraph{The Focal Plane Package:} After the tip-tilt mirror, the light is fed into the FPP via a reimaging lens. Using a beam distributor unit, the light is guided to the broadband filter imager (BFI), the narrowband filter imager (NFI) as well as to the spectropolarimeter (SP) and the correlation tracker. The SP is explained in Section~\ref{sec:polarimeter}, while the properties of the filtergraphs are summarized in the following.

\paragraph{Narrowband Filter Imager:} The NFI uses the same CCD camera as the BFI and was designed to provide full Stokes polarimetry in a range of spectral lines. To this end, a range of filters (FWHM = 0.6 nm) can be accessed to pre-select a spectral window via a wheel in under three seconds. A Lyot filter with a bandwidth of 9.5 pm at 630 nm is used to scan through various lines and the nearby continuum. The image has a size of 264\arcsec\,\,by 166\arcsec\,\,and exposure times range from 0.1 second to 1.6 second. The properties of the NFI are summarized in Table~\ref{tab:hinode_nfi}.

Dopplergrams are obtained either in fast or in normal mode, i.e. images taken at two or four wavelengths spaced equally within the line. To minimize necessary computer power, onboard memory calculates sums, differences and ratios of these images, i.e. $\rm{r}=(\lambda_1+\lambda_2-\lambda_3-\lambda_4)/(\lambda_1-\lambda_2-\lambda_3+\lambda_4)$. These are then converted to velocity, i.e. v(r), via a lookup table. Line of sight magnetograms are constructed from the I over V ratio of eight NFI filtergrams in a magnetic sensitive line. They have an RMS noise of approximately 10$^{15}$ Mx pixel$^{-1}$ and a cadence of 20 seconds. If NFI images are not only obtained at multiple wavelengths within the line, but also at different positions of the PMU, all Stokes parameters may be reconstructed by adding and subtracting the respective images. The Stokes parameters can be used to construct vector magnetograms with cadences ranging from 1.6 seconds to 4.8 seconds, depending on the size of the image (readout time of the CCD) and the exposure time \citep{2008SoPh..249..167T}.

\begin{table}[h]
\begin{center}
	\begin{tabular}{cccc}
		\hline
		\hline
		\\[-2ex]
		{Center [nm]} & {Spectral Line [nm]}  & {g$_{\rm{eff}}$} & {Scientific Purpose}\\
		\hline
		{517.2} & {Mg I b 517.27} & {1.75} & {Doppler and magnetograms}\\
		{} & {} & {} & {in lower chromosphere}\\
		{525.0} & {Fe I 524.71}& {2.00} & {Photospheric magnetograms}\\
		{--''--} & {Fe I 525.02} & {3.00} & {Together with Fe I 525.06}\\
		{--''--} & {Fe I 525.06} & {1.50} & {line ratio magnetograms}\\
		{557.6} & {Fe I 557.61} & {0.00} & {Photospheric Dopplergrams}\\
		{589.6} & {Na D 589.6} & {1.50} & {Weak and chromospheric fields}\\
		{630.2} & {Fe I 630.15} & {1.67} & {Photospheric magnetograms}\\
		{--''--} & {Fe I 630.25} & {2.50} & {Photospheric magnetograms}\\
		{--''--} & {Ti I 630.38} & {0.92} & {Sunspot umbra magnetograms}\\
		{656.3}& {H$_{\alpha}$ 656.28} & {} & {Chromospheric Structures}\\
		\hline	
	\end{tabular}
	\caption{Properties of filters of the NFI and their scientific purpose \citep{2008SoPh..249..167T}.}
	\label{tab:hinode_nfi}
\end{center}
\end{table}

After the launch of HINODE, it was realized that the NFI contains blemishes which degrade the image in some parts of the FOV. They are caused by air bubbles that move around in the fluid of the Lyot filter if the line is scanned. To suppress the image degradation, four out of eight tuning elements had to be blocked, and observation over extended periods of time is only possible for a small number of wavelengths within a singe spectral line. Rapidly switching between different lines is impossible, and Doppler shifts due to orbital motion cannot be corrected during the eclipse season \citep{2008ASPC..397....5I}.

\paragraph{Broadband Filter Imager:} The main component of the BFI is a filter wheel equipped with six interference filters that allow to image a FOV covering 218\arcsec\,\,by 109\arcsec. Typical exposure times range from 0.03 seconds to 0.8 seconds at a cadence of less than 10 seconds. The spectral properties of the different filters and the solar features that can be observed to address a range of scientific questions are summarized in Table~\ref{tab:hinode_bfi}.

\begin{table}[h]
\begin{center}
	\begin{tabular}{cccc}
		\hline
		\hline
		\\[-2ex]
		{Center [nm]}&{FWHM [nm]} & {Spectral Feature} & {Scientific Purpose}\\
		\hline
		388.35 & {0.7} & {CN}& {Magnetic network}\\
		396.85 & {0.3} & {Ca II H}& {Chromospheric heating}\\
		430.50 & {0.8} & {CH}& {Magnetic elements}\\
		450.45 & {0.4} & {blue continuum}& {Photospheric temperature}\\
		555.05 & {0.4} & {green continuum}& {--''--}\\
		668.40 & {0.4} & {red continuum}& {--''--}\\
		\hline
	\end{tabular}
	\caption{Properties of filters of the BFI and their scientific purpose \citep{2008SoPh..249..167T}.}	
	\label{tab:hinode_bfi}
\end{center}
\end{table}

\section{The Spectropolarimeter}
\label{sec:polarimeter}

The SP is build to introduce as little polarization as possible in order to minimize crosstalk between the Stokes parameters. It consist of the PMU, an entrance slit and an Echelle grating in an off-axis paraboloidal Littrow design optimized for a wavelength of 630~nm. The reimaging lens at the entrance of the FPP forms an image of the Sun at the spectrograph slit. The latter is 12~$\mu$m wide and can be displaced by a stepper motor to enable raster scans. After the slit, the light is reflected onto the grating by a Littrow collimating mirror with a focal length of 925~mm. This configuration is used to achieve a compact design and to avoid stray light from the additional surface of a focusing lens. The Echelle grating with 79 grooves per mm works in high order (36) and imparts very little polarization on the light. On return from the grating, the two orthogonal polarization stages of the dispersed light are spatially separated by a polarizing beam splitter (PBS) and projected onto different regions of the CCD. This so-called dual beam polarimeter approach is used to suppress motion-induced crosstalk in order to achieve a high S/N ratio \citep{2001ASPC..236...33L,Lites2010}.


\begin{table}[h]
\begin{center}
	\begin{tabular}{lll}
		\hline
		\hline
		\\[-2ex]
		{Science Requirement}& {As-Built} & {Scientific Purpose}\\
		\hline
		{Spectral Resolution} & {2.400 pm} & {Line width in umbra}\\
		{Spectral Sampling} &{2.155 pm pixel$^{-1}$} & {Resolve Stokes profiles}\\
		{Spectral Coverage} & {0.239 nm (normal)} & {Cover both Fe I lines @ 630.2 nm}\\
		{} & {0.477 nm (max)} & {plus high velocity events}\\
		{Spatial Coverage} & {162\arcsec\,N-S} & {Size of an active region}\\
		{} & {328\arcsec\,E-W} & {}\\
		{S/N (I$_{\rm{c}}$)} & {10$^{-3}$ in 4.8 sec} & {Active region fields}\\
		{Spatial Sampling} & {0.1585\arcsec\, along slit} & {Critical sampling of SOT}\\
		{} & {0.1476\arcsec\, slit step} & {resolution @ 630.2 nm}\\
		{Cadence} & {10 Hz} & {Avoid motion-induced crosstalk}\\
		\hline
	\end{tabular}
	\caption{Properties and scientific purpose of the SP \citep{2009ASPC..415..323C,Lites2010}.}	
	\label{tab:hinode_sp}
\end{center}
\end{table}

The measurement of the intensity of light that has passed through adequate optical elements allows to determine the degree of polarization, cf. Section~\ref{sec:pollight}. In the SP onboard HINODE, these elements are the PMU and the PBS. The PMU utilizes a bifringent Quartz-Sapphire waver that retards the extraordinary polarization plane 
with respect to the ordinary, thereby transforming circular into linear polarized light. A waver with a specific thickness is used to obtain equal retardance at 517~nm and 630~nm \citep{Elmore2004}. The waver rotates in the PMU at 5/8~Hz, causing a modulation of the polarization. Since the CCD operates with a 10~Hz frame transfer rate, 16 pictures are taken during a full revolution of the PMU, which corresponds to two basic modulation cycles \citep{1987ApOpt..26.3838L}. In normal mode, pictures from three full revolutions (six modulation cycles) are accumulated to achieve a polarization precision of 10$^{-3}$. The CCD measures not only a sinusoidal intensity variation, which is caused by the modulated polarization and the PBS, but also a phase shift between the two orthogonal polarization stages of light. These harmonic variations allow to extract all Stokes parameters, using a so-called demodulation scheme. Various schemes exist \citep{1987ApOpt..26.3838L,Beck2002,2008SoPh..249..233I}, but they all work by subsequently adding and subtracting images depending on the parameter and the rotational phase of the modulator.

\paragraph{Polarization Calibration:} Instrumental polarization is minimized by the design of the SP, but it can never be fully avoided. Mirrors, lenses and polarizers change the polarization state of light, lead to crosstalk and spurious polarimetric signals. Within the framework of the  M\"uller calculus \citep{collet}, this can be expressed as ${\bf{S'}}={\bf{M}}{\bf{S}}$, where {\bf{S}} represents the Stokes parameters describing the incident light, which is modified into {\bf{S'}} by the response matrix {\bf{M}} of the measuring device. Therefore, it is necessary to calibrate the instrument by subsequently measuring {\bf{M}}, determining its inverse {\bf{M$^{-1}$}} and calculating ${\bf{S}}={\bf{M^{-1}}}{\bf{S'}}$ to be able to eliminate instrumental polarization during the data reduction. This was conducted prior to the deployment of HINODE by feeding sunlight into the SOT through a range of polarizers that introduce a well-known state of polarization. The deviation of the polarization from the initial state of polarization measured by the SP allows to determine the elements of {\bf{M}}. The elements of the response matrix were determined multiple times with an accuracy sufficient to suppress crosstalk between the Stokes parameters below a statistical noise level of 10$^{-3}$ \citep{2008SoPh..249..233I}.

\section{Data Reduction and Image Reconstruction}
\label{sec:reduc}

Disturbances and errors occurring during data acquisition as well as instrumental polarization may be corrected a-posteriori in the data reduction process. These errors involve, for example, intensity variations over the field of view due to different properties of the individual pixels of the CCD or imperfect optical elements. They are usually corrected during the flat-field procedure, where a gain-table is constructed for each pixel of the CCD \citep{2006PhDT........30B}. Additionally, the electron and read-out noise of the CCD is taken into account by subtracting the dark current from each observation. For HINODE data this calibration process is automatized and only points important for the further analysis shall be discussed here.

\begin{figure}[h!]
	\centering
		\includegraphics[width={\textwidth}]{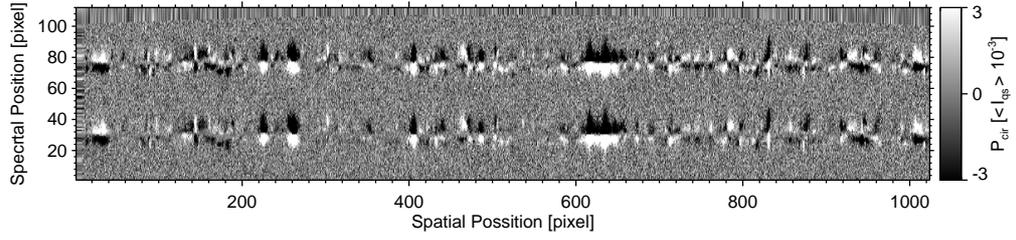}
		\caption{Stokes V spectrum along the slit saturated at the $3\sigma$ noise level. Note the top of the image, where a pattern of black and white stripes form a semicircle.}
		\label{fig:spec_shift}
\end{figure}

\paragraph{Spatial and Spectral drift:} A small focal length, which is due to the compact design of the HINODE SP, leads to a convex deformation of the spectral images (cf.~Fig.~\ref{fig:spec_shift}). Furthermore, thermal variations occur within the SP as the zenith angle of the Sun changes during the orbit of HINODE. These variations cause a drift of the image on the CCD. The shift occurs both: Spatially along the slit as well as in spectral direction, perpendicular to the slit. Another factor which adds to the displacement in the spectral direction is the orbital velocity of HINODE, which causes significant Doppler shifts (cf.~Section~\ref{sec:measuredoppler}).

\begin{figure}[h!]
	\centering
		\includegraphics[width={0.8\textwidth}]{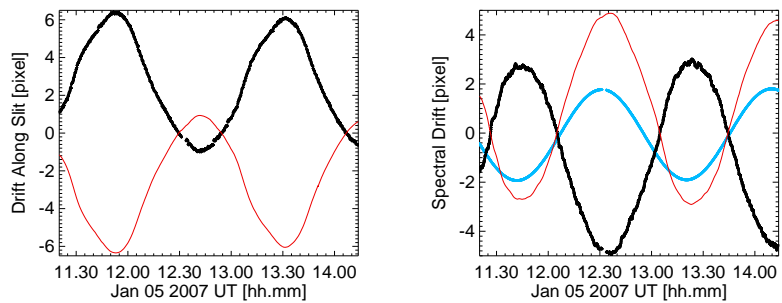}
		\caption{Left: Thermal drift of image along the slit of the SP. Measurements (black) and correction curve (red). Right: Same as left, but spectral drift due to thermal variation and orbital motion of HINODE (blue).}
		\label{fig:shift_temp}
\end{figure}

In the left plot of Fig.~\ref{fig:shift_temp}, the thermal variation of the image across the slit is shown exemplarily for the time of observation of  Spot 04 (cf.~Table~\ref{tab:data}). The black curve marks the measured shift due to thermal expansion and shrinkage of the SP. The red curve is derived from a fit to the data and used to correct for the shift. The plot on the right depicts the shift in spectral direction. Here, the Doppler shift is indicated by the blue curve.

Due to these drifts, the CCD does not always record the entire image. As an example, a Stokes V image from a QS region of data set Spot 04 is shown in Fig.~\ref{fig:spec_shift}. In the spatial dimension, black and white stripes are present on the left side of the image for $1 < \rm{pixel} < 10$. In the spectral dimension these stripes form a semicircle for $100 < \rm{pixel} < 112$. The stripe pattern is visible because the value of the last measurement available is interpolated to the edge of the image. This is of particular interest in the case of Stokes V, from which the total net circular polarization (NCP) is calculated. If these spurious signals are considered in the calculation of the NCP, they spoil the measurement as they are either positive or negative on the blue or red side of the profile.

\subsection*{Measures of Polarization}
\label{sec:measure_poli}

From the Stokes profiles, observables that allow 
to study e.g. the photospheric temperature or the strength and orientation of the magnetic field, can be computed.

Continuum images are constructed by computing an average value of Stokes I in the continuum and normalizing it to the average QS value. 

\begin{equation}
\rm{Continuum\,\,Intensity} =: \rm{I} = \frac{\rm{I}_{\rm{c}}}{< \rm{I}_{\rm{qs}} >}
\end{equation}

Total circular polarization (P$_{\rm{cir}}$) and total linear polarization (P$_{\rm{lin}}$) allow an estimate of the magnetic field component parallel and orthogonal to the LOS. They are computed from Stokes I, Q, U and V as follows:

\begin{equation}
\rm{Total\,\,Circular\,\,Polarization} =: \rm{P}_{\rm{cir}} = \int_{\lambda_1}^{\lambda_2} \frac{| \rm{V(\lambda)} |} {\rm{I}_{\rm{c}}} \rm{d}\lambda
\end{equation}

\begin{equation}
\rm{Total\,\,Linear\,\,Polarization} =: \rm{P}_{\rm{lin}} = \int_{\lambda_1}^{\lambda_2} \frac{\sqrt{\rm{Q}^2(\lambda)+\rm{U}^2(\lambda)}} {\rm{I}_{\rm{c}}} \rm{d} \lambda
\end{equation}

\noindent Note that in the case of $\rm{P}_{\rm{cir}}$, the integration of $| \rm{V(\lambda)} |$ leads to an increase of noise. To avoid this, the integration may be performed separately for both lobes.

Within the weak field limit, i.e. the two lobes of the Stokes V signal are not completely separated, the total polarization (P$_{\rm{tot}}$) may serve as a proxy of the magnetic field strength. It is calculated as:

\begin{equation}
\rm{Total\,\,Polarization} =: \rm{P}_{\rm{tot}} = \int_{\lambda_1}^{\lambda_2} \frac{\sqrt{\rm{Q}^2(\lambda)+\rm{U}^2(\lambda)+\rm{V}^2(\lambda)}} {\rm{I}_{\rm{c}}} \rm{d} \lambda
\end{equation}

\subsection*{Measures of Stokes Asymmetries}
\label{sec:measure_asym}

The deviation of Stokes I from a symmetric Voigt profile or the deviation of Stokes V from complete antisymmetry 
is commonly referred to as asymmetries of the respective Stokes profile. Theses asymmetries are of special interest as they contain additional information about unresolved structures within the resolution element, the variation of the atmospheric parameter with $\tau$ or a combination of both.
Asymmetries in Stokes I profiles may be measured, for example, by using the bisector method (cf.~Section~\ref{sec:bisec}). In the case of Stokes V profiles
, measures of asymmetry include:

\begin{equation}
\rm{Amplitude\,\,Asymmetry} =: \delta \rm{a} = \frac{\rm{a}_{\rm{r}} - \rm{a}_{\rm{b}}}{\rm{a}_{\rm{r}} + \rm{a}_{\rm{b}}}
\label{eq:ampasym}
\end{equation}
with $\rm{a}_{\rm{r}}$ and $\rm{a}_{\rm{b}}$ representing the amplitude of the red and the blue lobe, respectively. Even though Stokes V is antisymmetric, the amplitude of the respective lobes is considered to be a positive quantity. Note that $\delta\rm{a} \not= 0$ does not necessarily imply gradients along the LOS.

\begin{equation}
\rm{Total\,\,Net\,\,Circular\,\,Polarization} =: \mathcal{N} = \int_{\lambda_1}^{\lambda_2}{\frac{\rm{V}(\lambda)}{\rm{I}_{\rm{c}}}}\rm{d}\lambda
\label{eq:ncp}
\end{equation}

\noindent with $\lambda_1 < \lambda_2$ defining the wavelength interval which includes the entire spectral line. If [$\lambda_1, \lambda_2$] includes more than one spectral line, $\mathcal{N}$ may be referred to as Total Broad Band Polarization. Note that $\mathcal{N}$ is commonly expressed as an equivalent width with the unit of a length. Contrary to $\delta\rm{a}$, non vanishing values of $\mathcal{N}$ imply at least gradients of the velocity along the LOS \citep{1978A&A....64...67A}. Furthermore, the amplitude of $\mathcal{N}$ is influenced by the gradients with height of other atmospheric parameters such as the zenith angle and the azimuth of the magnetic field \citep{1996SoPh..164..191L}.

\begin{equation}
\rm{Area\,\,Asymmetry} =: \mathcal{A} = \frac{\rm{A}_{\rm{r}} - \rm{A}_{\rm{b}}}{\rm{A}_{\rm{r}} + \rm{A}_{\rm{b}}} = \mp1 \cdot \int_{\lambda_1}^{\lambda_2}\frac{{\rm{V}(\lambda)}} {\rm{|V}(\lambda)|} \rm{d}\lambda
\label{eq:areaasym}
\end{equation}

\noindent where $\rm{A}_{\rm{r}}$ and $\rm{A}_{\rm{b}}$ represent the area\footnote{Similar to amplitude (cf.~Equation \ref{eq:ampasym}), the area of the red and blue lobe of Stokes V is considered a positive quantity.} of the red and the blue lobe of Stokes V respectively. The factor $\mp1$ represents the polarity of the magnetic field with respect to the LOS. The first expression of $\mathcal{A}$ works only for regular Stokes V profiles with two lobes, while the second formulation allows to include profiles with three lobes as shown in the right panel in Fig.~\ref{fig:sketchstok}. $\mathcal{A}$ can be seen as $\mathcal{N}$ normalized to the area of the Stokes V profile. Thus, $\mathcal{A}$ is a measure of the degree of asymmetry of the Stokes V profile, while $\mathcal{N}$ is also sensitive to the apparent strength of the magnetic field. 

\section{Measurement of Doppler Velocity}
\label{sec:measuredoppler}


Spectral lines contain information about the atmospheric parameters -- i.e. the physical conditions -- within the line forming region (LFR). One of these parameters is the velocity of the solar plasma. The component of the motion occurring parallel to the line of sight can be measured directly because frequency and length of any wave are influenced by the relative motions between its emitter and receiver. This phenomenon is known as the Doppler effect. It may be expressed as:
\begin{equation}
\lambda' = \lambda - \frac{\rm{v}}{\nu}
\end{equation}
with $\lambda'$ being the wavelength at the receiver, while the source moves with a certain velocity v and emits a wave with length $\lambda$ and frequency $\nu$. In the case of light, it is possible to write: 
\begin{equation}
\frac{\delta\lambda_{\rm{dop}}}{\lambda} :=\frac{\lambda-\lambda'}{\lambda} = \frac{\rm{v}}{\rm{c}}
\label{eq:doppler}
\end{equation}
since $\lambda\nu=\rm{c}$, with c representing the speed of light. The quantity $\delta\lambda_{\rm{dop}}$ is thus a measure of the velocity between the reference frames of emitter and receiver.

Due to plasma velocities within the LFR, solar absorption lines are displaced in wavelength when compared to the respective line measured in vacuum. However, relative motions of the receiver (in this case the HINODE SP in its orbit around Earth), the rotation and even the gravitational potential of the Sun cause non-negligible Doppler shifts that have to be taken into account in the calibration and data reduction procedures (cf.~Section~\ref{sec:vcalib}).


Different parts of an absorption line form in different layers of the solar atmosphere (cf.~Section~\ref{sec:radtrans}). The line-core, the darkest part of the absorption line, forms at the top of the LFR, while the line-wings contain information about the lower LFR close to the solar surface. Thus, the velocity derived from $\delta\lambda_{\rm{dop}}$ significantly depends on the procedure used to deduct the Doppler shift of the absorption line. In the following, a range of methods to calculate $\delta\lambda_{\rm{dop}}$ are introduced, drawing attention to their advantages and drawbacks.

\subsection*{Center of Gravity}

This method calculates the barycenter of a spectral line. It is a robust and fast way to estimate line shifts. Since it takes the entire spectral line  into account, it is insensitive to any line asymmetries and yields only an average value of the velocity within the LFR.

\subsection*{Fit of Line-Core}
\label{sec:fitcore}

Velocities in the higher atmospheric layers are visible as a shift of the core of the absorption line. The most simple way to calculate the wavelength position of the core is to take the position of the measurement with minimal intensity. Due to the finite spectral resolution of the instrument, this position does not necessarily coincide with the actual position of the core. A better approach is to fit a second order polynomial to a range of measurements around the line minimum and approximate the line-core to sub pixel accuracy. 

However, the value of the minimum of the fit depends on the number of points taken into account. This is shown exemplarily in the upper part of Table~\ref{tab:core}, where an average Stokes I profile from the QS was fitted with different amounts of data points. For example, the polynomial fit to Fe I 630.15 nm changes by 0.14 pixel -- corresponding to a velocity of roughly 0.14 km s$^{-1}$ -- if 9 instead of 3 data points are considered. Due to the spectral resolution of the HINODE SP, the core of the line is usually sampled by 7 measurements. For a larger number of data points, the fit is thus more and more influenced by the wings of the line, which cannot be approximated by a simple polynomial fit. Drawback of this method is its insensitivety to the lower layers of the LFR and a lower quality of the fit in the presence of magnetic fields, which broaden or even split the line-core.


\begin{table}[h]
\begin{center}
	\begin{tabular}{ccc}
		\hline
		\hline
		\\[-2ex]
		{} &{Fe I 630.15 nm}&{Fe I 630.25 nm}\\
		{\# Data Points}&{Core Position [pixel]} & {Core Position [pixel]}\\
		\hline
		\multicolumn{3}{c}{Polynomial Fit to Line-Core}\\
		\hline	
		1 & {29.00} &  {75.00}\\
		3 & {29.06} &  {75.28}\\
		5 & {28.99} &  {75.12}\\
		7 & {28.94} &  {75.14}\\
		9 & {28.92} &  {75.12}\\
		11 & {28.92} &  {75.11}\\
		13 & {28.92} &  {75.11}\\
		\hline
		\multicolumn{3}{c}{Gaussian Fit to Line}\\
		\hline
		{13} & {28.91} & {75.19}\\	
		{17} & {28.91} & {75.19}\\
		{21} & {28.91} & {75.19}\\
		\hline
	\end{tabular}
	\caption{Influence of the number of data points considered on the position of the line-core as approximated by a polynomial and a Gaussian fit.}	
	\label{tab:core}
\end{center}
\end{table}

Another possibility is to approximate the line-core by a single Gauss\footnote{An absorption line profile is actually a Voigt function, i.e. the convolution of a Lorentz and a Gauss function (cf. Section~\ref{sec:radtrans}).} profile. However, this method yields wrong results if the line is asymmetric, and it completely fails if the number of data points considered is insufficient. The results for a Gauss fit to a symmetric Stokes I profile of the QS are shown in the lower part of Table~\ref{tab:core}. A minimum of 13 data points is necessary to reach convergence, but the position of the line-core is not sensitive to the number of data points considered in the fit. The results from the Gauss fit and from a polynomial fit with 7 data points differ by not more than 0.05 pixel (0.05 km s$^{-1}$).

\subsection*{Bisector}
\label{sec:bisec}

\begin{figure}[h!]
	\centering
		\includegraphics[width={\textwidth}]{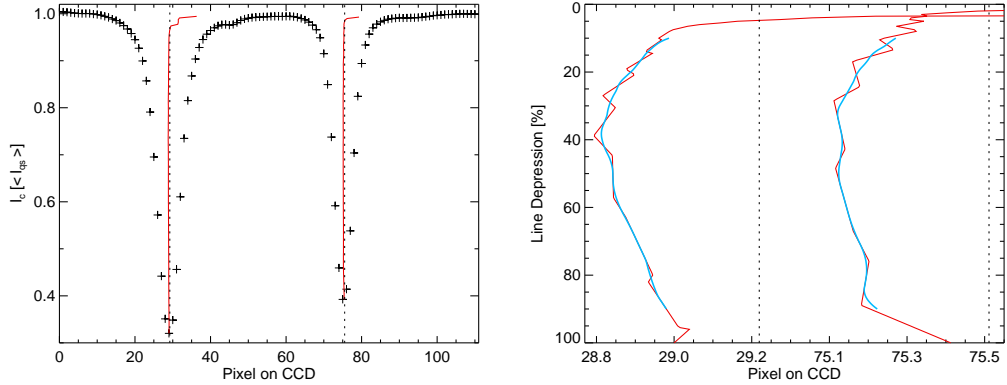}
		\caption{Left: Bisector (red) for an averaged profile (black crosses) of Fe I 630.15 (left) and Fe I 630.25 (right) from the QS. Dashed lines represent the position where the vacuum wavelength of the respective line would be located on the CCD camera. Right: Bisectors (red) for the same QS profile but in a narrow spectral range. The blue curves represent mean values averaged in an interval of $\pm$5\% line depression.}
		\label{fig:bisec_expl}
\end{figure}

The line representing the central position of an absorption profile at all line depths is called bisector. One example is depicted on the left side of Fig.~\ref{fig:bisec_expl}, where the bisector of an average QS profile is drawn in red. If the absorption profile is symmetric, the bisector is a straight line through the axis of symmetry. The right panel of Fig.~\ref{fig:bisec_expl} shows only a narrow spectral window around the core of both lines. It illustrates that the bisector is not only shifted towards the blue, 
but that it is also bent towards lower wavelengths at intermediate line depths resembling the shape of a C. This discrepancy is due to the convective blueshift (CBS), which will be discussed in Section~\ref{sec:averageQS}.

Another feature illustrated in Fig.~\ref{fig:bisec_expl} is the zigzag pattern of the bisector. It is a numerical artifact due to the finite spectral resolution of the HINODE SP, which makes it necessary to interpolate -- in first order linearly -- between the measurements to calculate the central position for all line depths. The numerical problems due to the interpolation can be overcome by computing mean bisector values. The blue line represents such a mean bisector, for which all values in an interval of $\pm$5\% line depression have been combined. However, this method fails close to the line-core -- at line depression values larger than 90\% -- where the exact position of the bisector does not coincide with the values found in Section~\ref{sec:fitcore}. For small line depressions values ($\rm{I}_{\rm{c}} \ge 0.98$), the bisector yields incorrect results as well, since it starts to bend towards higher wavelengths. This is caused by:
\begin{itemize}
\item[a)] The other iron line, which decreases $\rm{I}_{\rm{c}}$ between both lines.
\item[b)] A weak blend present between pixel 40 and 50.
\item[c)] Variations of $\rm{I}_{\rm{c}}$ itself, e.g. between pixel 90 and 105.
\end{itemize}
Extensive tests have shown that, 
the bisector method yields reliable Doppler shifts only between 10\% and 90\% line depression. Furthermore, it is necessary to average the bisector position for a range of line depression values to compensate for the problems arising from the finite spectral resolution.

\paragraph{Flow Velocity in the Deep Photosphere}

\begin{figure}[h!]
	\centering
		\includegraphics[width={\textwidth}]{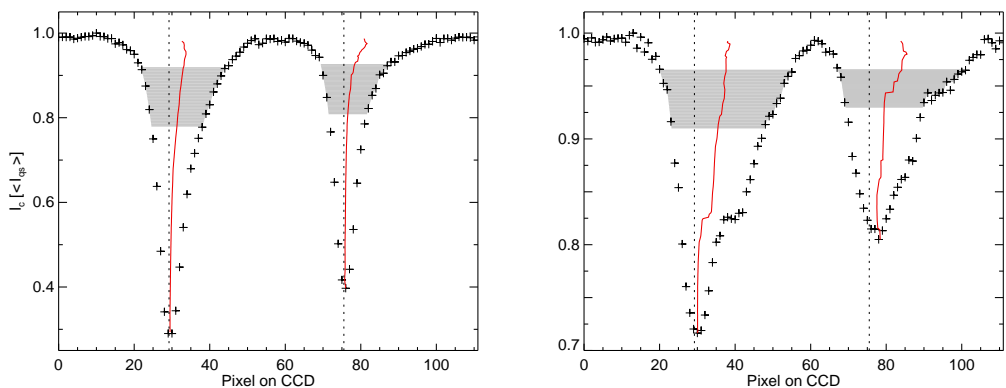}
		\caption{Left: Asymmetric Stokes I profile (black crosses) with computed bisector between 70\% and 90\% line depression (red). Right: Same as on the left, but an example of extreme asymmetry where a second component, a line satellite, is present in the Stokes I profile.}
		\label{fig:bisec_wing}
\end{figure}

Absorption lines usually contain asymmetries, and the shape of the bisector differs from a straight line. As an example, two penumbral profiles are shown in Fig.~\ref{fig:bisec_expl}. In the left panel, the profiles of both Fe lines show a gradually decreasing asymmetry with increasing line depression. The bisector is shifted towards the red in the line-wing and approaches the vacuum wavelength close to the core. The second profile on the right side of Fig.~\ref{fig:bisec_wing} represents a case of extreme asymmetry. In both Fe I lines, there is a shoulder on the red flank, which causes the bisector to be shifted towards higher wavelengths at low line depression values. Close to the line-core, the bisector of the Fe I 630.15 line approaches the vacuum wavelength. In the case of Fe I 630.25, however, it remains significantly displaced from the vacuum position.

These examples demonstrate that the strongest shifts of a solar absorption line occur close to the continuum of the line, indicating that the strongest velocities are present in the deep photosphere. Therefore, it is necessary to compute the position of the line at very small depression values and to use an averaging interval large enough to compensate for the finite spectral resolution. The gray shaded area in Fig.~\ref{fig:bisec_wing} shows the range between 10\% and 30\% line depression, which will be used to calculate the position of the line-wing. Any offset between this position and the position of a special feature of the solar spectrum (cf.~Section~\ref{sec:vcalib}) is interpreted as a Doppler shift caused by the velocities of the plasma. The Doppler maps, e.g. Fig.~\ref{fig:alldop}, are finally obtained by computing the velocities that correspond to the Doppler shifts at any given spatial position.

\paragraph{Molecular Lines in Umbral Profiles}

The atmospheric temperatures in the umbra are low enough to allow for the formation of molecules. The molecular absorption bands mask the two iron lines to an extent which makes it unreliable to compute a bisector and to derive velocities for the line-wing. Thus, all umbral profiles - i.e. $\rm{I}_{\rm{c}} < 0.33 \cdot \rm{I}_{\rm{qs}}$ - are excluded form the calculation of line shifts, and the Doppler velocity in the umbra is manually set to zero (cf.~Fig.~\ref{fig:alldop}).

\subsection*{Double Gaussian Fits}
\label{sec:multigauss}
\begin{figure}[h]
	\centering
		\includegraphics[width={\textwidth}]{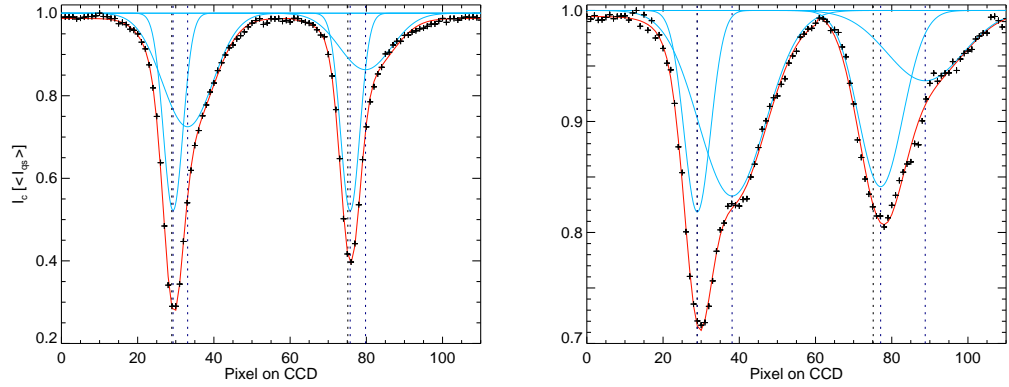}
		\caption{Asymmetric Stokes I profile (black crosses), multiple Gauss functions (solid blue) and their superposition (solid red). The vacuum wavelength of the line is indicated by the dashed black line, while the dashed blue lines mark the center of each Gauss function.}
		\label{fig:gauss_vdop}
\end{figure}

Asymmetric Stokes I profiles 
may be explained by a superposition of a major component and a strongly shifted second component, i.e. a line satellite \citep{Wiehr:1995p379}. This is shown in Fig.~\ref{fig:gauss_vdop}, where the same profiles as in Fig.~\ref{fig:bisec_wing} are evaluated with a double Gaussian fit. The asymmetries of the line are reasonably well reproduced by the superposition of the Gauss functions. 

The shift of the line satellites and bisectors with respect to the position of the core of the line fitted with 7 data points is listed in Table~\ref{tab:comp_bisec_multi}. In the case of a double Gaussian fit, the component with the larger shift was considered as the line satellite. 
The results show that the shift of the bisector is sensitive to the range of line depression values taken into account in the averaging process. The closer this range is to the continuum, the smaller the difference to the shift of the line satellite deduced from a double Gaussian fit becomes. In these examples, the difference between both methods is around 15\% for the Fe I 630.15 line, with a bisector averaged between 10\% and 20\% line depression. In the case of Fe I 630.25, though, this difference is always larger than 55\%. This shows that the bisector --  which is an average of the two line components -- underestimates the shift of the line satellite and can only serve as a lower limit. Even though a double Gaussian fit yields more realistic values, the bisector method will be used in the following.  This is because the double Gaussian fit technique is less robust, time-consuming and its convergence is strongly dependent on the initial conditions. 


\begin{table}[h!]
\begin{center}
	\begin{tabular}{ccccc}
		\hline
		\hline
		\\[-2ex]
		\multicolumn{5}{c}{Shift of Average Bisector [Pixel\footnotemark]}\\
		\hline	
		{Line Depression [\%]} & {Fe I 630.15}&{Fe I 630.25} & {Fe I 630.15}&{Fe I 630.25}\\
		\hline
		{10 -- 20} & {8.03} & {7.44} & {3.30} & {2.20}\\
		{10 -- 30} & {7.52} & {6.07} & {2.89} & {1.78}\\
		{10 -- 40} & {6.95} & {5.36} & {2.50} & {1.48}\\
		{20 -- 30} & {7.03} & {4.72} & {2.47} & {1.35}\\
		{20 -- 40} & {6.42} & {4.32} & {2.10} & {1.12}\\
		{30 -- 40} & {5.79} & {3.89} & {1.73} & {0.89}\\
		\hline
		\multicolumn{5}{c}{Shift of Line Satellite [Pixel$^{3}$]}\\
		\hline
		{} & {9.14} & {11.70} & {3.81} & {4.03}\\	
		\hline
	\end{tabular}
	\caption{Top: The bisector is calculated for the two asymmetric profiles in Fig.~\ref{fig:gauss_vdop} and averaged over different ranges of line depression. It is then compared to the line-core fitted with 7 data points. Bottom: Position of line satellite inferred from a fit with two Gauss functions compared to the core of the respective line.}	
	\label{tab:comp_bisec_multi}
\end{center}
\end{table}
\footnotetext{Due to the spectral sampling of the HINODE SP (cf. Table~\ref{tab:hinode_sp}) the pixel scale may be  approximately converted into a Doppler velocity scale of the unit km~s$^{-1}$. Thus, the velocities found using a double Gaussian fit are clearly supersonic.}


\subsection*{Inversions} 
An inversion is a sophisticated method to extract information about the atmospheric parameters within the LFR from absorption profiles. To this end, the radiative transfer equation is solved for a model atmosphere which yields synthetic line profiles. The synthetic profiles are compared to the observed profiles, and in an optimization process, the model atmosphere is altered to minimize the difference between observed and synthetic profile. Even though the obtained atmosphere is the best fit to the observation, it is not necessarily the real atmosphere, since the solution is not unique. A detailed explanation of this method can be found in Section~\ref{sec:inversion}.

\section{Velocity Calibration}
\label{sec:vcalib}

In order to calculate the Doppler velocities of the solar plasma using Equation~\ref{eq:doppler}, the line shifts of Table~\ref{tab:comp_bisec_multi} have to be converted into $\delta\lambda$, taking the spectral sampling of the HINODE SP (cf.~Table~\ref{tab:hinode_sp}) into account. However, these line shifts are only relative to the position of the line-core of an average QS profile that does not necessarily coincide with the vacuum wavelength of that line (cf.~Fig.~\ref{fig:bisec_expl}). A common method to calibrate the absolute wavelength scale is to use absorption lines of the terrestrial atmosphere \citep{Rezaei:2006p157} or to project a spectral line with a known wavelength into the measured spectrum \citep{1984A&A...134..134K}. However, since HINODE is a space-borne observatory and the SP does not use any reference line, absolute wavelength calibration is only possible using features from the pure solar spectrum. In the following, two methods for absolute wavelength calibration with pure solar spectra are explained, and the results are compared to theoretical values. 

\subsection*{Average Quiet Sun Profile}
\label{sec:averageQS}

The line-core of an average QS profile may serve as a reference to calibrate the wavelength scale. However, the wavelength of a solar photospheric line, whether averaged over sufficiently large areas or obtained with poor spatial resolution, does not correspond to the vacuum wavelength, even if the spectrum is corrected for solar-terrestrial motions and gravitational redshift (cf.~right panel in Fig.~\ref{fig:bisec_expl}). This discrepancy is the CBS, and its value depends on the spectral line \citep{1974kptp.book.....P}, the heliocentric angle \citep{1982ARA&A..20...61D,1988A&AS...72..473B} and even on the line depression, cf.~\citep{1984SoPh...93..219B} and Section~\ref{sec:bisec}.

\begin{figure}[h!]
	\centering
		\includegraphics[width={0.8\textwidth}]{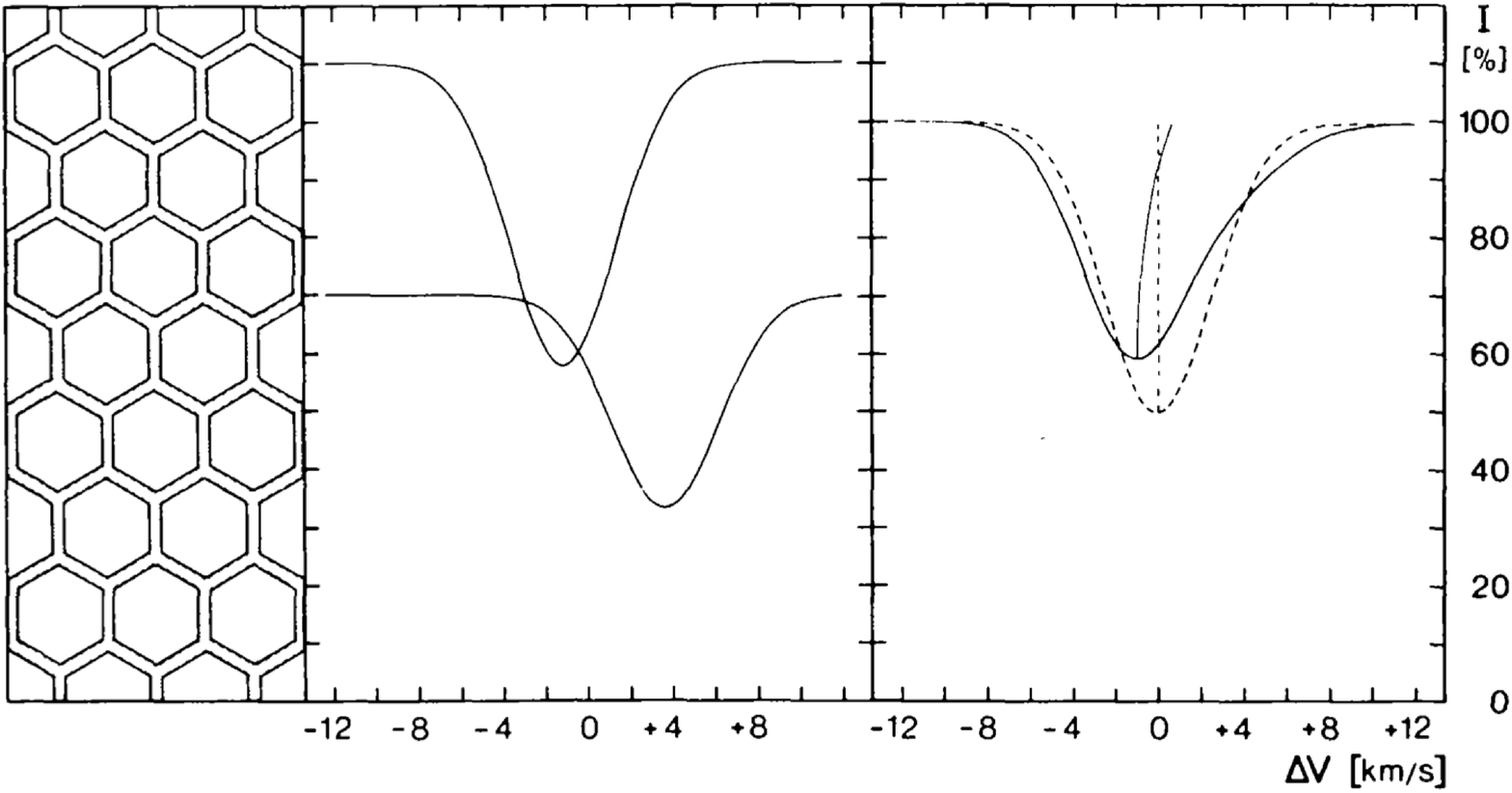}
		\caption{Left: Solar granulation idealized by a honeycomb structure. The hexagon represents the bright granules in between the intergranulum. Middle: Intense profile from the bright granules and weak profile from the dark intergranulum. Right: Superposition of both profiles leads to a shifted and asymmetric profile. Adopted from \citep{1982ARA&A..20...61D}.}
		\label{fig:cbs}
\end{figure}

It has been shown that the CBS of photospheric lines is due to the superposition of bright and blue shifted profiles from the granulum with dark and red shifted spectra found in the intergranulum. Because of their larger intensity, the blue shifted profiles contribute more to an average QS profile, displacing the latter towards lower wavelengths (cf.~\citet{1982ARA&A..20...61D,2004suin.book.....S} and Fig.~\ref{fig:cbs}). For the line-cores of Fe I 630.15 and Fe I 630.25, the CBS at disk center was determined by \citet{2006PhDT........30B} using the SIR code (cf.~Section~\ref{sec:inversion}) with a two-component model atmosphere of \citet{Borrero:2002p209}. The corresponding shifts were calculated to be $-0.18$ pixel and $-0.26$ pixel, which is equal to a blueshift of $-0.39$ pm and $-0.55$ pm or a velocity of $\rm{-0.19}$ km s$^{-1}$ and $\rm{-0.26}$ km s$^{-1}$ respectively. This offset has to be taken into account when using an average QS profile to calibrate the absolute wavelength scale.

Another effect that can spoil the quality of the calibration is the radially outward oriented moat flow near sunspots (cf.~\citet{1972SoPh...25...98S,2008ApJ...679..900V} and Section~\ref{sec:theo_penum}). Because of the residual heliocentric angle in some data sets, an offset is introduced in the average QS profile if only profiles from one side of the spot vicinity are taken into account. This effect is not negligible, and experiments have shown that it reaches values of 0.22 pixel for the two iron lines, depending on the amount and the spatial position of the profiles used to obtain an average QS profile.

\begin{figure}[h!]
	\centering
		\includegraphics[width={\textwidth}]{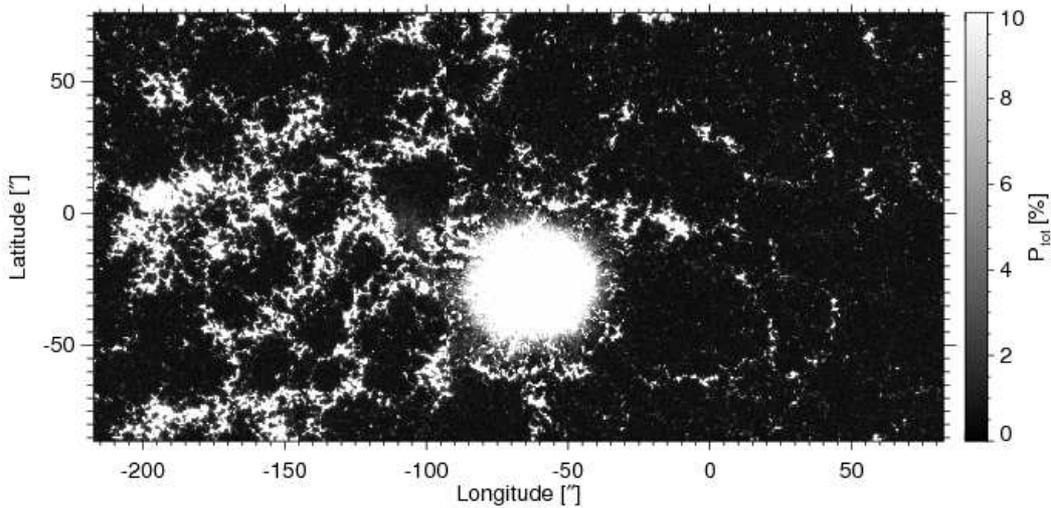}
		\caption{Total polarization in Spot 04 and in the QS, which serves as a mask to exclude profiles with a split core from the average QS profile.} 
		\label{fig:polmap}
\end{figure}

Furthermore, it has to be considered that Fe I 630.15 and Fe 630.25 have a g-factor of 1.67 and 2.5 respectively. Therefore, magnetic fields are able to lift the degeneracy of atomic levels and cause the profile to broaden or even split in several components (cf. Section~\ref{sec:zeeman}). To avoid contamination with split profiles, all pixels above a certain magnetic field strength have to be excluded from the averaging process. To this end, P$_{\rm{tot}}$ is used, which may serve as a proxy for the strength of the magnetic field. In Fig.~\ref{fig:polmap}, this is shown exemplarily for the data set of Spot 04, where the white areas represent pixels with $\rm{P}_{\rm{tot}} > 0.1 \cdot max(\rm{P}_{\rm{tot}})$ that were not considered. A comparison of the map of P$_{\rm{tot}}$ with the continuum intensity yields that the magnetic field in the QS is concentrated in the intergranular lanes. Thus, there is a tendency to exclude redshifted profiles from the averaging process, thereby introducing a systematic bias. However, close investigation yields that this bias is not more than 0.04 pixel for the two lines in all data samples.


\paragraph{Average Value of Line-Core in High Resolution Spectra:} Following the arguments above, the CBS should vanish if an average value of individual line-core position is calculated from data with reasonably high spatial resolution. However, when performing this experiment using Stokes I profiles from data set QS 01, a residual blueshift of $-$0.14 pixel and $-$0.16 pixel is obtained for Fe 630.15 and Fe 630.25. Possible explanations for this behavior include: 1) The finite spectral resolution of HINODE, which may be insufficient to resolve intergranular lanes correctly, and different sizes of up- and downflow regions, which occupy 57.6\% and 42.4\% of the QS area respectively. 2) If a velocity field free of divergence is assumed, mass continuity implies v$_{\rm{IG}}\rho_{\rm{IG}} = \rm{v}_{\rm{G}}\rho_{\rm{G}}$, with velocity and density in the intergranular (v$_{\rm{IG}}$ and $\rho_{\rm{IG}}$) regions as well as in granular (v$_{\rm{G}}$ and $\rho_{\rm{G}}$) regions. Different densities in these region, however, lead to unequal velocities as well. 3) Another issue is the corrugation $\tau$-surface, which shifts the sensitivity to different geometrical heights in the granular and intergranular regions. 

This experiment illustrates the limitations in using an average QS profile for calibration purposes and demonstrate the need to verify the results obtained above by a different method.

\subsection*{Average Umbral Profile}
\label{sec:umbcalib}

\citet{1977ApJ...213..900B} showed that vertical plasma motions do not occur in umbrae of sunspots observed at disk center. Consequently, an average umbral profile is shifted to the red when compared to the line-core of an average QS profile. This does not only allow to substantiate the values given by \citet{2006PhDT........30B}, but it also provides a possibility to calibrate the wavelength scale absolutely. 

The umbral magnetic field, however, is so strong that the core in both Fe lines is split in several components, which makes it impossible to determine the position of the line-core correctly. An alternative is to calculate the zero crossing of an average Stokes V profile. This feature, located midways between the two lobes, corresponds to the line-core of the respective average Stokes I profile. It can be calculated with reasonable precision in antisymmetric Stokes V profiles. Strict antisymmetry of Stokes V implies not only local thermodynamic equilibrium, but also points to the absence of velocity gradients in the line-forming region above the umbra \citep{1978A&A....64...67A} as well as to the absence molecular blends.

\begin{table}[h]
\begin{minipage}{\textwidth}
\begin{center}
	\begin{tabular}{ccccccc}
		\hline
		\hline
		\\[-2ex]
			{Data Set} & \multicolumn{2}{c}{CBS of QS [pixel]} & \multicolumn{2}{c}{RMS pixel]}\\
			{} & {Fe I 630.15} & {Fe I 630.25} & {Fe I 630.15} & {Fe I 630.25}\\
		\hline
		{Spot 01} & {-0.23}& {-0.29} & {0.09} & {0.08}\\
		{Spot 02} & {-0.24}& {-0.31} & {0.09} & {0.08}\\
		{Spot 03} & {-0.20}& {-0.29} & {0.10} & {0.09}\\
		{Spot 04} & {-0.23}& {-0.36} &  {0.14} & {0.10}\\
		\hline
	\end{tabular}
	\caption{Wavelength calibration using umbral Stokes V profiles. The RMS values denote the variation of the center position of umbral Stokes V profiles with $|\delta \rm{a} |< 0.01$.}
	\label{tab:calib}
\end{center}
\end{minipage}
\end{table}

To this end, $\delta \rm{a}$ (cf.~Section~\ref{sec:measure_asym}) was used as a measure of the degree of antisymmetry of Stokes V profiles. The amplitudes of the two lobes were determined from a parabola fitted to the respective extrema, including four adjacent measurements. Only umbral Stokes V profiles with $|\delta \rm{a} |< 0.01$ were used for an average Stokes V profile, from which the zero crossing position was then determined.

The results are summarized in Table~\ref{tab:calib} and show that the average QS is indeed blueshifted with respect to the average center position of umbral Stokes V profiles. Within the listed RMS errors, the values for the CBS agree with the ones reported by \citet{2006PhDT........30B}. Thus, an absolute wavelength calibration using antisymmetric profiles with $|\delta \rm{a} |< 0.01$ yields correct values.

The advantage of this procedure is that it allows for velocity calibration in observation off center \citep{Rezaei:2006p157}, which is not the case if an average QS profile is used. This is because the value of the convective blueshift (CBS) shows a center-to-limb variation (CLV), which has not been measured yet for Fe I 630.15 with an accuracy suitable for calibration. The values reported by \citet{1985SoPh...99...31B} represent only the average of an ensemble of twenty solar lines, and even though simulation of solar convection has recently been used to model the CLV of the CBS \citep{2011A&A...528A.113D}, it still needs to be confirmed by observation.

\subsection*{Error Discussion}
\label{sec:err}
The biggest challenge in the calibration process is to determine the exact position of the line-core of the average QS profile. Due to the finite spectral resolution, the core is sampled by not more than 7 measurements. Using, for example, only 5 data points to fit the core in the data set of Spot 04 changes the CBS of the QS in Fe 630.15 by 0.05 pixel ($+$0.05 km s$^{-1}$). In the QS, significant magnetic fields are predominately found in the intergranular downflow regions. As a result, more pixels showing downflows are excluded from the averaging, thereby introducing a bias to the calibration (cf.~Fig.~\ref{fig:polmap}). This bias leads to an overall blueshift of the maps between -0.01 and -0.04 pixel ($-0.01$ km s$^{-1}$ and $-0.04$ km s$^{-1}$) in all data samples. Spectral drift due to thermal instabilities is corrected during the calibration process. In all the data samples, it is within the range of 0.05 pixels ($\pm$0.05 km s$^{-1}$). Furthermore, Table~\ref{tab:calib} indicates that the values of the CBS of the QS inferred from model calculations and the values obtained from measurements differ by not more than 0.1 pixel ($\pm$0.1 km s$^{-1}$). Thus, the precision of the velocity calibration is $\pm0.1$ km s$^{-1}$.

\subsection*{Differences between Fe I 630.15 and Fe I 630.25}
\label{sec:diff_lines}

Differences between the velocities derived from the shift of an average bisector of the wing of the two Fe lines can arise from the unequal line parameters. 
One example is the formation height of the line-core. In the case of Fe I 630.15, it is $\rm{log}\,(\tau)=-2.9$, which corresponds to a geometrical height of 250 km above the solar surface. The core of Fe I 630.25 forms at $\rm{log}\,(\tau)=-2.0$, which is about 50~km lower \citep{1985SoPh...99...31B}. Thus, the wing of the second line seems to be more suited to measure velocities in the deep photosphere.

However, due to its higher Land\'e g factor, Fe I 630.25 is more sensitive to disturbances by magnetic fields and instrumental crosstalk. 
Radiative transfer calculations \citep{schliche:gross-doert-code} show that the line-wing of Fe 630.15 is most sensitive to atmospheric regions around log$\,(\tau)=-0.6$ regardless of the strength of the magnetic field. In the case of Fe I 630.25, though, the largest contribution to the wing of the line is in atmospheric layers around log$\,(\tau)=-0.5$ if no field is present and changes to layers around log$\,(\tau)=-0.8$ in the case of a magnetic field strength of 2 kG. Since the strength of the magnetic field decreases with height, its influence on the wing of Fe I 630.25 becomes even more important \citep{2007ApJ...659.1726K}.

Large differences in the calculated velocities ($\delta\rm{v_{dop}} > 0.5\,\rm{km\,s}^{-1}$) also occur in the case of extreme asymmetries. This can be seen in the right panel of Fig.~\ref{fig:bisec_wing}. In the case of Fe I 630.15, the line satellite fully contributes to the average bisector,  
while in Fe I 630.25, part of the average bisector is calculated without the second component. The result is a smaller shift 
in the latter line, causing a difference of 1.5 km s$^{-1}$ in the calculated velocities. In some cases, the asymmetries are below 10\% line depression and remain completely undetected in Fe I 630.25. However, only the amplitude of the velocities, but neither the morphology of up- and downflow patterns nor the direction of the flow are affected by these deviations.
\section{Summary}
\label{sec:concl_data}

The spectropolarimetric data used in this work was obtained by HINODE. Its spatial resolution is 0.\arcsec32 and it was sampled with 2.15~pm per pixel in the spectral dimension. From these data, different maps have been computed to infer the properties of the solar atmosphere in the penumbrae of sunspots. While the calculation of e.g. continuum intensity and total polarization is straightforward, the computation of the Doppler velocity is more sophisticated.

The velocity of the solar plasma can be measured along the line of sight via the Doppler effect from the shift of Fe I 630.15 and Fe I 630.25 respectively. In this Chapter, a range of possible measuring methods were introduced and their limitations and shortcomings discussed. 
The shift of an average bisector between 10\% and 30\% of line depth was finally used to calculate the Doppler velocity in the deep photosphere. 

Two independent procedures were used to calibrate the wavelength scale of the HINODE spectropolarimeter. The first method used the line-core of an average Stokes I profile of the quiet Sun, which was corrected for the convective blueshift. In the second approach, the center position of umbral Stokes V profiles with an amplitude asymmetry of less than 1\% was used to define a frame of rest on the solar surface. Within the uncertainties of $\pm$0.1~km~s$^{-1}$, both methods yield the same results.

\chapter{Observation of the Penumbral Velocity Field at Disk Center}
\label{ch:center}

This Chapter is dedicated to the investigation of the vertical component of the Evershed flow on spatial scales of 0.\arcsec3 ($\approx$ 240 km). Section~\ref{sec:global} discusses the radial dependency of the penumbral velocity field using azimuthally averaged values and summarizes the results of statistical studies of its global properties. It is followed by a comparison of the morphology of individual up- and downflow patches in Section~\ref{sec:morphology}. The predictions of the gappy penumbral model with respect to the plasma flows are studied globally in Section~\ref{sec:rolltype} and on a local scale, within individual filaments, in Section~\ref{sec:filaandflow}. The degree of correlation between continuum intensity and Doppler velocity is investigated in Section~\ref{sec:corr_center}, and the properties of bright penumbral downflows, which could account for the results of the correlation study, are provided in Section~\ref{sec:brightdownflow}. To conclude, the results of this chapter, especially the constrains they put on penumbral models, are summarized in Section~\ref{sec:concl_obs_center}.

\section{Global Velocity Field}
\label{sec:global}

\begin{figure}[h!]
	\begin{center}
		\includegraphics[width=1\textwidth]{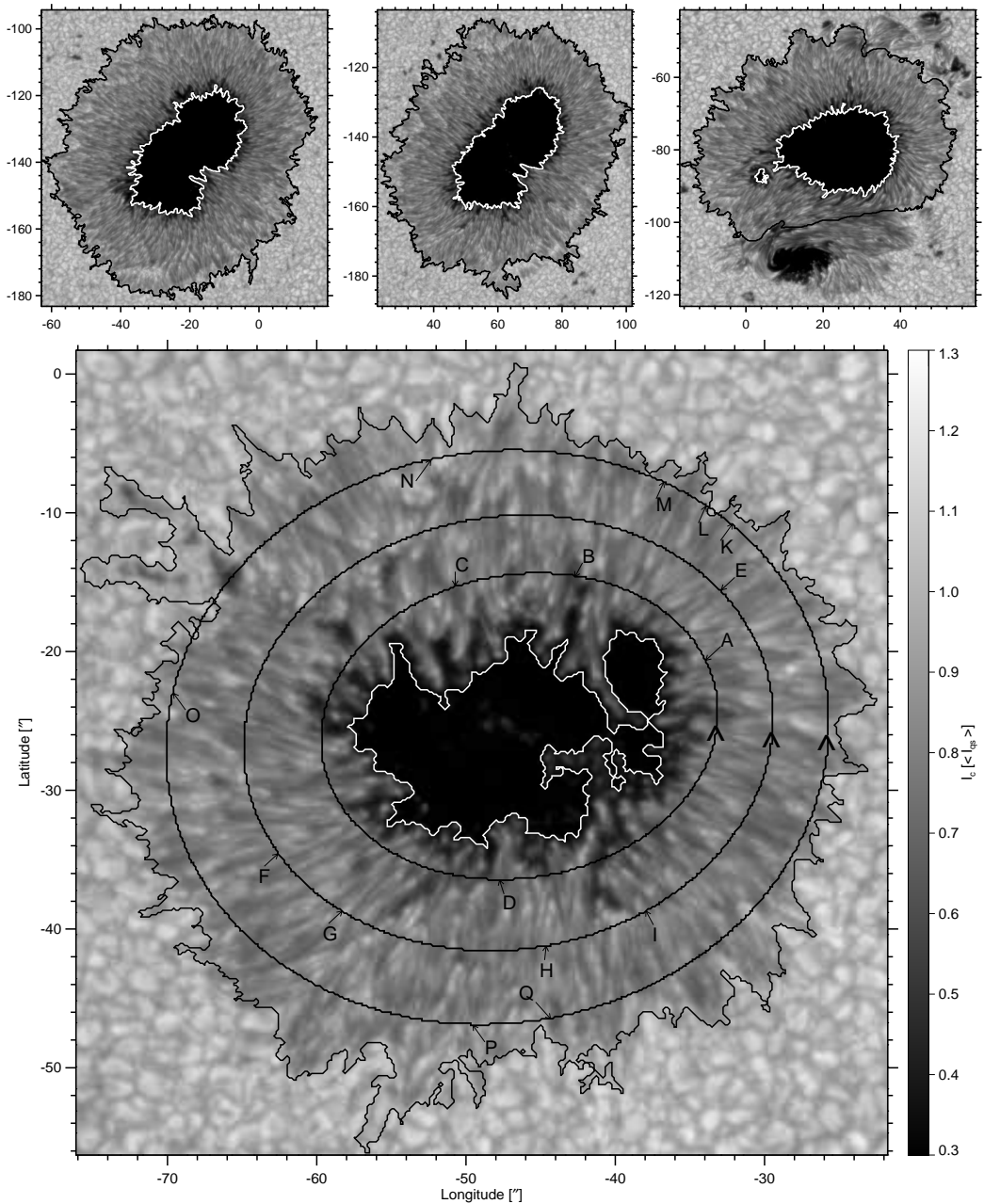}
		\caption{Photospheric continuum intensity in sunspots close to disk center. The behavior of I$_{\rm{c}}$ and v$_{\rm{dop}}$ is studied along the ellipse drawn in the lower map. Special features are indicated by letters and discussed in detail in Section~\ref{sec:morphology} and~\ref{sec:rolltype}.}
		\label{fig:alli}
		      \end{center}
\end{figure}

\begin{figure}[h!]
	\begin{center}
		\includegraphics[width=1\textwidth]{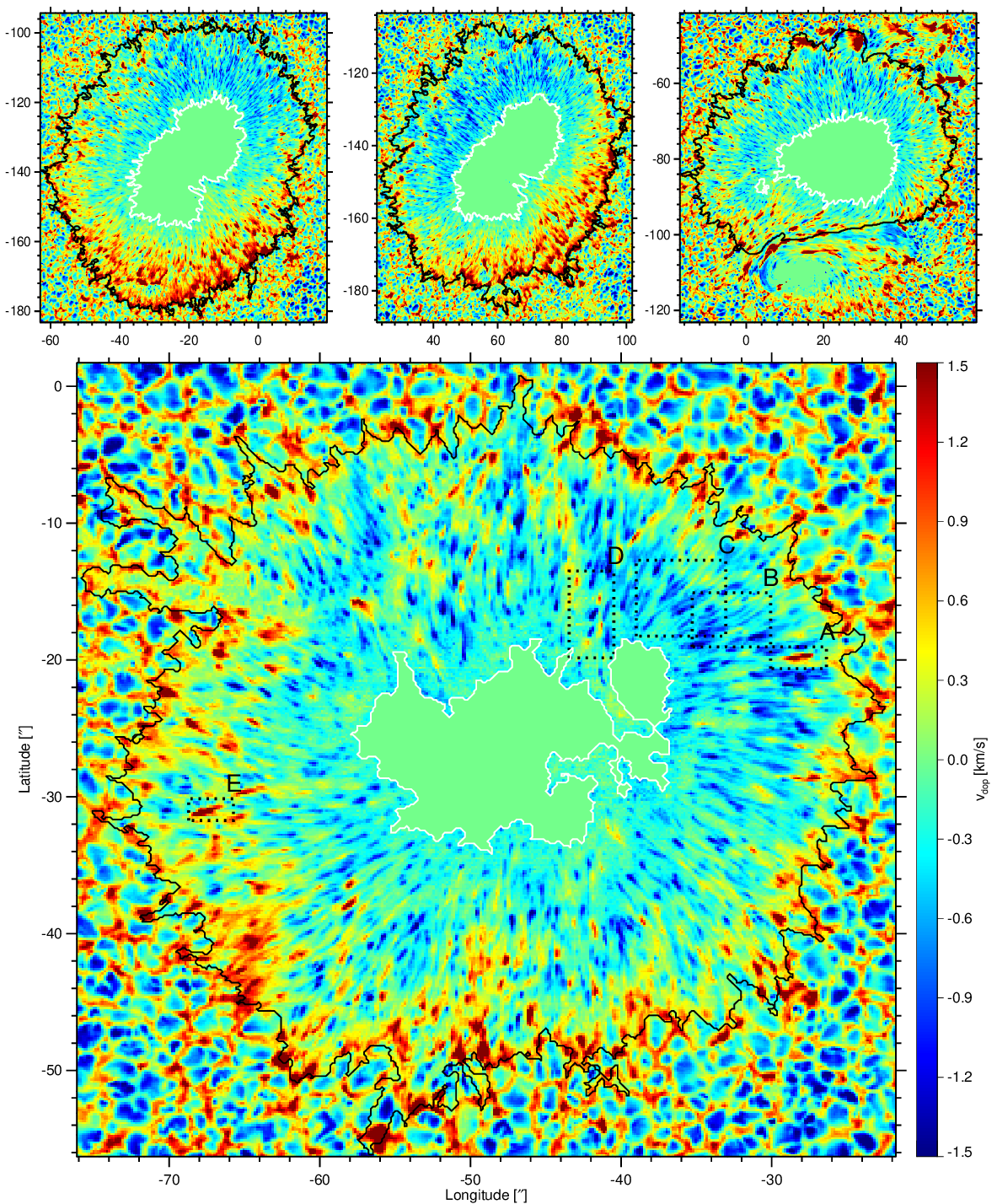}
		\caption{Same as Fig.~\ref{fig:alli} but photospheric Doppler velocities derived from the shifts in the line-wing of Fe I 630.15 nm. Upflows of increasing velocity are shown in green blue and violet, downflows of increasing strength are illustrated in green, yellow and red. Regions indicated as A -- E define areas studied in detail in Sections~\ref{sec:morphology}, \ref{sec:filaandflow} and \ref{sec:brightdownflow}.}
		\label{fig:alldop}
		      \end{center}
\end{figure}

All spectropolarimetric data used in this study were recorded with the SOT of the HINODE observatory and a exposure time of 4.8~s per slit position. The maps obtained with this instrument contain a gradient in time over the field of view. This is because the entrance slit of the instrument is successively scanned across the extended object, and the individual spectra are combined a-posteriori in the data reduction process. The gradient in time is approximately 29~s per second of arc perpendicular to the orientation of the slit, resulting in a time difference of about 30~min between the left and the right edge of the large image\footnote{It should be noted that the rotation of the Sun yields 9.\arcsec5 per hour on the equator and the central meridian. This effect alters the position of the solar feature on the disk, and has not been taken into account for the calculation of the solar coordinates in e.g. Fig. \ref{fig:alli} and \ref{fig:alldop}.} in Fig.~\ref{fig:alldop}.

\paragraph{Interpretation of Penumbral Doppler Maps:} Measuring the vertical component of the EF is only possible if a sunspot is observed exactly at disk center. For residual heliocentric angles, projection effects cause an azimuthal asymmetry between the center side penumbra (CSP), which is blueshifted, and the limb side penumbra (LSP), which is redshifted (cf. small inlets in the top row of Fig.~\ref{fig:alldop}). This is because the Doppler effect allows for measurements of velocity only parallel to the LOS. Thus, the amplitude of the vertical component of the EF decreases $\propto \cos(\Theta)$ and becomes contaminated with the projection of the horizontal component $\propto \sin(\Theta)$. Since the strength and the orientation of the EF is a-priori unknown, it is impossible to remove the contamination for a specific pixel a-posteriori. Nevertheless, even then it is feasible to extract the vertical component of the EF along  azimuthally averaged paths in the penumbra \citep{Schlichenmaier:2000p225}. 

The dominant contribution to the EF is a radial and horizontal outflow. Thus, the penumbral velocity field perpendicular to the line of symmetry\footnote{The line of symmetry connects the center of a sunspot with the center of the solar disk.} does not suffer from projection effects and allows for quantitative measurements of the vertical velocities at all heliocentric angles. For moderate heliocentric angles blueshifts in the LSP and redshifts in the CSP can be considered as real up- or downflows, as only their amplitude but not their direction is altered by projection effects, given that no reverse EF exist in the photosphere.

\paragraph{Description of Observational Data:} Table~\ref{tab:data} summarizes the data sets investigated in this chapter\footnote{A complete list of all the calibrated data sets available, including information on e.g. date of observation and heliocentric angle ($\Theta$), can be found in Appendix~\ref{ch:list}.}. The names given in the first column will be used throughout this work instead of the AR number of the National Oceanic and Atmospheric Administration (NOAA). The last column gives the $\mu$ values, i.e. cos($\Theta$),  of the penumbral and QS areas respectively. 

\begin{table}[h]
\begin{center}
	\begin{tabular}{cccc}
		\hline
		\hline
		\\[-2ex]
		{Name}&{NOAA} & {Date of} & {Heliocentric Angle}\\
		{}&{Active Region} & {Observation} & {cos$(\Theta$)}\\
		\\[-2ex]
		\hline		
		\\[-2ex]
		Spot 01 & {10923} & {Nov 14$^{\rm{th}}$ 2006} & {0.982 - 0.996}\\
		Spot 02 & {10923} & {Nov 14$^{\rm{th}}$ 2006} & {0.980 - 0.994}\\
		Spot 03 & {10930} & {Dec 11$^{\rm{th}}$ 2006} & {0.991 - 0.999}\\
		Spot 04 & {10933} & {Jan 05$^{\rm{th}}$ 2007} &{0.996 - 1.000}\\
		QS 01 & {...} & {Mar 10$^{\rm{th}}$ 2007} & {0.970 - 1.000}\\
		QS 02 & {...} & {Sep 06$^{\rm{th}}$ 2007} & {0.988 - 1.000}\\
		\hline
	\end{tabular}
		\caption{Data samples used in this study.}
	\label{tab:data}
\end{center}
\end{table}

Fig.~\ref{fig:alli} displays the continuum intensity (I$_{\rm{c}}$) of the four sunspots under study. Spot 01 to Spot 03 are depicted in the small inlets in the top row, while the large frame below shows Spot 04. The axis labels give the position on the solar disk in a heliocentric coordinate\footnote{Even though these coordinates do not correspond to the solar longitude and latitude, they may be derived from them.} system \citep{hinode:keywords}. The white and black contours mark the border between umbral and penumbral as well as between penumbra and QS respectively. Spot 03 contains a highly twisted satellite sunspot in the Southern periphery ($\delta$-configuration), which makes it necessary to draw the contour of the outer penumbra by hand, since it cannot be detected via pattern recognition.

Fig.~\ref{fig:alldop} shows the same sunspots, but this time the Doppler velocity is plotted. All maps are saturated at $\pm$ 1.5~km~s$^{-1}$, and the velocities in the umbra ($\rm{I}_{\rm{c}} < 0.33 \cdot \rm{I}_{\rm{qs}}$) are set to zero (cf. Section~\ref{sec:bisec}). The inner and outer penumbral boundaries are indicated by the white and black contours, respectively.

\paragraph{Extreme and Average Penumbral Velocities:} Table~\ref{tab:vdop_comp} summarizes the extreme and average values of the penumbral velocity field in all data samples as well as the amount of up- and downflows.  For reasons of comparison, two maps showing quiet Sun granulation at disk center -- QS 01 and QS 02 -- are evaluated in the same way. Contrary to the maximal upflow speed, the maximum downflow velocity shows variation of up to 80\%. Note that the former changes on a temporal scale of hours. It peaks around 9~km~s$^{-1}$ in the first observation (Spot 01), and 9 hours later, when the spot crossed the central meridian (Spot 02), the maximum downflow speed is about 7~km~s$^{-1}$.

When compared to the penumbra, the maximal downflow velocity (3~km~s$^{-1}$) in the QS is significantly lower, while the maximal upflow velocity ($-$3~km~s$^{-1}$) is about 50\% higher.


\begin{table}[h!]
\begin{center}
	\begin{tabular}{cccccccc}
		\hline
		\hline
		\\[-2ex]
		{Name} & \multicolumn{2}{c}{Extreme} & \multicolumn{2}{c}{Amount}  &  \multicolumn{2}{c}{Average}\\
		{of} & \multicolumn{2}{c}{v$_{\rm{dop\:630.15}}$ [km~s$^{-1}$]} & \multicolumn{2}{c}{v$_{\rm{dop\:630.15}}$ [\%]} &  \multicolumn{2}{c}{$\langle {\rm{v}_{\rm{dop}} \rangle}$ [km~s$^{-1}$]}\\
		{Data Set} & {Up} & {Down} & {Up} & {Down} & {Fe I 630.15} & {Fe I 630.25}\\
		\\[-2ex]
		\hline		
		\\[-2ex]
		{Spot 01} & {-2.17} & {9.02} & {48} & {52} & {0.14} & {0.11} \\
		{Spot 02} & {-2.26} & {7.19} & {52} & {48} & {0.08} & {0.11} \\
		{Spot 03} & {-2.43} & {8.02} & {60} & {40} & {-0.01} & {-0.01}\\
		{Spot 04} & {-2.50} & {5.10} & {57} & {43} & {0.09} & {0.08}\\
		{QS 01} & {-3.10} & {3.19} & {58} & {42} & {-0.16} & {-0.13}\\
		{QS 02} & {-3.13} & {2.93} & {55} & {45} & {-0.11} & {-0.10}\\
		\hline
	\end{tabular}
	\caption{Extreme up- and downflow velocities in the penumbrae as well as in the QS as computed from Fe I 630.15. Additionally, the amount of up- and downflows in the data sets is noted. The average velocity of the data sets is inferred from both Fe lines.}
	\label{tab:vdop_comp}	
\end{center}
\end{table}

\paragraph{Penumbral Redshift:} On a spatial average, mean values are obtained which correspond to a downflow velocity around 0.1~km~s$^{-1}$ for the penumbrae (except for the $\delta$-sunspot in the third data sampe). This finding is at odds with blueshifted penumbrae found in earlier investigations \citep{Schlichenmaier:2000p225}. 

It could be argued that this redshift is caused by the larger heliocentric angle of the LSP when compared to the CSP. This configuration favors redshifts as the contamination of up- and downflows with the horizontal component of the Evershed flow is larger on the LSP. This argument, however, is disproven by the fact that mean values computed for the penumbral quadrants perpendicular to the line of symmetry are significantly redshifted as well.

Another, more plausible explanation for this discrepancy is the irregular shape of the outer penumbra, which makes it difficult to define the boundary between penumbra and QS. Since most downflows are located there, the mean value depends on the criteria chosen to define the outer penumbral boundary. 

\begin{figure}[h!]
	\centering
		\includegraphics[width={0.7\textwidth}]{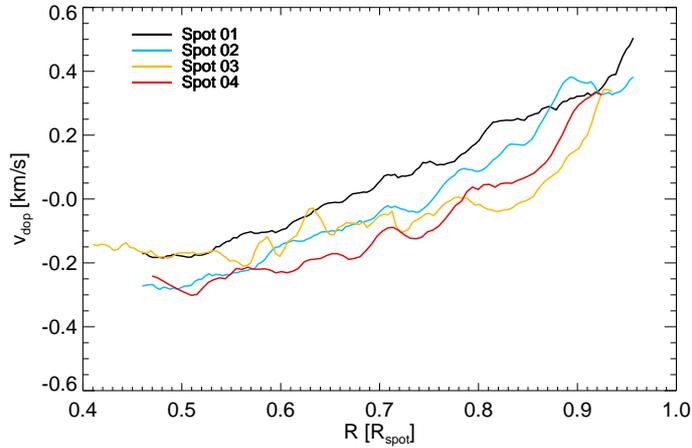}
		\caption{Radial dependency of azimuthally averaged up- and downflow velocity inferred from Fe I 630.15.}
		\label{fig:vdop_r}
\end{figure}

\paragraph{Radial Dependency of Azimuthally Averaged Velocity:} Fig.~\ref{fig:vdop_r} illustrates the mean values of the vertical velocity, which has been averaged along azimuthal paths, as a function of normalized radial distance from the center of the sunspot. It was ensured that only penumbral pixel were considered in the averaging process.

The plot shows  that in the inner penumbra, around 0.5 R$_{\rm{spot}}$, there is an upflow with velocity amplitudes between -0.15~km~s$^{-1}$ and -0.30~km~s$^{-1}$ is present. The mean upflow decreases with radial distance and changes its direction for 0.7 R$_{\rm{spot}} \le \rm{R}  \le 0.8$ R$_{\rm{spot}}$, where it becomes a downflow. In the outer penumbra, around 0.95 R$_{\rm{spot}}$, the average downflow reaches values between 0.33~km~s$^{-1}$ and 0.45~km~s$^{-1}$.

\begin{figure}[h!]
	\centering
		\includegraphics[width={\textwidth}]{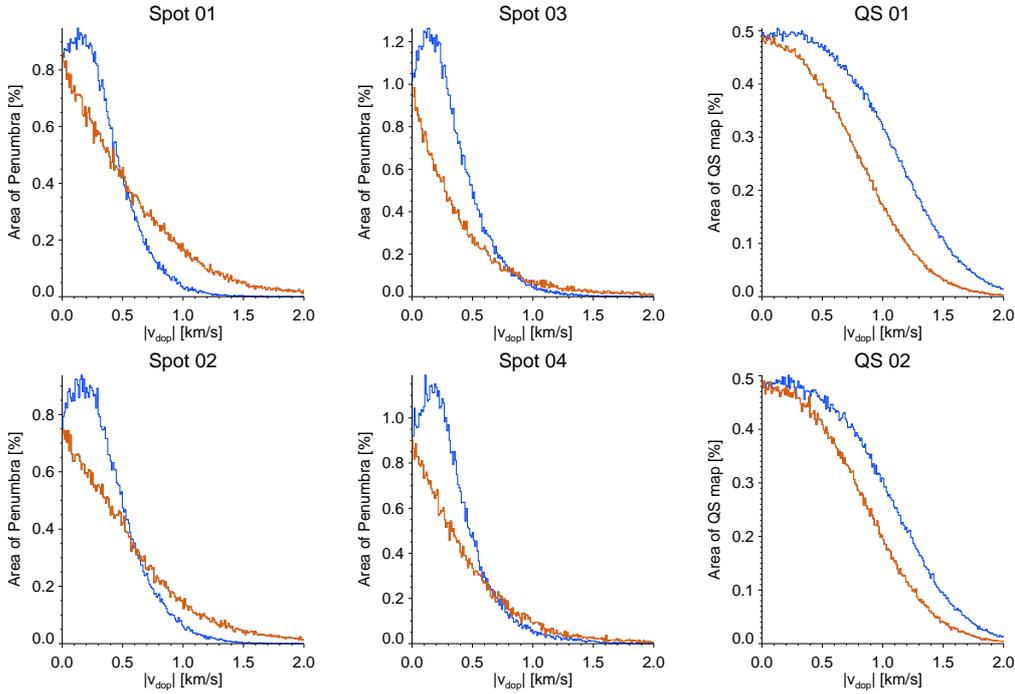}
		\caption{Histogram of $\rm{v}_{\rm{dop}}$ in penumbral (left and middle column) and quiet Sun (right column) observation. In all the plots, the error on upflows (blue) and downflows (red) is $\pm$0.1~km~s$^{-1}$.}
 		\label{fig:histo-sunspot}
\end{figure}

\paragraph{Probability Distribution of Up- and Downflows:} 
To investigate the penumbral velocity field in a statistical manner, probability distribution functions (PDFs) are plotted in Fig.~\ref{fig:histo-sunspot}. 
To visualize their skewness, i.e. the different amount of up- and downflows with a specific strength, only absolute values of velocity are plotted on the abscissae. 
Due to the uncertainty in the wavelength calibration (cf. Section~\ref{sec:err}), no unambiguous flow direction can be derived for $|\rm{v}_{\rm{dop}}| < 0.1$~km~s$^{-1}$.


For all sunspot data and low velocities, $|\rm{v}_{\rm{dop}}| < $ 0.4~km~s$^{-1}$, the area occupied by upflows is larger than that of downflows. The area showing blueshifts slightly increases and peaks around $-$0.2~km~s$^{-1}$, which seems to be a preferred velocity for penumbral upflows. For higher velocities, this area decreases gradually and is equal in size to the downflow area at velocities of 0.5~km~s$^{-1} \le |\rm{v}_{\rm{dop}}| \le 0.7$~km~s$^{-1}$. In Spot 03, the two lines intersect around 0.8~km~s$^{-1}$, which could be attributed to its $\delta$-configuration. The amount of penumbral upflow regions with $\rm{v}_{\rm{dop}} \le -1.5 \rm{km\,s}^{-1}$ is negligible.

The penumbral area showing downflows, by contrast, decreases monotonously towards higher velocities, shows no maximum and dominates for $|\rm{v}_{\rm{dop}}| > $ 0.8~km~s$^{-1}$. At a velocity of $\rm{v}_{\rm{dop}} \ge 1.5$~km~s$^{-1}$, a significant part of the penumbra is still occupied by downflows. The PDFs representing the velocity field of the QS have a different shape (cf. right column in Fig.~\ref{fig:histo-sunspot}). In the QS, the area occupied by upflows is always larger than the area showing downflows, regardless of velocity. Finally, the FWHM of the velocity PDFs in the QS are roughly 1~km~s$^{-1}$, while the FWHM of the penumbral velocity PDFs is approximately 0.5~km~s$^{-1}$. In a statistical sense, these observations are consistent with the results of magnetohydrodynamic simulations of sunspot penumbrae by \citet{2011ApJ...729....5R}, where the RMS velocities in the penumbra are about 50\% of that of the QS.



\section{Morphology of Up- and Downflows}
\label{sec:morphology}

To avoid contamination of the Doppler signal caused by the vertical flows with the horizontal component of the EF, only Spot 04 will be used for a quantitative discussion of the small-scale velocity field, as it shows the smallest heliocentric angle of all data sets. Nevertheless, the morphology of up- and downflow features is qualitatively compared to the other data samples.

If the full maps of $\rm{{v}}_{\rm{dop}}$ and $\rm{I}_{\rm{c}}$ are compared, it becomes apparent that upflows appear predominately, but not exclusively in the inner penumbra. In the Doppler map, upflows are visible as radially aligned, elongated patches separated by areas without any vertical plasma motion. They encompass peak velocities around $-$2.2~km~s$^{-1}$. A typical upflow area is about 2.\arcsec4 long and 0.\arcsec5 wide, resulting in a length/width ratio of about 5. 

Strong downflows with peak velocities of up to 5.0~km~s$^{-1}$ are generally located at the penumbra-QS boundary, while weaker downflows ($0\,\rm{km\,s}^{-1} \le \rm{v_{dop}} \le 2\,\rm{km\,s}^{-1}$) appear at all radial distances. None of the downflows appears elongated, they are roundish instead. The length of the downflow areas ranges from 1.\arcsec3 to 3.\arcsec0, while their width extends from 1.\arcsec3  to 2.\arcsec0. The downflow length/width ratio of 1 to 1.5 is therefore smaller than for the upflow areas. In the inner and mid penumbra, the upflows seem to correlate with the filamentary structure. This is no longer the case in the outer penumbra, where the downflow patches appear roundish and show no filamentary structure. A careful qualitative analysis of the other data sets yields the same picture.


\paragraph{Downflows in the Inner Penumbra:} There have been reports of downflows within the penumbral boundaries \citep{WestendorpPlaza:1997p238,delToroIniesta:2001p3738}, but HINODE data show localized up- and downflow patches at all radial distances.  

\begin{figure}[h!]
	\centering
		\includegraphics[width={0.87\textwidth}]{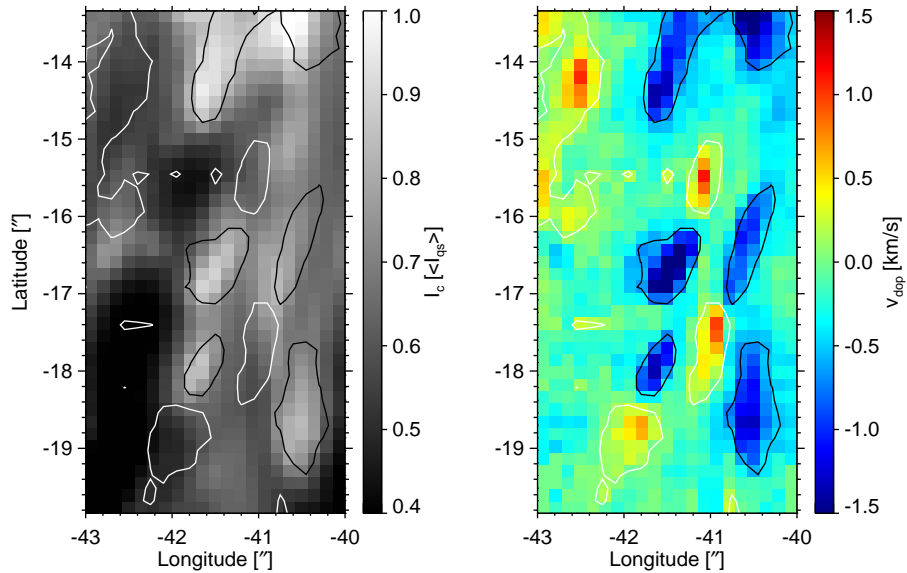}
		\caption{From left to right: $\rm{I}_{\rm{c}}$ and $\rm{{v}}_{\rm{dop}}$ in subsection D of Fig.~\ref{fig:alldop}. Black and white contours encircle flows with amplitudes of $-$0.6~km~s$^{-1}$ and 0.1~km~s$^{-1}$.}
		\label{fig:boxD}
\end{figure}

This is shown exemplarily in Fig.~\ref{fig:boxD}, i.e. an enlargement of the region D of Fig.~\ref{fig:alldop}, where strong downflows with $\rm{v}_{\rm{dop}} \ge 1$~km~s$^{-1}$ can be observed. The pattern of up- and downflow patches with amplitudes of $\rm{v}_{\rm{dop}} \ge \pm$1~km~s$^{-1}$ is anticorrelated with $\rm{I}_{\rm{c}}$. The up- and downflows stretching from $\rm{(x;y)}=(-42$\arcsec$;-19$\arcsec$)$ to $\rm{(x;y)}=(-41$\arcsec$;-15$\arcsec$)$ are remarkable in the sense that, in contrast to granulation, downflows do not exist around the hexagonal-shaped upflows, but appear patchy. There is no difference in shape between up- and downflows, and they appear radially aligned like pearls on a string -- cf.~\citet{2009ASPC..415..369F}.

\begin{figure}[h!]
	\centering
		\includegraphics[width={0.87\textwidth}]{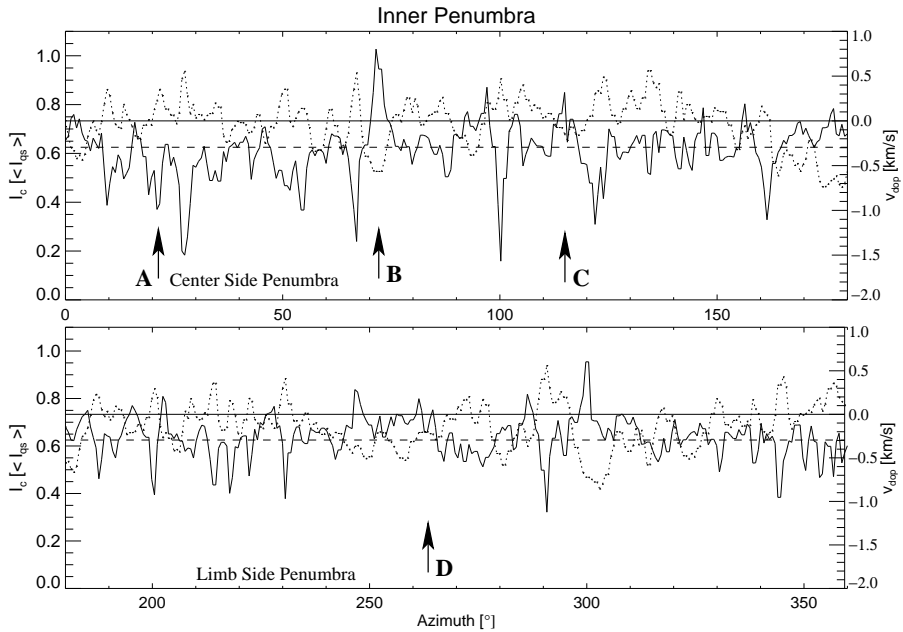}
		\caption{$\rm{I}_{\rm{c}}$ (dotted) and $\rm{v}_{\rm{dop}}$ (solid) in the penumbra along the small ellipse shown in Fig.~\ref{fig:alli}. The dashed line indicates the average velocity along that path.}
		\label{fig:inner}
\end{figure}

\begin{figure}[h!]
	\centering
		\includegraphics[width={0.87\textwidth}]{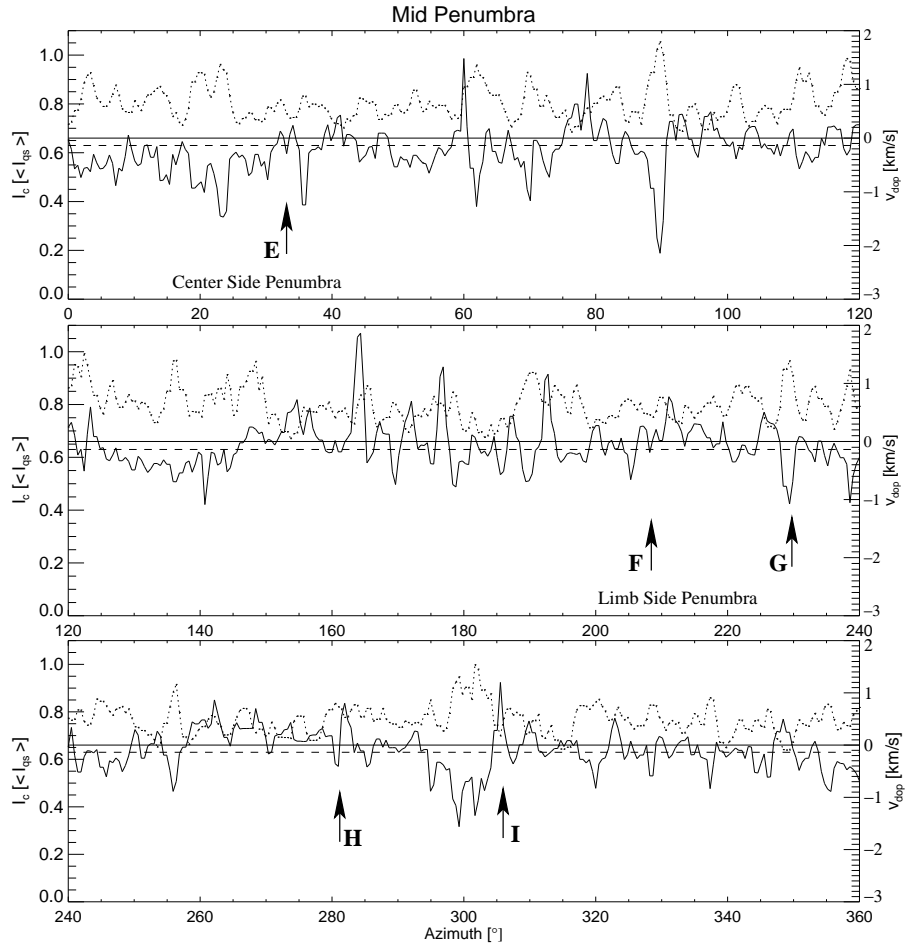}
		\caption{Same as Fig.~\ref{fig:inner}, but along the central path in the large map of Fig.~\ref{fig:alli}.}
		\label{fig:mid}
\end{figure}

\begin{figure}[h!]
	\centering
		\includegraphics[width={0.88\textwidth}]{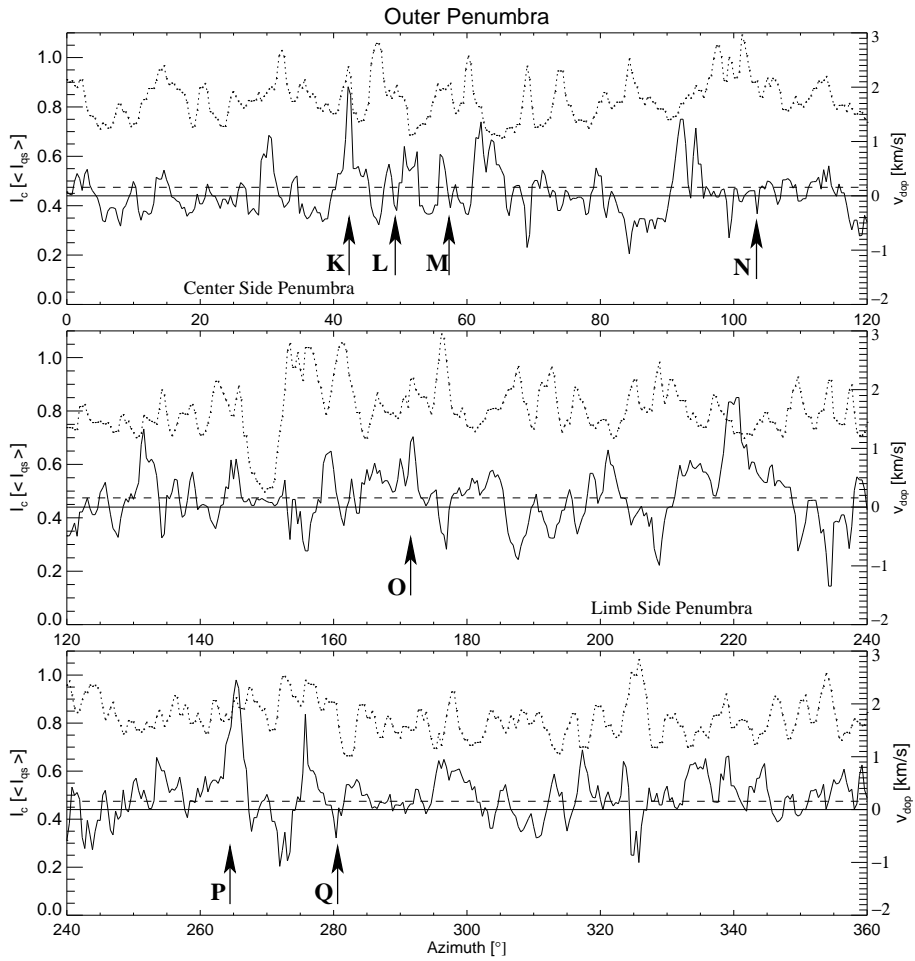}
		\caption{Same as Fig.~\ref{fig:inner}, but along the outer path in the large map of Fig.~\ref{fig:alli}.}
		\label{fig:outer}
\end{figure}

\paragraph{Flow Field and Continuum Intensity:} For this study, three ellipses were placed in the penumbra of Spot 04 (cf. Fig.~\ref{fig:alli}). The behavior of $\rm{I}_{\rm{c}}$ (dotted line with scale on the left ordinate) and $\rm{{v}}_{\rm{dop}}$ (solid line with scale on the right ordinate) along these azimuthal paths are plotted in Fig.~\ref{fig:inner}, Fig.~\ref{fig:mid} and Fig.~\ref{fig:outer}. The solid horizontal line marks zero velocity, while the average velocity along the ellipses are indicated by the dashed line. Prominent features, which will be discussed in the following, are indicated by letters. 

\paragraph{Ellipse in the Inner Penumbra:} 
Along this path, there is an average upflow of approx. $-$0.3~km~s$^{-1}$. The prominent spikes correspond to strong upflows with peak velocities of up to $-$1.5~km~s$^{-1}$. Sometimes downflows are visible, but they are weaker and broader when compared to the upflows. Areas of upflow often coincide with areas of increased $\rm{I}_{\rm{c}}$, and areas of downflow show a reduced $\rm{I}_{\rm{c}}$ (cf. position A \& B in Fig.~\ref{fig:inner}). This is in accordance with the idea that upflows transport hot plasma from lower layers to the surface, where it appears bright while cooling radiatively. Due to the release of energy, it becomes cooler and denser than the surrounding plasma ultimately sinking below the surface. 

\paragraph{Ellipse in the Mid Penumbra:} Strong upflow sites are seen less often in the mid penumbra, while more and more downflows appear. On average, the penumbra shows a net upflow of  $-$0.12~km~s$^{-1}$, and a correlation between strong upflows and enhanced $\rm{I}_{\rm{c}}$ is still ascertainable (cf. G \& I in Fig.~\ref{fig:mid}). Surprisingly, the strongest upflow area of Spot 04, which is $-2.2$~km~s$^{-1}$, is located in the middle penumbra. With 2.3~km~s$^{-1}$, the downflows show velocities of comparable strengths.

\paragraph{Ellipse in the Outer Penumbra:} Fig.~\ref{fig:outer} shows $\rm{{v}}_{\rm{dop}}$ and $\rm{I}_{\rm{c}}$ along the outer ellipse. Note that due to the irregular shape of the penumbral boundary, it is difficult to sample its outermost regions with a symmetric ellipse. As a result, not all pixel along the path are within the penumbral boundary (e.g. between positions K \& L or between N \& O). It is expected that the trend in the velocity field described below will be more prominent for  larger ellipses. 

Even though downflows are the dominant feature in the outer penumbra, upflow patches do not completely disappear. Downflows reach amplitudes of 2.5~km~s$^{-1}$, while the maximal upflow velocity is $-$1.3~km~s$^{-1}$. On average, the outer penumbra exhibits a downflow of almost 0.18~km~s$^{-1}$. It is interesting that there is no distinct anticorrelation between $\rm{I}_{\rm{c}}$ and v$_{\rm{dop}}$ anymore. Position K, O \& P in Fig.~\ref{fig:outer} indicate places with strong downflows that are co-spatial with a local increase of $\rm{I}_{\rm{c}}$. Sometimes, these downflows show an $\rm{I}_{\rm{c}}$ which is even higher than the $\rm{I}_{\rm{c}}$ of upflows (cf. I$_{\rm{c}}$ and v$_{\rm{dop}}$ around position K in Fig.~\ref{fig:outer}). Downflows appearing bright in continuum images will be investigated in greater detail in Section~\ref{sec:brightdownflow}.

\section{Search for Overturning Convection Within Penumbral Filaments}
\label{sec:rolltype}

In the framework of the gappy penumbral model, energy is transported by overturning convection along the penumbral gaps, which are areas without magnetic field below $\tau=1$. Thus, upflows would be present in the center or the dark-core of the filament, while a downflow should be observable adjacent to it at the position of the lateral brightening \citep{Scharmer2006}. Consequently, a down-up-down flow pattern, with a peak to peak distance between the downflows of 0.\arcsec5, ought to be visible along the azimuthal paths running perpendicular to the PFs.

The plots in Fig.~\ref{fig:inner}, \ref{fig:mid} and \ref{fig:outer}, however, show that strong up- and downflow patches have a width of more than 0.\arcsec5 (cf. positions A, B, G, I, K and P), and adjacent to them, the flow does not change its direction, but only declines in strength (cf. positions A, K \& P). The predicted flow pattern is not found on a scale of 0.\arcsec5. For weaker flows, all features that show down-up-down flow pattern on a scale of 1.\arcsec0 have been identified (positions C, D, E, L, M, N and Q). A close inspection of the map of $\rm{{v}}_{\rm{dop}}$ shows that this particular flow pattern appears were the path deviates from an elliptical shape (positions C, E, L, M, \& N). Due to the finite resolution of the ellipses, up- and downflows at different radial distances are compared. The remaining positions Q \& D do not show a bright filament at the upflow position, but appear rather dark in $\rm{I}_{\rm{c}}$. A careful analysis of the other data samples does not yield a down-up-down flow pattern on scales of 0.\arcsec5 that could be interpreted as an indication of overturning convection in penumbral filaments.

\section{Flow Field in Bright and Dark Penumbral Filaments}
\label{sec:filaandflow}

There are two reasons why it is not straightforward to use HINODE SP data 
to infer the velocity field within penumbral filaments, especially those with a dark-core. The first obstacle is the lifetime of PFs. Since it can be anything between 10 minutes, i.e. the lifetime of a typical granule, to 5 hours \citep{Bray:1958p3735,Loughhead:1964,Solanki:2003p2072}, the feature might have changed during the acquisition of SP data. The second complication lies in the vagueness of the definition 
itself. According to \citet{2004A&A...424.1049S}, dark-cored PFs consist of two bright PFs (the lateral brightening) and a dark feature in between the two (the dark-core), which move synchronously and follow the same trajectory. 

\paragraph{Investigation Procedure:} Because of these complications, the following  method of operation was applied to identify dark-cored PFs:
\begin{itemize}
\item[a)]{Identification of (dark-cored) PFs in maps of $\rm{P}_{\rm{tot}}$, where they are best visible \citep{BellotRubio2007}.}
\item[b)]{Cross-check of identified features in maps of $\rm{I}_{\rm{c}}$ and $\rm{P}_{\rm{tot}}$.}
\item[c)]{Identification of filaments under study in a time-series of HINODE BFI-pictures recorded simultaneously during the SP data acquisition.}
\item[d)]{If the results from a) b) \& c) are positive, the respective maps of $\rm{I}_{\rm{c}}$ and $\rm{{v}}_{\rm{dop}}$ are used to study the flow field within single PFs.}
\end{itemize}

\begin{figure}[h!]
	\centering
		\includegraphics[width={\textwidth}]{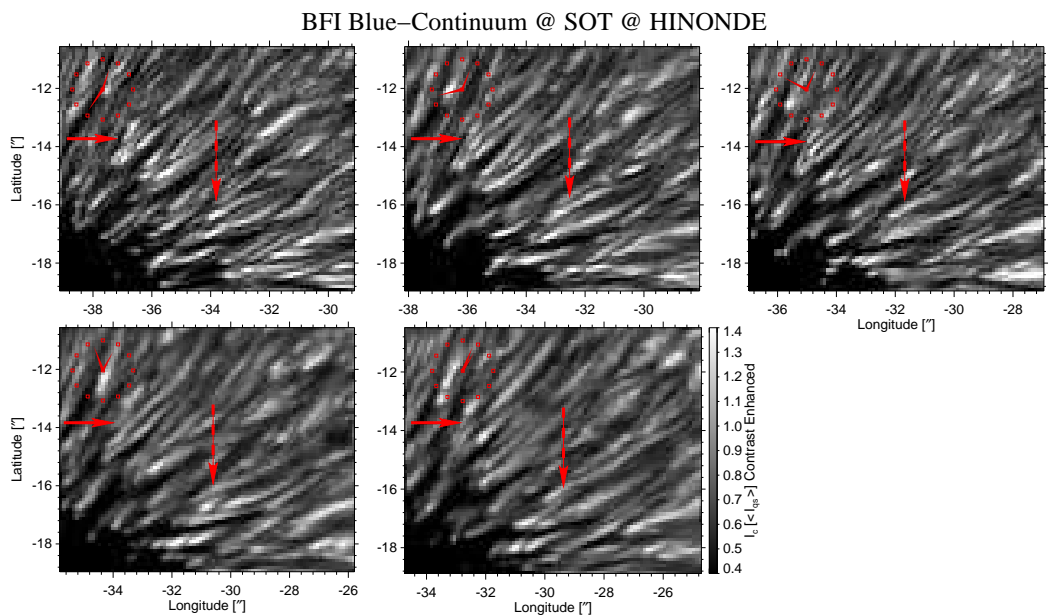}
		\caption{Contrast enhanced HINODE BFI maps, including the subfields B \& C of Fig.~\ref{fig:alli}. The images were recorded in the blue-continuum at 450.45 nm with a spatial resolution of 0.\arcsec2 and a cadence of 7~minutes. The arrows mark two PFs under study.}
		\label{fig:timeseries_enhanced}
\end{figure}

\paragraph{BFI Time-Series:} The penumbral region\footnote{The solar rotation causes a change in longitude during the data acquisition. Since this is not taken into account properly in the SP maps, the longitudes differ slightly in SP and BFI images.} around the area of the subfields denoted B and C in Fig.~\ref{fig:alldop} is shown in Fig.~\ref{fig:timeseries_enhanced}. The time-series was recorded by the HINODE BFI in the blue-continuum channel at 450.45 nm. The spatial resolution is of the order of 0.\arcsec2. The frame in the lower left corner is the one closest in time to the SP data set, while the other images were recorded beforehand and afterwards.

For the better identification of the dark-cored PFs, a standard IDL low pass filter was applied to artificially enhance the contrast. At the same time, it was ensured that only those features that are also ascertainable in the original data were studied. Despite the large cadence of 7 minutes in the BFI data set, a dark-cored PF (indicated by the solid red arrow) can be identified for more than 30 minutes during the SP data acquisition. However, some PFs that show a dark-core in the SP data set can be identified only in a single frame of the BFI time-series (cf. the structure below the dashed arrow).

\begin{figure}[h!]
	\centering
		\includegraphics[width={0.7\textwidth}]{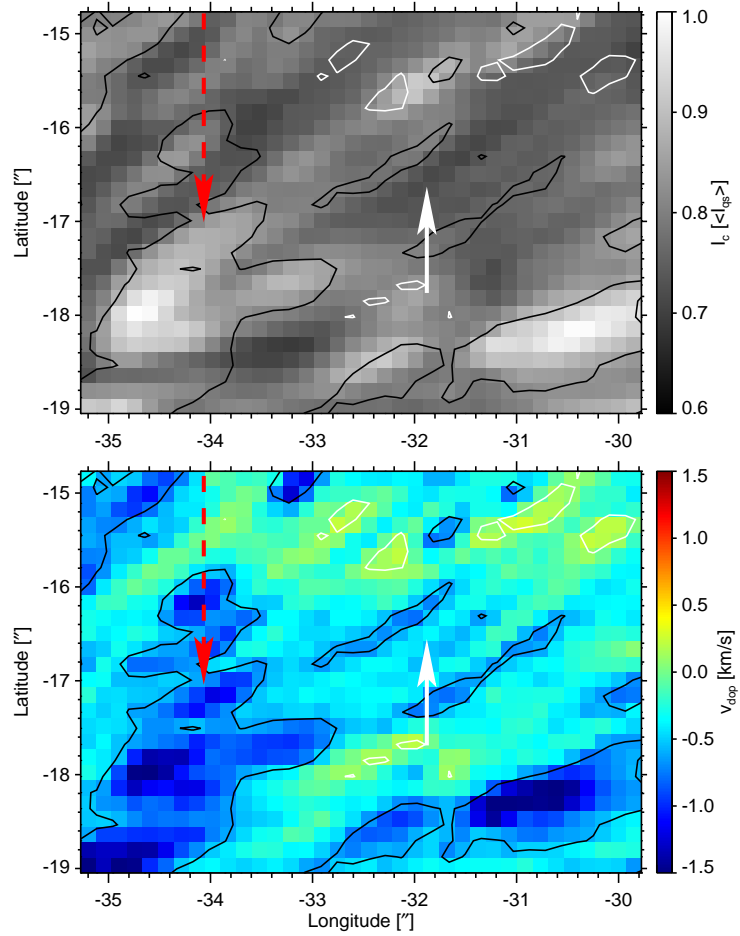}
		\caption{Top: $\rm{I}_{\rm{c}}$ map of subfield B of Fig.~\ref{fig:alldop}. Bottom: Respective map of v$_{\rm{dop}}$. Black (white) contours represent upflows (downflows) with amplitudes of $-$0.6~km~s$^{-1}$ (0.1~km~s$^{-1}$). The dashed red arrows point out the same feature as in Fig.~\ref{fig:timeseries_enhanced}.}
		\label{fig:boxB}
\end{figure}

\paragraph{Type I, Example 1:} Fig.~\ref{fig:boxB} displays $\rm{I}_{\rm{c}}$ and $\rm{v}_{\rm{dop}}$ of subfield B overlaid with a contour plot. Black represents upflows with  v$_{\rm{dop}} = -0.6$~km~s$^{-1}$ and white downflows with v$_{\rm{dop}} = 0.1$~km~s$^{-1}$. The dashed red arrows in Fig.~\ref{fig:timeseries_enhanced} and~\ref{fig:boxB} indicate the same PF, which is located between $\rm{(x;y)}=(-34.$\arcsec$6;-18.$\arcsec$)$ and $\rm{(x;y)}=(-30.$\arcsec$0;-15.$\arcsec$0)$ in the SP data set. 

The white arrow indicates a lane of weak upflow with amplitudes of $-0.25$~km~s$^{-1}$. It stretches from $\rm{(x;y)}=(-33.$\arcsec$8;-17.$\arcsec$4)$ to $\rm{(x;y)}=(-31.$\arcsec$2;-15.$\arcsec$8)$ and shows a significant lower $\rm{I}_{\rm{c}}$ when compared to the brighter structures above and below. Since the upflow velocity of v$_{\rm{dop}}=-0.6$~km~s$^{-1}$ is significantly higher, $\rm{I}_{\rm{c}}$ and v$_{\rm{dop}}$ are anti-correlated in this filament. However, the fact that the filament is visible in only one BFI image (cf. dashed red arrow in Fig.~\ref{fig:timeseries_enhanced}) indicates its limited lifetime and raises doubts as to whether this lane forms the dark-core of a PF.

\begin{figure}[h!]
	\centering
		\includegraphics[width={0.7\textwidth}]{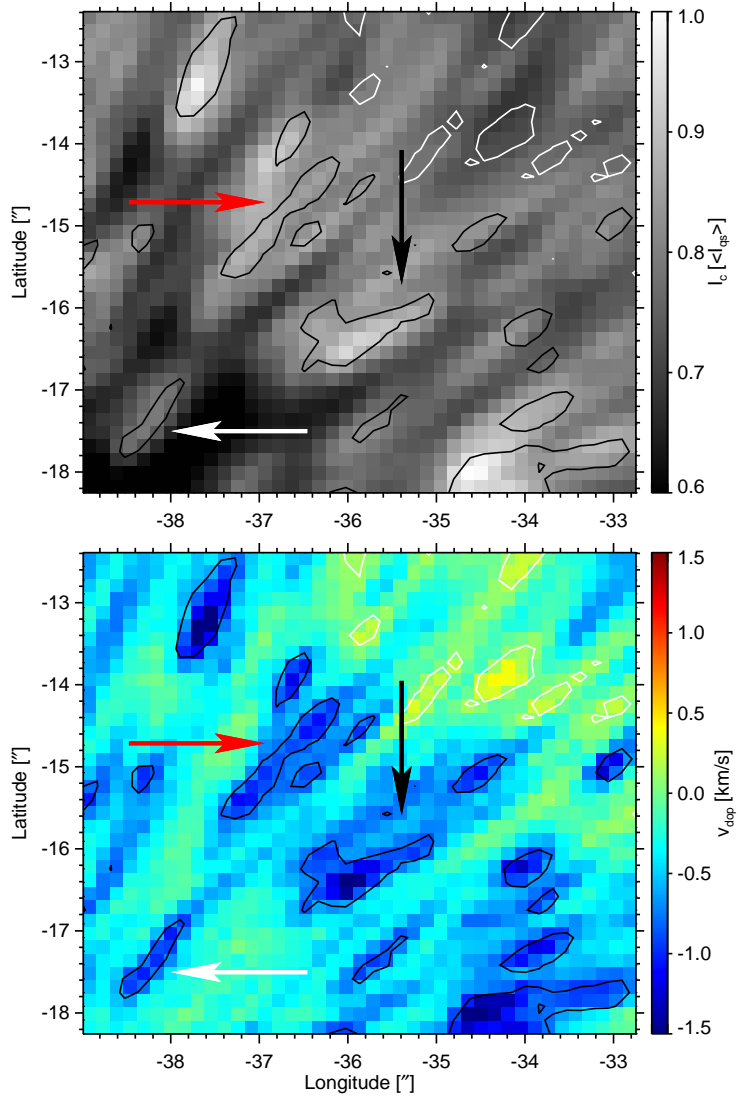}
		\caption{Same as Fig.~\ref{fig:boxB}, but subfield C of Fig.~\ref{fig:alldop}. Contours represent v$_{\rm{dop}}=-0.8$~km~s$^{-1}$ (black) and v$_{\rm{dop}}=0.1$~km~s$^{-1}$ (white).}
		\label{fig:boxC}
\end{figure}

\paragraph{Type I, Example 2:} In Fig.~\ref{fig:boxC}, a dark-cored PF is indicated by the solid red arrow (cf. also Fig.~\ref{fig:timeseries_enhanced}). It stretches from $\rm{(x;y)}=(-37.$\arcsec$5;-16.$\arcsec$0)$ to $\rm{(x;y)}=(-35.$\arcsec$8;-12.$\arcsec$4)$ and fulfills the \citet{2004A&A...424.1049S} criteria.

The map of v$_{\rm{dop}}$ illustrates that the strongest upflows occur at the bright head, i.e. at $\rm{(x;y)}=(-37.$\arcsec$5;-16.$\arcsec$0)$, and the lateral brightening of the PF. The upflow becomes weaker with increasing radial distance from the spot center. It stops and turns into a downflow around $\rm{(x;y)}=(-35.$\arcsec$8;-12.$\arcsec$4)$. There is also upflow in the dark-core, but it peaks around $-$0.7~km~s$^{-1}$ and weakens more quickly compared to the flow in the lateral brightening, before stopping $\rm{(x;y)}=(-35.$\arcsec$4;-14.$\arcsec$0)$.

\paragraph{Type I, Example 3:} The black arrow in Fig.~\ref{fig:boxC} points at another PF extending from $\rm{(x;y)}=(-36.$\arcsec$0;-16.$\arcsec$4)$ to $\rm{(x;y)}=(-34.$\arcsec$4;-15.$\arcsec$0)$. Around the bright head, i.e. located at $\rm{(x;y)}=(-36.$\arcsec$0;-16.$\arcsec$4)$, strong upflows with amplitudes larger than $-$0.8~km~s$^{-1}$ are measurable. As in the PF discussed before, the upflow diminishes towards the outer penumbra. The region with weaker upflow, i.e. v$_{\rm{dop}}=-0.4$~km~s$^{-1}$, from $\rm{(x;y)}=(-36.$\arcsec$0;-16.$\arcsec$2)$ to $\rm{(x;y)}=(-35.$\arcsec$0;-15.$\arcsec$4)$ appears less intense in the map of $\rm{I}_{\rm{c}}$. Again, this feature is unstable and fluctuates significantly in the BFI time-series. It does not fulfill the \citet{2004A&A...424.1049S} criteria for dark-cored PFs. 

\paragraph{Type II, Example 1:} The white arrow in Fig.~\ref{fig:boxC} points at the head of a PF located between coordinates $\rm{(x;y)}=(-38.$\arcsec$4;-17.$\arcsec$6)$ and $\rm{(x;y)}=(-35.$\arcsec$0;-13.$\arcsec$8)$ in Fig.~\ref{fig:boxC}. The bright head of the PF extends into the umbra, showing upflow velocities of $-$1 km s$^{-1}$. With increasing radial distance, I$_{\rm{c}}$ diminishes and a dark PF develops. This configuration is also observable during the 30 minutes of the BFI time-series (cf. Fig.~\ref{fig:timeseries_enhanced}). What is remarkable is the upflow within the dark PF. It peaks around $-$0.6 km/s at position $\rm{(x;y)}=(-37.$\arcsec$3;-16.$\arcsec$4)$ and weakens gradually along the dark PF. At the end of the latter, the flow returns into the solar surface with a maximal velocity of 0.4 km/s at position $\rm{(x;y)}=(-35.$\arcsec$8;-14.$\arcsec$6)$. Additionally, the upflow in the lateral brightening is weaker than in the dark area itself (compare I$_{\rm{c}}$ and v$_{\rm{dop}}$ at e.g $\rm{(x;y)}=(-37.$\arcsec$2;-16.$\arcsec1).

\paragraph{Type II, Example 2:} Another but less prominent example is seen between $\rm{(x;y)}=(-37.$\arcsec$2;-18.$\arcsec$2)$ and $\rm{(x;y)}=(-34.$\arcsec$0;-15.$\arcsec$4)$. Again, a bright feature that harbors an upflow is located in the umbra. Along the dark PF, there is a weak upflow that diminishes radially outwards. At $\rm{(x;y)}=(-34.$\arcsec$6;-16.$\arcsec$2)$, no upflow is present anymore. However, due to the uncertainty of 0.1~km~s$^{-1}$ in the maps of v$_{\rm{dop}}$, it is not possible to identify a downflow at the end of the dark PF.

\begin{figure}[h!]
	\centering
		\includegraphics[width={0.7\textwidth}]{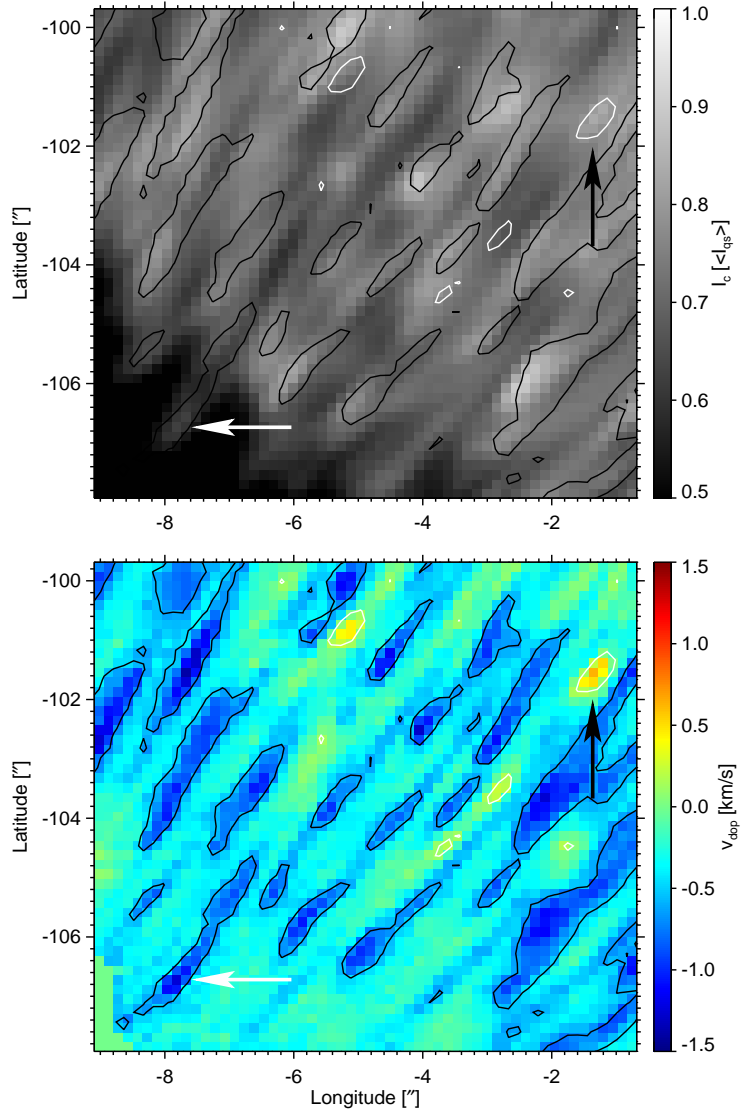}
		\caption{Same as Fig.~\ref{fig:boxB}, but what is shown is an enlargement of the inner CSP of Spot 01. Contours are v$_{\rm{dop}}=-0.6$~km~s$^{-1}$ (black) and v$_{\rm{dop}}=0.1$~km~s$^{-1}$ (white).}
		\label{fig:boxF}
\end{figure}


\paragraph{Type II, Example 3:} To show a third example of upflows within the dark-core of a PF, Fig.~\ref{fig:boxF} depicts an enlargement of the inner CSP of Spot 01. The white arrow points at a local enhancement of I$_{\rm{c}}$ at the umbral end of the dark-cored PF, which extends from $\rm{(x;y)}=(-6.$\arcsec$9;-106.$\arcsec$6)$ to $\rm{(x;y)}=(-5.$\arcsec$5;-102.$\arcsec$7)$. Contrary to the structures described in Fig.~\ref{fig:boxC}, the dark PF in the umbra is completely surrounded by bright areas. Additionally, this PF fulfills the \citet{2004A&A...424.1049S} criteria, as it can be identified for more than 30 minutes in the corresponding BFI images (not shown here). The Doppler map shows that upflows with amplitudes of at least v$_{\rm{dop}} = -0.6$~km~s$^{-1}$ occur at the head of the PF and the umbral side of the dark-core, i.e. between $\rm{(x;y)}=(-6.$\arcsec$9;-106.$\arcsec$6)$ and $\rm{(x;y)}=(-6.$\arcsec$4;-104.$\arcsec$6)$. The upflow speed at the lateral brightening is weaker $\rm{(x;y)}=(-5.$\arcsec$6;-102.$\arcsec$7)$. As in the filaments discussed before, the upflow speed decreases along the dark-core and turns into a downflow at its end, i.e. at position $\rm{(x;y)}=(-7.$\arcsec$2;-105.$\arcsec$0)$. Even though the downflows are hardly above the level of significance (only two pixel show v$_{\rm{dop}} >$ 0.1~km~s$^{-1}$), the actual amplitude is probably higher since $\Theta$ in Spot 01 is three times larger than $\Theta$ in Spot 04. As a result, the amplitude of downflows in the CSP are decreased by projection effects, while the amplitudes of the blueshift in the dark-core are increased.

\paragraph{Interpretation:} The previous analyses yield two families of PF: Type I PFs shows an anticorrelation of I$_{\rm{c}}$ and v$_{\rm{dop}}$, i.e. the upflow in the dark-core is weaker when compared to the lateral brightening of the PF. In type II PFs, the correlation is not distinct. Even though upflows are present in their bright heads as well, the upflow in the dark areas is stronger when compared to their bright surrounding. The upflow velocity within the dark-core of the PF decreases radially outwards, sometimes turning into a downflow at its end.

Thus, at a spatial resolution of 0.\arcsec3, it is not possible to decide whether I$_{\rm{c}}$ and v$_{\rm{dop}}$ are correlated in PFs. Furthermore, it is unclear if a transformation, or even an oscillation, exists between type I and type II PF.

\section{Bright Penumbral Downflows}
\label{sec:brightdownflow}
In the QS, granules are co-spatial with strong upflows of hot plasma. It is this excess of heat which leads to their surplus brightness. However, areas of increased I$_{\rm{c}}$ accompanied by downflows are known, for example, from bright points in the intergranular network \citep{Dunn:1973p4352,Mehltretter:1974p4351}. Recent measurements reveal that 7.5\% of all bright points are associated with v$_{\rm{dop}} > 1$~km~s$^{-1}$ \citep{Riethmuller:2010p4353}. It is agreed that network bright points are caused by strong magnetic fields, leading to a more transparent\footnote{As a result of the lateral pressure balance, the gas pressure outside the magnetic element has to balance the sum of gas and magnetic pressure inside the flux element. This leads to a reduction of the gas pressure inside the magnetic element.} plasma than in the mean photosphere. This allows photons to escape from deeper, and thus hotter sub-photospheric layers \citep{Spruit:1976p4356,Spruit:1977p4345}. 

Examples showing an anticorrelation between upflows and enhancements in I$_{\rm{c}}$ may be found in the penumbra as well, e.g. the dark-core of Type II PFs in Section~\ref{sec:filaandflow}. Additionally, downflows with an increase of I$_{\rm{c}}$ exist as well, e.g. Fig.~\ref{fig:boxF}. Here, the black arrow points to a region that harbors a downflow of v$_{\rm{dop}} \approx$ 0.4~km~s$^{-1}$, while the corresponding map of I$_{\rm{c}}$ does not show a local minimum at $\rm{(x;y)}=(-6.$\arcsec$4;-104.$\arcsec$6)$. These bright penumbral downflows (BPDs) are commonly observed in HINODE SP data and shall be discussed on the basis of the following two examples. 

\begin{figure}[h!]
	\centering
		\includegraphics[width={\textwidth}]{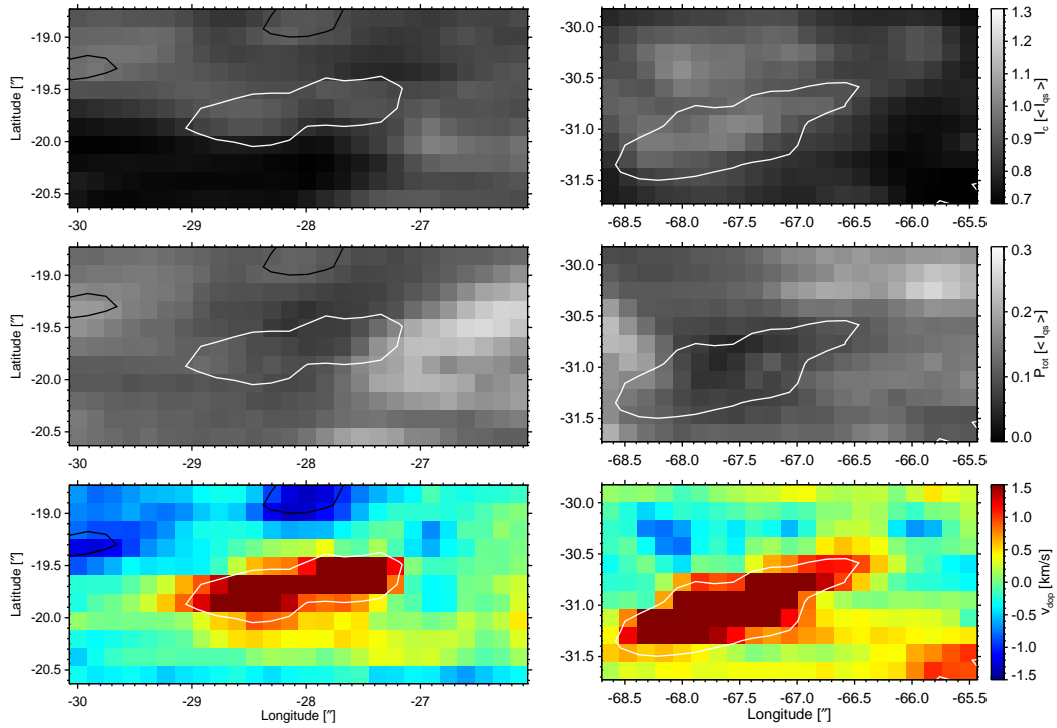}
		\caption{The left and right columns represent enlargements of the two subfields marked A and E in Fig.~\ref{fig:alldop}. The rows show from top to bottom: I$_{\rm{c}}$, P$_{\rm{tot}}$ and v$_{\rm{dop}}$. The black and white contours represent blue and redshifts with respective velocities of $\pm$1~km~s$^{-1}$.}
		\label{fig:twobox}
\end{figure}

\paragraph{Example 1:} The left and right columns of Fig.~\ref{fig:twobox} show enlargements of areas A and E (cf. bottom image in Fig.~\ref{fig:alldop}). The top row depicts I$_{\rm{c}}$, which is overlaid with contours marking up- and downflow areas with $|\rm{v}_{\rm{dop}}| = 1$~km~s$^{-1}$. The panels of the middle row show P$_{\rm{tot}}$, the percentage of polarized photons per pixel. This quantity may be used as a proxy of the magnetic field strength, given that more photons will be polarized if the magnetic field strength increases. Finally, the bottom row shows the Doppler velocity saturated at $\pm1.5$~km~s$^{-1}$. These pictures already indicate that downflow regions may also correlate with enhanced values of I$_{\rm{c}}$ without showing enhancements P$_{\rm{tot}}$ at this location. 

At $\rm{(x;y)}=(-28.$\arcsec$5;-19.$\arcsec$7)$, for example, a continuum intensity slightly below that of the average QS is obtained (I$_{\rm{c}} = 0.92 \cdot \rm{I}_{\rm{qs}}$). By contrast, the magnetic field is not really strong (P$_{\rm{tot}} = 9,0\%$), whereas the plasma returns back into the solar surface with v$_{\rm{dop}} = 2.36$~km~s$^{-1}$. 
It is surprising that the pixel at $\rm{(x;y)}=(-28.$\arcsec$0;-19.$\arcsec$3)$, which corresponds to a strong upflow of v$_{\rm{dop}} = -1.30$~km~s$^{-1}$ and a slightly larger magnetic field strength (P$_{\rm{tot}} = 11,5\%$), appears even less intense in continuum (I$_{\rm{c}} = 0.91\cdot\rm{I}_{\rm{qs}}$). 

\paragraph{Example 2:} A similar behavior is found in the feature depicted in the right column. I$_{\rm{c}}$ is of the same size as in the QS, while v$_{\rm{dop}} > 1$~km~s$^{-1}$ is measured and P$_{\rm{tot}}$ shows a local minimum. If the pixel at coordinates $\rm{(x;y)}=(-67.$\arcsec$5;-30.$\arcsec$9)$ is taken as an example, an I$_{\rm{c}}$ of $1.02 \cdot \rm{I}_{\rm{qs}}$ is obtained. At the same time, P$_{\rm{tot}}=6.0$\% is measured, while the redshift corresponds to v$_{\rm{dop}} = 2.36$~km~s$^{-1}$. The brightest pixel in the vicinity is located at $\rm{(x;y)}=(-68.$\arcsec$0;-30.$\arcsec$4)$ and shows I$_{\rm{c}} = 1.04 \cdot \rm{I}_{\rm{qs}}$, P$_{\rm{tot}} = 4,6\%$ and v$_{\rm{dop}} = -0.78$~km~s$^{-1}$. 

\paragraph{Interpretation:} 1.5 \% of the penumbral area in Spot 04 shows I$_{\rm{c}} > \rm{I}_{\rm{qs}}$. In 51\% of these cases, a significant upflow (v$_{\rm{dop}} < $ -0.1~km~s$^{-1}$) is measurable, while 36\% show a downflow (v$_{\rm{dop}} > $ 0.1~km~s$^{-1}$) at these locations. These numbers demonstrate that BPDs are a non negligible contributor to the penumbral brightness.

If the excess of I$_{\rm{c}}$ in BPDs is due to an opacity effect, which is in turn caused by concentrations of magnetic field, an increase of P$_{\rm{tot}}$ is to be expected. However, this increase in P$_{\rm{tot}}$  is not measured, which makes it unlikely that the BPDs and network bright points are caused by the same physical process.

A second mechanism to increase I$_{\rm{c}}$ in combination with downflows involves magnetohydrodynamic waves inside a flux-tube (cf. Sections~\ref{sec:siphon} and \ref{sec:buoyant}). Under certain conditions, theses waves can form shock fronts inside the tube, which compress the plasma in the downflow part of the loop. Due to the compression, the plasma is heated and appears bright in maps of I$_{\rm{c}}$.

\section{Correlations at Disk Center}
\label{sec:corr_center}

From the studies of Section~\ref{sec:morphology}, it can be inferred that enhancements in I$_{\rm{c}}$ are co-spatial with strong upflows, cf. also \citet{2004A&A...415..717T,Rimmele:2006p215,2006ASPC..358...96M,2010mcia.conf..186I}. However, Sections~\ref{sec:filaandflow} and \ref{sec:brightdownflow} show a range of examples where the opposite is true. To quantify the correlations between I$_{\rm{c}}$ and v$_{\rm{dop}}$ statistically, Spearman's rank correlation coefficient (r$_{\rm{S}}$) is taken as a measure\footnote{Since the underlying distribution of v$_{\rm{dop}}$ is not Gaussian (Section~\ref{sec:global}), r$_{\rm{S}}$ is a better measure for a possible correlation (cf. Appendix~\ref{ch:correlation}).}. Because upflows are defined as negative Doppler velocities, a value of r$_{\rm{S}}= -1$ would indicate a perfect correlation. r$_{\rm{S}}$ is computed based on the fluctuation of v$_{\rm{dop}}$ and I$_{\rm{c}}$ around the following mean values:

\begin{itemize}
\item{The entire penumbral area =: r$_{\rm{S}}$(A)}
\item{Along azimuthal cuts of different radii =: r$_{\rm{S}}$(R)}
\item{A small local area of 3.\arcsec2 by 3.\arcsec2 =: r$_{\rm{S}}$(L)}
\end{itemize}

\paragraph{Global Correlation Coefficients:} In the case of r$_{\rm{S}}$(A), values ranging from $-$0.02 in Spot 03 to $-$0.17 in Spot 01 are found. For r$_{\rm{S}}$(R), the extremum is $-$0.35 in Spot 04 and the minimum is $-$0.19 in Spot 01, while r$_{\rm{S}}$(L) shows a maximum of $-$0.36 in Spot 04 and a minimum of $-$0.30 in Spot 03. This demonstrates that on a global scale, there is no correlation between I$_{\rm{c}}$ and v$_{\rm{dop}}$ in penumbrae. Even though r$_{\rm{S}}$(R) and r$_{\rm{S}}$(L) are larger for the entire penumbra than r$_{\rm{S}}$(A), it is still too small to consider I$_{\rm{c}}$ to be correlated with upflows. This is in accordance with the conclusion of \citet{Schlichenmaier:2005p219}.

\begin{figure}[h!]
	\centering
		\includegraphics[width={\textwidth}]{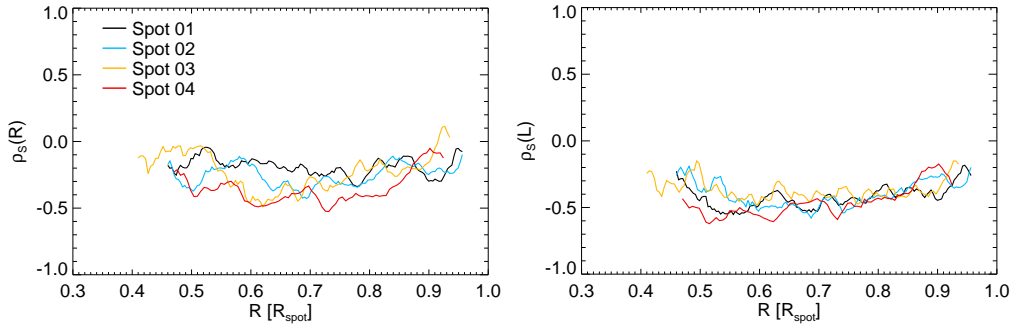}
		\caption{Left: Radial dependency of Spearman's rank correlation coefficient (r$_{\rm{S}}$) computed from the fluctuation of I$_{\rm{c}}$ and v$_{\rm{dop}}$ around an azimuthal mean value. Right: Same as left, but r$_{\rm{S}}$ computed from the fluctuation of the quantities around a local mean value.}
		\label{fig:corr_LR_cent}
\end{figure}

\paragraph{Radial Dependency of Correlation Coefficients:} In the case of r$_{\rm{S}}$(R) and r$_{\rm{S}}$(L), it is possible to calculate r$_{\rm{S}}$ as a function of the penumbral radius. The respective behavior is shown in Fig.~\ref{fig:corr_LR_cent}. The left panel shows that r$_{\rm{S}}$(R) does not only vary with radial distance\footnote{Note that it is not possible to reject the H$_0$ hypothesis (cf. Appendix~\ref{ch:correlation}) on a 3$\cdot\sigma$ confidence level for all radial cuts. This is especially the case in the inner penumbra of Spot 03.}, but also changes significantly between different penumbrae. Maximal and minimal values of r$_{\rm{S}}$(R) range from 0 to $-$0.53. In the case of Spot 04, a general trend may be assigned to r$_{\rm{S}}$(R) in the sense that it increases for $0.45 \le \rm{R}_{\rm{spot}} \le 0.60$, reaches an extremum around R$_{\rm{spot}} = 0.65$ and decreases for larger radii. This is interesting insofar that simulations by \citet{2011ApJ...729....5R} show not only the same behavior between vertical velocities and continuum intensity, but that the extremum is also located at R$_{\rm{spot}} = 0.62$. The different shape of the other curves may be attributed to the fact that the respective spots were observed with a residual heliocentric angle, which may spoil the analysis.

As illustrated by the right panel of Fig.~\ref{fig:corr_LR_cent}, the largest correlation coefficients are found along certain azimuthal paths of r$_{\rm{S}}$(L) -- i.e. if the fluctuations of I$_{\rm{c}}$ and v$_{\rm{dop}}$ are correlated around a local mean value. In the inner penumbra, r$_{\rm{S}}$(L) reaches values\footnote{In Appendix~\ref{ch:correlation}, it is shown that r$_{\rm{S}} = - 0.78$ for I$_{\rm{c}}$ and v$_{\rm{dop}}$ in the granulation of the QS.} of $-$0.62 in Spot 04. Besides a smaller scatter, the radial dependency of r$_{\rm{S}}$(L) shows the same general trend in all sunspots, even though the extremum is shifted towards the inner penumbra. A more detailed discussion about the interpretation of correlation coefficients is given in Section~\ref{sec:corr_limb}.

\section{Summary}
\label{sec:concl_obs_center}

The vertical component of the Evershed flow was investigated by means of spectropolarimetric observations of several sunspots close to disk center ($\Theta \le 11^{\circ}$). 
The results quantify the difference between the velocity field in the quiet Sun and in the penumbra. The maximal upflow velocity of the latter is weaker, while the downflow amplitudes are larger. For $|\rm{v}_{\rm{dop}}| < $ 0.8~km~s$^{-1}$, a larger fraction of the penumbra is occupied by upflows, but for $|\rm{v}_{\rm{dop}}| > $ 0.8~km~s$^{-1}$, downflows prevail. This behavior is reflected by the probability density functions of the velocity field which are significantly skewed for the penumbra but symmetric for the quiet Sun.

A morphological study reveals that upflows prevail in the inner penumbra. They appear elongated and follow the filamentary structure. The outer penumbra, by contrast, is dominated by downflows, which do not show a strand-like structure, but look roundish. Even though the distribution of up- and downflows is not absolute (downflows are found in the inner penumbra too) a study utilizing azimuthal mean values yields similar results. The upflow patches in the inner penumbra and the downflow locations in the outer penumbra could be interpreted as the sources and the sinks of the Evershed flow, even though an attribution to individual flow channels is challenging. No indication of overturning convection, i.e. a down-up-down flow pattern, was found on scales of the width of penumbral filaments.

Two types of (dark-cored) penumbral filaments were described in a case study. In the first type, the upflows in the lateral brightenings are stronger when compared to the dark-core. In the second type, the bright head of the penumbral filaments exhibits a strong upflow, which is also present in the dark-core, but diminishes with radial distance and turns into a downflow at the end of the penumbral filament. The lateral brightenings show no vertical plasma velocities.

Furthermore, penumbral downflows, which are accompanied by local enhancements of continuum intensity were identified. It was shown that these features are co-spatial with local minima of total polarization, which makes it unlikely that they are caused by an opacity effect due to strong magnetic fields. It is speculated that bright penumbral downflows are the result of shocks inside flux-tubes.

Spearmans ranked correlation coefficient ($\rho_{\rm{S}}$) was calculated for continuum intensity and Doppler velocity. It was found that 
$\rho_{\rm{S}}$ is not large enough to consider both quantities as being correlated neither globally nor locally. It was proposed that this is due to bright penumbral downflows or the second type of dark-cored penumbral filaments. 
Nevertheless, the radial dependency of $\rho_{\rm{S}}$ in observation is similar to that of sunspot simulations.

\paragraph{Implications for Penumbral Models:} The results of this Chapter are in favor of the penumbral flux-tube models, while they possess a range of problems for the gappy penumbral model. The main problem lies in the large number of downflows in the outer penumbra, which are proposed by the former scenario to guide hot plasma into the photosphere in the inner penumbra and cause it to submerge again in the outer penumbra. Additionally, the flow pattern in (dark-cored) penumbral filaments is different to that predicted by the gappy model. 

Whether this is due to the fact that a) overturning convection does not exist at all, b) occurs below the surface, or whether c) this plasma flow exists on scales not resolvable by HINODE is subject to speculation only. In the first two cases, the gappy model has to be rejected (or modified) as it does not explain, or will never provide, observational evidence. Finally, the argument of limited spatial resolution always applies, but from simulation of dark-cored penumbral filaments it is known that a down-up-down flow pattern should be observable on scales of 0.\arcsec5.

The investigation of the intensity and velocity field within individual penumbral filaments yields a diverse picture. Even though they sometimes show upflows in the dark-core, the velocity amplitude diminishes along their dark-cores and, in contrast to the propositions of the gappy scenario, turns into a downflow at its end.


\chapter{Observation of the Horizontal Component of the Evershed Effect}
\label{ch:limb}

This Chapter summarizes observation of the penumbral velocity field at large heliocentric angles. Section~\ref{sec:outflow} describes the general appearance of the Evershed flow alongside its extreme values. An interpretation of the radial behavior of azimuthally averaged values is given in Section~\ref{sec:rad_ave}. The morphology of individual flow channels, especially those with a line shift opposite to the general shift of the center- and the limb-side penumbra, are discussed in Section~\ref{sec:morph_out}. In Section~\ref{sec:incl_down}, these counterexamples are used to estimate the inclination of individual downflow channels. Finally, the correlation between continuum intensity and Doppler velocity is investigated in Section~\ref{sec:corr_limb}, and the results are summarized in Section~\ref{sec:concl_obs_limb}.

\section{The Evershed Outflow}
\label{sec:outflow}

Since plasma flows perpendicular to the LOS do not contribute to the Doppler shift of an absorption line, it is necessary to observe sunspots at large heliocentric angles to study the horizontal\footnote{In principle, it is impossible to measure the purely horizontal component of the EF. This would require observation of sunspots located exactly at the solar limb, where projection effects cause an extreme foreshortening and objects on the solar surface appear one-dimensional. Further complication arises from the Wilson depression of the umbra which results in a larger LSP when compared to the size of the CSP (cf.~\citet{Loughhead:1964} and Section~\ref{sec:theo_umbra}).} component of the EF. 
\begin{table}[h]
\begin{center}
	\begin{tabular}{ccccc}
		\hline
		\hline
		\\[-2ex]
		{Name}&{NOAA} & {Date of} & \multicolumn{2}{c}{cos$(\Theta$) of Penumbra}\\
		{}&{Active Region} & {Observation} & {Center} & {Limb}\\
		\\[-2ex]
		\hline		
		\\[-2ex]
		Spot 05 & {10923} & {Nov 10$^{\rm{th}}$ 2006} & {0.680 - 0650} & {0.637 - 0.585}\\
		Spot 06 & {10923} & {Nov 18$^{\rm{th}}$ 2006} & {0.665 - 0.637} & {0.626 - 0.594}\\
		Spot 07 & {10930} & {Dec 15$^{\rm{th}}$ 2006} & {0.625 - 0.595} & {0.583 - 0.564}\\
		Spot 08 & {10933} & {Jan 09$^{\rm{th}}$ 2007} & {0.695 - 0.677} & {0.670 - 0.652}\\
		\hline
	\end{tabular}
		\caption{Characteristics of data samples used in this study.}
	\label{tab:data_limb}
\end{center}
\end{table}

\begin{figure}[h!]
\begin{center}
    \includegraphics[width=0.9335\textwidth]{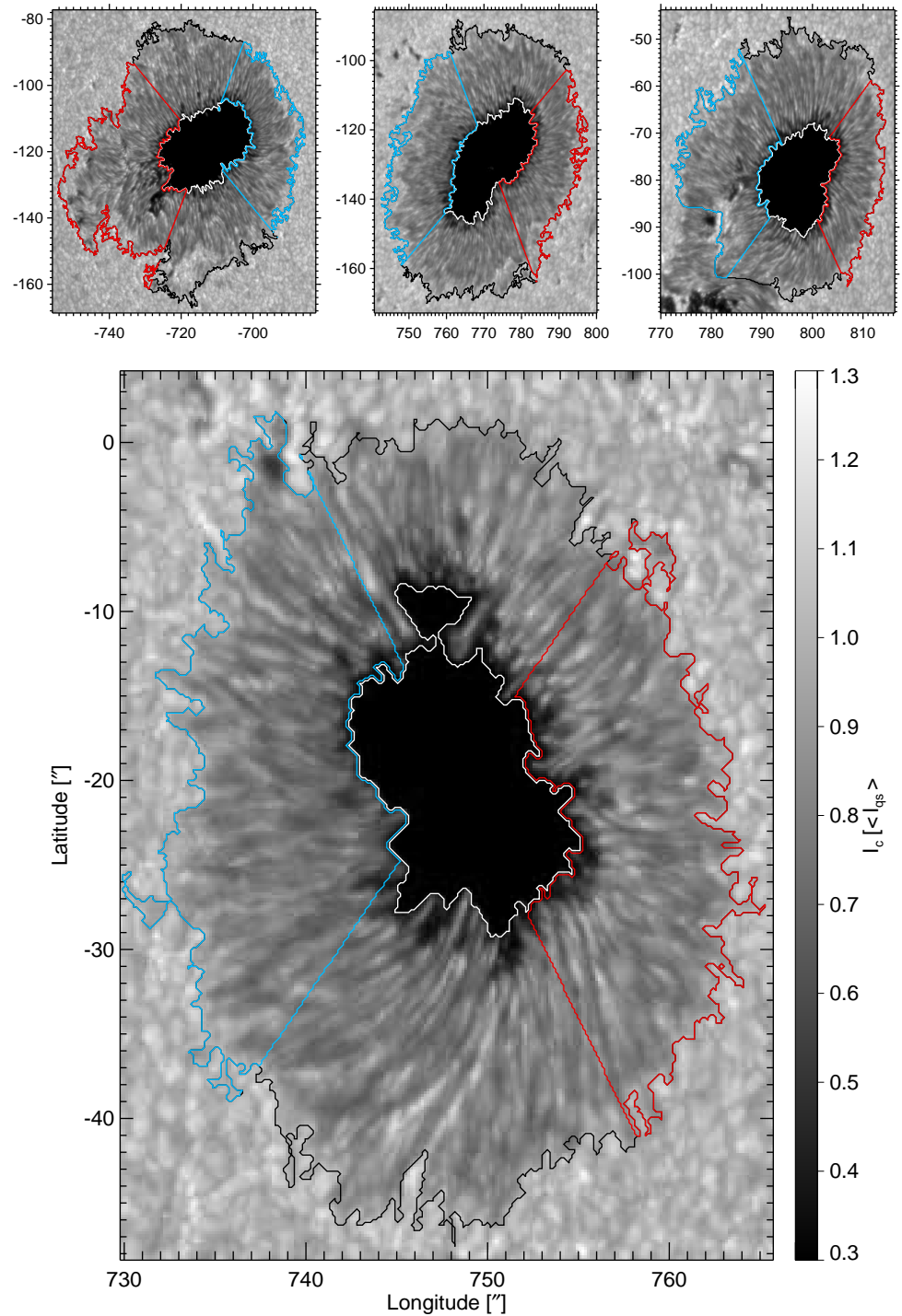}
\caption{Photospheric continuum intensity in sunspots at $\Theta\approx50^{\circ}$. The umbra and penumbra are indicated by the white and black contours. The red and blue contours indicate the areas defined as CSP and LSP.}
\label{fig:allimap_limb}
		\vspace{-40pt}
\end{center}
\end{figure}

\begin{figure}[h!]
\begin{center}
    \includegraphics[width=0.9335\textwidth]{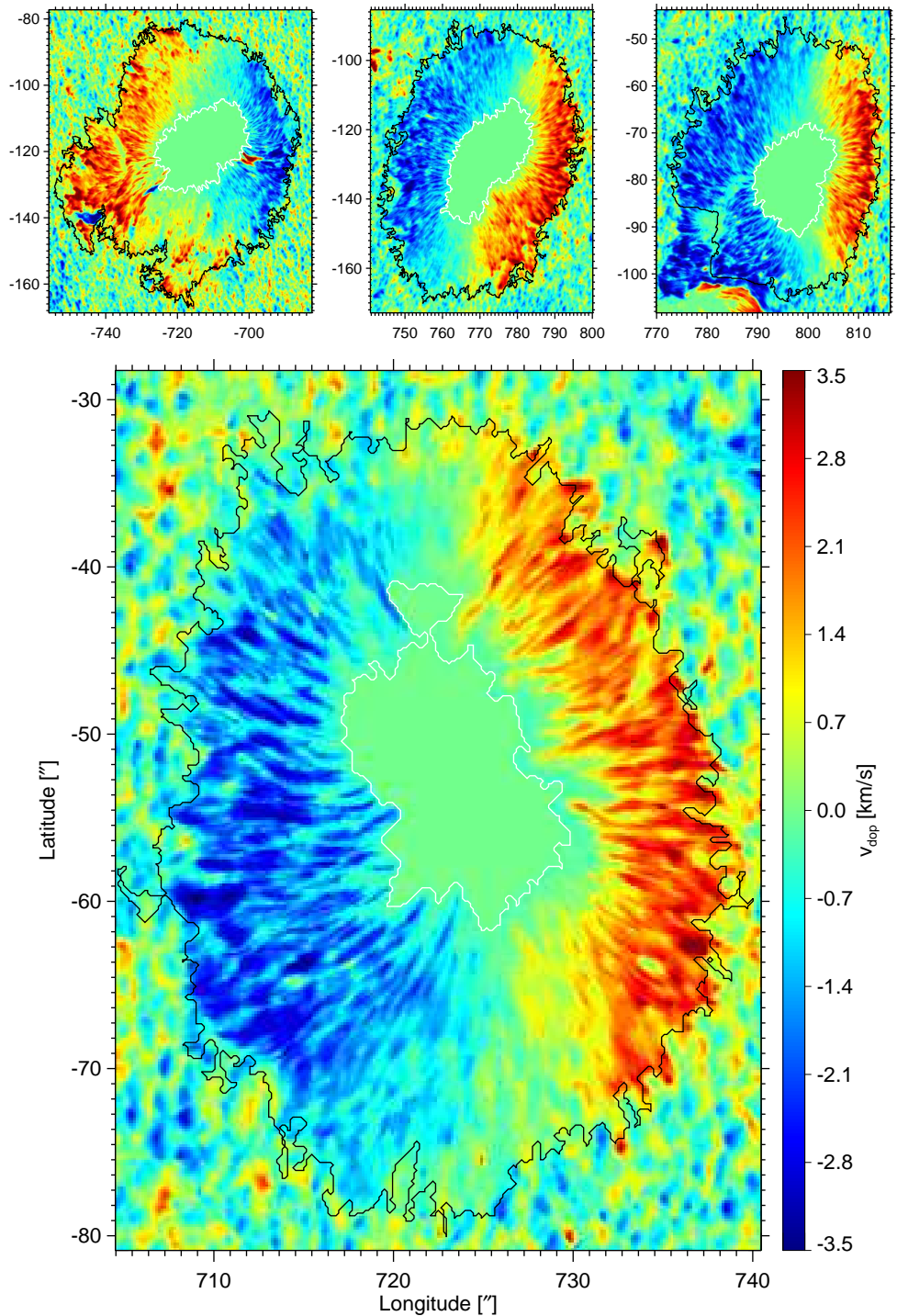}
\caption{Same as Fig.~\ref{fig:alli}, but photospheric Doppler velocities derived from the shifts in the line-wing of Fe I 630.15 nm. Motion towards the observer is shown from blue to green, motion away from the observer is indicated from green to red.}
\label{fig:alldop_limb}
		\vspace{-40pt}
\end{center}
\end{figure}

\paragraph{Description of Observational Data:} In this survey, four sunspots that were observed at a heliocentric angle of $\Theta \approx 50^{\circ}$ are evaluated. The characteristics of the data sets and their denotation are given in Table~\ref{tab:data_limb}. The CSP and the LSP are investigated separately because v$_{\rm{dop}}$ is oppositely oriented in the respective areas. Since $\Theta$ is different in the CSP and the LSP, their respective $\mu$-values are given individually as well.

The maps of Fig.~\ref{fig:allimap_limb} show I$_{\rm{c}}$ plotted between 30\% and 130\% of I$_{\rm{qs}}$. From left to right, the images in the top row represent Spot 05, Spot 06 and Spot 07, while Spot 08 is shown in the large picture. These sunspots are the same as the ones investigated in Section~\ref{sec:global}, but observed in the Eastern (negative longitude) or Western (positive longitude) hemisphere. The white and black contours mark the inner and outer penumbral boundaries, while the CSP and LSP are indicated in blue and red. CSP and LSP are defined as the area enclosed by the penumbral boundaries and two lines drawn from the center of the umbra in the direction of the line of symmetry at an angle of $\pm60^{\circ}$. Even though Spot 07 is recorded two days after Spot 03, the $\delta$-configuration with a little satellite sunspot is still present in the Southern periphery.

Photospheric Doppler velocities are plotted in Fig.~\ref{fig:alldop_limb}. As in Section~\ref{sec:global}, they are derived from the shift of the wing of Fe I 630.15. These velocities were calibrated using umbral profiles according to the method described in Section~\ref{sec:umbcalib}. All maps are saturated at $\pm$3.5~km~s$^{-1}$, and the velocities in the umbra ($\rm{I}_{\rm{c}} < 0.33 \cdot \rm{I}_{\rm{qs}}$) are set to zero (cf.~Section~\ref{sec:bisec} and \ref{sec:global}). The white and black contour lines represent the inner and outer penumbral boundaries.

\begin{table}[h!]
\begin{center}
	\begin{tabular}{ccccc}
		\hline
		\hline
		\\[-2ex]
		
		{Name} & \multicolumn{2}{c}{Extreme} &\multicolumn{2}{c}{Average}\\ 
		{of} & \multicolumn{2}{c}{v$_{\rm{dop\:630.15}}$ [km~s$^{-1}$]} &  \multicolumn{2}{c}{$\langle {\rm{v}_{\rm{dop}} \rangle}$ [km~s$^{-1}$]}\\
		{Data Set} & {Up} & {Down} & {Center} & {Limb}\\
		\\[-2ex]
		\hline		
		\\[-2ex]
		{Spot 05} & {-4.16} & {6.87} & {-1.15} & {1.26}\\
		{Spot 06} & {-3.85} & {5.00} & {-1.50} & {1.76}\\
		{Spot 07} & {-4.15} & {4.64} & {-1.63} & {1.60}\\
		{Spot 08} & {-3.90} & {4.40} & {-1.36} & {1.41}\\
		\hline
	\end{tabular}
	\caption{Extreme and average velocities in the CSP and LSP of sunspots at $\Theta\approx50^{\circ}$.}
	\label{tab:vdop_limb}	
	\vspace{-15pt}	
\end{center}
\end{table}

\paragraph{Extreme and Average Velocities:} Table \ref{tab:vdop_limb} summarizes the extreme and ave\-rage values of v$_{\rm{dop}}$ measured in the CSP and the LSP. The maximal blueshifts in all the data sets is larger by a factor of two when compared to the values measured at disk center (cf.~Table \ref{tab:vdop_comp}). The maximal redshift, though, is weaker by roughly 30\%. In Spot 07 and Spot 08, the maximal v$_{\rm{dop}}$ is alike while extreme redshifts show higher values in Spot 05 and Spot 06. Similar to the behavior at disk center, the maximal amplitudes of blueshifts are of equal strength and the maximal redshifts vary by 50\% between Spot 05 and Spot 08. On a spatial average, the Doppler shift of the CSP and LSP is of opposite sign, but almost identical.

\section{Radial Dependency of Average Velocity}
\label{sec:rad_ave}

\begin{figure}[h!]
\begin{center}
    \includegraphics[width=0.7\textwidth]{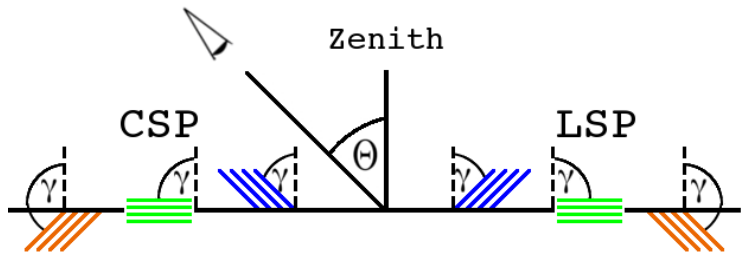}
\caption{Sketch of the penumbral velocity field under the projection of the heliocentric angle $\Theta$. The inclination $\gamma$ of the flows appears smaller, i.e. $\gamma - \Theta$, in the CSP and larger, i.e. $\gamma + \Theta$, in the LSP. The velocity field is axially symmetric around the zenith.}
\label{fig:inclupdown}
\end{center}
\end{figure}

Elliptical paths of increasing radii were placed in the penumbra to study the radial dependency of the horizontal EF. 
To ensure a good sampling for asymmetric sunspots, slightly different ellipses were used for the CSP and the LSP. For the interpretation of the results, it is useful to remember the geometry of the penumbral flows, especially under the projection of the heliocentric angle (cf.~Fig~\ref{fig:inclupdown}).

\begin{figure}[h!]
\begin{center}
    \includegraphics[width=\textwidth]{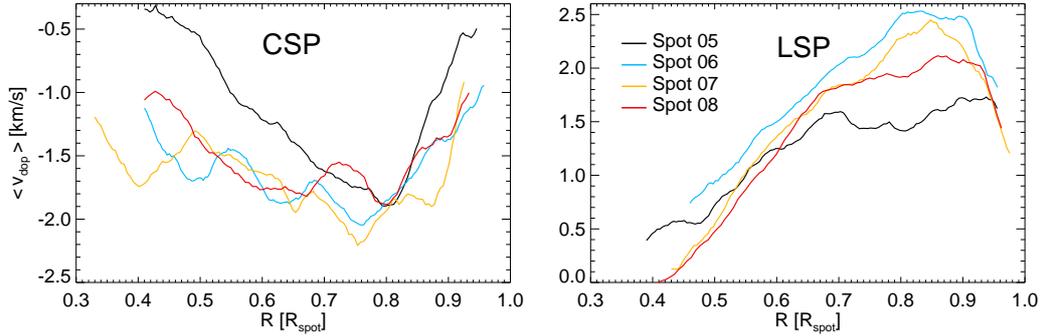}
\caption{Radial dependence of average values of Doppler velocity along elliptical cuts through the CSP and the LSP of four different spots.}
\label{fig:vdopave_limb}
\end{center}
\end{figure}

\paragraph{Center Side Penumbra:}  The left side of Fig.~\ref{fig:vdopave_limb} represents the radial dependency of the average velocity in the CSP. In the inner CSP, upflows are clearly visible as blueshifts because they appear almost parallel to the LOS (cf.~Fig.~\ref{fig:inclupdown}). With a blueshift of $-$1.1~km~s$^{-1}$, the average velocity in the inner CSP is comparable in all data sets. An exception is Spot 05, where a strongly redshifted area is found in the inner CSP (cf.~Section~\ref{sec:morph_out}). With increasing radial distance, the average velocity grows and reaches values of 1.8~km~s$^{-1}$ to 2.1~km~s$^{-1}$ at $\rm{R} = 0.8\cdot\rm{R}_{\rm{spot}}$. However, the increase of velocity with distance is not as smooth as in the LSP, and fluctuations of the order of 0.3~km~s$^{-1}$ are measurable (cf.~blue line in the plot on the right of Fig.~\ref{fig:vdopave_limb}). For $\rm{R} > 0.8\cdot\rm{R}_{\rm{spot}}$, the average blueshift decreases due to the domination of downflows in the outer penumbra.

\paragraph{Limb Side Penumbra:} The right panel of Fig.~\ref{fig:vdopave_limb} shows that the average velocity vanishes in the inner LSP in Spot 07 and Spot 08. This is because the inner penumbra is dominated by upflows (cf.~Section~\ref{sec:global}) that appear perpendicular\footnote{The zenith angle of of a typical upflow channel is 60$^{\circ}\pm5.5^{\circ}$ when determined from spectral inversion (cf.~Section~\ref{sec:invofupdown}).} to the LOS under the projection of the heliocentric angle (cf.~Fig.~\ref{fig:inclupdown}). Since the Doppler effect measures line shifts only along the LOS, upflows do not contribute to the velocity signal in the inner LSP. The residual velocity in Spot 05 and Spot 06 could be attributed to the irregular shape of the inner LSP that cannot be sampled correctly by an ellipse. For larger radii, the average Doppler velocities increases linearly, assumes a maximum of 2~km~s$^{-1}$ to 2.5~km~s$^{-1}$ at 85\% radial distance and decreases again for larger radii. 

While the growth in redshift with increasing radial distance can be be attributed to the fact that the horizontal EF appears more parallel to the LOS, the subsequent decrease is unexpected. This is because the outer penumbra is dominated by downflows (cf.~Section~\ref{sec:global}) that appear more parallel to the LOS than the horizontal EF (cf.~Fig.~\ref{fig:inclupdown}). There are a number of non exclusive explanations for this behavior:
\begin{itemize}
\item{The horizontal outflow ends in the outer penumbra and is on average much stronger when compared to the downflow.}
\item{The penumbral boundaries are not determined correctly, and granular upflows mix with the penumbral downflows, thereby decreasing the average penumbral redshift, cf.~\citet{BellotRubio:2004p4357}.}
\item{Part of the horizontal outflow continues along the magnetic field lines into the superpenumbral canopy \citep{Solanki:1994p4039,Rezaei:2006p157}. This decreases the average redshift, as this flow is directed upwards, i.e. more perpendicular to the LOS, in the outer penumbra.}
\item{The zenith angle in the downflow channels is larger than 140$^{\circ}$, and projection effects cause a decrease in redshift. This possibility will be investigated in detail in Section~\ref{sec:incl_down}.}
\end{itemize}



\section{Morphology of Evershed Outflow}
\label{sec:morph_out}



The horizontal component of the EF is clearly visible as a blueshift in the CSP and a redshift in the LSP. Perpendicular to the line of symmetry, the lineshift vanishes, indicating a radially outwards directed plasma flow. The filamentary structure of the EF is much more present in observation at large heliocentric angles (cf.~Fig.~\ref{fig:alldop_limb}) when compared to observation at disk center (cf.~Fig.~\ref{fig:alldop}).

\paragraph{Center Side Penumbra:} Narrow channels with v$_{\rm{dop}} < $ $-$2.5~km~s$^{-1}$ can be seen in the inner CSP, e.g. between $\rm{(x;y)}=(716$\arcsec$;-53$\arcsec$)$ and $\rm{(x;y)}=(721$\arcsec$;-56$\arcsec$)$ 
Under the projection of the heliocentric angle, the upflows in the inner CSP are almost parallel to the LOS. This explains the larger velocity when compared to the observation at disk center. The flow channels have a length ranging from 2\arcsec\,\,to 5\arcsec, and they are less than 1\arcsec\,\,in width. In the outer CSP, the appearance of areas with v$_{\rm{dop}} < $ $-$1.5~km~s$^{-1}$ is rather roundish, e.g. the patch between $\rm{(x;y)}=(707.$\arcsec$5;-58.$\arcsec$5)$ and $\rm{(x;y)}=(712$\arcsec$;-61.$\arcsec$5)$. The length of these areas ranges from 2\arcsec\,\,to 5\arcsec, while their width varies between 1\arcsec\,\,and 3.5\arcsec. 

\paragraph{Limb Side Penumbra:} With respect to the filamentary structure, the general appearance of the LSP coincides with that of the CSP. However, the redshifts in the inner LSP are weaker when compared to the blueshift of the inner CSP. This is due to the geometry of the upflow channels in the inner LSP (cf.~Fig.~\ref{fig:inclupdown}), which appear almost perpendicular to the LOS. In the mid LSP, the filamentary structure of the EF is visible. Flow channels with v$_{\rm{dop}} > $ 1.5~km~s$^{-1}$ are 3.\arcsec5 to 5.\arcsec5 long and less than 1\arcsec\,\,wide -- cf.~Spot 08 between $\rm{(x;y)}=(732$\arcsec$;-54$\arcsec$)$ and $\rm{(x;y)}=(737$\arcsec$;-60$\arcsec$)$. In the outer LSP the areas with v$_{\rm{dop}} > $ 1.5~km~s$^{-1}$ become more roundish, and the filamentary structure disappears. The strong downflow between $\rm{(x;y)}=(735.$\arcsec$5;-62$\arcsec$)$ and $\rm{(x;y)}=(737$\arcsec$;-64.$\arcsec$5)$, for example, shows redshifts corresponding to v$_{\rm{dop}} > $ 3.6~km~s$^{-1}$ and is of almost circular shape.


\paragraph{Counterexamples:} Even though the CSP shows a blueshift and the LSP a redshift, patches of opposite lineshifts can be found in both penumbral sides.

In the inner CSP, redshifts are located at $\rm{(x;y)}=(715.$\arcsec$5;-$54\arcsec$)$ and correspond to Doppler velocities of 0.1~km~s$^{-1}$. In the mid CSP, redshifts are located at $\rm{(x;y)}=(711$\arcsec$;-$49\arcsec$)$, $\rm{(x;y)}=(711$\arcsec$;-$53.\arcsec$5)$ and $\rm{(x;y)}=(715$\arcsec$;-$65.\arcsec$5)$ with a Doppler velocity of approximately 0.6~km~s$^{-1}$. Other examples are found at $\rm{(x;y)}=(717.$\arcsec$5;-$64\arcsec$)$ and $\rm{(x;y)}=(713.$\arcsec$5;-$43\arcsec$.5)$, with velocities exceeding 1~km~s$^{-1}$. The most intense redshift in the CSP shows a velocity of 1.8~km~s$^{-1}$ and resides at $\rm{(x;y)}=(708.$\arcsec$5;-$57\arcsec$)$. These examples indicate that $\gamma > 90^{\circ} + \Theta$ within some downflow channels in the outer penumbra. 

Blueshifts in the LSP appear close to the umbra, e.g. at $\rm{(x;y)}=(708.$\arcsec$5;-$57\arcsec$)$, $\rm{(x;y)}=(708.$\arcsec$5;-$57\arcsec$)$ and $\rm{(x;y)}=(708.$\arcsec$5;-$57\arcsec$)$ with a velocity of $-$0.6~km~s$^{-1}$. The mid and outer LSP is void of blueshifts, except for one patch with a LOS velocity of $-$0.2~km~s$^{-1}$ located at $\rm{(x;y)}=(711$\arcsec$;-$56.\arcsec5$)$.

\begin{figure}[h!]
\begin{center}
    \includegraphics[width=0.8\textwidth]{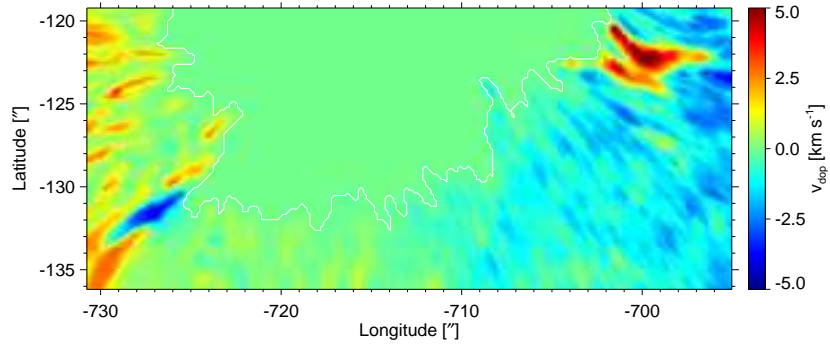}
\caption{Enlargement of Spot 05 with blue- and redshifts in the inner LSP and CSP.}
\label{fig:counterex_spot05}
\end{center}
\end{figure}

\paragraph{Possible Signs of Reconnection:} Another very remarkable counterexample can be seen in Spot 05 (cf.~Fig.~\ref{fig:counterex_spot05}). In the region around $\rm{(x;y)}=(-700$\arcsec$;$-124\arcsec$)$, strong redshifts are present in the inner CSP, while the inner LSP shows an area of strong blueshift around $\rm{(x;y)}=(-727$\arcsec$;-133$\arcsec$)$. The measured amplitudes correspond to $|\rm{v}_{\rm{dop}}| > $ 4~km~s$^{-1}$. 

If it is assumed that these line shifts are not due to an inverse\footnote{Even though an inverse Evershed flows has not been observed at photospheric layers it existence can not be ruled out.} Evershed flow, the respective up- and downflow velocity reaches 6~km~s$^{-1}$. Additional evidence for vertical plasma motion comes from \citet{2011arXiv1101.0751L,2011ApJ...727...49L} who investigated the redshifted area in the CSP in detail. Using a two-component Milne-Eddington type inversion they find velocities of up to 8.5~km~s$^{-1}$ . 

The mechanism to produce such strong downflows in the inner penumbra is still unknown. Even though some downflow patches in Spot 04 show velocity amplitude of comparable size (cf.~Fig.~\ref{fig:alldop}), they are all located in the outer penumbra. \citet{2011ApJ...727...49L} follow the argumentation of \citet{2008ApJ...686.1404R} that reconnection processes of highly twisted and interlaced penumbral filaments provide the energy to accelerate the plasma to the measured velocities.

\section{The Inclination of the Downflow Channels}
\label{sec:incl_down}

If it is assumed that the solar plasma follows the magnetic field lines, the zenith angle ($\gamma$) of penumbral downflows can be inferred from spectropolarimetric observation. \citet{BellotRubio:2003p206,BellotRubio:2004p4357} and \citet{Langhans:2005p9} found $\gamma \approx 100^{\circ}$ in the outer penumbra from inversion of Stokes profiles and the observation of sunspots at various $\Theta$. \citet{Ichimoto:2007p178} deducted $\gamma \approx 120^{\circ}$ for the downflow channels from the distribution of opposite polarity patches in the outer penumbra. However, as it will be demonstrated in the following, the redshifts on the CSP of Spot 08 imply $\gamma~>~135^{\circ}$ in some downflow channels.

\begin{figure}[h!]
\begin{center}
    \includegraphics[width=0.7\textwidth]{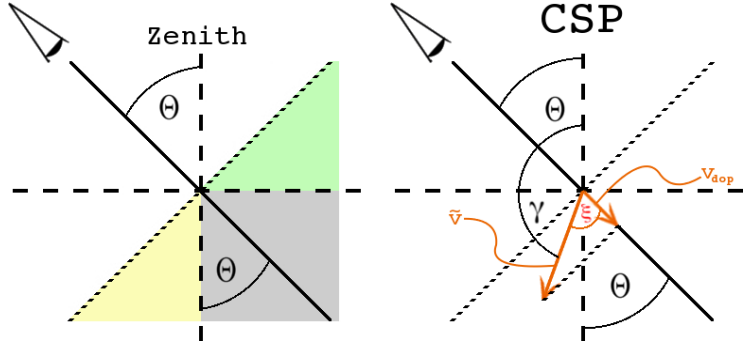}
\caption{Left: Possible values of $\gamma$ under the projection of $\Theta$. If $\gamma$ is within the colored regions redshifts are visible in the CSP. Right: Configuration of downflow in the CSP.}
\label{fig:incl_down_def}
\end{center}
\end{figure}

\paragraph{Geometrical Considerations:} From the plot on the left side of Fig.~\ref{fig:incl_down_def}, it is evident that redshifts on the CSP occur only if $\gamma$ lies within the colored regions, i.e. $\Theta + 90^{\circ} < \gamma < \Theta + 270^{\circ}$. Since the flow into the superpenumbral canopy is a continuation of the EF \citep{Solanki:1994p4039,Rezaei:2006p157}, it is directed radially outwards and cannot produce redshifts on the CSP. Thus, it is expected that flows with $270^{\circ} < \gamma < 270^{\circ} + \Theta$, i.e. the green region, are not causing the redshifts seen in the CSP. If it is assumed that the EF is directed radially outwards, it follows $180^{\circ} < \gamma < 270^{\circ} + \Theta$, and the gray region is discarded.

\paragraph{Velocity Amplitude Considerations:} The right side of Fig.~\ref{fig:incl_down_def} illustrates that the difference in v$_{\rm{dop}}$, i.e the measured Doppler velocity, is caused by the projection of v, i.e. the absolute velocity, onto the LOS. Fig.~\ref{fig:incl_down_def} yields:

\begin{eqnarray}
\gamma_{\rm{CSP}} & = & 180^{\circ} + \Theta_{\rm{CSP}} + \xi \,\,\,\,\,;\,\,\,\,\, -90^{\circ} \le \xi \le 90^{\circ} \nonumber \\
& = & 180^{\circ} + \Theta_{\rm{CSP}}  \pm \rm{cos}^{-1} \left( \frac{\rm{v_{\rm{dop}}}} {\rm{v}} \right)
 \label{eq:CSP}
\end{eqnarray}

\begin{figure}[h!]
\begin{center}
    \includegraphics[width=0.8\textwidth]{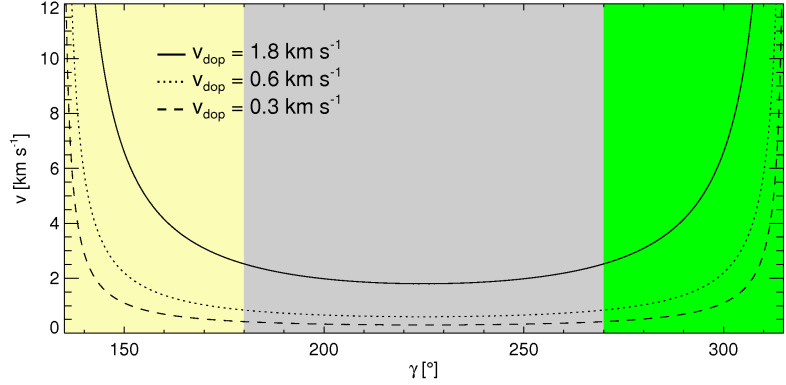}
\caption{Possible angles and absolute velocities of the downflow channels in the CSP of Spot 04. The colors represent the same range of angles as in Fig.~\ref{fig:incl_down_def}.}
\label{fig:incl_down_v_dop}
\end{center}
\end{figure}

\noindent In Fig.~\ref{fig:incl_down_v_dop}, the absolute velocity is plotted as a function of the inclination of the downflow channel. The solid line corresponds to a measured v$_{\rm{dop}}$ of 1.8~km~s$^{-1}$, while the dotted line represents v$_{\rm{dop}} = 0.6$~km~s$^{-1}$ and the dashed  represents v$_{\rm{dop}} = 0.3$~km~s$^{-1}$, i.e. the  v$_{\rm{dop}}$ values measured in the CSP of Spot 08. The yellow region illustrates redshifts caused by radially outwards directed downflows.

\paragraph{Results:} If it is assumed that the EF is a radial outwards directed flow, it is possible to put constraints on the configuration of the downflow channels causing the redshifts in the CSP of Spot 08.

Since $\Theta = 45^{\circ}$ at these locations, an inclination of $135^{\circ} < \gamma < 180^{\circ}$, i.e. the yellow region in the plot on the left side of Fig.~\ref{fig:incl_down_def}, is obtained from geometrical considerations.

If it is further assumed that absolute plasma velocities do not exceed\footnote{\citet{Ichimoto:2007p178} found a maximal v$_{\rm{dop}}$ of 7~km~s$^{-1}$, while \citet{2001ApJ...549L.139D} 
found vertical velocities of 
16~km~s$^{-1}$ from inversion of Stokes profiles. The highest v$_{\rm{dop}}$ obtained in this work is 12~km~s$^{-1}$, cf.~Section~\ref{sec:multigauss}.} 12~km~s$^{-1}$, Fig.~\ref{fig:incl_down_v_dop} yields $143^{\circ} < \gamma < 180^{\circ}$ in the case of v$_{\rm{dop}} = 1.8$~km~s$^{-1}$ and $137^{\circ} < \gamma < 180^{\circ}$ for v$_{\rm{dop}} = 0.6$~km~s$^{-1}$.

\section{Correlations at Large Heliocentric {Angles}}
\label{sec:corr_limb}

For penumbrae at large heliocentric angles, it not reasonable to calculate a global correlation coefficient for I$_{\rm{c}}$ and v$_{\rm{dop}}$. Both quantities increase with radial distance from the center of the spot, thereby introducing an artificial correlation, e.g. \citep{1994ApJ...430..413S} and Section~\ref{sec:outflow}. Thus, a correlation coefficient was computed based on the fluctuations of the respective quantities around:

\begin{itemize}
\item{The mean along azimuthal cuts of different radii =: r$_{\rm{S}}$(R)}
\item{A local mean value =: r$_{\rm{S}}$(L)}
\end{itemize}

\noindent Furthermore, the CSP and the LSP have to be investigated individually, as the opposite velocity signals would otherwise cancel each other. Due to the definition of blue- and redshifts, a value of r$_{\rm{S}} = 1$ corresponds to a flow in the dark structures in the CSP, while it indicates a flow in the bright structures on the LSP.

\paragraph{Average Correlation Coefficients:} A calculation of the average r$_{\rm{S}}$(R) in the LSP yields values ranging from $-$0.08 in Spot 08 to $-$0.21 in Spot 06. In the CSP, values between 0.01 and 0.11 are obtained in Spot 05 and Spot 06 respectively. For r$_{\rm{S}}$(L), the values of the LSP vary between $-$0.37 in Spot 06 to $-$0.26 in Spot 05, while the CSP shows correlation coefficients between 0.05 in Spot 05, and 0.13 in Spot 08. On a global scale, r$_{\rm{S}}$(R) and r$_{\rm{S}}$(L) are thus too small to conclude that the Evershed outflow is concentrated in the dark filaments.

There is a tendency for r$_{\rm{S}}$(R) to be smaller than r$_{\rm{S}}$(A), which indicates that correlation exists rather on a local  scale. The correlation coefficients of the CSP are always smaller when compared to the LSP. This could be due to the Wilson depression, which results in a larger angle between LOS and the zenith on the CSP than on the LSP. Thus, on the CSP,  v$_{\rm{dop}}$ is measured in a slightly higher atmospheric layer when compared to the LSP.

\begin{figure}[h!]
\begin{center}
    \includegraphics[width=\textwidth]{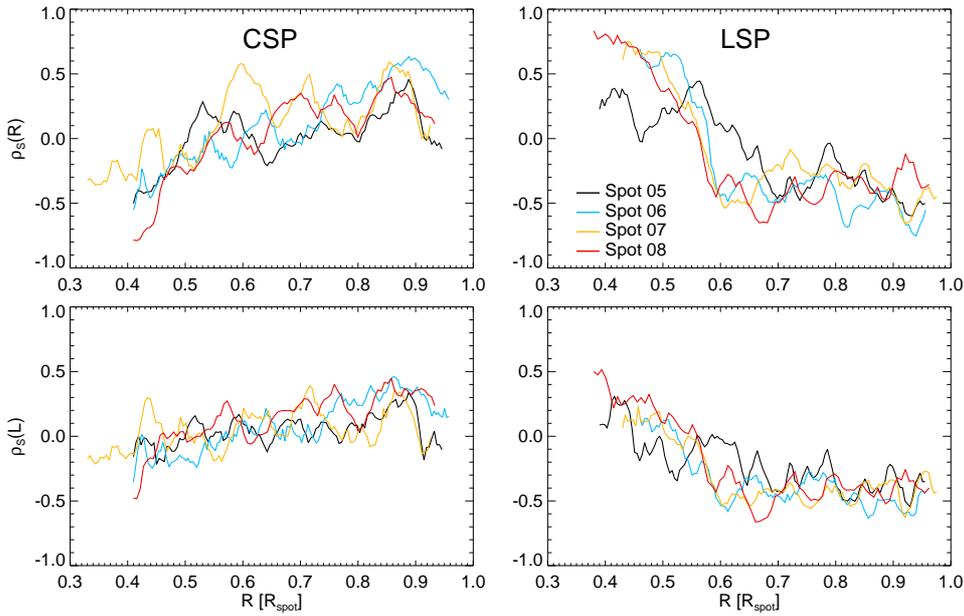}
\caption{Spearman's rank correlation coefficient (r$_{\rm{S}}$) for I$_{\rm{c}}$ and v$_{\rm{dop}}$ along elliptical cuts of different radii. Top: Radial dependency of r$_{\rm{S}}$ computed from the fluctuation of the quantities around their azimuthal mean value in the CSP and the LSP. Bottom: Same as top, but r$_{\rm{S}}$ computed from the fluctuation of the quantities around their local mean value. }
\label{fig:corr_Ic_vdop_limb}
\end{center}
\end{figure}

\paragraph{Radial Dependency of Correlation Coefficients:} In the top row of Fig.~\ref{fig:corr_Ic_vdop_limb}, the radial dependency of r$_{\rm{S}}$(R) is plotted for the LSP and the CSP respectively. The correlation coefficient\footnote{Note that the null hypothesis (H$_0$) cannot be rejected on a 3$\sigma$ confidence level (cf.~Appendix~\ref{ch:correlation}). This is especially the case for r$_{\rm{S}}$(R) and r$_{\rm{L}}$(R) in the CSP, where the significance drops below 1$\sigma$ for some radial cuts.} assumes large positive values (r$_{\rm{S}}(\rm{R})\ge 0.7$) in the inner LSP, indicating that the flow is present in the bright structures. Between 50\% and 60\% radial distance r$_{\rm{S}}$(R) drops to values of around $-$0.5. Even though the fluctuations of r$_{\rm{S}}$(R) are large for $0.6 \le \rm{R}_{\rm{Spot}} \le 1$ on average, there is a tendency for the flow in the mid and outer LSP to be situated in the dark structures. Note that the behavior of r$_{\rm{S}}$(R) in the LSP of Spot 05 is significantly different from that of the other data samples. This is probably due to the irregular shape of the LSP.

In the inner CSP, the correlation is not as distinct as in the inner LSP, and r$_{\rm{S}}$(R) shows a large scatter at 40\% radial distance. Only Spot 08 provides a clear indication for a correlation between Doppler velocity and continuum intensity. Similar to the LSP, r$_{\rm{S}}$(R) reverses its sign between 50\% and 60\% radial distance. However, the variance of r$_{\rm{S}}$(R) in the mid and outer CSP is larger when compared to the LSP. The correlation coefficient in the LSP assumes sporadically r$_{\rm{S}}(\rm{R}) \ge 0.5$, but the general trend shows a tendency for the flow to be present in the bright structures in the inner CSP and in the dark structures in the mid and outer CSP. 

The bottom row of Fig.~\ref{fig:corr_Ic_vdop_limb} shows r$_{\rm{S}}$(L) computed from the fluctuation of I$_{\rm{c}}$ and v$_{\rm{dop}}$ around their local averages. The radial dependence of r$_{\rm{S}}$(L) in the LSP and the CSP shows a similar behavior as r$_{\rm{S}}$(R), but the amplitudes and fluctuations are smaller. This is in contrast to the behavior at disk center, where r$_{\rm{S}} (\rm{L}) \ge \rm{r}_{\rm{S}} (\rm{R})$. In the CSP, $|\rm{r}_{\rm{S}}(\rm{L})|$ is always below 0.5 at all radial distances, which is too small to infer a definite correlation between I$_{\rm{c}}$ and v$_{\rm{dop}}$. Nevertheless, the general trend of flows present in the bright structures of the inner penumbra and in dark structures in the mid and in the outer penumbra is still ascertainable.

\paragraph{Results:} The results found in this study are in accordance with \citet{Ichimoto:2007p178}, who report on a correlation of I$_{\rm{c}}$ and v$_{\rm{dop}}$ in the inner but on an anticorrelation in the outer penumbra.

\paragraph{Discussion:} Other reports on a correlation of I$_{\rm{c}}$ and v$_{\rm{dop}}$ in penumbrae at large heliocentric angles are contradictory (cf.~Section~\ref{sec:theo_penum}). \citet{Wiehr:1989p2899} as well as \citet{1990ApJ...355..329L} do not find a correlation, while \citet{Wiehr:1992p3002}, \citet{1994ApJ...430..413S} and \citet{Langhans:2005p9} deduce an anticorrelation (flow in dark structures) from visual inspection. \citet{Title:1993p216} report on high Doppler velocities in the dark structures inferred from a correlation analysis along selected paths through the penumbra. \citet{Schlichenmaier:2005p219} find a correlation only for short azimuthal traces that cover a few filaments, but no correlation for larger paths through the penumbra. 

Part of the confusion arises from the lack of spatial resolution of the investigated data sets and from the different methods used to calculate the Doppler velocity. \citet{Wiehr:1994p412} as well as \citet{Rimmele:1995p2411} find an anticorrelation as long as they compare both quantities at a spatial resolution better than 0.\arcsec5\,\,and in similar atmospheric layers. In \citet{Wiehr:1992p3002,Title:1993p216,Ichimoto:2007p178}, I$_{\rm{c}}$ and v$_{\rm{dop}}$ are compared along radial cuts. A visual inspection easily yields an anticorrelation between I$_{\rm{c}}$ and v$_{\rm{dop}}$. This result, however, might be biased by the fact that the human eye tends to focus on very prominent features. The same argument holds for a visual inspection of two dimensional maps \citep{1994ApJ...430..413S,Langhans:2005p9}.

The most reliable way to study a correlation is to compute correlation coefficients. If Pearson's correlation coefficient r$_{\rm{P}}$ is used (cf.~Appendix \ref{ch:correlation}), a linear relationship is assumed a-priori, which is not always justified. Additional problems arise from a lack of standardization in the interpretation of the results. If, for example, r$_{\rm{S}} = 0.5$, only 25\% of the variance of the data can be explained by the correlation \citep{LSachs:1999}.

If the results are interpreted in this context, it could be argued that the respective quantities cannot be called correlated if the correlation coefficient does not exceed a certain value, e.g. r$_{\rm{S}} \ge 0.5$ \citep{JCohen:2003}. From that perspective, the studies of \citet{Title:1993p216}, \citet{Schlichenmaier:2005p219} and \citet{Ichimoto:2007p178} as well as the results of Section~\ref{sec:corr_center} and \ref{sec:corr_limb} do not contradict each other.

\section{Summary}
\label{sec:concl_obs_limb}

Observations of sunspots at large heliocentric angles ($\Theta \approx 50^{\circ}$) were used to study the horizontal component of the Evershed flow. The velocity field shows a filamentary structure comparable to the maps of continuum intensity, with blueshifts in the center and redshifts in the limb side penumbra. In the inner center side penumbra, the strong blueshifts are confined to narrow channels, which become roundish with increasing radial distance form the center of the spot. Except for the inner limb side penumbra, which exhibits no Doppler shift, large redshifts cover the mid and outer limb side penumbra. Similar to the center side penumbra, the Doppler velocity is confined to narrow channels in the mid limb side penumbra, but occurs in more circular patches further outwards. When compared to observations at disk center, the maximal amplitudes of the Doppler velocity are larger for blueshifts, but smaller for redshifts.

Ellipses of different radii were placed in the center and in the limb side penumbra to study the velocity field averaged along azimuthal paths. The dependency of these average values on the radial distance can be explained, by taking into account the configuration of up- and downflow channels under the projection of the heliocentric. The fact that the Doppler shift vanishes in the inner limb side penumbra was interpreted as evidence that the flow occurs perpendicularly to the line of sight. This implies a zenith angle ($\gamma$) of approximately 40$^{\circ}$ for the upflows in the inner penumbra. 

Even though the center and in the limb side penumbra show predominately blue- and redshifts, there is a range of counterexamples. Patches of redshifts in the mid and outer center side penumbra were used to calculate the zenith angle within the downflow channels. With $\gamma > 135^{\circ}$, the zenith angle obtained in this study is larger than all values previously reported.

Spearman's ranked correlation coefficient ($\rho_{\rm{S}}$) was calculated for the intensity and the Doppler velocity field. The results show that $\rho_{\rm{S}}$ is positive in the inner, but negative in the outer penumbra. However, $\rho_{\rm{S}}$ is not large enough to consider the continuum intensity and the Doppler velocity as being correlated. This is neither true for fluctuations around an azimuthal, nor for fluctuations around a local mean value. These results were discussed in the context of previous reports on e.g. correlations of strong outflows in the dark penumbral filaments, and it was proposed that contradictory results depend on unequal evaluation procedures or a lack of standardization for the interpretation.

\chapter{Asymmetries and Crossover Profiles}
\label{ch:asym}
Asymmetries in Stokes profiles are of great interest, as they contain valuable information on the gradient with height of atmospheric parameters. The shape of the bisector of Stokes I profiles is used in Section~\ref{sec:gradv} to compare the stratification of the velocity field in the penumbra and in the quiet Sun. The implications of crossover Stokes V profiles for penumbral models are described in Section~\ref{sec:hiddenopp}. Asymmetries in Stokes V are quantified in Section~\ref{sec:implications}, where it is furthermore argued that all penumbral plasma flows are magnetized. Section~\ref{sec:invofupdown} demonstrates the difficulties of accounting for line asymmetries in spectral inversions by means of typical Stokes profiles from a blue- and a redshifted region. Finally the results of this Chapter are summarized in Section~\ref{sec:concl_asym}.

\section{Height Dependency of Penumbral and Quiet Sun Velocity Fields}
\label{sec:gradv}
The Doppler maps presented in Chapters~\ref{ch:center} and~\ref{ch:limb} quantify the plasma flows occurring just above the solar surface. Doppler shifts at different line depths that contain information on the velocity field at different atmospheric heights were neglected.

In the following, the asymmetry of typical penumbral Stokes I profiles will be characterized and qualitatively compared to QS profiles. Additionally, the entire velocity field of a penumbral region and a QS region of comparable size is statistically analyzed at different line depths. 

\begin{figure}[h!]
	\centering
		\includegraphics[width={0.9\textwidth}]{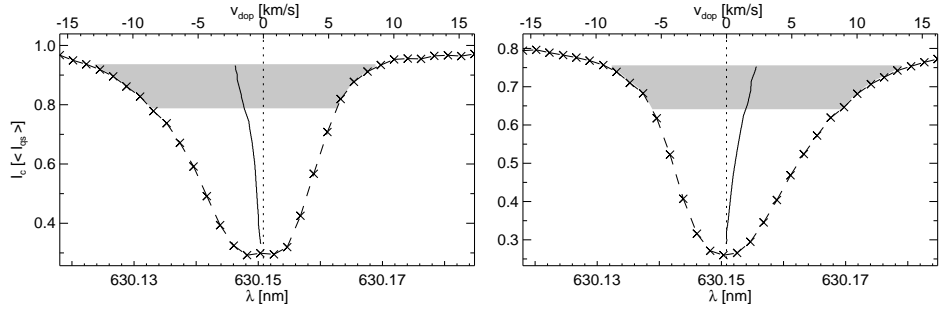}
		\caption{Stokes I profile of Fe I 630.15 nm for a typical blue- (left) and redshifted (right) region in the penumbra. The bisector (solid black) is plotted between 10\% and 90 \% line depression. The line-wing (shaded gray) is used to compute the Doppler maps.}
		\label{fig:bisec_pen}
\end{figure}

\paragraph{Penumbra:} Fig.~\ref{fig:bisec_pen} shows the absorption line profile of Fe I 630.15 from a typical\footnote{The blueshifted profile shows a broader and split line-core. These minima belong to the blue- and the redshifted $\sigma^+$ and $\sigma^-$ components of the atomic transition. This indicates a larger magnetic field strength in the upflow area when compared to the downflow patch, since the energy gap between atomic levels is proportional to the strength of an external magnetic field (Zeeman effect). This finding is in accordance with the fact that the latter appears at greater radial distances from the umbra, where the overall magnetic field strength is lower (cf.~Section~\ref{sec:theo_penum}).} up- (left) and downflow (right) region from Spot 04. The bisector (solid black) is evaluated between 10\% and 90 \% line depression (cf.~Section~\ref{sec:bisec}), and the Doppler velocity, which is equivalent to a respective shift in wavelength, is plotted at the top. It can be seen that the maximal shift occurs in the line-wing (shaded gray). With increasing line depth, the absorption profile becomes more symmetric and the bisector monotonously approaches the vacuum wavelength. Close to the line-core, no significant Doppler shift is measurable.

\paragraph{Quiet Sun:} Different results are obtained, if the same analysis is performed with typical QS profiles. The left side of Fig.~\ref{fig:bisec_pen} shows a blueshifted profile, which was measured in a granule of data set QS 01. The plot on the right side depicts a redshifted profile, which was observed in the intergranulum of the same data set. It can be seen that the QS profiles are more symmetric than penumbral profiles, which is reflected by the shape of the bisectors. In contrast to the penumbra, QS profiles show a significant Doppler shift at all line depression values, even at the core of the line.

\begin{figure}[h!]
	\centering
		\includegraphics[width={0.9\textwidth}]{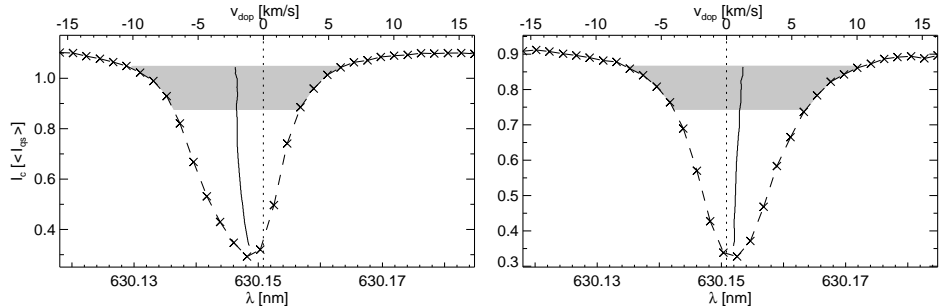}
		\caption{Same as in Fig.~\ref{fig:bisec_pen}, but for a red- (left) and blueshifted (right) region in the QS.}
		\label{fig:bisec_qs}
\end{figure}

\paragraph{Interpretation:} The asymmetry of individual penumbral profiles from up- and downflow regions can be best understood by assuming that the amplitude of v$_{\rm{dop}}$ decreases with geometrical height. This picture is further supported by the mo\-no\-to\-nous shift of the bisector and the fact that line profiles from the center of an up- or downflow region are more asymmetric when compared to profiles from the edge of such a flow patch. It is unlikely that the distribution of lateral unresolved structures is at the origin of these asymmetries. Thus, the amplitude of penumbral plasma flows are strong in the lower atmospheric layers and decreases with height. In the higher layers of the LFR, velocities are largely suppressed.

In the QS, the so-called convective overshoot \citep{1990ARA&A..28..263S,2000A&A...355..381K} results in plasma motions within the LFR. The superadiabatic stratification of the convection zone causes hot parcels of plasma to rise buoyantly from below the solar surface. Even though the buoyancy vanishes in the convectively stable layers of the solar photosphere, these parcels continue to rise due to the conservation of momentum. During its further ascent, the plasma cools radiatively and its velocity is reduced by gravity. Eventually, the cool plasma returns below the surface because of its larger density. 
The shape of the bisectors, and especially the shift of the line-core position of QS profiles implies that the vertical flows are of comparable strength in low atmospheric layers, whereas the plasma in the QS ascents much higher into the atmosphere when compared to the penumbra.

\begin{figure}[h!]
	\centering
		\includegraphics[width={\textwidth}]{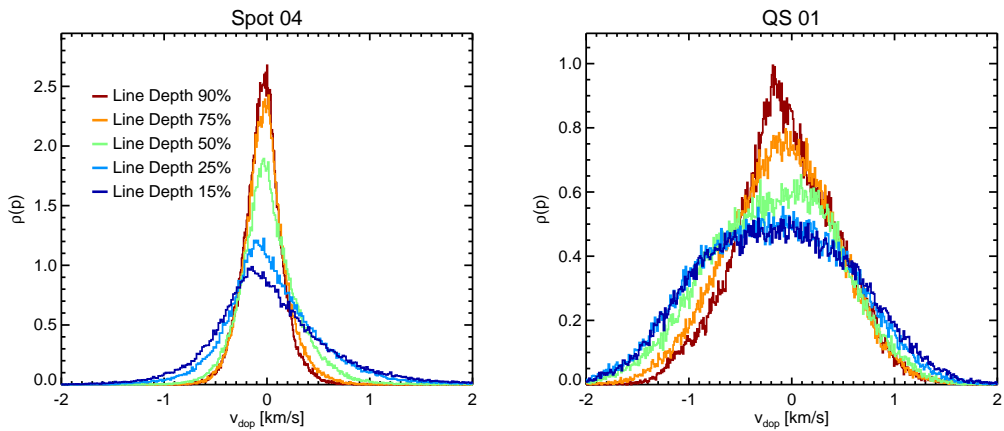}
		\caption{Left: PDFs for penumbral Doppler maps computed from line shift of bisector averaged for $\pm$5\% line depression values around the indicated line depth. Right: Same as left, but for an equal area of QS at disk center.}
		\label{fig:bisec_height}
\end{figure}

\paragraph{Statistical Analysis:} To obtain statistical results on the gradient of the velocity field with height, Doppler maps were constructed from the line shift of an average bisector computed from an interval of $\pm$5\% around different line depths. From these maps, PDFs were computed for the velocity field of the penumbra (cf.~left side of Fig.~\ref{fig:bisec_height}) and in an area of equal size of the QS data set (cf.~right side of Fig.~\ref{fig:bisec_height}).

\paragraph{Results:} The characteristics of these PDFs, i.e. mean and RMS values as well as the skewness, are listed in Table~\ref{tab:PDFs_compare}. These parameters allow a comparison of the velocity field at different atmospheric heights above the penumbra and the QS.




\begin{table}[h!]
\begin{center}
	\begin{tabular}{ccccccc}
		\hline
		\hline
		\\[-2ex]
		{Line} & \multicolumn{2}{c}{Mean [km~s$^{-1}$]} & \multicolumn{2}{c}{RMS [km~s$^{-1}$]} & \multicolumn{2}{c}{Skewness}\\
		{Depth [\%]} & {Spot 04} & {QS 01} & {Spot 04} & {QS 01} & {Spot 04} & {QS 01}\\
		\\[-2ex]
		\hline		
		\\[-2ex]
		{90} & {-0.03} & {-0.04} & {0.19}& {0.49} & {0.73} & {0.10}\\
		{75} & {0.01} & {-0.10} & {0.22}& {0.54} & {0.94} & {0.01}\\
		{50} & {0.04} & {-0.17} & {0.30}& {0.63} & {1.22} & {-0.28}\\
		{25} & {0.07} & {-0.18} & {0.49}& {0.71} & {1.23} & {-0.02}\\
		{15} & {0.06} & {-0.14} & {0.61}& {0.73} & {1.33} & {0.04}\\
		\hline
	\end{tabular}
	\caption{Characteristics of PDFs for the Doppler maps of Spot 04 and QS 01. The maps are computed from an average bisector around the indicated line depth.}
	\label{tab:PDFs_compare}	
\end{center}
\end{table}

At large line depression values, which correspond to high atmospheric layers, the penumbra of Spot 04 shows a blueshift. With decreasing line depth, however, an increasing redshift is measured, reaching maximal values of 0.07~km~s$^{-1}$ at 25\% line depression. The map of QS 01 shows a slight blueshift at large line depths. With decreasing line depression values, the blueshift increases before it decreases again for very small line depths. This behavior is reasonable since the bisector of an average QS profile exhibits a C-shape (cf.~Fig.~\ref{fig:bisec_expl}). 

The RMS values of both maps, demonstrate that the amplitudes of the velocity field decrease with atmospheric height. The decrease between the lower and upper LFR is about 70\% in the map of Spot 04, while it amounts to 30\% in the map of QS 01. Between 15\% and 25\% line depression, the drop in RMS amplitude is 20\% in Spot 04 and 2\% in QS 01. 

The skewness, a measure of the asymmetry of the histogram, is also different in the two data samples. In Spot 04, it is positive at all line depths, indicating a larger number of high downflow velocities (cf.~Fig.~\ref{fig:histo-sunspot}). Since the skewness decreases monotonously with height, there is a tendency for the high downflow velocities to appear in the lower atmospheric regions. Compared to Spot 04, the histogram of QS 01 shows a smaller skewness at all line depths, and no distinct trend can be observed with atmospheric height.


\paragraph{Conclusion:} The statistical analysis confirms the picture obtained from the investigation of individual bisectors. The RMS velocities drop by 70\% between the bottom and the top of the LFR in the penumbra. In the quiet Sun, however, this decrease is only 30\%. Thus, there is a distinct difference between the velocity gradient with height above the penumbra and the QS. It is most likely that the strong and inclined magnetic field in the penumbra causes the vertical velocity to ease much faster with height when compared to the QS. 

\section{Crossover Profiles and Hidden Opposite Polarity}
\label{sec:hiddenopp}

\begin{figure}[h!]
\begin{center}
    \includegraphics[width=\textwidth]{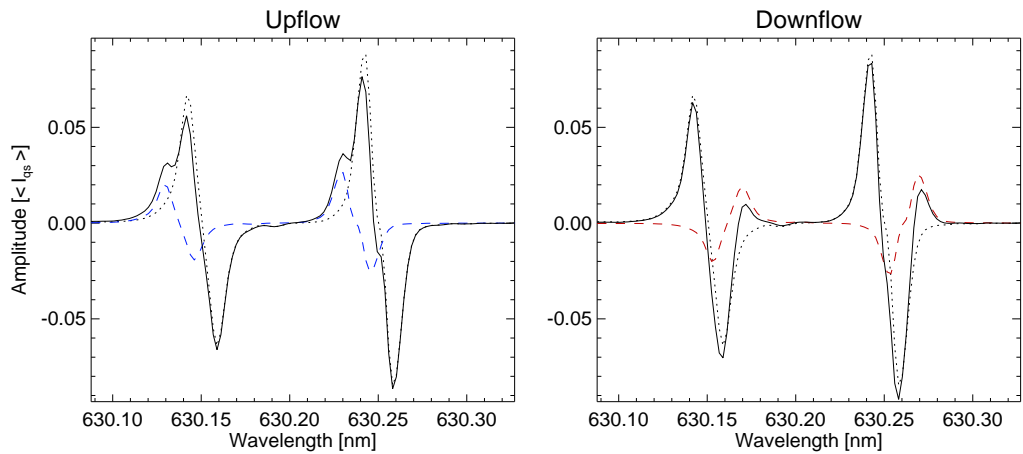}
\caption{Left: Plot of two antisymmetric Stokes V profiles -- unshifted (dotted) and blueshifted (dashed) -- with different amplitudes and the same polarity as well as their superposition (solid). Right: Superposition (solid) of two antisymmetric Stokes V profiles -- unshifted (dotted) and redshifted (dashed) -- of different amplitudes but opposite polarity.}
\label{fig:sketchstok}
\end{center}
\end{figure}

Asymmetries are not only measurable in Stokes I, but in the other Stokes parameters as well. Stokes V profiles, for example, often deviate from antisymmetry and show a third lobe on the red side of the regular profile (cf.~right plot in Fig.~\ref{fig:sketchstok}). These asymmetries, can be understood in the framework of a multilayer atmosphere, which 
is of special interest for the discussion of adequate penumbral models. \citet{Spruit:2006p2} argue that neither the idea of turbulent pumping (cf.~Section~\ref{sec:siphon}), nor the flux tube model (cf.~Section~\ref{sec:buoyant}) are correct descriptions of the penumbra, since both scenarios require large areas of opposite magnetic polarity in the outer penumbra, which are not visible in magnetograms with high spatial resolution, e.g. \citet{Langhans:2005p9}. In the following it will, however, be demonstrated that a large number of Stokes V profiles from downflow regions in the outer penumbra indeed show a component with opposite polarity.

\paragraph{Evidence for a Multilayer Atmosphere:} %
\noindent \citet{Wiehr:1995p379} showed that the asymmetries in Stokes I are due to so-called line satellites, i.e. an additional line component that shows Doppler shifts of different strengths. In Section~\ref{sec:multigauss}, it was demonstrated that even strongly asymmetric Stokes I profiles can be reproduced reasonably well with a double Gaussian fit. Furthermore, it has been argued that the asymmetries in Stokes I are due to the gradient with height of the velocity field (cf.~Section~\ref{sec:gradv}). In the simplest scenario, this gradient is approximated by a stack of two atmospheric layers. Incorporating Stokes V profiles in the analysis allows to infer basic 	 characteristics of the magnetic field in these layers.

\paragraph{Crossover Profiles:} Fig.~\ref{fig:sketchstok} shows an asymmetric Stokes V profile. In the upflow example, there is a shoulder on the blue wing of an atisymmetric profile, while the downflow profile contains a third lobe on the red side of a regular Stokes V profiles. Profiles with this kind of asymmetries are called crossover profiles. They were observed in the spectra of stars \citep{Babcock:1951p2083} as well as in solar spectra of sunspots \citep{Grigorjev:1972p175}.

Crossover profiles can be understood as the superposition of a Stokes V profile from a steady but magnetized atmospheric layer with a blue or redshifted profile from an atmospheric layer with magnetic fields and strong plasma velocities (cf.~Fig.~\ref{fig:sketchstok}). Furthermore, it is necessary to superimpose Stokes V profiles of the same polarity to reproduce the crossover profile of the upflow, while the asymmetries of the downflow profile require the redshifted profile (dashed line in right panel of Fig.~\ref{fig:sketchstok}) to be of opposite polarity.

\paragraph{Implication for Penumbral Observations:} These two examples demonstrate that it is important to use solar observation with  high spatial and high spectral resolution. The study of \citet{Langhans:2005p9}, for example, was conducted at the Swedish Solar Tower Telescope, using the Lockheed Solar Optical Universal Polarimeter (SOUP). This Lyot-type filtergraph measures the intensities of left and right circular polarized light (I$^{\rm{LCP}}$ and I$^{\rm{RCP}}$) with a bandpass of FWHM 7.2 pm at $\pm5$ pm from the line-core \citep{2001ApJ...553..449B,Langhans:2007p16}. In the case of Fe I 630.25, magnetograms are constructed like:

\begin{equation}
\rm{M}=\frac{\rm{I}_{\rm{blue}}^{\rm{LCP}}-\rm{I}_{\rm{blue}}^{\rm{RCP}}}{\rm{I}_{\rm{blue}}^{\rm{LCP}}+\rm{I}_{\rm{blue}}^{\rm{RCP}}}
\label{eq:magneto}
\end{equation}

\noindent with $\rm{I}_{\rm{blue}}^{\rm{LCP}}$ and $\rm{I}_{\rm{blue}}^{\rm{RCP}}$ being the intensities measured by SOUP detuned $-5$ pm to the blue with respect to the line-core. Since the third lobe in Stokes V profiles from downflow regions 
is not detectable within a redshift of 10 pm from the zero crossing, a significant amount of flux that actually returns into the Sun within the penumbral boundary remains undetected.

\paragraph{Estimates of Hidden Opposite Polarity:} To estimate the amount of hidden opposite polarity, three experiments were conducted:

\begin{itemize}
\item[1)]{Magnetograms were constructed according to Equation~\ref{eq:magneto}, using HINODE SP data of Spot 04. A linear interpolation was applied between the spectral measurements to infer the polarity of Stokes V at $-$5 pm from the line-core of Fe I 630.25.} 
\item[2)]{Stokes V profiles were analyzed to determine the number of local extrema above the 3$\cdot\sigma$ noise level. This procedure was spoiled by the so-called magneto-optical effect, represented by the antisymmetric $\rho$ terms in the absorption matrix (cf. Equations~\ref{eq:rhoq}, \ref{eq:rhou} and \ref{eq:rhov}) which causes additional extrema around the zero crossing in Stokes V. This is especially the case in profiles from the inner penumbra, where the magnetic field is strong. To exclude the magneto-optical effect from the study, the signal of an adjacent $\pm3$ pm interval around the local extrema is taken into account. Only if the mean value of the interval is of the same polarity as the extrema itself is it considered as a lobe.}
\item[3)]{To ensure that the additional extrema occur on the red side of the antisymmetric profile, the signal strength of Stokes V was averaged around 16$\pm$3 pm from the line-core.}
\end{itemize}

A cross-check between the second and the third study yields differences only in places where antisymmetric Stokes V profiles with a polarity opposite to that of the umbra are measured. Thus, only the first and second survey will be compared in the following. 

\begin{figure}[h!]
\begin{center}
    \includegraphics[width=\textwidth]{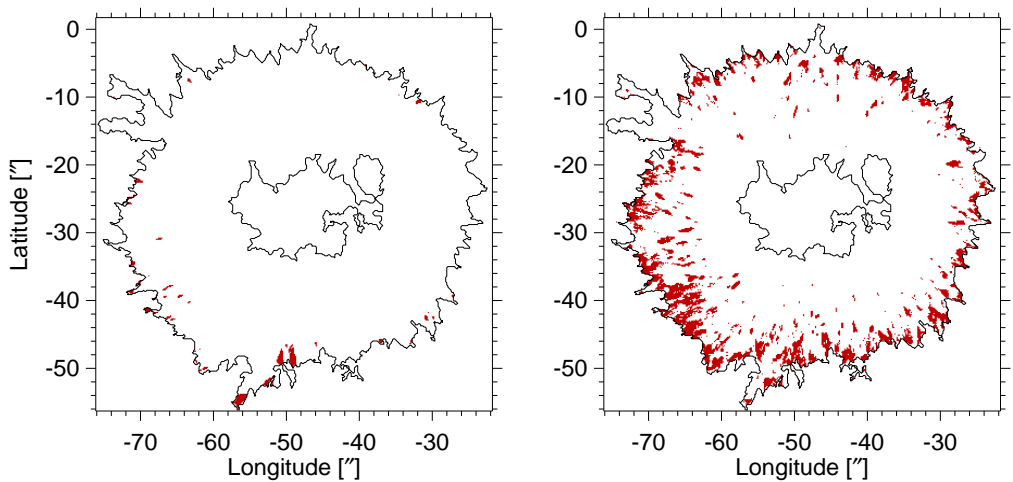}
	\caption{Left: Penumbra of Spot 04 outlined by the black countours. The red areas mark regions where a magnetogram constructed at $-$5 pm from the line-core detects fields of opposite polarity in downflow regions. Right: Same as left, but the red areas mark places where the Stokes V profile contains an additional lobe on the red side.}
\label{fig:lobes}
\end{center}
\end{figure}

\paragraph{Results:} The red areas in the left panel of Fig.~\ref{fig:lobes} indicate downflows with v$_{\rm{dop}} > 0.1$~km~s$^{-1}$, where Stokes V at $-$5 pm shows a polarity opposite to that of the umbra. In the plot on the right, red areas denote downflows as well, but the Stokes V profiles contain an additional lobe on the red side. A comparison already shows that the majority of magnetic fields with opposite polarity is hidden in classical magnetograms. A quantitative investigation shows that using method 1), only 3\% of the penumbral downflows show Stokes V profiles with a polarity opposite to that of the umbra. Applying method 2) it is found that 36\% of all penumbral downflows contain Stokes V profiles with hidden opposite polarity. 
 
It is necessary to note that this represents a lower limit only. Most probably, the magnetic field returns back into the solar surface in all penumbral downflows. The reason why HINODE SP measures opposite polarities in only 40\% of the penumbral downflows is that the second and shifted component is of smaller amplitude compared to the major and unshifted component (cf.~Fig.~\ref{fig:lobes}). Thus, depending on the amplitude of the shifted component, the additional lobe on the red side is only visible for large Doppler velocities. Smaller velocities, though, lead to both a smaller amplitude and a smaller area of the red lobe. Since the algorithm used is not very sophisticated, a large number of asymmetric profiles remain undetected, especially for small downflow velocities.

\paragraph{Conclusion:} Taking these arguments into account, the study of \citet{Langhans:2005p9} cannot be considered as evidence that the outer penumbra lacks areas of opposite magnetic polarity as stated by \citet{Spruit:2006p2}. 

\section{Properties of the Magnetic Flow}
\label{sec:implications}

\citet{Spruit:2006p2} and \citet{Scharmer:2006p25} argue that the filamentary structure of the penumbra is due to convection in regions of weak or no magnetic field (cf.~Section~\ref{sec:gaps}). 
However, following the arguments from Section~\ref{sec:hiddenopp}, crossover profiles 
contradict the idea of penumbral plasma motions occurring in regions void of magnetic fields. Nevertheless, it could be argued 
that the technique described in Section~\ref{sec:hiddenopp} works only for downflows and suffers from the fact that large values of v$_{\rm{dop}}$ are necessary to identify the additional lobe.


In the following, it shall be demonstrated that the additional lobe on the blue and red sides of the Stokes V profile is a continuous effect of v$_{\rm{dop}}$, implying that all measurable penumbral plasma flows occur in a magnetic field. To this end, a mean Stokes V profile was calculated for different classes of velocity. In other words, all Stokes V profiles for which the line shift of the wing of the corresponding Stokes I profile amounts to, e.g. $-$1.32 km s$^{-1} < \rm{v}_{\rm{dop}} < -1.31$ km s$^{-1}$, were averaged. 

\begin{figure}[h!]
\begin{center}
    \includegraphics[width=\textwidth]{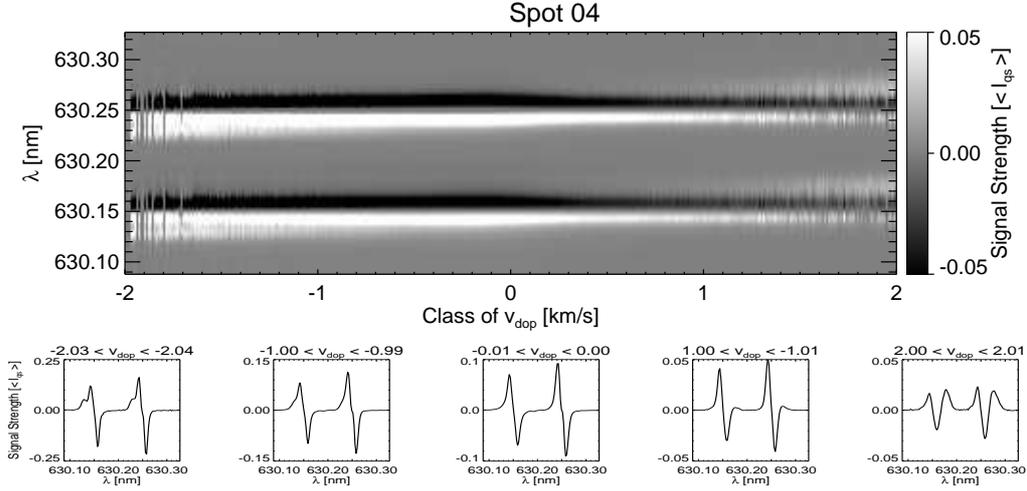}
\caption{Top: Average Stokes V profiles of different classes of v$_{\rm{dop}}$. White represents the positive lobe, black the negative lobe. The picture is saturated at an intensity of 5\% of that of the QS. Bottom: Examples of profiles at various classes of Doppler velocity.}
\label{fig:meanv_dop}
\end{center}
\end{figure}

\paragraph{Vertical Component:} Fig.~\ref{fig:meanv_dop} shows the results for Spot 04, representing the vertical component of penumbral plasma flows. For a comparison, the average Stokes V profiles of the velocity classes $-$2, $-1$, 0, 1 and 2~km~s$^{-1}$ are plotted in the bottom row. The crossover effect is clearly visible in Stokes profiles with large velocities. The additional lobes in the crossover profiles cause the asymmetry in the plot at the top of Fig.~\ref{fig:meanv_dop}.

For strong upflows, the white area, i.e. the positive lobe, is much wider than the black area, i.e the negative lobe. With decreasing upflow velocity, the width of the positive lobe decreases continuously. At a zero Doppler shift, both lobes have the same width and the profile is antisymmetric. But at a velocity class of $\pm$0.1~km~s$^{-1}$, an asymmetry in the width of the positive and negative lobe is already detectable in the average Stokes V profile above the $3\sigma$ noise level\footnote{This accounts for more than 80\% of all penumbra up- or downflows.}. With increasing downflow velocity, the negative lobe becomes narrower and weaker, while an additional lobe appears on the red side of the profile. This can be seen in Fig.~\ref{fig:meanv_dop} where the intensity of the black line decreases while an additional white stripe appears approximately $+$25~pm from the line-core of both lines. 

\begin{figure}[h!]
\begin{center}
    \includegraphics[width=\textwidth]{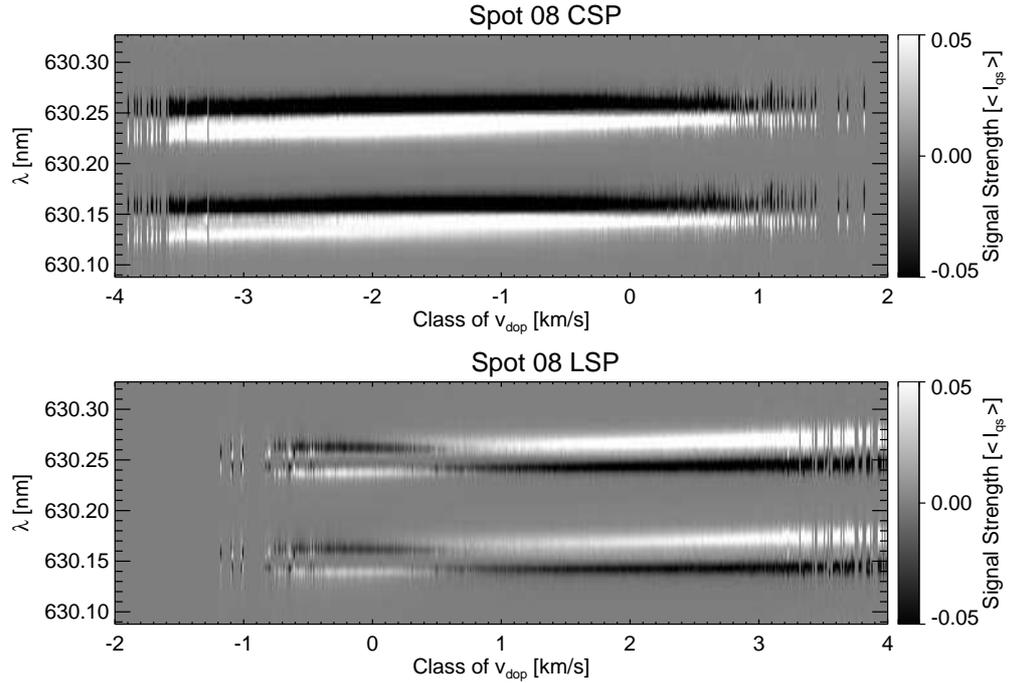}
\caption{Top: Average Stokes V profiles of different classes of v$_{\rm{dop}}$ in the CSP of Spot 08. Bottom: Same as top but LSP of Spot 08.}
\label{fig:meanv_dop2}
\end{center}
\end{figure}

\paragraph{Horizontal Component:} To study the horizontal component of penumbral plasma motion, the same investigation is performed on Spot 08. Fig.~\ref{fig:meanv_dop2} shows the individual results for the LSP and the CSP. The explanation of the shape of Stokes V profiles is not straightforward because, under the projection of $\Theta = 50^{\circ}$, the line shifts are not purely due to the horizontal outflow of plasma. Due to the overall geometry of the penumbral magnetic field, the latter appears perpendicular to the LOS in the inner LSP, causing the so-called magnetic neutral line (MNL). Along the MNL, Stokes V profiles have a small amplitude but large asymmetries \citep{SanchezAlmeida:1992p160}. Beyond the MNL, in the mid and outer LSP, the polarity of the magnetic field is reversed. Despite these complications, a conclusive interpretation of Stokes V profiles is still possible. 


\paragraph{Center Side Penumbra:} The CSP is dominated by blushifts, which is reflected by the large number of negative velocity classes. Comparable to Spot 04, the width of the positive lobe grows with increasing blueshift. However, for Fe I 630.15 and blueshifts larger than $-$2~km~s$^{-1}$, the amplitude of the additional lobe becomes larger than the amplitude of the positive lobe of the regular Stokes V profile. This is not the case in Fe I 630.25, where the zero crossing position shifts instead towards a lower wavelength. In contrast to Spot 04, the redshifts on the CSP do not show a positive lobe on the red side of the regular Stokes V profile, but an increasing width of the negative lobe. This is surprising as it implies that the magnetic field in the redshifted component is of the same polarity as the umbra. \citet{2010A&A...524A..20K} suggested that downflows with such a configuration are causes by magnetic reconnection and chromospheric activity.

\paragraph{Limb Side Penumbra:} Small red- and blueshifts ($-$0.5~km~s$^{-1} < \rm{v}_{\rm{dop}} < 0.5$~km~s$^{-1}$) are located in the inner LSP along the MNL. The average Stokes V profiles have the same polarity as the umbra, but a smaller amplitude. For increasing blueshifts, the width of the positive lobe grows and for increasing redshifts, an additional positive lobe is present on the red side of the profile. With increasing redshift, the positive lobe growth in width and the zero crossing shifts towards higher wavelengths in both lines. What is remarkable is the weak signal of reversed polarity around the zero crossing of Fe 630.25. Radiative transfer calculations in a model atmosphere show that this can be attributed to the magneto-optical effect, which causes a splitting of the lobes in Fe 630.15, but two lobes of reversed polarity in Fe 630.25 \citep{schliche:gross-doert-code}. Redshifts larger 0.5~km~s$^{-1}$, are located beyond the MNL. Here, the profiles are of opposite polarity as the umbra and have a larger amplitude. As before, the positive lobe growth in width with increasing redshift, while the zero crossing shifts towards higher wavelengths in both lines.

\paragraph{Conclusion:} This study shows that the appearance of the additional lobe in Stokes V is a continuous effect of the shift of the wing of Stokes I. This can only be explained if it is assumed that all penumbral plasma motions occur within magnetic fields. 
These finding is in accordance with the results obtained in Section~\ref{sec:invofupdown} and~\ref{sec:hiddenopp} as well as by, e.g. \citet{Rezaei:2006p157} and \citet{Borrero:2008p143}, but contradict the idea of plasma motion occurring in regions void of magnetic field as proposed by \citet{Spruit:2006p2} and \citet{Scharmer:2006p25}. 

\section{Inversion of Up- and Downflow Profiles}
\label{sec:invofupdown}


Even though it has previously been argued that the velocity field changes with height (Section~\ref{sec:gradv}) and that all penumbral flows occur within magnetic fields (Section~\ref{sec:implications}), which are of opposite polarity in downflow channels (Section~\ref{sec:hiddenopp}), no quantitative results were obtained. The behavior of atmospheric parameters (e.g v$_{\rm{dop}}$, strength and inclination of the magnetic field) as a function of $\tau$ can be inferred within the LFR by performing a so-called inversion of all Stokes parameters.

\paragraph{Inversion of HINODE SP Data:} Stokes inversion based on response functions (SIR) uses a least square iteration process to minimize the differences between observed Stokes profiles and Stokes profiles calculated from a model atmosphere in which the atmospheric parameters change with $\tau$ (cf.~Section~\ref{sec:inversion}). The drawback of inversions is that even though the model atmosphere results in the best fit to the observation, it does not necessarily represent the real atmosphere, since it is not derived from an unique solution \citep{1992ApJ...398..375R}. 

Another, more severe shortcoming, is the impossibility to invert all penumbral profiles obtained with the HINODE SP with sufficient accuracy. This is either because the SIR code does not converge properly and remains insensitive to the large asymmetries, especially in crossover profiles, or it yields unphysical solutions. Therefore a case study on individual profiles has been performed, which will be discussed in the following by two representative up- and downflow profiles.

\paragraph{Inversion Procedure:} In the initial atmosphere, the distribution of T was randomized around the values of the Harvard Smithsonian Reference Atmosphere (HSRA), while the values for the electron pressure (p$_{\rm{e}^-}$) where kept as in the HSRA model. Furthermore, the values for v$_{\rm{dop}}$, magnetic field strength (B), inclination ($\gamma_{\rm{mag}}$) and azimuth ($\phi_{\rm{mag}}$) were randomized as well. 

In the first cycle of the inversion, all atmospheric parameters (except p$_{\rm{e}^-}$) were allowed to change constantly with $\tau$. In the second cycle, linear gradients were added to the distribution of B and $\gamma_{\rm{mag}}$. For v$_{\rm{dop}}$, it was necessary to allow for seven degrees of freedom to be able to fit the asymmetries of Stokes V profiles. In fact, it is impossible to model the crossover character of upflow profiles with less nodes in v$_{\rm{dop}}$. This is feasible for the redshifted profiles, but only at the cost of increasing the nodes in B and $\gamma_{\rm{mag}}$. The reason for this lies in the mode of operation of the code itself. SIR distributes the nodes equally with $\tau$ (cf.~Section~\ref{sec:inversion}), which makes it difficult to model a flow mainly present in the lower layers of the atmosphere. 

To ensure that the code did not approach a local minimum of the $\chi^2$ surface, the inversion was performed 100 times for the same pixel. From the resulting pool of solutions, the fit with a minimal $\chi^2$ was selected.

\begin{figure}[h!]
	\centering
		\includegraphics[width={0.9\textwidth}]{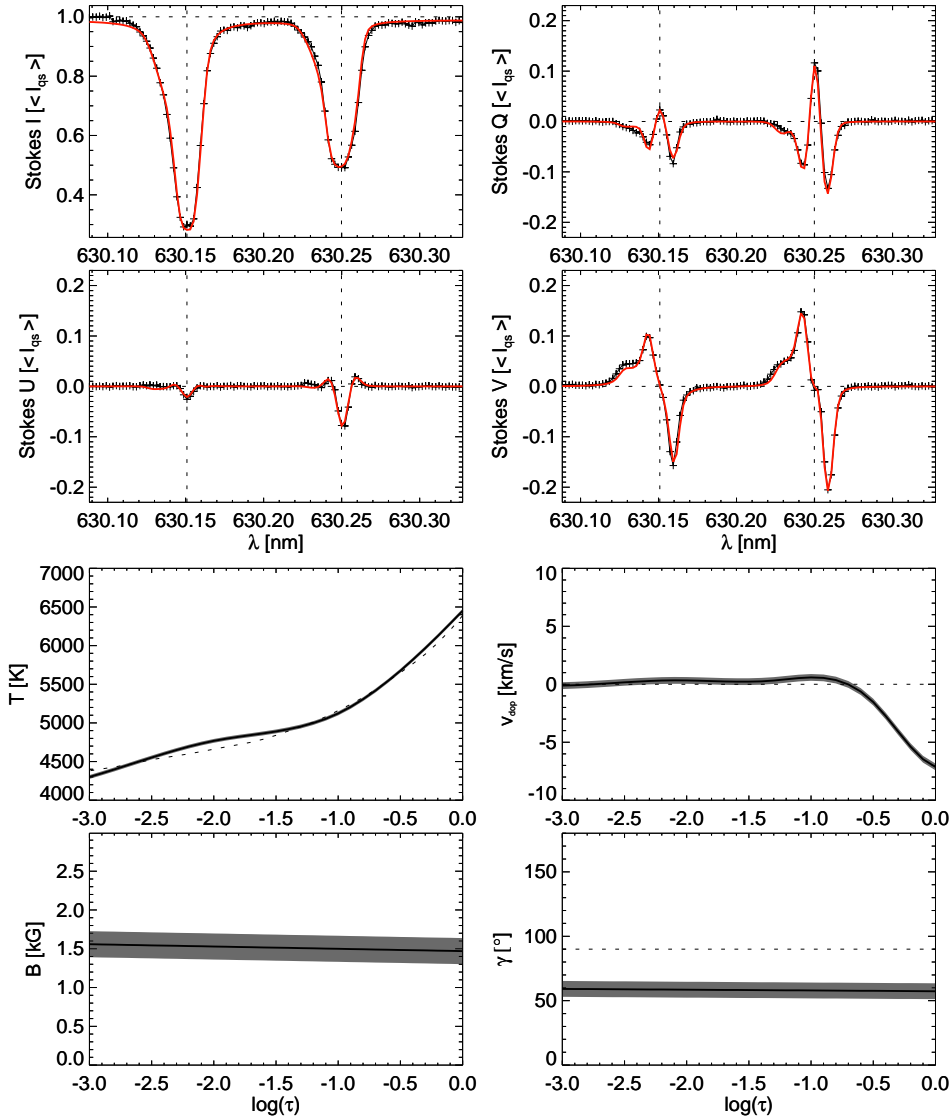}
		\caption{Top part clockwise: Stokes I,  Q,  V and U profiles (black) of an upflow region. Note the crossover effect in the Stokes V profile. The red lines indicate the result of radiative transfer calculations within the atmosphere plotted below. Bottom part clockwise: T, v$_{\rm{dop}}$, $\gamma_{\rm{mag}}$ and B (black) with errors (shaded gray) at different $\tau$ values. The dashed lines indicate the HSRA stratification of T, plasma at rest in v$_{\rm{dop}}$ and 90$^{\circ}$ for $\gamma_{\rm{mag}}$, respectively.}
		\label{fig:profileup}
\end{figure}

\paragraph{A Typical Upflow:} The observed Stokes I profile -- the same as in the left panel of Fig.~\ref{fig:bisec_pen} -- is indicated by the black crosses in the upper left panel of Fig.~\ref{fig:profileup}. Since the entire spectral range of the HINODE SP is used in the inversion, both Fe lines are plotted. The respective vacuum wavelength of the transition is indicated by the dashed vertical lines. The red line indicates the result of radiative transfer calculations within an atmosphere that yields a minimal divergence between observation and fit. The other panels in the top half of Fig.~\ref{fig:profileup} show Stokes Q, Stokes V and Stokes U (black crosses) together with the best fit (solid red). The crossover effect in Stokes V is visible as a positive shoulder on the blue wing of the profile. 

\paragraph{Resulting Atmosphere:} The two bottom rows of Fig.~\ref{fig:profileup} illustrate the model atmosphere, for which the radiative transfer calculations result in a profile with the least deviation from the observation. Depicted are T, v$_{\rm{dop}}$, B and $\gamma_{\rm{mag}}$ (solid black), including the respective error (shaded grey), 
while other parameters like micro and macroturbolence as well as p$_{\rm{e}^-}$ and $\phi_{\rm{mag}}$ are not plotted. This is because they are either not inverted (p$_{\rm{e}^-}$) or remain constant along the LOS ($\phi_{\rm{mag}}$, micro- and macroturbolence). The variation of the atmospheric parameters along the LOS is plotted only between log($\tau)=-3$ and log($\tau)=0$ because the two lines are significantly sensitive at these $\tau$ values. 

In this model, T increases monotonously from 4300~K at log($\tau)=-3$ to 6450~K at log($\tau)=0$, with an error of $\pm$30~K. Except for a little plateau around {log$(\tau)=-2$}, this distribution is similar to that of the HSRA (cf. dashed line in respective plot of Fig.~\ref{fig:profileup}). The strongest flows occur in the low layers of the atmosphere, where the largest amplitudes of v$_{\rm{dop}}$ are present. Between log$(\tau)=-0.7$ and the surface, the amplitude increases from v$_{\rm{dop}} = 0$~km~s$^{-1}$ to v$_{\rm{dop}} = -7$~km~s$^{-1}$ while -- within the uncertainties -- almost no v$_{\rm{dop}}$ is measured in the layers above ($-3.0 \le \rm{log}(\tau) \le -0.7$). B and $\gamma_{\rm{mag}}$ remain constant throughout the LFR with values of 60$^{\circ}\pm5.5^{\circ}$ and 1.5$\pm$0.16~kG respectively.

\begin{figure}[h!]
	\centering
		\includegraphics[width={0.9\textwidth}]{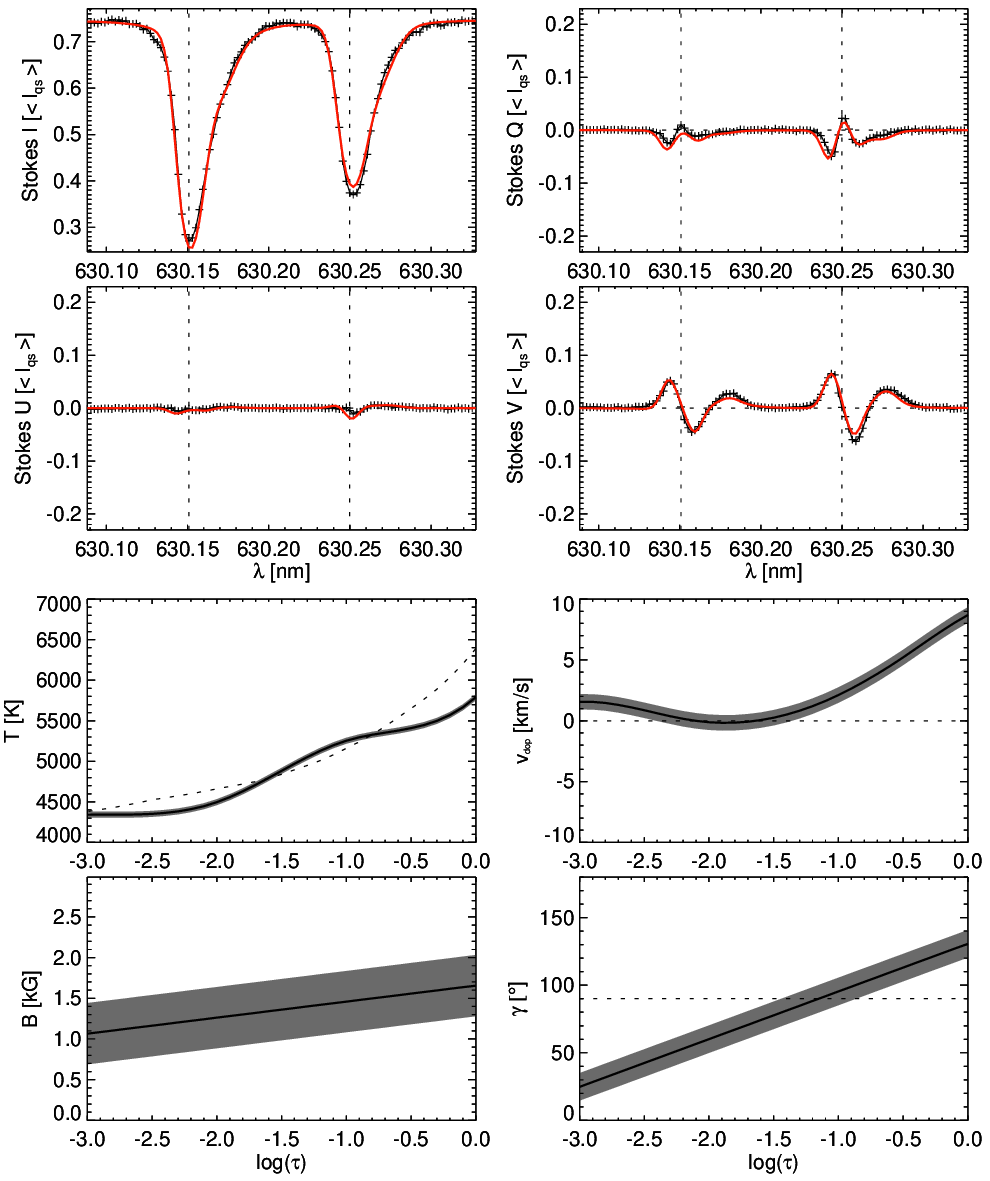}
		\caption{Same a Fig.~\ref{fig:profileup} but for a downflow profile.}
		\label{fig:profiledown}
\end{figure}

\paragraph{A Typical Downflow:} The upper left panel of Fig.~\ref{fig:profiledown} shows Stokes I (black crosses) of a typical downflow region (cf. right panel of Fig.~\ref{fig:bisec_pen}). As mentioned before, the red line represents the best fit to all Stokes parameters. When comparing these profiles to the upflow case (cf.~Fig.~\ref{fig:profileup}), it becomes apparent that Stokes I is shifted towards higher wavelengths, especially in the wing of the line. Furthermore, the value of I$_{\rm{c}}$ is significantly below I$_{\rm{qs}}$, and the maximal amplitudes of Stokes Q, U and V are lower than in the upflow profile. The crossover effect in this Stokes V profile manifests itself as an additional positive lobe on the red side of the line.

\paragraph{Resulting Atmosphere:} In the downflow channel, T increases monotonously from 4350~K at log($\tau)=-3$ to 5300~K at log($\tau)=0$, with an error of roughly $\pm$40 K (cf. Fig.~\ref{fig:profiledown}). At lower layers, this atmosphere is significantly cooler when compared to the HSRA (T$_{{\rm{log}(\tau)}}(-3) = 4400$ and T$_{{\rm{log}(\tau)}}(0) = 6400$), which explains the 25\% drop in I$_{\rm{c}}$. Furthermore, the distribution of T shows a little plateau around {log($\tau)=-1$}, a feature not present in the HSRA. For v$_{\rm{dop}}$, strong flows are found in the low layers of the atmosphere. Their amplitude increases monotonously from v$_{\rm{dop}} = 0$~km~s$^{-1}$ at log$(\tau)=-1.5$ to large values of v$_{\rm{dop}} = 8.5$~km~s$^{-1}$ at the continuum. There is also a slight downflow in the higher layers (v$_{\rm{dop}} = 1.5$~km~s$^{-1}$ for $-3 \le \rm{log}(\tau) \le -2.5$), but no flow for $-2.5 \le \rm{log}(\tau) \le -1.5$. The uncertainty in v$_{\rm{dop}}$ is about $\pm0.5$~km~s$^{-1}$. B increases from 1.1~kG to 1.6~kG between log($\tau)=-3$ and log($\tau)=0$. However, the relatively large error of $\pm$0.35 kG does not allow to rule out a constant B throughout the LOS. In contrast to the upflow example, 
$\gamma_{\rm{mag}}$ increases from 25$^{\circ}$ at log($\tau)=-3$ to 130$^{\circ}$ at log($\tau)=0$, with an error of $\pm$10$^{\circ}$. This means that the magnetic field reverses its polarity somewhere between log($\tau)=-1.5$ and log($\tau)=-0.9$, an atmospheric height reached by the strong downflows of the lower layers.

\paragraph{Discussion:} Even though the inversion results have only been discussed expli\-citly for two examples, it should be stressed that similar results were obtained for numerous other up- and downflow profiles. Nevertheless, there is no guarantee that an inversion of other profiles with the same initial atmosphere and number of nodes successfully models crossover Stokes V profiles.  

Within the limitations of this method, the results in this Section yield a scenario in which strong gradients of atmospheric parameters, especially of the v$_{\rm{dop}}$, occur within the LFR. In upflows, the atmosphere at the continuum is significantly hotter than in downflows\footnote{For an exception see the bright penumbral downflows in Section~\ref{sec:brightdownflow}}. The upper layers (log($\tau)~=~-3$) of the atmosphere show a similar temperature in both cases. The asymmetries in Stokes I 
are caused by a v$_{\rm{dop}}$ of at least $\pm$7~km~s$^{-1}$ around the continuum layers of the atmosphere. At the same time, the atmosphere above log($\tau)~=~-1.5$ remains rather steady. The configuration of the magnetic field shows significant differences between up- and downflow regions with rather constant B and $\gamma_{\rm{mag}}$ throughout the LFR in the former and a polarity reversal of $\gamma_{\rm{mag}}$ around log$(\tau)=-1.2$ in the latter case.

The inversion results are in accordance with the outcome of Section~\ref{sec:gradv}, where it was demonstrated that the penumbral plasma flows dominate in the deep photosphere. Additional support comes from the results presented in Section~\ref{sec:morph_out} and \ref{sec:rad_ave}, where $\gamma$ of upflow channels in the inner LSP was estimated to be around 50$^{\circ}$. Furthermore, the studies of Sections~\ref{sec:hiddenopp} and \ref{sec:implications} show that a magnetic component of opposite polarity is present in penumbral downflow channels.

 

Up to now, there have been only a few reports of spectral inversions with gradients along the LOS for high resolution data of penumbrae at disk center. 

\citet{Jurcak:2010p4739}, for example, performed a study on crossover profiles from penumbral downflows 
and found a v$_{\rm{dop}}$ above 3~km~s$^{-1}$ in the lower layers of the atmosphere. However, they did not report on a polarity reversal of the magnetic field within the LFR.

\citet{BellotRubio:2003p206,BellotRubio:2004p4357} investigated a sunspot that was artificially transformed into a local reference frame. 
From a multicomponent Milne-Eddington type atmosphere (i.e. no gradients along the LOS), they obtained $\gamma_{\rm{mag}}$ for a flux tube and a background component. Since their study was done in a different spectral line and with lower spatial resolution, a comparison between their results and the atmosphere represented in e.g. Fig.~\ref{fig:profileup} is not straightforward. 
At some places in the inner penumbra where a lot of upflows are located, their flux tube and background component shows a $\gamma_{\rm{mag}}$, variing between 50$^{\circ}$ and 70$^{\circ}$. Assuming that the flux tube component is located in the deep photosphere and that the background component maps the higher atmospheric layers\footnote{\citet{Borrero:2004p140} showed that a two-component inversion with a Milne-Eddington atmos\-phere and a one-component inversion with gradients along the LOS yield similar results.}, an atmosphere comparable to that of Fig.~\ref{fig:profileup} seems plausible.


\section{Summary}
\label{sec:concl_asym}


Asymmetries in Stokes I were investigated to compare the gradient with height of the velocity field above the penumbra and the quiet Sun. For representative penumbral profiles, it was found that the line-wing is significantly shifted, while the line-core is located at the vacuum wavelength. In the quiet Sun, the shift of the line-wing is of comparable strength. However, in contrast to penumbral profiles, the line-core of quiet Sun profiles also shows a significant shift. A statistical analysis of velocity maps computed at different line depression values yields similar results: The amplitudes of the penumbral velocity field diminish much faster with hight when compared to the quiet Sun. These findings demonstrate that the solar plasma in the quiet Sun reaches into higher atmospheric layers than plasma in the penumbrae of sunspots.

By interpreting crossover Stokes V profiles in the framework of a multilayer atmosphere, it is shown that at least 40\% of all penumbral downflows contain a magnetic component with opposite polarity. It is demonstrated that these hidden opposite polarities are not present in classical magnetograms, 
which therefore, can not be used as an argument against neither the flux tube model, nor the scenario of turbulent pumping.

Furthermore, Stokes V profiles have been sorted and averaged according to the line shift of the respective Stokes I profiles. This procedure allows an estimate of the shape of Stokes V profiles and illustrates that the crossover effect occurs increasingly with Doppler velocity. The results prove that both the vertical and the horizontal component of the Evershed flow occur in a magnetized atmosphere. This is in accordance with the ideas of penumbral flux tubes, but it is at odds with the predictions of the gappy model.

To obtain quantitative values of the atmospheric parameters of penumbral up- and downflows, spectral inversions were preferment by means of two representative profiles from a blue- and a redshifted regions. Because of the difficulties in accounting for line asymmetries in spectral inversions, it was ensured that the crossover effect in Stokes V profiles was modeled correctly. The outcome of the inversion confirms the results presented so far, i.e. strong penumbral plasma flows only in the deep photosphere, penumbral magnetic fields exceeding kilogauss field strength throughout the line forming region and reversed magnetic polarities in the atmospheric region occupied by penumbral downflows.

\chapter{Net Circular Polarization}
\label{ch:ncp}

In this Chapter, it will be investigated how the total net circular polarization ($\mathcal{N}$), that is a measure of asymmetries in Stokes V, may be used to refine penumbral models. In Section~\ref{sec:corr_dop_ncp}, $\mathcal{N}$ is used to put constraints on the geometry of penumbral up- and downflows. The center to limb variation of $\mathcal{N}$ is studied in Section~\ref{sec:clv_ncp}, where it is also compared to previous investigations. Section~\ref{sec:pol_rev} discusses not only the effects which may create the pattern of mixed polarities of $\mathcal{N}$ in the center side penumbra, but it also studies this polarity reversal on a larger scale. Section~\ref{sec:stat_ana} works out the implications of this reversal for the geometry of the penumbral magnetic field, using spectral inversion. The results of this Chapter are finally discussed in Section~\ref{sec:concl_ncp}.

\section{Penumbral Net Circular Polarization at Disk Center}
\label{sec:corr_dop_ncp}

\begin{figure}[h!]
\begin{center}
    \includegraphics[width=\textwidth]{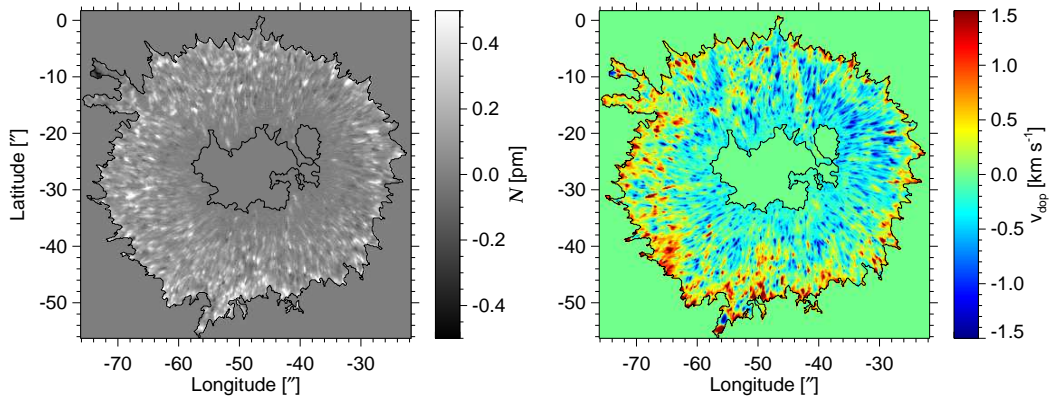}
\caption{Map of $\mathcal{N}$ (left) and v$_{\rm{dop}}$ (right) inferred from Fe I 630.15 in Spot 04.}
\label{fig:ncp_vdop}
\end{center}
\end{figure}

\citet{Ichimoto:2008p194} report that up- and downflows correlate with positive $\mathcal{N}$ at disk center. As an example, the penumbra of Spot 04 is shown in Fig.~\ref{fig:ncp_vdop}. Indeed, places with high values of $|\rm{v}_{\rm{dop}}|$ correspond to large values of $\mathcal{N}$ -- cf. upflow at $\rm{(x;y)}=(-47$\arcsec$;-$10\arcsec$)$ and downflow at $\rm{(x;y)}=(-58$\arcsec$;-$38\arcsec$)$. The outcome of a statistical evaluation, however, yields a more diverse picture.

\paragraph{Global Correlation Coefficients:} For this study, an analysis similar to the one in Section~\ref{sec:corr_center} was performed. In the entire penumbra, r$_{\rm{S}}(\rm{A})=0.23$ was computed for v$_{\rm{dop}}$ and $\mathcal{N}$. This value is to small to consider both quantities as correlated. 

In a next step, r$_{\rm{S}}$ was calculated for $\mathcal{N}$ as well as up- and downflows separately. In the case of v$_{\rm{dop}} < 0$~km~s$^{-1}$, a r$_{\rm{S}}(\rm{A})=-0.16$ was obtained, while v$_{\rm{dop}} > 0$~km~s$^{-1}$ yields r$_{\rm{S}}(\rm{A})=0.39$. This shows that even though r$_{\rm{S}}$ remains too small for a significant correlation, it is higher in the case of downflows.

\begin{figure}[h!]
\begin{center}
    \includegraphics[width=\textwidth]{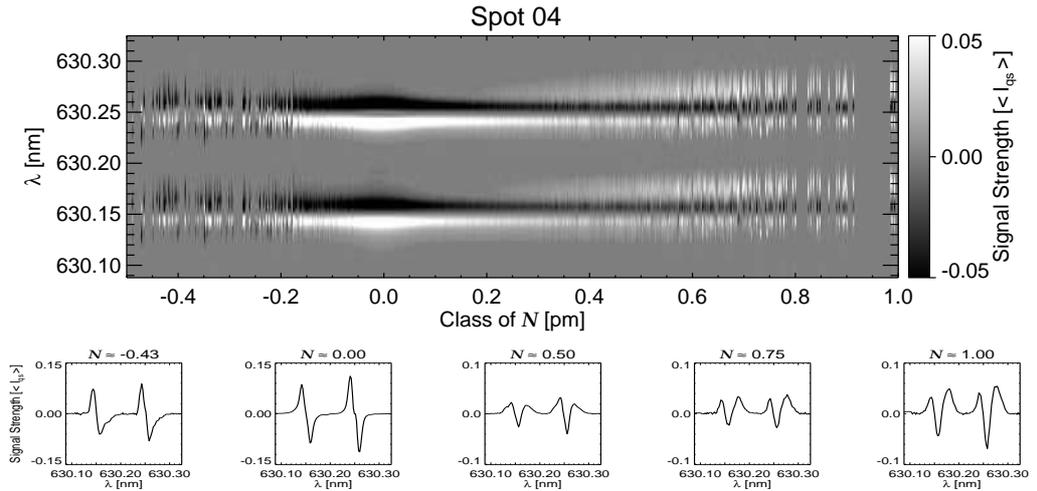}
\caption{Top: Average Stokes V profiles of different classes of $\mathcal{N}$. White represents the positive lobe, black corresponds to the negative lobe. The picture is saturated at an intensity of 5\% of that of the QS. Bottom: Profiles at various classes of $\mathcal{N}$.}
\label{fig:meanv_ncp}
\end{center}
\end{figure}

\paragraph{Shape of Stokes V Profiles:} The tendency of downflow profiles to show higher values of $\mathcal{N}$ is also evident from the shape of Stokes V. Similar to the investigation in Section~\ref{sec:implications}, all profiles within a specific $\mathcal{N}$-range were averaged, and the resulting mean profiles were plotted as a function of $\mathcal{N}$. As an example, the results for Spot 04 are plotted in Fig.~\ref{fig:meanv_ncp} alongside a number of selected profiles.

For $\mathcal{N} \approx 0$, the Stokes V profiles are antisymmetric, and both lobes have the same amplitude\footnote{This is not trivial as $\mathcal{N} = 0$ does not imply $\delta\rm{a}=0$. A profile with a narrow but strong positive lobe and a shallow but wide negative lobe could have $\mathcal{N} = 0$ as well.}. With increasing negative $\mathcal{N}$, the blue lobe becomes wider. These profiles are caused by downflow patches with the same polarity as the umbra (cf.~\citep{2010A&A...524A..20K} and Section~\ref{sec:implications}). For increasing positive\footnote{For penumbrae at disk center, the PDFs of $\mathcal{N}$ (not shown) are highly skewed. In Spot 04, for example, 84\% of all Stokes V profiles show $\mathcal{N} > 0$.} $\mathcal{N}$, the profiles start to show the crossover effect. This is evident as an additional white area at $+$25~pm from the line-core of both lines, if $\mathcal{N} > 0.2$~pm. Since profiles from up- and downflows with the same $\mathcal{N}$ were averaged, these profiles also show a shoulder on the blue lobe (cf. profile with $\mathcal{N} = 0.5$~pm in lower row of Fig.~\ref{fig:meanv_ncp}). For $\mathcal{N} > 0.6$~pm, however, the lobe on the blue side of the profile diminishes, while the additional lobe on the red becomes larger in amplitude than the lobes from the regular profile (cf. right two plots in lower row of Fig.~\ref{fig:meanv_ncp}).

The fact that the profiles for large $\mathcal{N}$ look like crossover profiles from downflow regions is in accordance with the findings of the correlation analysis. To understand these results, the generation of $\mathcal{N}$ shall be described in a multilayer atmosphere.


\paragraph{Ingredients:} %
\noindent \citet{1975A&A....41..183I} explained non-vanishing values of $\mathcal{N}$ as the result of gradients of v$_{\rm{dop}}$ and B along the LOS. \citet{1978A&A....64...67A} showed that already a constant B and a gradient with height of v$_{\rm{dop}}$ within the LFR are sufficient to obtain $\mathcal{N} \not= 0$. Nevertheless, the amplitude of $\mathcal{N}$ is altered not only by the variation of v$_{\rm{dop}}$ and B, but also by the difference of $\gamma_{\rm{mag}}$ and to a minor extent also of $\phi_{\rm{mag}}$ in distinct atmospheric regions \citep{1996SoPh..164..191L}. The contribution of these so-called $\scriptstyle{\Delta}\textstyle$v$_{\rm{dop}}$, $\scriptstyle{\Delta}\textstyle$B and $\scriptstyle{\Delta}\textstyle\gamma_{\rm{mag}}$ effects to the amount of $\mathcal{N}$ have been investigated in a two-layer atmosphere by \citet{Solanki:1993p210}.

\begin{figure}[h!]
\begin{center}
    \includegraphics[width=0.9\textwidth]{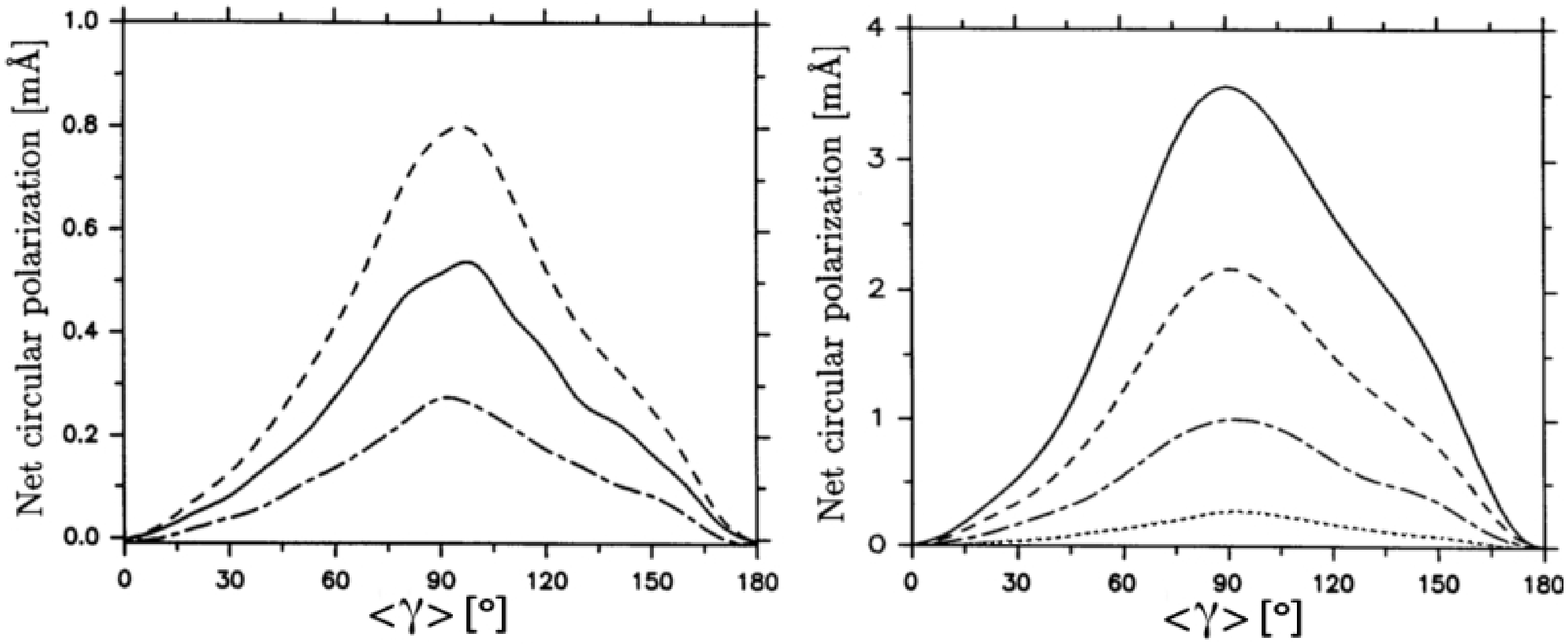}
\caption{Left: $\mathcal{N}$ of Fe I 630.25 as a function of $<$$\gamma$$>$ for $\rm{B}=1.5$~kG, $\scriptstyle{\Delta}\textstyle\gamma=10^{\circ}$ and v$_1=\rm{v}_2=1$~km~s$^{-1}$ (dot-dashed), v$_1=\rm{v}_2=2$~km~s$^{-1}$ (solid) and v$_1=\rm{v}_2=3$~km~s$^{-1}$ (dashed). Right: Same as left, but for $\rm{B}=1.5$~kG, v$_1=\rm{v}_2=1$~km~s$^{-1}$ and $\scriptstyle{\Delta}\textstyle\gamma=10^{\circ}$ (dotted), $\scriptstyle{\Delta}\textstyle\gamma=20^{\circ}$ (dot-dashed), $\scriptstyle{\Delta}\textstyle\gamma=30^{\circ}$ (dashed) and $\scriptstyle{\Delta}\textstyle\gamma=40^{\circ}$ (solid). Adopted from \citep{Solanki:1993p210}.}
\label{fig:ncp_solanki}
\end{center}
\end{figure}

The left panel of Fig.~\ref{fig:ncp_solanki} shows $\mathcal{N}$ as a function of the average zenith angle of both atmospheric regions, i.e. $<$$\gamma$$> =  \frac{1}{2} (\gamma_1+\gamma_2)$. It can be seen that $\mathcal{N}$ is largest if $<$$\gamma$$>$ is perpendicular\footnote{This is not at odds with the fact that Stokes V vanishes if the magnetic field is observed perpendicularly, since $\gamma_1 \not= \gamma_2 \not= 90^{\circ}$ within the individual atmospheric layers.} to the LOS. The respective plots show the impact of v on the generation of $\mathcal{N}$ in a two-layer atmosphere with B$_1=\rm{B}_2=1.5$~kG and $\scriptstyle{\Delta}\textstyle\gamma_{\rm{mag}}=10^{\circ}$. Even though v$_1=\rm{v}_2$ for all plots, v$^{\rm{dop}}_1 \not= \rm{v}^{\rm{dop}}_2$ since $\scriptstyle{\Delta}\textstyle\gamma \not=0^{\circ}$, and thus $\mathcal{N}$ increases with v. The right panel of Fig.~\ref{fig:ncp_solanki} demonstrates that $\scriptstyle{\Delta}\textstyle\gamma_{\rm{mag}} = | \gamma_1-\gamma_2 |$, i.e. the difference of the zenith angle in both atmospheric regions contributes more to $\mathcal{N}$ than $\scriptstyle{\Delta}\textstyle$v. In fact, the lowest curves in the right and left panels refer to an atmosphere with B$_1=\rm{B}_2=1.5$~kG, v$_1=\rm{v}_2 = 1$~km~s$^{-1}$ and $\scriptstyle{\Delta}\textstyle\gamma =10^{\circ}$. For $\scriptstyle{\Delta}\textstyle\gamma =20^{\circ}$, $\scriptstyle{\Delta}\textstyle\gamma =30^{\circ}$ and $\scriptstyle{\Delta}\textstyle\gamma =40^{\circ}$, $\mathcal{N}$ is larger than in all plots on the left.

\paragraph{Interpretation:} To understand the different correlation of $\mathcal{N}$ in up- and downflow channels, it is thus important to estimate $\scriptstyle{\Delta}\textstyle\gamma_{\rm{mag}}$ and $<$$\gamma$$>$.

From Section~\ref{sec:rad_ave}, $\gamma_1 \approx 40^{\circ}$ was obtained in the inner CSP, while the spectral inversion of a typical upflow profile yielded $\gamma_1 \approx \gamma_2 \approx 60^{\circ}$ (cf.~Section~\ref{sec:invofupdown}). For downflows, there is indirect evidence that $\gamma_1 > 90^{\circ}$ (cf.~Section~\ref{sec:hiddenopp}), which was confirmed for the deep atmospheric layers by spectral inversions (cf.~Section~\ref{sec:invofupdown}). The distribution of opposite polarity patches in penumbrae off disk center allowed \citet{Ichimoto:2007p178} to infer $\gamma_1 \approx 120^{\circ}$ in downflows. From a multicomponent Milne-Eddington inversion, \citet{BellotRubio:2003p206} found $\gamma_1 \approx 70^{\circ}$ and $\gamma_2 \approx 30^{\circ}$ for the flux tube and background components in the inner penumbra, respectively. Furthermore, they obtained $\gamma_1 \approx 100^{\circ}$ and $\gamma_2 \approx 55^{\circ}$ for the outer penumbra.

\begin{figure}[h!]
\begin{center}
    \includegraphics[width=0.9\textwidth]{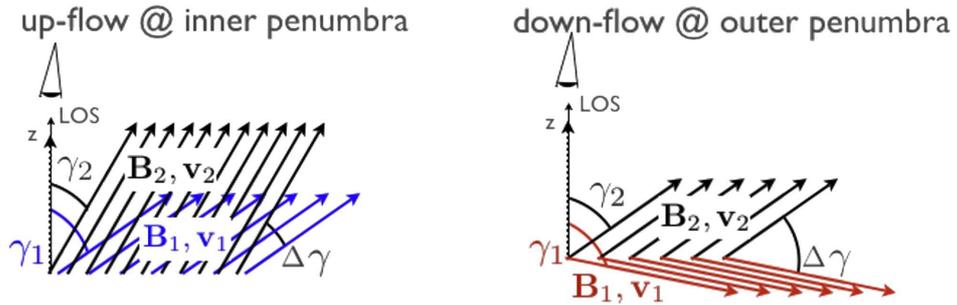}
\caption{Schematic configuration of penumbral up- (left) and downflow channels (right).}
\label{fig:ncp_updown}
\end{center}
\end{figure}

Despite the large uncertainties of the magnetic field configuration in up- and downflow channels, the scenario depicted in Fig.~\ref{fig:ncp_updown} seems plausible. Such a configuration explains the tendency for $\mathcal{N}$ to show larger values in downflow channels when compared to upflows, because:
\begin{itemize}
\item{Upflows have a smaller $\gamma$ when compared to downflows.}
\item{The predominant occurrence of up- and downflows in the inner and outer penumbra results in a larger $<$$\gamma$$>$ in downflows.}
\item[$\Rightarrow$]{These conditions consequently result in a larger $\scriptstyle{\Delta}\textstyle\gamma$ in downflows when compared to upflows.}
\end{itemize}

\noindent The effects described so far always lead to a positive $\mathcal{N}$ and do not explain the 16\% of negative $\mathcal{N}$ in Spot 04. The impact of the other parameters on $\mathcal{N}$ in general, and the $\scriptstyle{\Delta}\textstyle$B-effect in particular, are studied in Section~\ref{sec:pol_rev} and \ref{sec:stat_ana}, where a possible explanation is given for the pattern of positive and negative $\mathcal{N}$ in the CSP of sunspots at large heliocentic angles.

\section{Center to Limb Variation of Penumbral Net Circular Polarization}
\label{sec:clv_ncp}

In the previous Section, it was shown that the configuration of B and $\gamma_{\rm{mag}}$ in individual up- and downflows has an impact on $\mathcal{N}$. Additionally, $\mathcal{N}$ can also be used to put constrains on the global properties of the penumbral magnetic field. \citet{Schlichenmaier:2002p722} and \citet{Muller:2002p3983}, for example, showed that the effects of anomalous dispersion have to be taken into account to understand the azimuthal properties of $\mathcal{N}$ in penumbral observation within the visible and infrared regime of the EMS. Furthermore, \citet{Solanki:1993p210} and \citet{MartinezPillet:2000p223} restrained the atmospheric parameters of penumbral models by means of the center to limb variation (CLV) of $\mathcal{N}$. In the following, the CLV of $\mathcal{N}$ will be derived from high resolution HINODE SP data and compared to these models.

\begin{figure}[h!]
\begin{center}
    \includegraphics[width=0.85\textwidth]{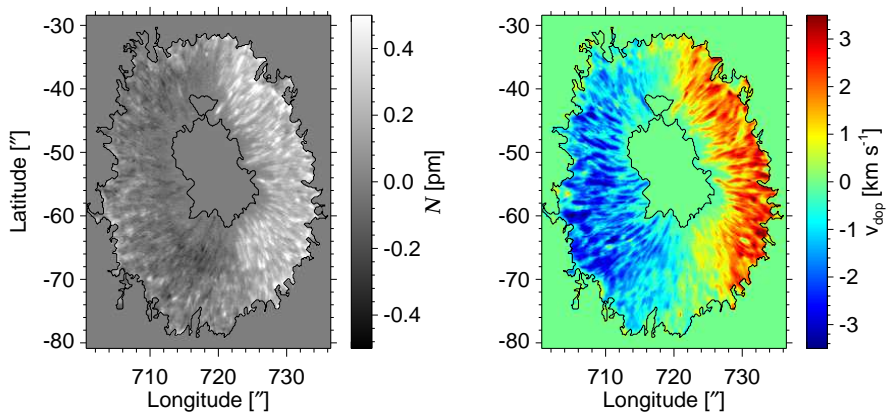}
\caption{Map of $\mathcal{N}$ (left) and v$_{\rm{dop}}$ (right) inferred from Fe I 630.15 in Spot 08.}
\label{fig:ncp_vdop_limb}
\end{center}
\end{figure}

\paragraph{Observation at large heliocentric angles:} Fig.~\ref{fig:ncp_vdop_limb} shows $\mathcal{N}$ (left) and v$_{\rm{dop}}$ (right) in Spot 08. $\mathcal{N}$ assumes large positive values on the LSP, while it is of smaller but positive and negative sign in the CSP.

\paragraph{Setup of Study:} All data samples listed in Table~\ref{tab:data_limb2} have been used for this survey. To compare the results with the studies mentioned above, $\mathcal{N}$ was extracted for the LSP and the CSP individually, using a similar procedure as in Chapter~\ref{ch:limb}. For this analysis, however, the sector was defined by a cone angle of $\pm$2$^{\circ}$ measured from the center of the umbra along the line of symmetry.

\begin{figure}[h!]
\begin{center}
    \includegraphics[width=0.9\textwidth]{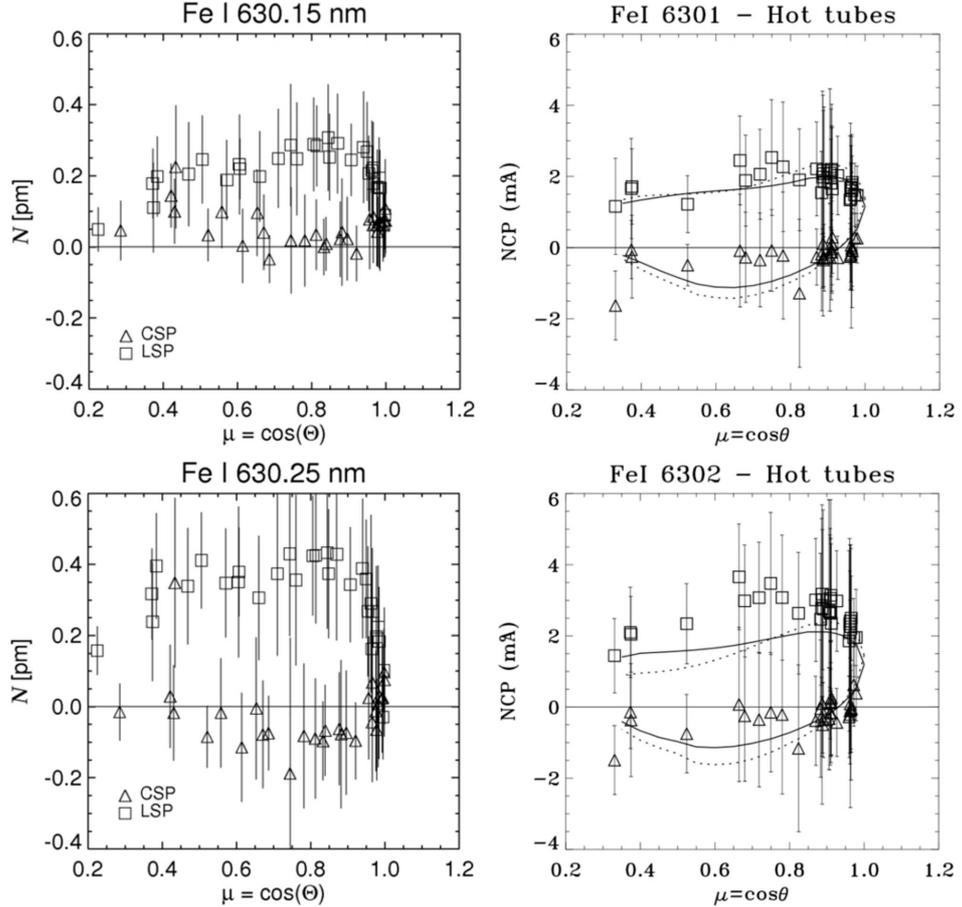}
\caption{Left column: Mean value of $\mathcal{N}$ in the CSP (triangles) and LSP (squares) for various heliocentric angles as well as Fe I 630.15 (top) and Fe I 630.25 (bottom). The error bar represents the standard deviation. Right column: Same as left, but study of \citet{MartinezPillet:2000p223}, where the error bars represent three times the standard deviation and the dashed/solid curves show model calculations for flux tubes of different diameters.}
\label{fig:clv_ncp}
\end{center}
\end{figure}

\paragraph{Description of Results:} Mean values and the standard deviation\footnote{The error bars of the mean values are smaller than the symbols. Thus, the variability of $\mathcal{N}$ around the respective mean values is plotted instead.} of $\mathcal{N}$ are plotted for the CSP and the LSP in the left column of Fig.~\ref{fig:clv_ncp}. Due to the smaller g-factor, $\mathcal{N}$ is weaker for Fe I 630.15 when compared to Fe I 630.25. In the first line, it reaches a maximum of 0.3~pm at $\mu = 0.83$, and in the second line, it assumes a maximum of 0.43~pm in a number of measurements of the LSP around $\mu = 0.8$. 

In contrast to the LSP, where $\mathcal{N}$ is always positive, its behavior on the CSP is more diverse. In Fe I 630.15, it fluctuates around zero with slightly negative values for $0.6<\mu<0.95$. In Fe I 630.25, $\mathcal{N}$ assumes values of approximately $-$0.03~pm for $0.3~<~\mu~<~0.5$, $-$0.1~pm for $0.5~<~\mu~<~0.98$ and positive values around 0.05~pm close to disk center. Due to the irregular shape of some sunspots, there are a number of outliers, e.g. the large positive values for the CSP around $\mu = 0.43$.

\begin{figure}[h!]
\begin{center}
    \includegraphics[width=0.8\textwidth]{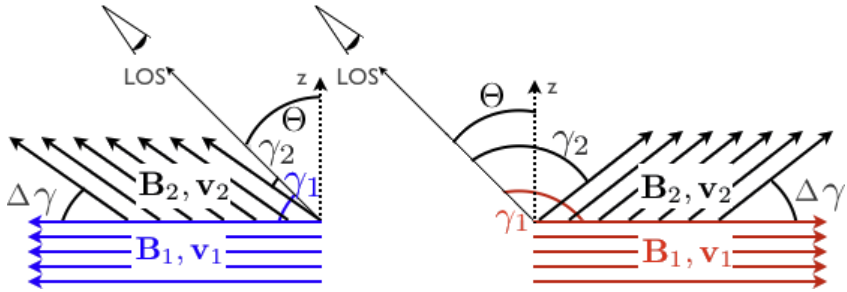}
\caption{Configuration of v, B, and $\gamma$ in the CSP (left) and in the LSP (right).}
\label{fig:config_cenlimb}
\end{center}
\end{figure}

\paragraph{Geometrical Considerations:} The large values of $\mathcal{N}$ in the LSP are comprehensible 
if the overall geometry of the atmospheric parameters is considered under the projection of the heliocentric angle. From Fig.~\ref{fig:config_cenlimb}, it can be seen that not only $<$$\gamma$$>$, but also $\scriptstyle{\Delta}\textstyle\gamma$ is smaller in the CSP when compared to the LSP. 

To calculate $<$$\gamma$$>=\frac{1}{2}(\gamma_1+\gamma_2)+\Theta$ 
and $\scriptstyle{\Delta}\textstyle\gamma=|\gamma_2-\gamma_1|$ in order to derive $\mathcal{N}$ from Fig.~\ref{fig:ncp_solanki}, $\Theta_{\rm{CSP}} \approx 50^{\circ}$ is used (cf.~Section~\ref{sec:outflow}), while the values for $\gamma_1$ and $\gamma_2$ are taken from \citet{BellotRubio:2003p206}. The results are summarized in Table~\ref{tab:N_delta_gam} and demonstrate that the $\scriptstyle{\Delta}\textstyle\gamma$-effect vanishes on the CSP, while it is strong on the LSP. This causes the larger $\mathcal{N}$ in the LSP when compared to the CSP.

\begin{table}[h!]
\begin{center}
	\begin{tabular}{ccccc}
		\hline
		\hline
		\\[-2ex]
		{Quantity} & {Inner Penumbra} & {Mid Penumbra} & {Outer Penumbra} \\
		\\[-2ex]
		\hline		
		\\[-2ex]
		{$\gamma_1$ [$^{\circ}$]} & {70} & {85} & {100}\\
		{$\gamma_2$ [$^{\circ}$]} & {30} & {45} & {55}\\
		{$\scriptstyle{\Delta}\textstyle\gamma$ [$^{\circ}$]} & {40} & {40} & {45}\\
		{$<$$\gamma$$>_{\rm{CSP}}$ [$^{\circ}$]} & {0} & {15} & {30}\\
		{$<$$\gamma$$>_{\rm{LSP}}$ [$^{\circ}$]} & {100} & {115} & {130}\\
		{$\mathcal{N}_{\rm{CSP}}$ [pm]} & {0.0} & {0.02} & {0.05}\\
		{$\mathcal{N}_{\rm{LSP}}$ [pm]} & {0.34} & {0.28} & {0.22}\\
		\hline
	\end{tabular}
	\caption{Values for $\scriptstyle{\Delta}\textstyle\gamma$ and $<$$\gamma$$>$ in the CSP and LSP corresponding to \citet{BellotRubio:2003p206} and $\mathcal{N}$ resulting from the $\scriptstyle{\Delta}\textstyle\gamma$-effect according to \citet{Solanki:1993p210}.}
	\label{tab:N_delta_gam}	
\end{center}
\end{table}

\paragraph{Comparison with Previous Studies:} This survey and the study of \citet{MartinezPillet:2000p223} yield similar results. However, the positive values of $\mathcal{N}$ in the CSP at small heliocentric angles were not that obvious in previous investigations. They are in accordance with the synthetic CLV curves of $\mathcal{N}$ from \citet{Solanki:1993p210} as well as \citet{MartinezPillet:2000p223}, but are not reproduced by \citet{Muller:2002p3983}\footnote{Their choice of the sign of $\mathcal{N}$ is opposite to that of the other studies. Hence, the CLV curves have to be interpreted inversely.}. The error bars in this study and the investigation of \citet{MartinezPillet:2000p223} represent 1$\cdot\sigma$ and 3$\cdot\sigma$ respectively. Thus, the fluctuations of $\mathcal{N}$ around its mean value are significantly larger in the high resolution data of HINODE. 

Differences between the CLV curves of $\mathcal{N}$ in this investigation and other reports are found primarily on the CSP for Fe 630.15 and on the LSP for Fe 630.25. For Fe 630.15, \citet{MartinezPillet:2000p223} obtained larger negative values on the CSP, while this study shows positive values for $\mu < 0.65$. For Fe 630.25 on the LSP, \citet{MartinezPillet:2000p223} reports smaller values, especially for $\mu < 0.6$ where differences of up to 100\% can be seen. A comparison with \citet{Solanki:1993p210} and \citet{Muller:2002p3983} yield even larger differences, indicating that the parameters in the flow channel and the ambient atmosphere need some fine tuning.

\paragraph{Discussion:} The basic behavior of the CLV curves of $\mathcal{N}$ in the two iron lines is in accordance with other studies. The positive values of $\mathcal{N}$ in the CSP, which are expected for $\mu \approx 1$, were measured for the first time. The differences between the CLV curves of $\mathcal{N}$ observed with HINODE and earlier studies indicate that the configuration of velocity and magnetic field in the penumbra is much more diverse than previously assumed.

The higher values of $\mathcal{N}$ in the LSP are due to projection effects, which result in a larger $\scriptstyle{\Delta}\textstyle\gamma$ and subsequently in a higher $\mathcal{N}$. Due to the high spatial resolution of the HINODE SP data, more detailed studies can be performed in the future. Since $\gamma$ is different for the flow channels in the inner, mid and outer penumbra, different CLV curves should be obtained if $\mathcal{N}$ is averaged in just these areas. The comparison of such an observation with artificial CLV curves obtained from a known atmosphere would help to put further constraints on penumbral models.

\section{Polarity Reversal of Net Circular Polarization}
\label{sec:pol_rev}

The scenario discussed in Sections~\ref{sec:corr_dop_ncp} and \ref{sec:clv_ncp} explain positive values of $\mathcal{N}$ as well as its different strengths in the CSP and the LSP. However, they do neither explain the negative $\mathcal{N}$ measured in 16\% of penumbral Stokes V profiles at disk center (cf.~Section~\ref{sec:corr_dop_ncp}), nor do they account for mixed polarities in the CSP (cf.~Fig.~\ref{fig:ncp_vdop_limb}).

\begin{figure}[h!]
\begin{center}
    \includegraphics[width=0.9\textwidth]{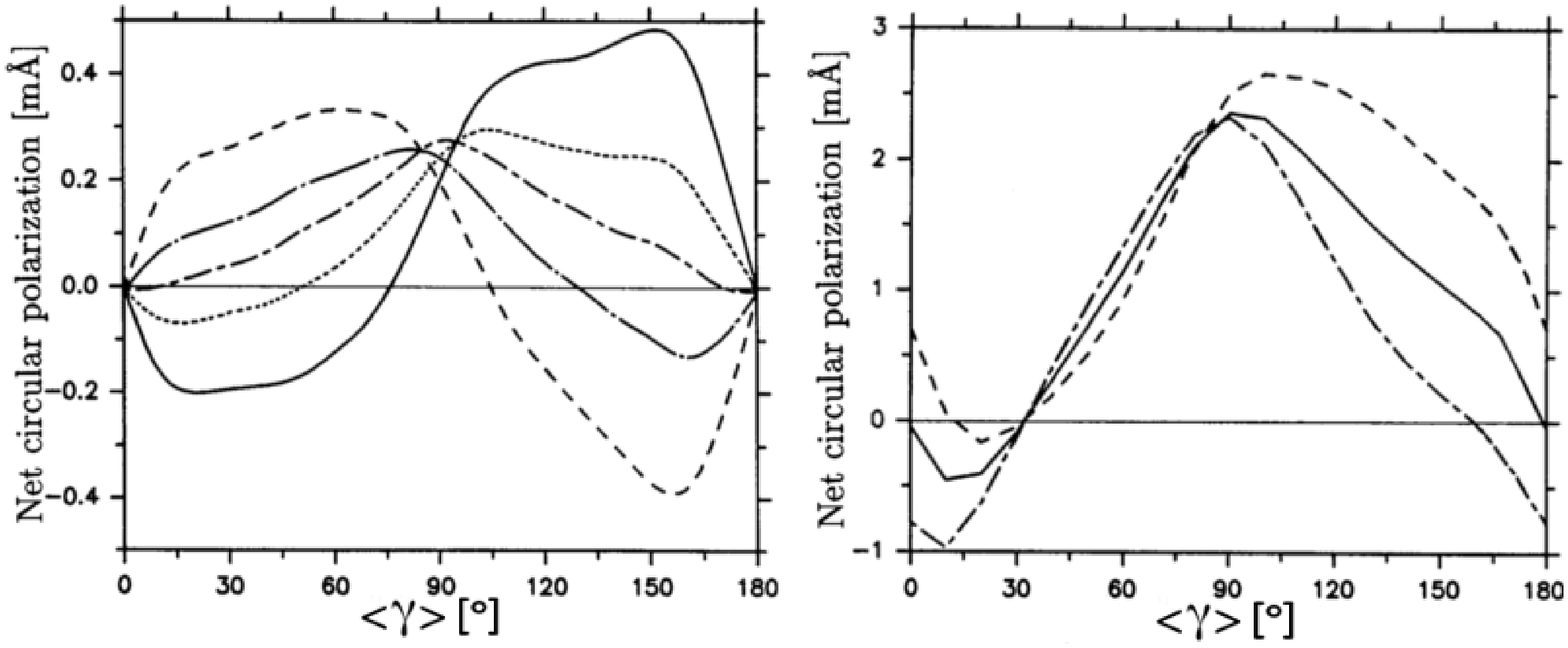}
\caption{Left: $\mathcal{N}$ of Fe I 630.25 versus $<$$\gamma$$>$ for various $\scriptstyle{\Delta}\textstyle$v. In all plots, $\rm{B}=1.5$~kG and $\scriptstyle{\Delta}\textstyle\gamma=10^{\circ}$. Besides, v$_1=0.5$~km~s$^{-1}$ and v$_2=1$~km~s$^{-1}$ (dashed), v$_1=0.8$~km~s$^{-1}$ and v$_2=1$~km~s$^{-1}$ (dot-long-dashed), v$_1=\rm{v}_2=1$~km~s$^{-1}$(dot-dashed), v$_1=0.8$~km~s$^{-1}$ and v$_2=1$~km~s$^{-1}$ (dotted) as well as v$_1=0.5$~km~s$^{-1}$ and v$_2=1$~km~s$^{-1}$ (solid). Right: Same as left, but for various $\scriptstyle{\Delta}\textstyle\rm{B}$. In all plots, $\scriptstyle{\Delta}\textstyle\gamma=20^{\circ}$, v$_1=1$~km~s$^{-1}$ and v$_2=0.8$~km~s$^{-1}$, while B$_1=\rm{B}_2=1$~kG (solid), B$_1=1.7$~kG and B$_1=1.5$~kG (dashed) as well as B$_1=1.5$~kG and B$_1=1.7$~kG (dot-dashed). Adopted from \citep{Solanki:1993p210}.}
\label{fig:ncp_solanki2}
\end{center}
\end{figure}

\paragraph{Negative $\mathcal{N}$:} Fig.~\ref{fig:ncp_solanki2} shows the influence of the $\scriptstyle{\Delta}\textstyle$v-effect (left) and of the $\scriptstyle{\Delta}\textstyle$B-effect (right) on $\mathcal{N}$ in the framework of the uncombed penumbral model \citep{Solanki:1993p210}. Both effects are capable of producing negative $\mathcal{N}$. 

The dot-dashed curves on the left side of Fig.~\ref{fig:ncp_solanki} and \ref{fig:ncp_solanki2} refer to the same atmospheric configuration. In all the other plots of Fig.~\ref{fig:ncp_solanki2}, v$_1\not=\rm{v}_2$, which  significantly alters the contribution of the $\scriptstyle{\Delta}\textstyle$v-effect on $\mathcal{N}$. The interval of $<$$\gamma$$>\in[0^{\circ}...90^{\circ}]$ is of special interest for the explanation of mixed polarities in the CSP. It can be seen that $\mathcal{N}$ becomes negative for v$_1 < \rm{v}_2$. This, however, is at odds with the results of Section~\ref{sec:gradv} and \ref{sec:invofupdown}, where it was demonstrated that the strongest plasma flows are present in the deep photosphere, which implies v$_1 > \rm{v}_2$. Furthermore, it does not explain the negative $\mathcal{N}$ in the inner CSP, since $<$$\gamma$$>=0$, which results in $\mathcal{N}=0$ for all v$_1\not=\rm{v}_2$. Thus, the $\scriptstyle{\Delta}\textstyle$v-effect 
is not at the origin of negative $\mathcal{N}$ in the CSP.

The right plot of Figure~\ref{fig:ncp_solanki2} demonstrates that the $\scriptstyle{\Delta}\textstyle$B-effect produces mainly positive  $\mathcal{N}$ on the CSP, except for $<$$\gamma$$>\in[0^{\circ}...30^{\circ}]$. The solid curve refers to B$_1=\rm{B}_2$, while the dashed curve represents B$_1 > \rm{B}_2$. Even though these two atmospheric configurations produce $\mathcal{N}<0$, they lead to $\mathcal{N}>0$ in the inner CSP, where $<$$\gamma$$>=0^{\circ}$ (cf.~Table~\ref{tab:N_delta_gam}). Only the dot-dashed curve referring to B$_1<\rm{B}_2$ yields $\mathcal{N}<0$ for $<$$\gamma$$>=0^{\circ}$. Thus, the only explanation which satisfies all the results previously discussed, i.e. $<$$\gamma$$>=0^{\circ}$, v$_1 > \rm{v}_2$ and $\mathcal{N}<0$, is a decrease of B with $\tau$ in the inner CSP.

\begin{figure}[h!]
\begin{center}
    \includegraphics[width=1\textwidth]{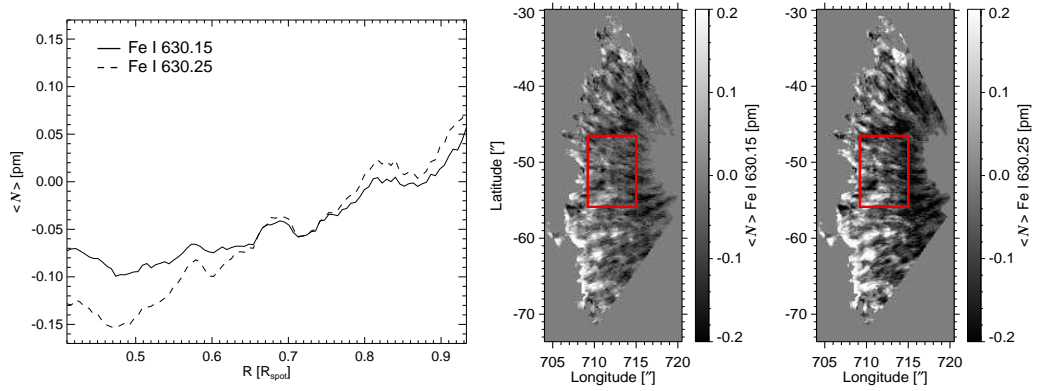}
\caption{Left: Average $\mathcal{N}$ along azimuthal paths of different radii in the CSP for the two Fe lines. Middle and right: Enlargement of the CSP of Spot 08. The red box indicates the area studied in Section~\ref{sec:stat_ana}.}
\label{fig:ncprev_csp}
\end{center}
\end{figure}

\paragraph{Polarity Reversal:} In this context, it is interesting that \citet{Tritschler:2007p217} reported a polarity reversal of $\mathcal{N}$ when averaged along azimuthal paths of different radii placed in the CSP. Fig.~\ref{fig:ncprev_csp} shows the result of the same study using HINODE SP data of the CSP of Spot 08. It is evident that  
$\mathcal{N}$ does not only become more and more positive with increasing distance from the center of the spot, but that it also reverses its polarity at 80\% of radial distance. On average, the inner and mid CSP are dominated by negative $\mathcal{N}$, while positive $\mathcal{N}$ is prevalent in the outer CSP.

\citet{Tritschler:2007p217} interpreted this polarity reversal as evidence for B$_1 < \rm{B}_2$ in the inner CSP\footnote{In their contribution, B$_1$ corresponds to the magnetic field strength in the flow channel and B$_2$ denotes the field strength in the background field.}. With increasing radius, B$_2$ diminishes, while B$_1$ increases slightly, resulting in B$_1 > \rm{B}_2$ and $\mathcal{N}>0$ in the outer CSP.

This is in accordance with the results presented before. However, the pattern of mixed polarities is visible throughout the CSP in the HINODE SP data, indicating that regions with B$_1 < \rm{B}_2$ exist in the mid and outer CSP as well.

\paragraph{Conclusion:} In a two-layer atmosphere, negative $\mathcal{N}$ may be produced by the $\scriptstyle{\Delta}\textstyle$v-effect if v$_1 \not= \rm{v}_2$ as well as by the $\scriptstyle{\Delta}\textstyle$B-effect. However, since the strong plasma flows occur in the deep atmospheric layers, only the $\scriptstyle{\Delta}\textstyle$B-effect with B$_1 < \rm{B}_2$ is capable of producing $\mathcal{N}<0$ in the inner CSP.

\section{Inversion}
\label{sec:stat_ana}

To test the assumption that the pattern of opposite polarities of $\mathcal{N}$ in the CSP is due to an alternating gradient in B, a spectral inversion was performed on 2400 Stokes profiles located within the red box of Fig.~\ref{fig:ncprev_csp}.

\paragraph{Setup:} The procedure is similar to the one explained in Section~\ref{sec:invofupdown}. To minimize the possibility for the algorithm to approach a local minimum of the $\chi^2$ surface, the inversion was performed ten times for the same profile, and the atmosphere with the minimal $\chi^2$ was selected. For the initial atmosphere of each of these inversions, the distribution of T was randomized around the values of the HSRA. In contrast to p$_{\rm{e}^-}$, the values for v$_{\rm{dop}}$, B, $\gamma_{\rm{mag}}$ and $\phi_{\rm{mag}}$ were randomized as well.

In the first cycle of the inversion, all atmospheric parameters (except p$_{\rm{e}^-}$) were allowed to change constantly with $\tau$. In the second cycle, linear gradients were added to the distribution of B and v$_{\rm{dop}}$. In contrast to the setup in Section~\ref{sec:invofupdown}, $\gamma_{\rm{mag}}$ was not allowed to change with $\tau$, and the number of free parameters in v$_{\rm{dop}}$ was significantly smaller. This is possible because:
\begin{itemize}
\item{Under the projection of $\Theta$, $\scriptstyle{\Delta}\textstyle\gamma$ is small and will not contribute significantly\footnote{To validate this premise, another inversion was performed a) with linear gradients in B, v$_{\rm{dop}}$ and $\gamma$ as well as a third inversion, b) with linear gradients v$_{\rm{dop}}$ and $\gamma$, but only constant values of B with $\tau$. While the second experiment yielded only slightly better results, it was not possible to reproduce the pattern of $\mathcal{N}$ in the CSP in the third attempt. This is taken as evidence that linear gradients in B and v$_{\rm{dop}}$ are sufficient for this investigation.} to $\mathcal{N}$ in the CSP.}
\item{The polarity of $\mathcal{N}$ shall be reproduced, which makes it necessary to model the gradient with height of the atmospheric parameters, but allows to neglect spectral features like the additional shoulder on the blue side of Stokes V.}
\end{itemize}

\begin{figure}[h!]
\begin{center}
    \includegraphics[width=0.8\textwidth]{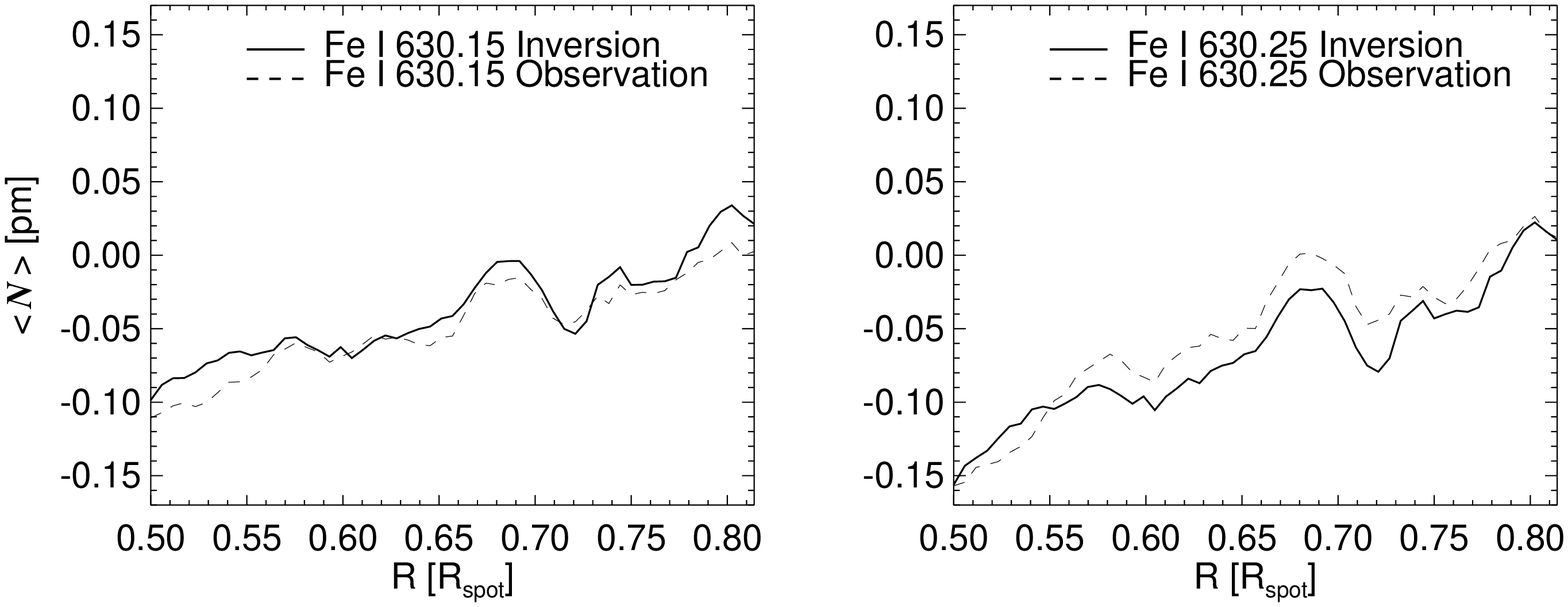}
\caption{Average $\mathcal{N}$ of Fe I 630.15 (left) and Fe I 630.25 (right) along azimuthal paths of different radii within the red box indicated in Fig.~\ref{fig:ncprev_csp}. Plotted are both the results from the observation (dashed) and inversion (solid).}
\label{fig:ncp_invobs_radial}
\end{center}
\end{figure}

\paragraph{Quality of Results:} Fig.~\ref{fig:ncp_invobs_radial} compares the radial dependency of $<$$\mathcal{N}$$>$, i.e. the average $\mathcal{N}$ along azimuthal paths, inferred from observation (dashed) and from inversion (solid) in Fe I 630.15 (left) and Fe I 630.25 (right). 

\noindent Despite the simplicity of the model, it is possible to reproduce not only the global trend of $<$$\mathcal{N}$$>$ with increasing radial distance, i.e. the polarity reversal, but also the fluctuations of $<$$\mathcal{N}$$>$, e.g. a local maximum at $\rm{R}=0.69\cdot\rm{R}_{\rm{spot}}$. In general, $<$$\mathcal{N}$$>$ is reproduced better for Fe I 630.15 than for Fe I 630.25. In the first case, $<$$\mathcal{N}$$>$ is overestimated by the inversion for $0.50\cdot\rm{R}_{\rm{spot}} < \rm{R} < 0.55\cdot\rm{R}_{\rm{spot}}$ and for $\rm{R} > 0.78\cdot\rm{R}_{\rm{spot}}$. In the second case, the inversion overestimates $<$$\mathcal{N}$$>$ for $\rm{R} < 0.55\cdot\rm{R}_{\rm{spot}}$ and underestimates it for $0.55\cdot\rm{R}_{\rm{spot}} < \rm{R} < 0.79\cdot\rm{R}_{\rm{spot}}$. The differences between observed and inverted $<$$\mathcal{N}$$>$ might be owed to the simplicity of the model. The overestimation of $\mathcal{N}_{630.15}$ and the underestimation of $\mathcal{N}_{630.25}$ seem plausible since both lines are inverted at the same time and $\rm{g}_{630.15} < \rm{g}_{630.25}$.

\begin{figure}[h!]
\begin{center}
    \includegraphics[width=0.9\textwidth]{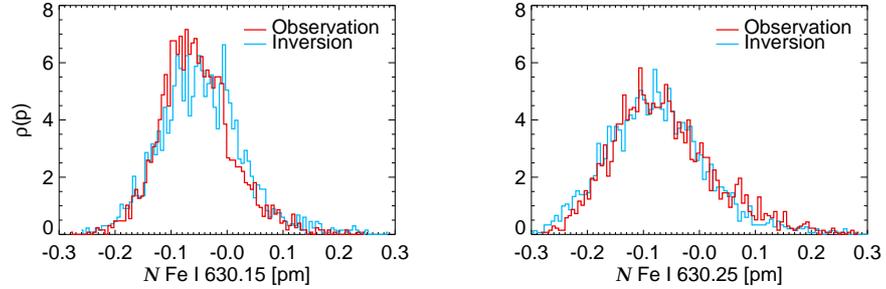}
\caption{PDFs of $\mathcal{N}$ from observation (red) and from inversion (blue) for both Fe lines within the red box of Fig.~\ref{fig:ncprev_csp}.}
\label{fig:histo_invobs}
\end{center}
\end{figure}

PDFs of $\mathcal{N}$ within the red box of Fig.~\ref{fig:ncprev_csp} are shown in Fig.~\ref{fig:histo_invobs}. The left plot shows observation (red) and inversion (blue) for Fe I 630.15, while the same is depicted on the right side, but for Fe I 630.25. It is evident that for Fe I 630.15, more Stokes V profiles have small negative values of $\mathcal{N}$ than in the case of Fe I 630.25. Even though both PDFs are skewed, the asymmetry is more apparent in the latter case.

\noindent In general, the histogram of inversion of Fe I 630.25 agrees better with the observation then Fe I 630.15. In the latter case, the amount of Stokes V profiles with $-0.13~{\rm{nm}} < \mathcal{N} < -0.01~{\rm{nm}}$ is larger in the inversion when compared to the observation, while the opposite is true for $-0.01~{\rm{nm}} < \mathcal{N} < 0.07~{\rm{nm}}$. In Fe I 630.25, the observed $\mathcal{N}$ is much better reproduced by the inversion, except for $-0.28~{\rm{nm}} < \mathcal{N} < -0.2~{\rm{nm}}$ and $0.06~{\rm{nm}} < \mathcal{N} < 0.14~{\rm{nm}}$, where the inversion over- and underestimates the amount of Stokes V profiles.

\begin{figure}[h!]
\begin{center}
    \includegraphics[width=0.9\textwidth]{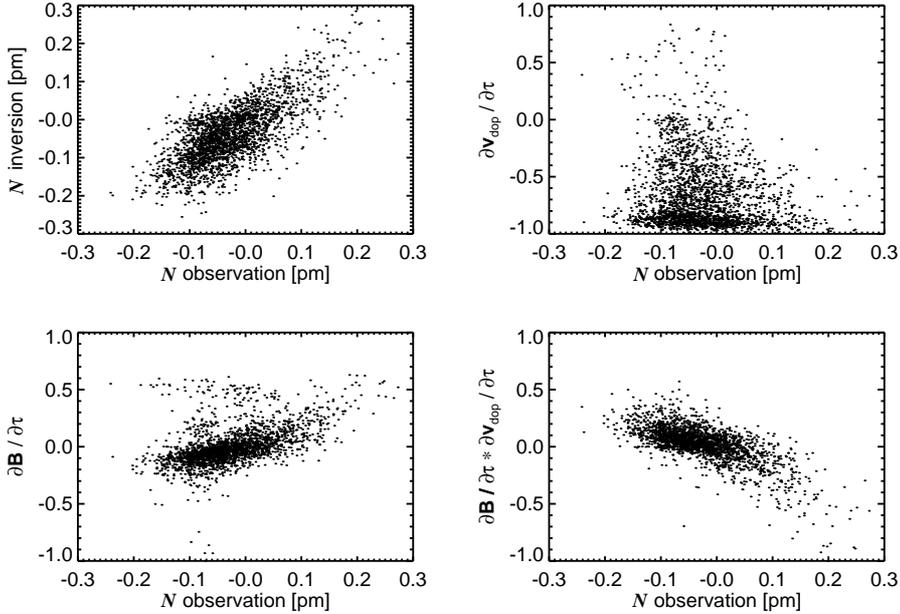}
\caption{Scatterplots of results from inversion. Top left: $\mathcal{N}$ from inversion versus $\mathcal{N}$ from observation. Top right: Gradient of v$_{\rm}$ with $\tau$ versus observed $\mathcal{N}$. Bottom left: Gradient of B with $\tau$ versus observed $\mathcal{N}$. Bottom right: Product of the gradients of v$_{\rm}$ and B with $\tau$ versus observed $\mathcal{N}$.}
\label{fig:inv_ncp_res}
\end{center}
\end{figure}

\paragraph{Interpretation:} The resulting gradient of the atmospheric parameters with respect to $\mathcal{N}$ is depicted in Fig.~\ref{fig:inv_ncp_res}. The top left panel shows a scatterplot of inverted versus observed $\mathcal{N}$. Since the inversion does not yield a perfect correlation between these quantities, r$_{\rm{S}}=0.67$ is obtained.

In the top right panel, the normalized gradient of v$_{\rm{dop}}$ is plotted with respect to the observed $\mathcal{N}$. It is evident that the gradient of v$_{\rm{dop}}$ with $\tau$ is negative in almost all profiles. This indicates that the amplitude of the plasma flow increases\footnote{In this study, blueshifts correspond to negative values of v$_{\rm{dop}}$.} with $\tau$, which is in accordance with the results of Section~\ref{sec:gradv} and \ref{sec:invofupdown}. Since a large range of $\mathcal{N}$ values is obtained for the same gradient of v$_{\rm{dop}}$, there is, however, no correlation between these quantities, and a small correlation coefficient of r$_{\rm{S} = -0.13}$ is derived.

The scatterplot in the bottom left panel illustrates the gradient of B with $\tau$ versus the observed $\mathcal{N}$. A trend is ascertainable in the sense that the polarity of $\mathcal{N}$ is related to the sign of the gradient of B with $\tau$. Furthermore, the strength of the gradient seems to account for the amplitude of the observed value of $\mathcal{N}$. A ranked correlation analysis yields r$_{\rm{S}}=-0.57$. 

In the bottom right panel, the product of the gradients of v$_{\rm}$ and B with $\tau$ is plotted with respect to the observed $\mathcal{N}$. Since 
sign$(\mathcal{N})= - \rm{sign} \left ( \rm{dB}/\rm{d}\tau \cdot \rm{dv}_{\rm{dop}}/\rm{d}\tau \right ) $ \citep{Solanki:1988p212}, it is not surprising that the correlation coefficient is r$_{\rm{S}}=0.64$ and is thus slightly larger than in the previous case. Nevertheless, the difference is only 10\% and shows that it is indeed the gradient in B with $\tau$ that is the dominant contributor to the polarity and the amplitude of $\mathcal{N}$.

\paragraph{Conclusion:} Given the simplicity of the model -- gradients with $\tau$ only in T,  v$_{\rm{dop}}$ and B -- the quality of the inversion results is astonishing. It is possible to reproduce not only the observed radial dependency of $<$$\mathcal{N}$$>$, but also the PDFs of $\mathcal{N}$, including their skewness.

The results from the inversion indicatet that $\mathcal{N}$ is a proxi of the gradient of B with $\tau$ in the CSP of sunspots observed at $\Theta \approx 50^{\circ}$. Furthermore, the pattern of positive and negative $\mathcal{N}$ is caused by the $\scriptstyle{\Delta}\textstyle$B-effect, since the dominant $\scriptstyle{\Delta}\textstyle\gamma$-effect is small under the projection of $\Theta$ and the amplitude of the plasma flow increases with $\tau$.
 
\section{Summary}
\label{sec:concl_ncp}

The influence of the gradients with height of various atmospheric parameters on asymmetries in Stokes V, i.e. the total net circular polarization ($\mathcal{N}$), was investigated by means of a two-layer model atmosphere. It was shown that at disk center, Sperman's ranked correlation coefficient for $\mathcal{N}$ and the Doppler velocity is larger in downflow channels than it is in upflow channels. This behavior was not only attributed to the larger value of the average zenith angle, but also to the larger difference between the zenith angles of the individual atmospheric components in downflow channels.

The center to limb variation of $\mathcal{N}$ was investigated using 30 images of three sunspots taken during their passage across the solar disk. The  outcome of this survey not only confirms but also extends the results of earlier studies. Various peculiarities of the center to limb variation curves of $\mathcal{N}$ are interpreted within the framework of a two-layer atmosphere. It was found that the fluctuation of $\mathcal{N}$ are larger when compared to previous studies with lower spatial resolution. Finally, ideas and proposals for future investigations were developed.

It was argued that 
only a decrease of the magnetic field strength with optical depth, can account for the negative $\mathcal{N}$ in the center side penumbra of sunspots at a heliocentric angel of approximately $50^{\circ}$ when assuming a two-layer atmosphere. Using HINODE data, it was possible to confirm the radial dependency of azimuthally averaged values of $\mathcal{N}$ in the center side penumbra in general and the sign reversal of $\mathcal{N}$ in particular.

By inverting 2400 Stokes profiles from the center side penumbra, it was demonstrated that a simple model atmosphere with linear gradients in Doppler velocity and magnetic field strength is sufficient to account not only for the distribution of $\mathcal{N}$ but also for its amplitude. The inversion results yield a scenario in which 
the velocity always increases with optical depth, while the gradient of the magnetic field strength with optical depth is either positive or negative thereby accounting for the pattern of positive and negative $\mathcal{N}$. It was concluded that $\mathcal{N}$ is a proxy of the gradients of the magnetic field strength with optical depth in center side penumbra of sunspots at large heliocentric angles ($\Theta \approx 50^{\circ}$).

\chapter{Conclusion}
\label{ch:outlook}

Spectropolarimetric data obtained with the HINODE space-borne observatory were used to study the horizontal and vertical components of the Evershed flow. The high spatial and spectral resolution of the data allows for a precise investigation of the velocity field on spatial scales of  240~km. The Doppler shift of the wing of Fe I 630.15 was used to infer the plasma velocities ($\rm{v}_{\rm{dop}}$) in the deep photosphere. For observation at disk center, the wavelength scale was calibrated\footnote{For observation of sunspots away from the center of the solar disk, procedure b) was applied.} using:
\begin{itemize}
\item[a)]{The line-core of an average Stokes I profile of the quiet Sun after the correction for the convective blue shift.}
\item[b)]{The center position of umbral Stokes V profiles with an amplitude asymmetry of less than 1\%.}
\end{itemize}
\noindent Within the uncertainties of $\pm$0.1~km~s$^{-1}$, both methods yield the same results. 

\paragraph{The Penumbral Velocity Field:} The vertical penumbral velocity field was investigated using HINODE observation of several sunspots close to disk center ($\Theta \le 11^{\circ}$). The horizontal contribution to the Evershed flow was studied in a range of sunspots located at large heliocentric angles ($\Theta \approx 50^{\circ}$). The results can be summarized as follows:
\begin{itemize}
\item[$\alpha$)]{The maximal upflow velocity in the penumbra is weaker when compared to the quiet Sun, while the downflow amplitudes are larger.}
\item[$\beta$)]{Probability density functions of the penumbral velocity field are positively skewed. This is because a larger fraction of the penumbra shows upflows  for $|\rm{v}_{\rm{dop}}| < $ 0.8~km~s$^{-1}$, while downflows prevail for $|\rm{v}_{\rm{dop}}| > $ 0.8~km~s$^{-1}$. By contrast, histograms representing the occurrence of velocity amplitudes in the quiet Sun are symmetric with upflows dominating for all velocities.}
\item[[$\gamma$)]{In both the penumbra and the quiet Sun, the strongest plasma flows are present in the deep photospheric layers. However, the gradient with height of the velocity field is larger in the penumbra when compared to the quiet Sun. This could be due to the large inclination of the penumbral magnetic field which hinders the plasma to reach the same atmospheric heights as in the quiet Sun.}
\item[$\delta$)]{The horizontal component of the Evershed flow shows a filamentary structure comparable to the one seen in images of the continuum intensity, with blue and redshift in the center and the limb side penumbra respectively.}
\item[$\epsilon$)]{A ranked correlation analysis for continuum intensity and the vertical component of the Evershed flow yields a radial dependency of the correlation coefficient ($\rho_{\rm{S}}$) similar to that of sunspot simulations. For the continuum intensity and the horizontal plasma flow, $\rho_{\rm{S}}$ is positive in the inner, but negative in the outer penumbra. However, $\rho_{\rm{S}}$ is not large enough to consider the Evershed flow as being connected to the penumbral continuum intensity.}
\end{itemize}

\noindent A morphological study of the global penumbral velocity field and a case study of local penumbral features reveal that:
\begin{itemize}
\item[1)]{Upflows appear elongated, follow the filamentary structure and prevail in the inner penumbra.}
\item[2)]{Downflows do not show a strand-like structure, but look roundish and dominate the outer penumbra.}
\item[3)]{Even though points 1) and 2) are confirmed by the radial dependency of azimuthal averages of the Doppler velocity, downflows may appear in the inner penumbra, too.}
\item[4)]{No overturning convection, i.e. a down-up-down flow pattern, occurs on the scale of the width of a penumbral filament.}
\item[5)]{Two families of (dark-cored) penumbral filaments exist:}
\begin{itemize}
\item{In one family of penumbral filaments, the upflows in the lateral brightenings are stronger when compared to the dark-core.}
\item{In the other family, the bright head of the penumbral filaments exhibits a strong upflow, which is also present in the dark-core, but diminishes with radial distance and turns into a downflow at the end of the penumbral filament. The lateral brightenings show no vertical plasma velocities.}
\end{itemize}
\item[6)]{Some penumbral downflows are accompanied by a local enhancement in continuum intensity associated with a local minimum in total polarization.}
\item[7)]{The upflow channels in the inner penumbra have a zenith angle of $\gamma \approx 40^{\circ}$.}
\item[8)]{The zenith angle is $\gamma > 135^{\circ}$ in some penumbral downflow channels.}
\end{itemize}

\paragraph{The Penumbral Magnetic Field:} Information on the configuration of the magnetic field were either derived from the interpretation of (crossover) Stokes V profiles in the framework of a multilayer atmosphere or from inversion using the SIR code. It was found that:
\begin{itemize}
\item[A)]{At least 40\% of all penumbral downflows contain a magnetic component with opposite polarity, which remains hidden in magnetograms obtained with instruments of limited spectral resolution.}
\item[B)]{The crossover effect increases continuously with the Doppler velocity proving that both the vertical and the horizontal components of the Evershed flow are magnetized.}
\item[C)]{Points A) and B) are in accordance with the results of an inversion of representative Stokes V profiles from up- and downflow channels.}
\item[D)]{The correlation coefficient between the total net circular polarization ($\mathcal{N}$) and the Doppler velocity is larger in downflow channels when compared to upflows. This is in accordance with the idea that the magnetized Evershed flow returns below the solar surface within the penumbra. In the framework of a two-layer model atmosphere, it is the average inclination of the magnetic field as well as the difference between its components which causes $\mathcal{N}$ to be larger in penumbral downflow channels when compared to upflows.}
\item[E)]{The radial dependency of azimuthally averaged values of $\mathcal{N}$ in the center side penumbra is due to an average decrease of the magnetic field strength with optical depth in the inner, but an increase of this strength with optical depth in the outer center side penumbra.}
\item[F)]{Results from spectral inversion indicate that the pattern of positive and negative $\mathcal{N}$ in the center side penumbra is associated with the sign of the gradient of the magnetic field strength with optical depth.}
\end{itemize}

\paragraph{Implications Regarding Penumbral Models:} Most of the results in this work are in accordance with the predictions of penumbral flux-tube models: 

\begin{itemize}
\item{The up- and downflow patches in the inner and outer penumbra can be interpreted as the sources and the sinks of the Evershed flow.}
\item{Downflows contain magnetic fields of opposite polarities.}
\item{Both the vertical and the horizontal components of the Evershed flow are magnetized.}
\item{One family of penumbral filaments shows an upflow at the head and a downflow at its end.}
\item{Bright penumbral downflows could be the result of shocks inside penumbral flux-tubes.}
\end{itemize}

\noindent Furthermore, the investigated HINODE data pose a range of problems for the scenario of penumbral gaps:

\begin{itemize}
\item{The proposed flow pattern within penumbral filaments is not observed.}
\item{All penumbral plasma flows occur in a magnetized atmosphere.}
\item{Magnetic fields with opposite polarity are not proposed by the gappy model.}
\end{itemize}

\noindent In the light of HINODE observation, flux-tube models thus appear as the more favorable scenario to explain the penumbra of sunspots.

\paragraph{Future Research:} The high resolution of HINODE data has shown that the penumbral atmosphere is much more diverse and structured on smaller scales than previously thought.

Thus, it seems reasonable to repeat a range of previously conducted studies with HINODE data. In the survey of \citet{MartinezPillet:2000p223}, for example, the center to limb variation of $\mathcal{N}$ was studied by means of average values of the center and the limb side penumbra. However, the geometry of the atmosphere of a) upflows in the inner, b) outflows in the mid and c) downflow channels in the outer penumbra has a significant influence on $\mathcal{N}$. The high resolution data of the HINODE SP allows to obtain the center to limb variation of $\mathcal{N}$ for these penumbral regions individually.


Additionally, the increasing diversity of the penumbral atmosphere on smaller scales makes it difficult to rule out the gappy penumbral model completely. It is possible that the predicted flow pattern in penumbral filaments is observable on smaller scales. The next generation of solar telescopes, i.e.  NST (1.6~m aperture, in operation), GREGOR (1.5~m aperture, commissioning 2011), ATST (4~m aperture, commissioning 2018) and EST (4~m aperture, commissioning $>$ 2020), will have a better spatial resolution, allowing to draw definite conclusions.

Finally, the recent progress in the magneto-convective modeling of entire sunspots \citep{2011ApJ...729....5R} will surely lead to an increase of knowledge about the penumbra in the near future. Solving the radiative transfer equation in such a simulated penumbral atmosphere yields observable quantities, which can be compared to observation and used in turn to fine-tune the computer models.

\appendix

\chapter{Estimation of Straylight in HINODE SP Data}
\label{ch:stray}

Straylight is the light that enters a resolution element from the surrounding FOV and all telescope operating at the diffraction limit have straylight issues, especially if extended objects like the Sun are observed. The major sources of straylight are scattering processes along the optical path of the instrument as well as the microroughness of mirrors and other optical elements. The latter contribution is described by the point spread function (PSF) of the telescope, and may be accounted for in the a-posteriori data reduction process. Even though the contribution from scattering processes can be minimized by using high quality optics and a limited number of mirrors, it can never be avoided completely. An advantage of space-borne observations is that a third source of straylight, i.e. scattering processes within Earth's atmosphere, is absent in such data.

Straylight may be estimated from observation of the solar limb. A scan across the edge of the solar disk does not shows not only show a drop of intensity at the solar solar limb, but also residual light at a distance of several seconds of arc. Since it can be assumed that no emission is present at greater distances from the solar disk\footnote{This is true for the continuum. Note, however, that some lines, e.g. Ca, are visible in emission at higher atmospheric layers.}, the residual intensity can be attributed to straylight. Commonly the superposition of a number of Gauss functions is fitted to the intensity profile to account for the solar atmosphere and quantify the degree of straylight. In the case of planetary transits, e.g. of Mercury or Venus, the shadow of the planet can be used as well \citep{2009A&A...501L..19M}.

In another approach \citet{1992SoPh..140..207M,1993A&A...270..494M} derived a scaling relation between Stokes V and I$_{\rm{c}}$ from calculations of polarized radiative transfer within Milne-Eddington type atmospheres. This relation allowed them to quantify the straylight contamination. Finally, \citet{2008A&A...484L..17D} compared the contrast in HINODE SP observation with magnetohydrodinamic simulation of solar convection. Besides instrumental effects due to imperfect alignment, they had to assume a straylight level of 4.7\% in order to match simulation and observation.

\begin{figure}[h!]
	\centering
		\includegraphics[width={\textwidth}]{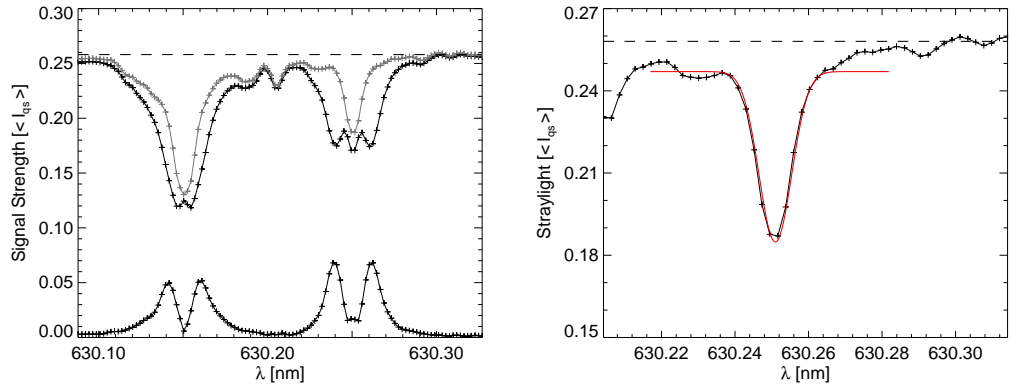}
		\caption{Left: Average Stokes I  (black) and absolute values of average Stokes V profile (gray) from the umbra with I$_{\rm{c}} < 0.33 \cdot \rm{I}_{\rm{qs}}$. The umbral continuum level is marked by the dashed line. Right: Difference profile (black) and fit to the straylight component (red).}
		\label{fig:straylight}
\end{figure}

\paragraph{Estimation of Straylight} In this work, Hinode SP observation of a sunspot at disk center is used to estimate straylight. Fe I 630.25 is suited best for this measurement, because it is a simple line triplet in the presence of a magnetic field and has a large Land\'e g factor of 2.5.  As a result, the line-core of Stokes I is fully split into three components (cf. left panel in Fig. \ref{fig:straylight}) if $\rm{B} > 1.5$~kG; a reasonable assumption for umbral magnetic fields. Apart from the red and blue shifted $\sigma^{\pm}$ components, Fe I 630.25 shows a third but unshifted dip within the line-core. This unshifted component is a combination of the $\pi$ transition and nonmagnetic straylight. If a sunspot is observed at disk center, where the umbral magnetic field appears parallel to the line of sight, the transversal Zemann effect disappears and the $\pi$ component vanishes. If it is assumed that the umbral magnetic field is strong enough to polarize all radiation, the residual dip, allows an estimate of straylight contamination.

Average Stokes I and P$_{\rm{tot}}$ profiles\footnote{P$_{\rm{tot}}$ is used instead of Stokes V to account for a possible contamination of P$_{\rm{lin}}$ due to the magneto-optical effect.} (black) from the umbra of Spot 04 are plotted in the left panel of Figure \ref{fig:straylight}. A superposition of both profiles, allows to remove the $\sigma^{\pm}$ components from the umbral Stokes I profile. The resulting profile (grey) is assumed to be completely due to straylight. 
The line-core of the residual component of Fe I 630.25 was determined by a Gauss fit with a constant baseline. This was necessary, because molecular blends are present in the wings of the umbral stokes V profile making it difficult for the fit to converge if the entire profile is considered. If the sum of constant baseline and amplitude of the fit is considered, the straylight contamination in HINODE SP data amounts to 7.6\%.

\paragraph{Application to HINODE SP Data} Since straylight is linked to the PSF of the instrument, it has a large contribution from the immediate surrounding FOV. The more distant surrounding, represented by the tail of the PSF, adds less to the straylight. However, the low I$_{\rm{c}}$ in the umbra, makes this kind of observation especially sensitive to the latter effect. This is because the tail of the PSF samples the penumbra and the QS where I$_{\rm{c}}$ is up to five times as large when compared to the umbra. Thus, the straylight contribution from the more distant surrounding is of significance in the umbra, which has to be taken into account when normalizing the Stokes Q, U and V profiles to I$_{\rm{c}}$.

\begin{figure}[h!]
	\centering
		\includegraphics[width={\textwidth}]{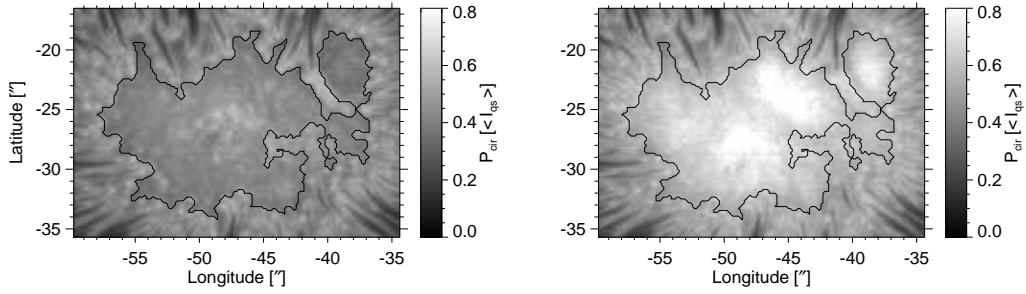}
		\caption{Map of P$_{\rm{cir}}$ in the umbra of Spot 04. Left: The maximal value of a Stokes V profile after normalization to the local continuum. Right: Same as left, but taking into account a global and constant straylight contamination of 8.7\%.}
		\label{fig:straymap}
\end{figure}

An example is give in Fig. \ref{fig:straymap}, where P$_{\rm{cir}}$
is plotted for the umbra of Spot 04. In the left panel of Fig. \ref{fig:straymap} P$_{\rm{cir}}$ is derived from the maximal value of the respective Stokes V profile after normalization to I$_{\rm{c}}$. It can be seen that P$_{\rm{cir}}$ assumes its maximal values at the head of PFs located in the inner penumbra and in some places in the inner umbra, while it appears less intense in the outer umbra and in DN -- e.g. at $\rm{(x;y)}=(-51$\arcsec$;-27.$\arcsec$5)$.

This is suspicious, because the places with a large P$_{\rm{cir}}$ in the umbra coincide with UDs and the sunspot is observed almost at disk center. Since the magnetic field is assumed to be weaker in UDs, and since the umbral fields appear parallel to the LOS \citep{Lites:1990p2995,1991Natur.350...45L,2002AN....323..165S}, it seems reasonable to assume that the DN and not the UDs should appear more intense in P$_{\rm{cir}}$. Especially the spatial correlation between local enhancements in I$_{\rm{c}}$ and P$_{\rm{cir}}$ suggest a contamination of straylight. If the Stokes V signal is normalized not only to I$_{\rm{c}}$ but is also corrected by a constant, mimicking straylight, theses peculiarities disappear. This can be seen in the right image of Fig. \ref{fig:straymap}), where the DN are the areas most prominent in P$_{\rm{cir}}$, while the drop of P$_{\rm{cir}}$ at the umbral boundary has disappeared.

\paragraph{Conclusion} Umbral maps of P$_{\rm{cir}}$ demonstrate that is necessary to correct the HINODE SP observation for instrumental straylight. The superposition of an average Stokes I and Stokes V profile observed in the umbra at disk center was used to estimate the level of straylight contamination. To this end it was assumed that the magnetic field is strong enough to not only separate the $\sigma^{\pm}$  components but also to completely polarize all photons. Since this implies I$_{\rm{c}} = \rm{P}_{\rm{cir}}$ within the line, the amplitude of the residual $\pi$ component present in the superimposed profile allows an estimation of straylight. 

Using this method a straylight contamination of 7.6\% was found for HINODE SP data. This is higher than the value of \citet{2008A&A...484L..17D}, who obtained a straylight contribution of 4.7\% but lower than the value of \citet{2009A&A...501L..19M}, who obtained 10\% and 11.5\% for the HINODE BFI at 555.0~nm and 668.4~nm, respectively. The discrepancy might be owed to the simplicity of the method and to the fact that the HINODE BFI was used in the study by \citet{2009A&A...501L..19M}.

\chapter{Correlation Analysis}
\label{ch:correlation}


To quantify a dependency between two variables it is possible to performed, e.g. a linear correlation analysis, first introduced by \citet{Pearson:1900p1860}, or a ranked correlation analysis according to \citet{Spearman:1987p1802}.

The Pearson correlation coefficient $\rho\left(\zeta,\xi\right)$ of two random variables $\zeta$ and $\xi$ with normal distribution is a measure of the linear dependence between $\zeta$ and $\xi$. It is defined as:

\begin{equation}
\rho \left( \zeta, \xi \right) := \frac{\mathcal{C} \left( \zeta,\xi \right) }{\sigma_\zeta\sigma_\xi}
\label{eq_1}
\end{equation}

\noindent with $\sigma$ being the standard deviaton, and $\mathcal{C} \left( \zeta,\xi \right) $ the covariance of $\zeta$ and $\xi$. In the case of two data samples $\zeta_{\rm{i}}$ and $\xi_{\rm{i}}$, with mean values $\mu_{\rm{x}} = \frac{1}{\rm{n}}\sum_{\rm{i}=0}^{\rm{n}}\rm{x}_{\rm{i}} $ and a variances
$\sigma_{\rm{x}}^2 = \frac{1}{\rm{n-1}}\sum_{\rm{i}=0}^{\rm{n}}(\rm{x}_{\rm{i}}-\mu_{\rm{x}})^2$, $\rho\left( \zeta,\xi \right) $ can be written as:

\begin{equation}
\rho \left( \zeta, \xi \right) =: \rm{r}({\zeta_{\rm{i}}, \xi_{\rm{i}}}) = \lim_{\rm{n}\to\infty} \frac{{\sum_{\rm{i}=0}^{\rm{n}}\left(\zeta_{\rm{i}}-\mu_{\zeta}\right)\left(\xi_{\rm{i}}-\mu_{\xi}\right)}}{ \sqrt{\sum_{\rm{i}=0}^{\rm{n}}\left(\zeta_{\rm{i}}-\mu_{\zeta}\right)^2} \sqrt{\sum_{\rm{i}=0}^{\rm{n}}\left(\xi_{\rm{i}}-\mu_{\xi}\right)^2}}
\end{equation}

\begin{figure}[h!]
	\centering
		\includegraphics[width=0.9\textwidth]{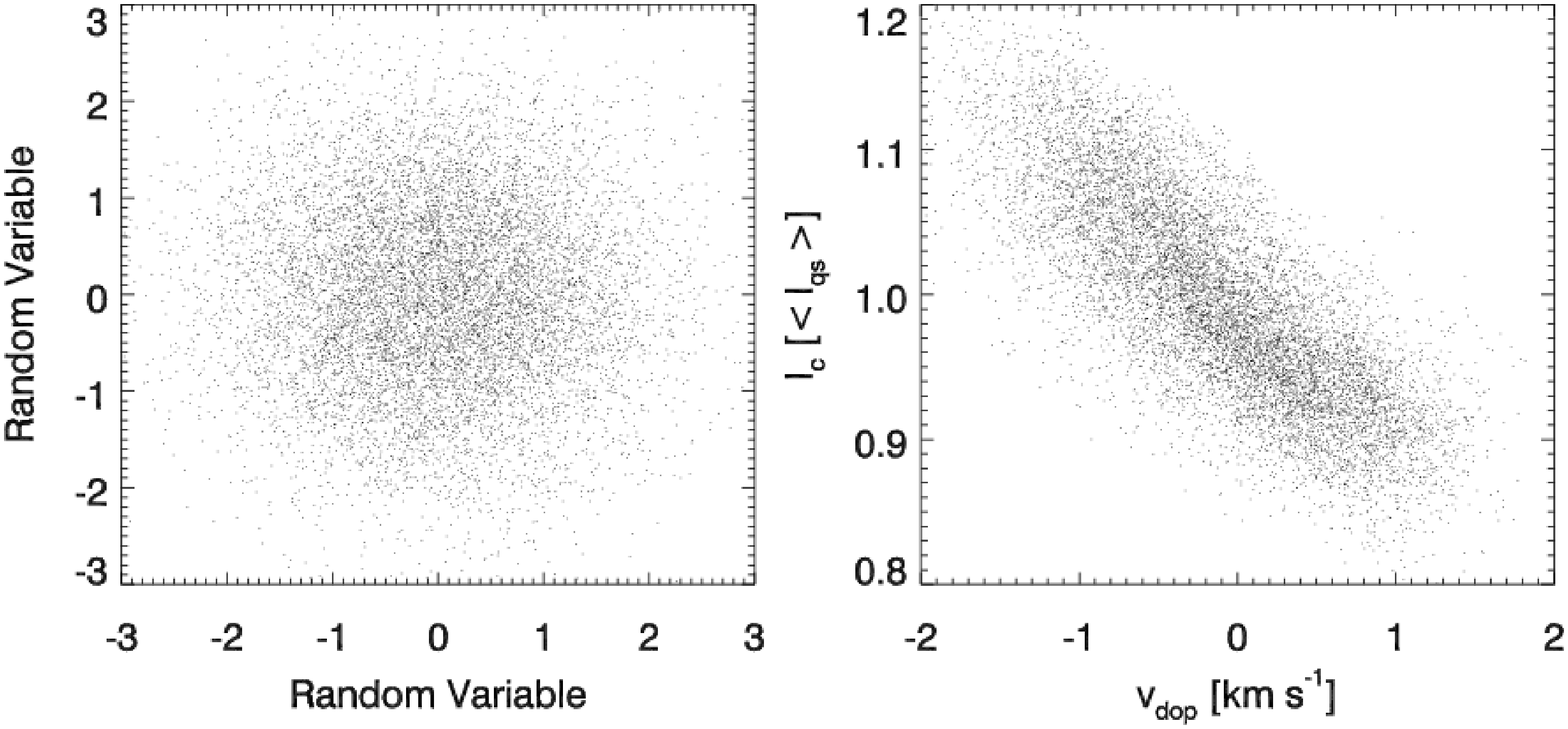}
		\caption{Scatterplots of two random variables (left) and I$_{\rm{c}}$ vs. v$_{\rm{dop}}$ (right) in the QS.}
		\label{fig:scatter}
\end{figure}

To give an example two scatterplots are shown in Fig.~\ref{fig:scatter}. The data cloud of two random variables appears circular (left side) because there is no linear correlation between these variables. This is reflected in a correlation coefficient of $\rm{r} = 0.04$. The data cloud on the right depicts I$_{\rm{c}}$ vs. v$_{\rm{dop}}$ from a section (16\arcsec~by 16\arcsec) of the QS at disk center. A linear dependency between I$_{\rm{c}}$ and v$_{\rm{dop}}$ is reflected in $\rm{r} = -0.778$.

If the underlying statistics of the random variables is not normal or unknown, it is more appropriate use non-parametric statistics, e.g Spearman's rank-correlation, to measure a correlation between the variables \citep{Henze:1979p1852,LSachs:1999}.

Spearman's correlation coefficient $\rho_{\rm{s}}(\zeta,\xi)$ is a measure of a monotonic\footnote{Keep in mind that correlations are not always linear.} dependency between two non-Gaussian distributed variables $\zeta$ and $\xi$. In analogy to equation (\ref{eq_1}) it is defined as: 
\begin{equation}
\rho_{\rm{s}} \left( \zeta, \xi \right) := \frac{\mathcal{C} \left[\mathcal{R}(\zeta),\mathcal{R}(\xi) \right] }{\sigma_{\mathcal{R}(\zeta)}\sigma_{\mathcal{R}(\xi)}}
\end{equation}

\noindent with $\mathcal{R}({\zeta})$ and $\mathcal{R}({\xi})$ being ranks of two variables $\zeta$ and $\xi$. For two non-binomial distributed data samples $\zeta_{\rm{i}}$ and $\xi_{\rm{i}}$, it can be estimated like:

\begin{equation}
\rho_{\rm{s}}(\zeta, \xi) =: \rm{r}_{\rm{s}}({\zeta_{\rm{i}}, \xi_{\rm{i}}}) = \lim_{\rm{n}\to\infty} 1-\frac{6 \sum_{\rm{i}=0}^{\rm{n}}\left[\mathcal{R}({\zeta_{\rm{i}})-\mathcal{R}({\xi_{\rm{i}}})}\right]^2}{\rm{n}(\rm{n}^2-1)}
\label{eq4}
\end{equation}
 
\noindent The number of tied ranks, i.e. $\mathcal{R}(\zeta_{\rm{i}}) - \mathcal{R}(\zeta_{\rm{j}}) = 0$ with  $\rm{i}\not=\rm{j}$, in either one of the data sets should be less than 20\% to ensure a reasonable estimate of $\rm{r}_{\rm{s}}$ using Equation~\ref{eq4}. Additionally, $\rm{r}_{\rm{s}}$ approaches $\rho_{\rm{s}}$ instead of $\rho$ for $\lim_{\rm{n}\to\infty}$ but the difference between $\rho_{\rm{s}}$ and $\rho$ is always smaller then 0.018 \citep{LSachs:1999}.

\paragraph{Significance of a correlation:} Significance is a measure to quantify how likely a result did not occur on chance. For a correlation it is a way to quantify how well r$({\zeta_{\rm{i}}, \xi_{\rm{i}}})$, derived from finite data samples, approximates the correlation $\rho(\zeta,\xi)$ of two variables. On way to test the significance of a correlation is to calculate the probability that the Null hypothesis H$_0$ -- i.e. a set of random numbers ${\zeta_{\rm{rdm}}, \xi_{\rm{rdm}}}$ with r$({\zeta_{\rm{rdm}}, \xi_{\rm{rdm}}}) \ge  \rm{r}({\zeta_{\rm{i}}, \xi_{\rm{i}}})$ -- is true. That probability is then used to define a level of confidence, for which H$_0$ can be rejected and r$({\zeta_{\rm{i}}, \xi_{\rm{i}}}) = \rho(\zeta,\xi)$ \citep{LSachs:1999}.

The level of confidence is often expressed in terms of standard deviations $\sigma$ of a normal distribution. 1$\sigma$ translates into a confidence level of 68.3\%, meaning that for 683 realizations out of 1000 permutations of ${\zeta_{\rm{rdm}}, \xi_{\rm{rdm}}}$ the Null hypothesis can be rejected. For 2$\sigma$, the level of confidence is 95.4\% and for 3$\sigma$ H$_0$ is wrong in 99.7\% of all cases.

\chapter{Calibrated HINODE SP Data}
\label{ch:list}

\begin{table}[h!]
\begin{center}
	\begin{tabular}{ccccc}
		\hline
		\hline
		\\[-2ex]
		{Name}&{NOAA} & {Date of} & \multicolumn{2}{c}{cos$(\Theta$) of Penumbra}\\
		{}&{Active Region} & {Observation} & {Center} & {Limb}\\
		\\[-2ex]
		\hline		
		\\[-2ex]	
		Spot 01 & {10923} & {Nov 14$^{\rm{th}}$ 2006} & \multicolumn{2}{c}{0.982 - 0.996\footnotemark}\\
		Spot 02 & {10923} & {Nov 14$^{\rm{th}}$ 2006} & \multicolumn{2}{c}{0.980 - 0.994$^1$}\\
		Spot 03 & {10930} & {Dec 11$^{\rm{th}}$ 2006} & \multicolumn{2}{c}{0.991 - 0.999$^1$}\\
		Spot 04 & {10933} & {Jan 05$^{\rm{th}}$ 2007} & \multicolumn{2}{c}{0.996 - 1.000$^1$}\\
		
		Spot 05 & {10923} & {Nov 10$^{\rm{th}}$ 2006} & {0.680 - 0650} & {0.637 - 0.585}\\
		Spot 06 & {10923} & {Nov 18$^{\rm{th}}$ 2006} & {0.665 - 0.637} & {0.626 - 0.594}\\
		Spot 07 & {10930} & {Dec 15$^{\rm{th}}$ 2006} & {0.625 - 0.595} & {0.583 - 0.564}\\
		Spot 08 & {10933} & {Jan 09$^{\rm{th}}$ 2007} & {0.695 - 0.677} & {0.670 - 0.652}\\
		Spot 09 & {10923} & {Nov 11$^{\rm{th}}$ 2006} & {0.818 - 0.796} & {0.789 - 0.741}\\
		Spot 10 & {10923} & {Nov 12$^{\rm{th}}$ 2006} & {0.887 - 0.870} & {0.864 - 0.834}\\
		Spot 11 & {10923} & {Nov 13$^{\rm{th}}$ 2006} & {0.979 - 0.968} & {0.979 - 0.970}\\
		Spot 12 & {10923} & {Nov 15$^{\rm{th}}$ 2006} & {0.964 - 0.952} & {0.963 - 0.951}\\
		Spot 13 & {10923} & {Nov 15$^{\rm{th}}$ 2006} & {0.961 - 0.950} & {0.947 - 0.934}\\
		Spot 14 & {10923} & { Nov 16$^{\rm{th}}$ 2006} & {0.883 - 0.862} & {0.862 - 0.838}\\
		Spot 15 & {10923} & {Nov 16$^{\rm{th}}$ 2006} & {0.847 - 0.822} & {0.822 - 0.798}\\
		Spot 16 & {10923} & {Nov 18$^{\rm{th}}$ 2006} & {0.571 - 0.539} & {0.527 - 0.492}\\
		Spot 17 & {10923} & {Nov 18$^{\rm{th}}$ 2006} & {0.536 - 0.503} & {0.488 - 0.455}\\
		Spot 18 & {10923} & {Nov 19$^{\rm{th}}$ 2006} & {0.443 - 0.412} & {0.400 - 0.360}\\
		Spot 19 & {10923} & {Nov 20$^{\rm{th}}$ 2006} & {0.301 - 0.266} & {0.244 - 0.208}\\

		Spot 20 & {10930} & {Dec 6$^{\rm{th}}$ 2006} & {0.441 - 0.423} & {0.404 - 0.368}\\
		Spot 21 & {10930} & {Dec 8$^{\rm{th}}$ 2006} & {0.750 - 0.734} & {0.724 - 0.698}\\
		Spot 22 & {10930} & {Dec 10$^{\rm{th}}$ 2006} & {0.979 - 0.974} & {0.971 - 0.961}\\
		Spot 23 & {10930} & {Dec 12$^{\rm{th}}$ 2006} & {0.979 - 0.971} & {0.967 - 0.960}\\
		Spot 24 & {10930} & {Dec 13$^{\rm{th}}$ 2006} & {0.903 - 0.884} & {0.879 - 0.866}\\
		Spot 25 & {10930} & {Dec 14$^{\rm{th}}$ 2006} & {0.791 - 0.766} & {0.756 - 0.736}\\
		Spot 26 & {10930} & {Dec 16$^{\rm{th}}$ 2006} & {0.434 - 0.405} & {0.389 - 0.363}\\
		\hline
		\end{tabular}

\end{center}
\end{table}
\footnotetext{Spot at disk center. No CSP and no LSP was define.}
\begin{table}
\begin{center}

		\begin{tabular}{ccccc}
		\hline
		\hline
		\\[-2ex]
		{Name}&{NOAA} & {Date of} & \multicolumn{2}{c}{cos$(\Theta$) of Penumbra}\\
		{}&{Active Region} & {Observation} & {Center} & {Limb}\\
		\\[-2ex]
		\hline
		Spot 27 & {10933} & {Jan 04$^{\rm{th}}$ 2007} & {0.968 - 0.961} & {0.961 - 0.952}\\
		Spot 28 & {10933} & {Jan 06$^{\rm{th}}$ 2007} & {0.987 - 0.981} & {0.980 - 0.975}\\
		Spot 29 & {10933} & {Jan 07$^{\rm{th}}$ 2007} & {0.925 - 0.916} & {0.912 - 0.903}\\
		Spot 30 & {10933} & {Jan 08$^{\rm{th}}$ 2007} & {0.838 - 0.826} & {0.821 - 0.809}\\
		\\
		QS 01 & {...} & {Mar 10$^{\rm{th}}$ 2007} & \multicolumn{2}{c}{0.970 - 1.000\footnotemark}\\
		QS 02 & {...} & {Sep 06$^{\rm{th}}$ 2007} & \multicolumn{2}{c}{0.988 - 1.000$^2$}\\
		\hline		
		\end{tabular}
		\caption{Data samples used throughout this work. The first column gives the name of the data set. The second column refers to the denotation of the NOAA, while the third column specifies the date of observation. The fourth and fifth column gives the maximal and minimal $\mu$ values of the CSP and the LSP respectively.}
	\label{tab:data_limb2}
\end{center}
\end{table}
\footnotetext{QS observation at disk center. Values represent extrema of $\mu$ within the FOV.}


\backmatter



\phantomsection 
\addcontentsline{toc}{chapter}{{\indent{}}Bibliography} 
\bibliographystyle{astronurl}
\bibliography{references}
\chapter*{Acronyms and Abbreviations
\label{ch:abbv}
\markboth{Abbreviations}{}}
\addcontentsline{toc}{chapter}{{\indent{}}Abbreviations}

\begin{small}
\begin{longtable}{ll}
{$\mathbb{1}$\dotfill} & {identity matrix}\\
{$\mathcal{A}$\dotfill} & {area asymmetry}\\
{A$_{\rm{ul}}$\dotfill} & {Einstein coefficient accounting for spontaneous emissions}\\
{$\delta$a\dotfill} & {amplitude asymmetry}\\
{A$_{\rm{b}}$\dotfill} & {area of blue lobe in Stokes V}\\
{A$_{\rm{r}}$\dotfill} & {area of red lobe in Stokes V}\\
{a$_{\rm{b}}$\dotfill} & {amplitude of blue lobe in Stokes V}\\
{a$_{\rm{r}}$\dotfill} & {amplitude of red lobe in Stokes V}\\
{a\dotfill} & {reduced variable accounting for radiative damping}\\
{AR\dotfill} & {active region}\\
{ATST\dotfill} & {Advanced Technology Solar Telescope}\\
{B\dotfill} & {magnetic field strength}\\
{$\scriptstyle{\Delta}\textstyle\rm{B}$\dotfill} & {difference between magnetic field strength in a two-layer atmosphere}\\
{{\rm B}\dotfill} & {magnetic field vector}\\
{B$_{\nu}$\dotfill} & {Kirchhoff-Plank function}\\
{B$_{\rm{ul}}$\dotfill} & {Einstein coefficient accounting for induced emissions}\\
{B$_{\rm{lu}}$\dotfill} & {Einstein coefficient accounting for photo absorption}\\
{BFI\dotfill} & {broadband filter imager}\\
{$\chi^2$\dotfill} & {merit function}\\
{$\mathcal{C}$\dotfill} & {covariance}\\
{c\dotfill} & {speed of light}\\
{C\dotfill} & {carbon}\\
{Ca\dotfill} & {calcium}\\
{CBS\dotfill} & {convective blue shift}\\
{CCD\dotfill} & {charged coupled device}\\
{CLV\dotfill} & {center to limb variation}\\
{CSP\dotfill} & {center side penumbra}\\
{CUD\dotfill} & {central umbral dots}\\
{DN\dotfill} & {dark nuclei}\\
{$\epsilon_{\nu}$\dotfill} & {emission coefficient}\\
{e$^{-}$\dotfill} & {electron}\\
{e\dotfill} & {electron charge}\\
{E$_{\rm{J,M}}$\dotfill} & {Energy of Zeeman sublevel with quantum numbers J and M}\\
{EC\dotfill} & {Evershed clouds}\\
{EE\dotfill} & {Evershed effect}\\
{EF\dotfill} & {Evershed flow}\\
{EIS\dotfill} & {extreme ultraviolet imaging spectrometer}\\
{EMS\dotfill} & {electromagnetic spectrum}\\
{EST\dotfill} & {European Solar Telescope}\\
{EUV\dotfill} & {extreme ultraviolet}\\
{$\mathcal{F}$\dotfill} & {Faraday-Voigt function}\\
{f-spot or f-leg\dotfill} & {following (eastern) spot or leg of $\Omega$-loop}\\
{Fe\dotfill} & {iron}\\
{FOV\dotfill} & {field of view}\\
{FPP\dotfill} & {focal plane package}\\
{FWHM\dotfill} & {full width at half maximum}\\
{$\Gamma$\dotfill} & {Gauss profile}\\
{$\gamma$\dotfill} & {zenith angle}\\
{$\gamma_{\rm{mag}}$\dotfill} & {zenith angle of magnetic field}\\
{$\scriptstyle{\Delta}\textstyle\gamma$\dotfill} & {difference of zenith angle in a two-layer atmosphere}\\
{$<$$\gamma$$>$\dotfill} & {mean zenith angle in a two-layer atmosphere}\\
{$\gamma^{\rm{rad}}$\dotfill} & {radiative damping constant}\\
{g or g$_{\rm{eff}}$\dotfill} & {Land\'e g-factor or effective Land\'e g-factor}\\
{$\eta_0$\dotfill} & {line to continuum absorption coefficient}\\
{$\mathcal{H}$\dotfill} & {Hamiltion operator}\\
{H\dotfill} & {hydrogen}\\
{h or $\hbar = \rm{h} \, / \, 2\pi$\dotfill} & {Plank's constant}\\
{H$_0$\dotfill} & {null hypothesis}\\
{He\dotfill} & {helium}\\
{I\dotfill} & {Stokes parameter representing intensity}\\
{I$_{\nu}$\dotfill} & {intensity of radiation field}\\
{$\mathbf{I}_{\nu}$\dotfill} & {intensity of radiation field in a magnetized atmosphere}\\
{I$_{\rm{c}}$\dotfill} & {local continuum intensity}\\
{I$_{\rm{qs}}$\dotfill} & {continuum intensity of the average quiet Sun}\\
{J or $\mathcal{J}$\dotfill} & {quantum number or operator of total angular momentum}\\
{$\kappa_{\nu}$\dotfill} & {absorption coefficient}\\
{$\kappa_{\nu}^{\rm{cont}}$\dotfill} & {absorption coefficient in continuum}\\
{$\kappa_{\nu}^{\rm{line}}$\dotfill} & {absorption coefficient in a spectral line}\\
{$\mathcal{K}$\dotfill} & {absorption matrix}\\
{$\tilde{\mathcal{K}}$\dotfill} & {line absorption matrix}\\
{K\dotfill} & {Kelvin}\\
{$\rm{k}_0$\dotfill} & {Boltzmann's constant}\\
{kG\dotfill} & {kilogauss}\\
{KH\dotfill} & {Kelvin-Helmholtz}\\
{$\Lambda$\dotfill} & {Lorentz profile}\\
{$\lambda$\dotfill} & {wavelength}\\
{$\lambda_0$\dotfill} & {vacuum wavelength of unsplit line}\\
{$\delta\lambda_{\rm{dop}}$\dotfill} & {wavelength shift due to the Doppler effect}\\
{L or $\mathcal{L}$\dotfill} & {quantum number or operator of total orbital angular momentum}\\
{l or u\dotfill} & {lower or upper atomic state participating in a transition}\\
{LB\dotfill} & {light bridge}\\
{LFR\dotfill} & {line forming region}\\
{LMSAL\dotfill} & {Lockheed Martin Solar and Astrophysics Laboratory}\\
{LOS\dotfill} & {line of sight}\\
{LSP\dotfill} & {limb side penumbra}\\
{$\mu$\dotfill} & {cos($\Theta$)}\\
{M\dotfill} & {eigenvalue of $\mathcal{J}$}\\
{{\bf M}\dotfill} & {M\"uller matrix for an optical device}\\
{m$_{\rm{A}}$\dotfill} & {mass of atom}\\
{m$_{\rm{e}}$\dotfill} & {mass of electron}\\
{mag\dotfill} & {magnitude}\\
{MF\dotfill} & {moatflow}\\
{MISMA\dotfill} & {micro structured magnetic atmospheres}\\
{MMF\dotfill} & {moving magnetic features}\\
{MNL\dotfill} & {magnetic neutral line}\\
{$\nu$\dotfill} & {frequency}\\
{$\tilde{\nu}$\dotfill} & {reduced variable accounting for Doppler and Zeeman frequency shifts}\\
{$\scriptstyle\Delta\displaystyle\nu_{\scriptscriptstyle\rm{D}\displaystyle}$\dotfill} & {Doppler broadening of line profile}\\
{$\mathcal{N}$\dotfill} & {total net circular polarization}\\
{$<$$\mathcal{N}$$>$\dotfill} & {azimuthal average of total net circular polarization}\\
{n$_{\rm{l}}$ or n$_{\rm{u}}$\dotfill} & {number densities of l or u}\\
{N\dotfill} & {nitrogen}\\
{Ne\dotfill} & {neon}\\
{NFI\dotfill} & {narrowband filter imager}\\
{NOAA\dotfill} & {National Oceanic and Atmospheric Administration}\\
{NST\dotfill} & {New Solar Telescope}\\
{$\mathcal{O}$\dotfill} & {evolution operator}\\
{O\dotfill} & {oxygen}\\
{$\omega$\dotfill} & {circular frequency}\\
{OTA\dotfill} & {optical telescope assembly}\\
{$\phi_{\rm{mag}}$\dotfill} & {azimuth of magnetic field}\\
{$\pi$ component\dotfill} & {components of Zeeman pattern with $\scriptstyle{\Delta}\textstyle\rm{M}=0$}\\
{p$_{\rm{e}^{-}}$\dotfill} & {electron pressure}\\
{p-spot or p-leg\dotfill}  & {preceding (western) spot or leg of $\Omega$-loop}\\
{P$_{\rm{cir}}$\dotfill} & {circular polarization}\\
{P$_{\rm{lin}}$\dotfill} & {linear polarization}\\
{P$_{\rm{mag}}$\dotfill} & {magnetic pressure}\\
{P$_{\rm{tot}}$\dotfill} & {total polarization or degree of polarization}\\
{PBS\dotfill} & {polarizing beam splitter}\\
{PDF\dotfill} & {probability density function}\\
{PF\dotfill} & {penumbral filaments}\\
{PG\dotfill} & {penumbral grains}\\
{PMU\dotfill} & {polarizing modulator unit}\\
{PSF\dotfill} & {point spread finction}\\
{PUD\dotfill} & {peripheral umbral dots}\\
{Q\dotfill} & {Stokes parameter representing linear polarization ($\rm{Q}=\rm{U}+45^{\circ}$)}\\
{QS\dotfill} & {quiet sun}\\
{$\rho$\dotfill} & {density}\\
{$\mathcal{R}$\dotfill} & {rank of a variable}\\
{R$_{\rm{spot}}$\dotfill} & {normalized sunspot radius}\\
{r$_{\rm{P}}$\dotfill} & {Pearson's correlation coefficient}\\
{r$_{\rm{S}}$\dotfill} & {Spearman's rank correlation coefficient}\\
{r$_{\rm{S}}(\rm{L})$\dotfill} & {r$_{\rm{S}}$ calculated for fluctuations around a local mean value}\\
{r$_{\rm{S}}(\rm{R})$\dotfill} & {r$_{\rm{S}}$ calculated for fluctuations around a radial mean value}\\
{RMS\dotfill} & {root mean square}\\
{$\sigma$\dotfill} & {standard deviation}\\
{$\sigma^{\pm}\dotfill$ component\dotfill} & {components of Zeeman pattern with $\scriptstyle{\Delta}\textstyle\rm{M}=\pm1$}\\
{S or $\mathcal{S}$\dotfill} & {quantum number or operator of total spin angular momentum}\\
{{\bf S}\dotfill} & {Stokes Vector}\\
{$\mathcal{S}_{\nu}$\dotfill} & {source function}\\
{$\SSS_{\nu}$\dotfill} & {source function in a magnetized atmosphere}\\
{S/N\dotfill} & {signal to noise}\\
{SIR\dotfill} & {Stokes inversion based on response functions}\\
{SOT\dotfill} & {solar optical telescope}\\
{SOHO\dotfill} & {Solar and Heliospheric Observatory (sattelite)}\\
{SOUP\dotfill} & {Lockheed Solar Optical Universal Polarimeter}\\
{SP\dotfill} & {spectropolarimseter}\\
{$\tau$\dotfill} & {optical depth}\\
{$\Theta$\dotfill} & {heliocentric angle}\\
{T\dotfill} & {temperature}\\
{TRACE\dotfill} & {Transition Region and Coronal Explorer (sattelite)}\\
{U\dotfill} & {Stokes parameter representing linear polarization}\\
{UD\dotfill} & {umbral dots}\\
{$\mathcal{V}$\dotfill} & {Voigt function}\\
{V\dotfill} & {Stokes parameter representing circular polarization}\\
{$\scriptstyle{\Delta}\textstyle\rm{v}$\dotfill} & {difference of velocity  in a two-layer atmosphere}\\
{v\dotfill} & {absolute velocity}\\
{v$_{\rm{dop}}$\dotfill} & {Doppler velocity}\\
{$<$v$_{\rm{dop}}$$>$\dotfill} & {azimuthal average of Doppler velocity}\\
{v$_{\rm{mic}}$\dotfill} & {microturbolence velocity}\\
{w$_{\rm{l}}$ or w$_{\rm{u}}$\dotfill} & {statistical weighting of l or u}\\
{XRT\dotfill} & {x-ray telescope}\\
{$\xi_{\rm{x}}$ or $\xi_{\rm{y}}$\dotfill} & {amplitude of the electric field in x or y direction}\\
\vspace{-30pt}
\end{longtable}
\end{small}

\chapter*{Curriculum Vitae
\label{ch:vita}
\markboth{Curriculum Vitae}{}}
\addcontentsline{toc}{chapter}{{\indent{}}Curriculum Vitae}

\noindent {\bf\Large{Morten Franz}}\\
\noindent born September 20$^{\rm{th}}$ 1979 in Gifhorn, Germany

\begin{cv}

  \begin{cvlist}{Education}

  \item[since 07/2007] Ph.D. studies in Physics, Kiepenheuer Institut for Solar Physics, Freiburg, Germany
  \item[04/2006--04/2007]  Binational Diploma in Physics, University of Freiburg, Germany and University of Adelaide, Australia
  \item[04/2004--03/2006]  Diploma studies in Physics, University of Freiburg, Germany
  \item[08/2003--12/2003]  IAESTE stipendiary, Catholic University, Porto Alegre, Brasil 
  \item[10/2000--07/2003]  Diploma studies in Physics, University of Freiburg, Germany

  \item[08/1997--07/1999]  Robert-Bosch Comprehensive School, Hildesheim, Germany
  \item[08/1996--07/1997]  Pleasent Valley High School, Chico (CA), USA
  \item[08/1990--07/1996]  Robert-Bosch Comprehensive School, Hildesheim, Germany
  \end{cvlist}

  \begin{cvlist}{Professional Affiliations}
  \item[12/2008]  Visiting scientist, Indian  Institute for Astrophysics, Bangalore, India
  \item[04/2008--05/2008]  Visiting scientist, Stanford-Lockheed Institute for Space Research and Lockheed Martin Solar and Astrophysics Laboratory, Palo Alto (CA), USA
  \item[since 07/2007]  Research scientist, Kiepenheuer Institut for Solar Physics, University of Freiburg, Germany
  \item[04/2006--10/2006]  DAAD-stipendiary, Center for Bio-Medical Engineering, University of Adelaide, Australia
  \item[05/2004--03/2006]  Research assistent, Dept. Atomic, Molacular and Optical Physics, Freiburg, Germany
  \item[09/2003] Visiting scholar, Microgravity Center, Catholic University, Porto Alegre, Brasil
  \item[08/2003--12/2003]  Trainee, Electric/Electronic Calibration and Tests Specialized Laboratories, Catholic University Porto Alegre, Brasil 
  \item[04/2002--07/2003]  Part-time employee, Student Union, Freiburg, Germany
  \item[07/2001--08/2001]  Part-time employee, Eon-Avacon, Sarstedt, Germany
  \item[07/1999--06/2000]  Civil servant, Diakonische Werke, Hildesheim, Germany
  \end{cvlist}

  \begin{cvlist}{Voluntary Affiliations}
  \item[since 06/2008]	Head of "DAAD-Freundeskreis", Freiburg, Germany,
  \item[09/2001--09/2005] 	Student self-administration, Freiburg, Germany
  \end{cvlist}

\end{cv}

\chapter*{Acknowledgment
\label{ch:ackno}
\markboth{Acknowledgment}{}}
\addcontentsline{toc}{chapter}{{\indent{}}Acknowledgment}

It would have been impossible to write this thesis without the help, support and participation of a lot of colleagues, friends and relatives.

First of all, I would like to thank my supervisors {\bf Wolfgang Schmidt} and {\bf Rolf Schlichenmaier}. Not only did they arouse my interest in solar physics, but also took the time to listen to my (sometimes silly) questions. Furthermore, they gave me the freedom to follow my own interests, develop my own research style and work on scientific topics not directly related to my PhD studies.

Additionally, I would like to mention {\bf Oskar Steiner}, with whom I spend hours (preferentially at night) chatting about solar physics and other topics. We had a lot of fun at various conferences in India, the US and Italy. Together with {\bf Nazaret Bello Gonzalez}, I shared the privilege to dig through terabyte of data from the Sunsrise mission. I thank her for all the presents (Baumkuchen) and for the shelter she provided me with in G\"ottingen and on Tenerife. {\bf Reza Rezai} is an inexhaustible source of information and helped me more than once to solve my problems without following a wrong path for too long. I still cherish the interesting time we had together in Dwingeloo. {\bf Juan Manuel Borrero} did not only manage to explain the SIR code to me, but also told me about various economy-related issues. Progress with this thesis would have been much slower without {\bf Christian Bethge}, who installed all the nifty tools on my computer. Furthermore, he kept the students at KIS together by organizing the daily lunch tours or the various movie nights. From {\bf Sven Bingert}, I did not only learn how to properly program in IDL, but also how to adjust my bicycle gearing.

I would like to thank {\bf Pia Zacharias},  {\bf J\"orn Warnecke}, {\bf Daniel Siegel}, {\bf Hardi Peter}, {\bf Rainer Hammer}, {\bf Markus Roth}, {\bf Christian Beck}, {\bf Hans-Peter Doerr}, {\bf Peter Calligari}, {\bf Torsten Waldmann}, {\bf Christian Nutto}, {\bf Jo Bruls}, {\bf Tanja Leist}, {\bf Eva Barkowsky}, {\bf Oskar von der L\"uhe}, {\bf Svetlana Berdyugina}, {\bf Helm\-hold Schleicher}, {\bf Anastasios Nesis}, {\bf Roland Fellman}, {\bf Frank Heidecke}, {\bf Mi\-cha\-el Weisssch\"adel} and {\bf Thomas Keller} for KIS movie nights, summer fairs and christmas parties, the help with all my little and large computer problems, taking me on bike tours, helping me with administrative matters and for sharing their knowledge on the (quiet) Sun, spectropolarimetry, high resolution observation, radiative transfer, cosmology, exoplanets, string theory, the VTT, IDL etc. with me. But above all, I would like to thank them for making the KIS the special place it is.

Im grateful to {\bf Alan Title} for hosting me at the LMSAL in Palo Alto. My time in Stanford was enriched by the daily lunch discussions with {\bf Mark Cheung}, {\bf Karel Schrijver}, {\bf Richard Shine} and {\bf Mandy Hagenaar}. I also had a great time during an intense hiking tour with {\bf Mark De Rosa} and {\bf Jesper Schou} and with the bike which I borrowed from {\bf Bart De Pontieu}. Furthermore, I have to thank {\bf Bruce Lites}, who shared his knowledge about the HINODE spectropolarimeter with me. Sorry for not taking you on the bike tour as promised. I hope you enjoy your life in your new home in Costa Rica.

Im also indebted to {\bf Siraj Hasan} for accommodating me at the guesthouse of the IIA during my visit to Bengaluru. This exchange would not have taken place without {\bf Vigeesh}, who stimulated my interest for India already by the fact that he does not have a family name and even further during the numerous lunch breaks we had together eating "H\"anchend\"oner".

Further thanks go to {\bf Valentin Mart\'inez Pillet} for his tremendous effort reducing the IMaX data and for his easy going attitude, which I enjoyed every time we met. It was a great pleasure to work with the nice people making the SUNRISE mission a success. In particular, I would like to acknowledge {\bf Sanja Danilovic}, {\bf Jose Carlos del Toro Iniesta}, {\bf Jose Antonio Bonet}, {\bf Luis Bellot Rubio}, {\bf Andreas Lagg}, {\bf Judith Palacios Hernandez}, {\bf Sami Solanki}, {\bf Tino Riethm\"uller}, {\bf Alex Feller} and {\bf Achim Gandorfer}.

Even though the administration denied us the additional lecture on the Eskimo roll in the hotel pool, I still had a very interesting and informative week at the USO summer school in Dwingeloo. Thank you {\bf Tayeb Aiouaz} and {\bf Rob Rutten} for the organizational effort.

\vspace{1cm}

Money makes the world go round, and therefore I am indebted to the {\bf German Research Foundation} for their financial support of this project.

\vspace{1cm}

Special thanks also go to my family {\bf Elke, Karl-Heinz} and {\bf Nele Franz} for all the faith they put in me throughout my entire life.

\vspace{1cm}

My girlfriend {\bf Andrea Moll} deserves the largest tribute. She did not only proof read this thesis several times, but also gave me strength during hours of self-doubt and stress and provided me with the necessary energy to keep on going.

\newpage
\clearpage
\thispagestyle{empty}

\begin{figure}[p!]
	\centering
		\includegraphics[width={\textwidth}]{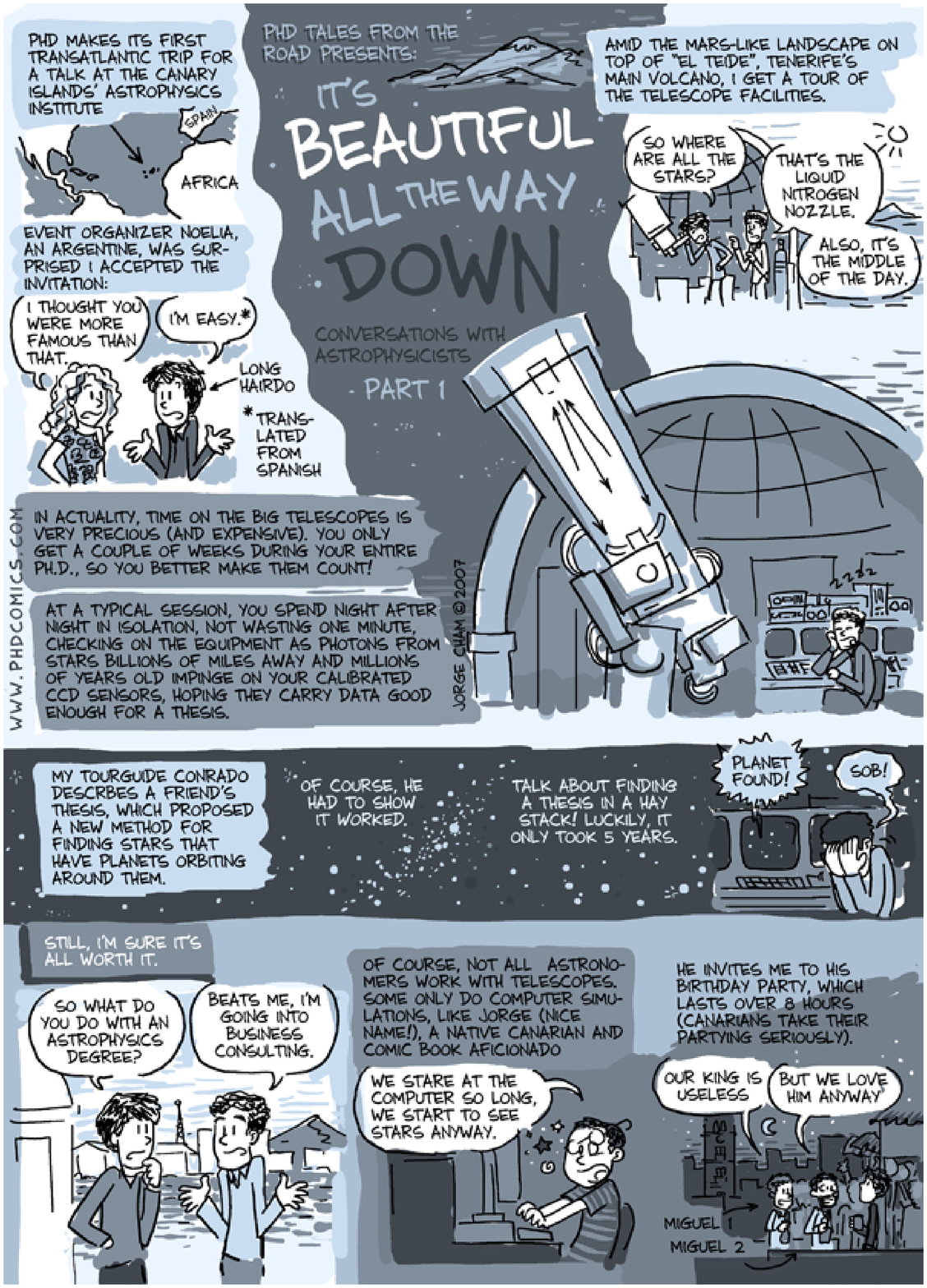}
\end{figure}

\newpage
\clearpage
\thispagestyle{empty}

\begin{figure}[p!]
	\centering
		\includegraphics[width={\textwidth}]{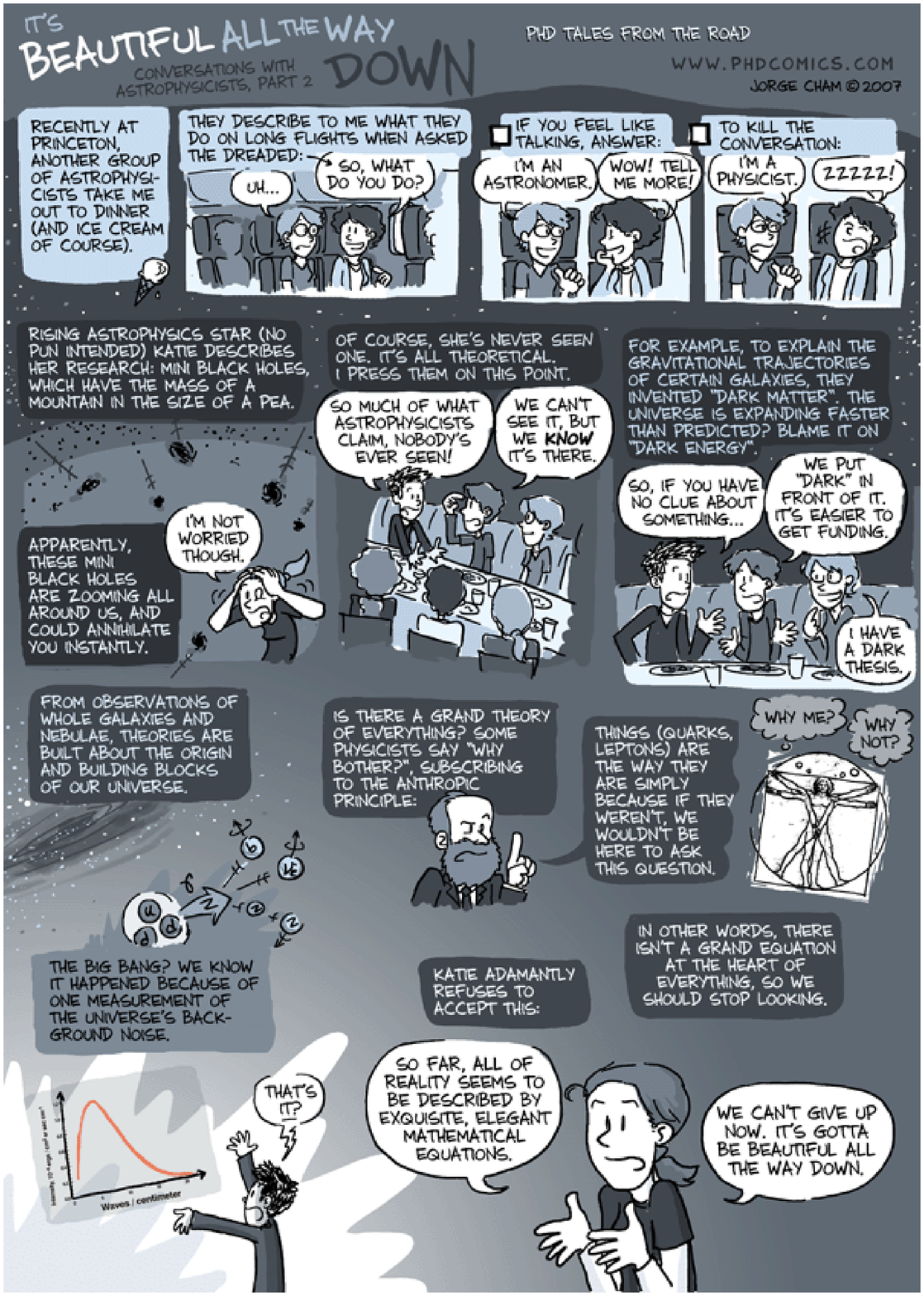}
\end{figure}

\cleardoublepage
\addcontentsline{toc}{chapter}{\numberline{}Index}
\printindex

\end{document}